\documentclass[a4paper,12pt]{article}
\pdfoutput=1
\usepackage[UKenglish]{babel}
\usepackage[toc,page]{appendix}
\usepackage[T1]{fontenc} 
\usepackage{amsmath}
\usepackage{amsfonts}
\usepackage{amssymb}
\usepackage{amsthm}
\usepackage{graphicx}
\usepackage[table]{xcolor}
\usepackage{physics}
\usepackage{tikz}
\usepackage{mathtools}
\usepackage{enumitem}
\usetikzlibrary{positioning}
\usepackage{fancyvrb}
\usepackage[bf]{caption}
\usepackage{subcaption}
\usepackage[bookmarksopen, bookmarksnumbered, bookmarksopenlevel=2]{hyperref}
\usepackage{cite}
\usepackage{booktabs}
\usepackage{cancel}
\usepackage{tabularx}
\usepackage[all]{xy}
\usepackage{tikz-cd}
\usepackage{dsfont}
\hypersetup{
    colorlinks=true,
    linkcolor=blue,
    linktoc=all,     
    filecolor=blue,      
    urlcolor=blue,
    citecolor=blue,
}
\usepackage{physics}
\usepackage{fancyvrb}
\usepackage{comment}
\usepackage{setspace}
\usepackage{rotating}
\usepackage{pdflscape}
\usepackage{caption}
\usepackage{adjustbox}

\usepackage{tabularray}
\UseTblrLibrary{diagbox} 
\usepackage[table]{xcolor}
	
\usepackage{color, colortbl}

\newtheorem{definition}{Definition}
\numberwithin{equation}{section}
\newcommand{\tikzcircle}[2][red,fill=red]{\tikz[baseline=-0.5ex]\draw[#1,radius=#2] (0,0) circle ;}

\def\SW{\text{SW}}

\newcommand{\fixme}[1]{{}}

\def\CC{{\mathcal C}}

\def\CF{{\mathcal F}}

\def\CM{{\mathcal M}}
\def\CN{{\mathcal N}}
\def\CO{{\mathcal O}}
\def\CP{{\mathcal P}}

\def\CT{{\mathcal T}}

\def\CZ{{\mathcal Z}}

\def\C{\mathbb{C}}

\def\Z{\mathbb{Z}}
\def\R{\mathbb{R}}
\def\SL{\mathrm{SL}}

\def\tilde{\widetilde}
\renewcommand{\bar}{\overline}

\def\^{{\wedge}}
\def\*{{\star}}

\makeatletter
\newcommand*{\shifttext}[2]{%
  \settowidth{\@tempdima}{#2}%
  \makebox[\@tempdima]{\hspace*{#1}#2}%
}
\makeatother

\makeatletter
\renewcommand{\paragraph}{%
  \@startsection{paragraph}{4}%
  {\z@}{1.5ex \@plus 1ex \@minus .2ex}{-0.4em}%
  {\normalfont\normalsize\bfseries}%
}
\makeatother

\usepackage{geometry}
\usepackage{changepage}
\usepackage{lipsum}

\newlength{\offsetpage}
{\end{adjustwidth}}

\def\TMF{{\text{TMF}}}
\def\MF{{\text{MF}}}
\def\SW{{\mathrm{SW}}}
\def\cp{{\mathbb{CP}}}

\begin{document} 

\begin{titlepage}
\begin{center}
\rightline{\small }

\vskip 15 mm

{\Large \bf
 E-Strings and Four-Manifolds
} 
\vskip 11 mm

Du Pei $^{1}$ and David H. Wu $^{2}$
\vskip 11 mm

\small ${}^{1}$ {\it Centre for Quantum Mathematics, University of Southern Denmark, \\Odense 5230, DK} \\[3 mm]
\small ${}^{2}$
{\it Jefferson Physical Laboratory, Harvard University, Cambridge, MA 02138, USA}\\[7mm]

 \href{mailto:dpei@imada.sdu.dk}{dpei@imada.sdu.dk}, \href{mailto:dwu@g.harvard.edu}{dwu@g.harvard.edu}

\end{center}
\vskip 17mm

\begin{abstract}

We investigate the physics of the E-string theory and its compactifications as well as their applications to four-dimensional topology. In particular, we compute the partition function of the topologically twisted theory on $M_4\times T^2$, where $M_4$ is a four-manifold. In a range of examples, we verify that this partition function, as a $q$-series, 1) has integral coefficients, 2) is modular, and 3) can be lifted to a topological modular form (TMF). Remarkably, the E-string theory ``knows'' about various subtle aspects of the world of smooth 4-manifolds, as the TMF-condition of the partition function is contingent on a collection of properties of 4-manifolds and their Seiberg--Witten invariants, including, notably, the simple-type conjecture. 
\end{abstract}

\vfill
\end{titlepage}

\newpage

\tableofcontents

\section{Introduction}
\label{sec:intro}

The present work aims to initiate a program that bridges the rich physics of 6d superconformal theories (SCFTs) with the equally rich world of smooth 4-manifolds, using the partition function of 6d theories on $M_4\times T^2$ as the central link. We focus on the case of the E-string theory \cite{Witten:1995gx,Ganor:1996mu,Seiberg:1996vs}, which, despite being one of the simplest 6d SCFTs, already exhibits remarkable power in uncovering deep structures in 4-manifold topology that merit further exploration.

\subsection{6d theories and 4-manifolds}

The introduction of gauge theory and quantum field theory into the study of smooth 4-manifolds has profoundly transformed the field of low-dimensional topology~\cite{Donaldson:1983wm,Witten:1988ze,Witten:1994cg}, fostering a rich exchange of ideas between mathematics and theoretical physics. A key element in this interplay is the construction of 4d topological quantum field theories (TQFTs) from the topological twist of 4d $\CN=2$ theories. Their correlation functions give rise to topological invariants of four-manifolds, offering powerful tools for their classification and analysis.

Despite the remarkable success achieved through this interplay, there are some intrinsic limitations to approaches rooted in 4d quantum field theories. One notable constraint is that, since the correlation functions are complex numbers, they are ill-suited for detecting certain ``nilpotent operations'' in the world of 4-manifold topology,\footnote{One such operation is the connected sum with $S^2\times S^2$, which is ``nilpotent'' in the sense that it kills exotic smooth structures when performed a sufficient number of times~\cite{Wall,GOMPF1984115}. To address such and related difficulties, there are developments from the mathematical side such as homotopy refinements of gauge-theoretic invariants~\cite{Furuta:2001,bauer2004stable}, which would be interesting to better understand from the physics perspective.}
 and there is a sentiment that the subject would benefit greatly from some new physics ideas to overcome these limitations.  

One proposal for such an idea is to use quantum field theories of higher dimensions, and it was argued in~\cite{Gukov:2018iiq} and further demonstrated in~\cite{Gukov:2025nmk} that these theories should be able to generate new types of invariants, which can be further packaged into a novel type of TQFTs. Such a TQFT would associate to a 3-manifold not a vector space, but a ``TMF-module,'' 
\begin{equation}
    \CZ\,:\quad \{\text{3-manifolds}\}\quad\rightarrow\quad \TMF\text{-mod}\, ,
\end{equation}
and to a 4-dimensional cobordism $W$ between two 3-manifolds $M_3$ and $M_3'$ a module homomorphism,
\begin{equation}
    \CZ[W]\;\in\; \mathrm{Hom}_{\TMF}\left(\CZ[M_3]\,,\CZ[M'_3]\right)\, .
\end{equation}
For a closed 4-manifold $M_4$, the new invariants $\CZ[M_4]$ are valued in $\pi_*(\TMF)$,\footnote{Just as the usual partition function can diverge, this assignment of TMF-classes to four-manifolds is also subject to ``non-compactness issues'' and often requires certain regularization or deformation. One way to make it well defined is to use global symmetries of the 6d theory, as will be discussed in detail later. For now, partition functions and TMF-classes should be understood with an appropriate regularization in mind.\label{foot:noncompact}} the ring of ``topological modular forms'' (see~\cite{douglas2014topological} for a comprehensive introduction to this subject and~\cite[Sec.~3]{Gukov:2018iiq} for a review from the physics perspective relevant for the present setting), whose richer structure is expected to make it better equipped to distinguish 4-manifolds.

Although this approach is conceptually promising, in the absence of powerful computational tools for these invariants at the present moment, they are only understood in some of the simplest cases.

One goal of the present work is to partially bridge this gap in technology by determining the image of $\CZ[M_4]$ in the ring of modular forms under the map
\begin{equation}\label{TMFtoMF}
    \pi_*(\TMF)\rightarrow \MF_{*/2}^\Z\, .
\end{equation}
The image can be understood as the partition function of the 6d theory on $M_4\times T^2$, for which there are more standard tools to carry out the computation.

\subsection{Properties of the partition function}

Given a 6d $(1,0)$ theory $\CT$, the partition function $Z_\CT[M_4\times T^2]$ can be understood from two different lower-dimensional perspectives,
\begin{equation}
    \begin{array}{ccccc}
\; & \;\text{6d $(1,0)$ theory $\CT$} &\quad\text{on}  &\text{ $M_4\times T^2$}  \; & \; \\
\text{\qquad\raisebox{10pt}{compactify on:}} & \text{\rotatebox[origin=c]{45}{$\xleftarrow{{\quad M_4\quad}}$}} \qquad\qquad\qquad& \; &\qquad \text{\rotatebox[origin=c]{-45}{$\xrightarrow{{\quad T^2\quad}}$}}& \; \\
\text{\shifttext{140pt}{ 2d $(0,1)$ theory}} & \; & \; & \; & \text{\shifttext{-120pt}{4d $\CN=2$ KK-theory  }} \\
\text{\shifttext{140pt}{ $\CT[M_4]$ on $T^2$}} & \; & \; & \; & \text{\shifttext{-120pt}{$\CT[T^2]$ on $M_4$}}
\end{array}
\end{equation}
The equality of the three partition functions,
\begin{equation}
    Z_\CT[M_4\times T^2]=Z^{(\rm 4d)}[M_4]=Z^{(\rm 2d)}[T^2]\,,
\end{equation}
gives two distinct ways to carry out the computation.
We will derive $Z_\CT[M_4\times T^2]$ via the ``right route'' as a partition function of the 4d effective theory,\footnote{One benefit of taking this route is that, as we will see later, the computation can be done almost uniformly for all $M_4$ subject to certain conditions. On the other hand, to carry out the computation the other way around, one will have to first understand the effective theory $\CT[M_4]$, which is in general difficult---the results we obtained here can in fact shed light on this issue---and has been only achieved in limited cases such as when $M_4=\Sigma_1\times \Sigma_2$ with appropriate fluxes~\cite[Sec.~4]{Gukov:2018iiq}.} while the relation to the 2d theory $\CT[M_4]$ on the left imposes a sequence of highly non-trivial constraints:

    \paragraph{Integrality.} $Z[M_4\times T^2]\in \Z(\!(q)\!)$ has an integral $q$-series. The coefficients can be regarded as the Witten indices of the supersymmetric quantum mechanics obtained from the compactification on $M_4\times S^1$ in sectors with different KK-momenta along $S^1$. This is a ``5d property'' as it does not require a lift to 6d, but holds true already for a 5d theory with a $U(1)$ symmetry (e.g.,~a 5d gauge theory with an instanton symmetry).
    
\paragraph{Modularity.} $Z[M_4\times T^2]\in \MF_*$ is a modular form. This originates from the action of the mapping class group $\SL(2,\Z)$ on the $T^2$ factor, which is expected to lead to equivalent systems. However, due to anomalies, the partition function turns out to be not strictly invariant under the full SL$(2,\Z)$ but can be a modular form with non-trivial level and weight.
(We review some relevant facts about modular forms in Appendix~\ref{sec:modular forms}.)

\paragraph{Topological modularity.} $Z[M_4\times T^2]$ is also expected to be a reduction of a topological modular form. In other words, it should admit a lift to $\pi_*(\TMF)$ along~\eqref{TMFtoMF}. This condition, motivated by a conjecture of Segal--Stolz--Teichner~\cite{Segal-1988,Stolz-Teichner-2004,Stolz:2011zj}, is a genuine 6d constraint, as it stems from the requirement that $\CT[M_4]$ is also a well-defined theory by itself.
\medskip

Together, these conditions are highly constraining and can provide a powerful check on the computation of the partition function. Furthermore, if $\CT$ possesses a flavor symmetry $G$, then each of the three constraints can be further refined and significantly strengthened by requiring the partition function to be 1) an integer $q$-series with coefficients being virtual characters of $G$, 2) a ``$G$-equivariant Jacobi form'' with given weight, level, and index, and 3) liftable to a $G$-equivariant topological modular form, $\CZ\in \widetilde{\TMF}^G$.\footnote{Here the reason that the notation $\widetilde{\TMF}^G$ is used as opposed to ${\TMF}^G$ is related to the potential non-compactness of the theory (cf.~Footnote~\ref{foot:noncompact}). Indeed, even when the partition function of the theory can be made finite with generic $G$-holonomies on $T^2$, the limit where the holonomies are turned off is in general divergent. Therefore one expects to obtain \textit{meromorphic} (as opposed to just weakly holomorphic) $G$-equivariant Jacobi forms. When considering the homotopy refinement, the lift of the partition function should thus be the \textit{meromorphic} version of $G$-equivariant topological modular forms. As the notation $\TMF^G$ is most commonly used for $G$-equivariant topological modular forms that are only weakly holomorphic, to emphasize this point, we will instead use the notation $\widetilde{\TMF}^G$, which is yet to be defined mathematically. And later by ``$G$-equivariant topological modular forms,'' we are actually always referring to the meromorphic/non-compact version $\widetilde{\TMF}^G$. One important difference is that $\TMF^G$ has a well-defined map to $\pi_*(\TMF)$, while $\widetilde{\TMF}^G$ in general does not. We thank an anonymous referee for emphasizing the importance of clarifying this point.\label{foot:tilde}}

To see how the three constraints can be useful, first note that, related to the integrality constraint, $Z_\CT[M_4\times T^2]$ often has a canonical normalization as it counts BPS states of the 6d theory on $M_4\times S^1$. This is in contrast to partition functions of 4d theories, whose normalization can be changed by modifying couplings to the background metric. One thing we will discuss in the paper is how such couplings in the 4d low-energy effective theory can be fixed from the 6d perspective.

Modularity, on the other hand, could offer deep insights into the topology of 4-manifolds and the behavior of their invariants. One way to harness its power is to first express the partition function as 
\begin{equation}\label{SumModular}
    Z_\CT[M_4\times T^2]=\sum_{\lambda}C_\lambda\cdot\SW(\lambda) +Z_{\text{Coulomb}}\,,
\end{equation}
via the 4d low-energy effective theory~\cite{Witten:1994cg,Moore:1997pc}, where the sum is over the Seiberg--Witten invariants labeled by spin$^c$ structures $\lambda$ (see Appendix~\ref{appendix:SW} for a brief review of the Seiberg--Witten invariants) and $Z_{\text{Coulomb}}$ is the ``contribution of the Coulomb branch.'' Modularity of this expression turns out to be a remarkably strong statement as the coefficients $C_\lambda$ are typically assembled from pieces that are not individually modular with the right weight or level. 

In this paper, we focus on the case where $\CT$ is the E-string theory. For various classes of 4-manifolds that we examine, we test the topological modularity of \eqref{SumModular}, which turns out to be highly non-trivial and hinges on a collection of properties obeyed by 4-manifolds and their Seiberg--Witten invariants. These range from well-established facts, such as how SW$(-\lambda)$ is related to SW$(\lambda)$, to more speculative assumptions, including the simple-type conjecture.

\subsection{Why E-strings?}

What makes the E-string theory the ideal starting point for studying 4-manifolds from the 6d perspective? Below we enumerate some of the rationales for selecting it as the first example in this program.
\begin{itemize}
\item The E-string theory is among the most extensively studied theories in 6d, with deep insights into its physics available through its embeddings in  M-, F-, and string theory. (See a brief review in Appendix~\ref{sec:More on the E-string theory}.) These embeddings provide powerful tools that greatly facilitate our analysis.

 \item Despite being arguably the simplest interacting SCFT in 6d, the E-string theory already exhibits remarkably rich dynamics. This ensures a non-trivial interplay between its physical content and the topology of smooth 4-manifolds.\footnote{This is in contrast with non-interacting theories, such as those studied in \cite[Sec.~2.8]{Gukov:2018iiq}, whose twisted partition functions are often determined by homology and insensitive to exotic smooth structures.}

    \item Related to the points above, the E-string theory and its compactifications admit a variety of limits in their parameter spaces, giving rise to a landscape of interesting lower-dimensional theories including gauge theories with various matter contents. The study of the E-string theory thus implicitly encompasses these descendant theories.

    \item The E-string theory is ``absolute,'' meaning that it does not have to reside on the boundary of a non-invertible 7d TQFT. As a result, the partition function $Z[M_4\times T^2]$ is expected to be a full SL$(2,\Z)$ modular form. This is in contrast with ``relative'' 6d theories, which, to give rise to a well-defined theory in 4d, require a choice of a ``polarization on $T^2$,'' thereby breaking the full SL$(2,\Z)$ into a subgroup.  (See~\cite{Gukov:2020btk,GHP2} for further discussions.)
\item The E-string theory can be defined canonically on $M_4\times T^2$ even when $M_4$ is not spin. Moreover, when $M_4$ is spin, the partition function can be made independent of the spin structure, leading to simplifications over the more general situation. We elaborate on this in Section~\ref{sec:topological twist}.
\item Another important feature of the E-string theory is its $E_8$ global symmetry. Although we do not fully exploit its power in the present work, we anticipate that incorporating it would lead to a more refined understanding of 4-manifold topology and sharper predictions.

\item Lastly, two copies of the E-string theory can be combined into a theory with $E_8\times E_8$ symmetry. This bigger theory, when compactified to 4d with generic holonomies, has a compact moduli space and is expected to define an invariant that is fully topological without any metric dependence, even when $b_2^+\leq 1$. The existence of this bigger theory suggests that the partition function of the E-string theory might also have full topological invariance.  This point is discussed in more detail in Section~\ref{sec:E8E8} and~\ref{sec:FullTop}.

\end{itemize}

\paragraph{Structure of the paper.} We begin with a discussion of the E-string theory and its compactifications in Section~\ref{sec:2}, with emphasis on the moduli spaces and the construction of the topologically twisted partition function. The computational strategy for evaluating this partition function is explained in Section~\ref{sec:3} (with additional technical details in Appendix~\ref{sec:details on the partition function}), where we also study some aspects of its behaviors and key properties. Section~\ref{sec:examples} presents a detailed study of several classes of examples across the geography of 4-manifolds, selected to demonstrate different facets of the intricate interplay between the 6d theory and topology. We conclude in Section~\ref{sec:5} with a discussion of promising directions for future research.

\section{E-string theory and its compactifications}
\label{sec:2}

The E-string theory is a 6d $(1,0)$ SCFT with $E_8$ global symmetry which can be realized in multiple ways via string/M/F-theory~\cite{Witten:1995gx,Ganor:1996mu,Seiberg:1996vs,Witten:1996qb,Morrison:1996na,Morrison:1996pp}. Here, we will briefly discuss the M-theory construction while deferring other approaches to Appendix~\ref{sec:E-string constructions}.

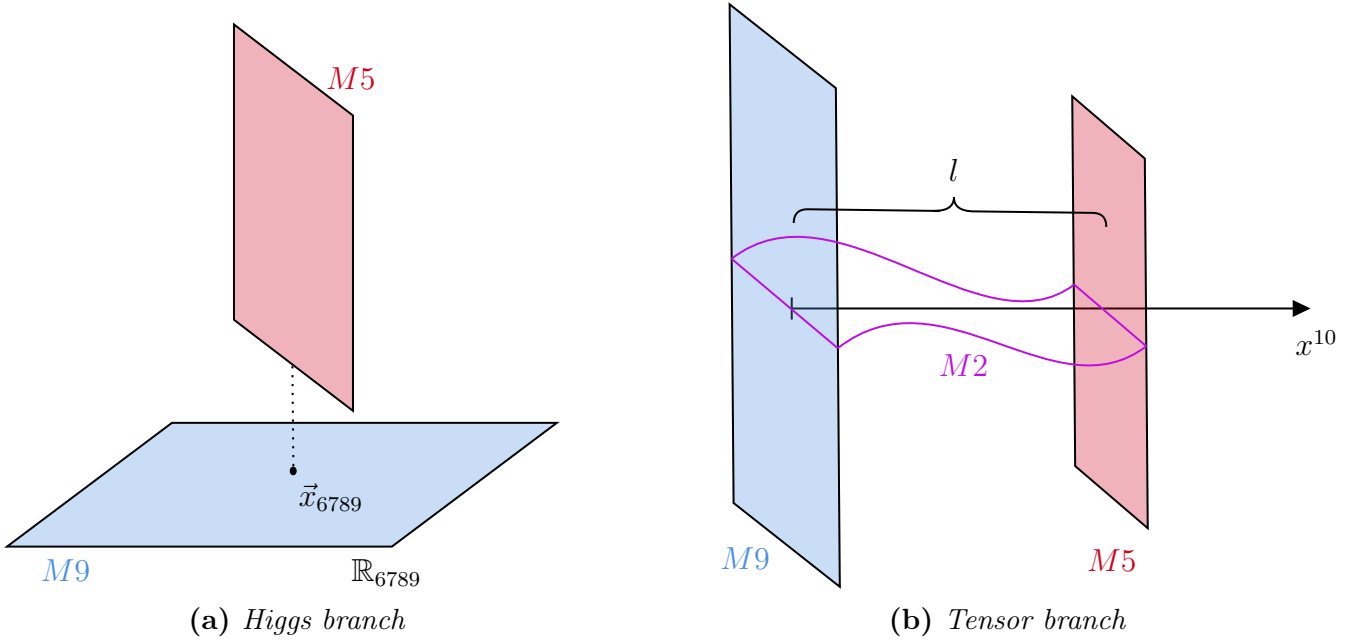
\begin{figure}[ht!]
    \centering
    \begin{subfigure}{.45\textwidth}
\tikzset{every picture/.style={line width=0.75pt}} 

\begin{tikzpicture}[x=0.75pt,y=0.75pt,yscale=-0.75,xscale=0.55]

\draw  [fill={rgb, 255:red, 74; green, 144; blue, 226 }  ,fill opacity=0.3 ] (221.7,296) -- (571,296) -- (421.3,379) -- (72,379) -- cycle ;
\draw  [dash pattern={on 0.84pt off 2.51pt}]  (331,258) -- (332,331.5) ;
\draw  [fill={rgb, 255:red, 208; green, 2; blue, 27 }  ,fill opacity=0.3 ] (386,89.88) -- (386,287.95) -- (278,226.87) -- (278,28.8) -- cycle ;

\draw  [fill={rgb, 255:red, 0; green, 0; blue, 0 }  ,fill opacity=1 ] (329.5,328.25) .. controls (329.5,327.01) and (330.51,326) .. (331.75,326) .. controls (332.99,326) and (334,327.01) .. (334,328.25) .. controls (334,329.49) and (332.99,330.5) .. (331.75,330.5) .. controls (330.51,330.5) and (329.5,329.49) .. (329.5,328.25) -- cycle ;

\draw (382,385.4) node [anchor=north west][inner sep=0.75pt]    {$\mathbb{R}_{6789}$};
\draw (100,385.4) node [anchor=north west][inner sep=0.75pt]    {$\textcolor[rgb]{0.29,0.56,0.89}{M9}$};
\draw (359,55.4) node [anchor=north west][inner sep=0.75pt]    {$\textcolor[rgb]{0.82,0.01,0.11}{M5}$};
\draw (334,334.9) node [anchor=north west][inner sep=0.75pt]    {$\vec{x}_{6789}$};
\end{tikzpicture}
\caption{Higgs branch}
    \end{subfigure}
    \hfill
    \begin{subfigure}{.45\textwidth}

\tikzset{every picture/.style={line width=0.75pt}} 

\begin{tikzpicture}[x=0.75pt,y=0.75pt,yscale=-1,xscale=1]

\draw [fill={rgb, 255:red, 189; green, 16; blue, 224 }  ,fill opacity=1 ]   (166,229) -- (422,229) ;
\draw [shift={(425,229)}, rotate = 180] [fill={rgb, 255:red, 0; green, 0; blue, 0 }  ][line width=0.08]  [draw opacity=0] (8.93,-4.29) -- (0,0) -- (8.93,4.29) -- cycle    ;
\draw [shift={(166,229)}, rotate = 180] [color={rgb, 255:red, 0; green, 0; blue, 0 }  ][line width=0.75]    (0,5.59) -- (0,-5.59)   ;
\draw  [fill={rgb, 255:red, 74; green, 144; blue, 226 }  ,fill opacity=0.3 ] (136.97,326.84) -- (135,76) -- (188.03,118.16) -- (190,369) -- cycle ;
\draw  [fill={rgb, 255:red, 208; green, 2; blue, 27 }  ,fill opacity=0.3 ] (307.48,308.26) -- (306.02,122.28) -- (342.25,153.57) -- (343.71,339.54) -- cycle ;
\draw [color={rgb, 255:red, 189; green, 16; blue, 224 }  ,draw opacity=1 ][line width=0.75]    (136,204) .. controls (188,163) and (261.14,252.09) .. (307,217) ;
\draw [color={rgb, 255:red, 189; green, 16; blue, 224 }  ,draw opacity=1 ]   (136,204) -- (189,249) ;
\draw [color={rgb, 255:red, 189; green, 16; blue, 224 }  ,draw opacity=1 ]   (307,217) -- (343,248) ;
\draw [color={rgb, 255:red, 189; green, 16; blue, 224 }  ,draw opacity=1 ][line width=0.75]    (189,249) .. controls (241,208) and (297.14,283.09) .. (343,248) ;

\draw   (323,188) .. controls (323.06,183.33) and (320.76,180.97) .. (316.09,180.91) -- (256.9,180.16) .. controls (250.23,180.07) and (246.93,177.7) .. (246.99,173.03) .. controls (246.93,177.7) and (243.57,179.99) .. (236.9,179.9)(239.9,179.94) -- (174.09,179.1) .. controls (169.42,179.04) and (167.06,181.34) .. (167,186.01) ;

\draw (416,238.4) node [anchor=north west][inner sep=0.75pt]    {$x^{10}$};
\draw (312,348.4) node [anchor=north west][inner sep=0.75pt]    {$\textcolor[rgb]{0.82,0.01,0.11}{M5}$};
\draw (238,250.4) node [anchor=north west][inner sep=0.75pt]  [color={rgb, 255:red, 189; green, 16; blue, 224 }  ,opacity=1 ]  {$M2$};
\draw (129,347.4) node [anchor=north west][inner sep=0.75pt]    {$\textcolor[rgb]{0.29,0.56,0.89}{M9}$};
\draw (243,152.4) node [anchor=north west][inner sep=0.75pt]    {$l$};

\end{tikzpicture}

\caption{Tensor branch}
\end{subfigure}
    \caption{\textbf{Phases of the E-string theory.} The E-string theory has two different phases, which can be understood from the M5-M9 system. (a) For the configuration on the left, the M5 is constrained to live (and ``dissolve'') on the M9, where it can be viewed as an $E_8$ instanton. This gives the Higgs branch of the E-string theory, and the SCFT is realized as the zero-size limit for the instanton at the origin of the Higgs branch. (b) The configuration on the right gives vacua on the tensor branch. In this phase, the E-strings, realized as M2-branes suspended between the M5 and M9, acquire tension proportional to the displacement $l$. From the viewpoint of this phase, the SCFT is at the tensionless limit $l\rightarrow0$.}
    \label{fig:The E-string theory's M-theory's brane configuration}
\end{figure}

In M-theory, the E-string theory can be obtained by having an M5-brane probing an ``end-of-the-world'' M9-brane (also known as the Ho\v{r}ava--Witten wall \cite{Horava:1995qa,Horava:1996ma}). The configuration of branes in the M-theory setup is specified in Table~\ref{table:The E-string theory's M-theory's brane configuration} and illustrated in Figure~\ref{fig:The E-string theory's M-theory's brane configuration}.
This can be viewed as the decompactification limit of M-theory on $S^1/\mathbb{Z}_2=I$ near one of the end points of the interval $I$, which gives rise to a well-defined quantum field theory with only a single copy of the $E_8$ as a global symmetry on the worldvolume of the M5-brane.\footnote{One can also engineer the rank-$Q$ E-string theory by considering $Q$ M5-branes stacked together approaching the M9-brane. It will be a 6d $(1,0)$ SCFT with $E_8\times SU(2)_F$ global symmetry. The focus of this paper is the $Q=1$ case, and we will use the term ``the E-string theory'' to implicitly refer to the rank-1 E-string theory. }

\begin{table}[h!]
    \centering
    \begin{tabular}{c|c|c|c|c|c|c|c|c|c|c|c}
         & $x^0$ & $x^1$ & $x^2$ & $x^3$ & $x^4$ & $x^5$ & $x^6$ & $x^7$ & $x^8$ & $x^9$ & $x^{10}$\\
        \hline
        M5 & $\times $& $\times $& $\times $& $\times $& $\times $& $\times $ & -- & -- & -- & -- & $l$ \\
        \hline
        M9 & $\times $ & $\times $& $\times $ & $\times $& $\times $& $\times $& $\times $ & $\times $& $\times $& $\times $ & $0$ \\
        \hline M2 & $\times $ & $\times $ & -- & -- & -- & -- & -- & -- & --& --& $\times $\\
        \hline
    \end{tabular}
    \caption{\textbf{Brane configuration for the E-string theory.} The theory is identified with the worldvolume theory of the M5-brane, with M2 giving rise to the ``E-strings.'' In the table, ``$\times$'' indicates space-filling along the dimension $x^i$ while ``--'' denotes that the brane sits at a particular point along $x^i$. $0$ and $l$ denote the origin and a point a distance $l$ away along $x^{10}$.}
    \label{table:The E-string theory's M-theory's brane configuration}
\end{table}

The namesake ``E-strings'' in the theory are excitations given by M2-branes suspended between the M5- and the M9-brane. From the point of view of the 6d theory, these two-dimensional objects are coupled to the self-dual two-form $B^+_{\mu\nu}$ in the tensor multiplet. As the tension of the strings is proportional to the distance between the two branes, the strings become ``tensionless'' at the superconformal point when they coincide. Partially due to the existence of these tensionless E-strings, the 6d theory admits no known Lagrangian descriptions.

\subsection{Moduli space of vacua}\label{sec:moduli space}

M5-branes on an M9-brane are small $E_8$ instantons in 10d~\cite{Witten:1995gx}. As a consequence, the Higgs branch of the E-string theory is expected to be the moduli space of one $E_8$ instanton, or, equivalently, the minimal nilpotent orbit of $E_8$~\cite{kronheimer1990instantons}. This is a hyper-Kähler space with quaternionic dimension 29 with the action of both the $E_8$ global symmetry and the $SU(2)$ R-symmetry. For higher-rank E-string theories, there is an action of another $SU(2)_F$ global symmetry.

Moving the M5-brane away from the M9-brane by a distance $l$ deforms the 6d theory away from the superconformal point. This corresponds to giving a vacuum expectation value (vev) to the scalar component $\langle \phi\rangle\sim l$ of the tensor multiplet. This vev parametrizes the tensor branch $\mathbb{R}^+$ of the theory. While the low-energy effective theory on the Higgs branch is given by 29 massless hypermultiplets, the theory on the tensor branch is an abelian tensor multiplet. 

When compactified to 5d on a circle of radius $R_6$, the Higgs branch is unchanged, while the tensor branch gives rise to the Coulomb branch of the 5d theory. Although the Coulomb branch is still $\mathbb{R}^+$, there are actually two singularities as opposed to only one in 6d~\cite{Ganor:1996pc}. At the origin, one finds an ``$E_8$ theory'' with the Higgs branch being the same one-instanton moduli space, while there is another ``$A_0$ point'' at $\phi=\frac{1}{\sqrt{2}R_6}$, where a hypermultiplet with non-trivial KK-momentum becomes massless.\footnote{For 5d theories, we adopt the convention coming from the ADE classification of massless rank-1 theories \cite{Seiberg-1996}. Recall that $A_{n-1}$ is the $U(1)$ gauge theory with $n$ charge-1 hypermultiplets, $D_n$ is an $SU(2)$ theory with $n$ flavors, while $E_{n+1}$ is the strong coupling limit of $D_n$. }

\begin{figure}
    \centering
    \includegraphics[width=0.85\linewidth]{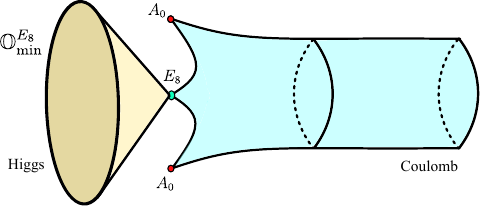}
    \caption{\textbf{Moduli space of the $T^2$-compactified E-string theory.} The Coulomb branch has a cigar shape with three singularities near the tip. The Higgs branch, which is the minimal nilpotent orbit of $E_8$, opens up at the $E_8$ singularity. }
    \label{fig:Moduli}
\end{figure}

The Higgs branch is again preserved upon compactification on another circle to $\mathbb{R}^4\times T^2$. It sits at the origin of the 4d Coulomb branch, which is now complexified by a periodic scalar given by the holonomy of the two-form field in 6d on $T^2$ (or alternatively, the holonomy of the 5d gauge field). The Coulomb branch has a cigar shape, which resembles $\mathbb{R}^+\times S^1$ for a large value of the Coulomb vev, but there are three singularities near the origin. One is an $E_8$ singularity where the Higgs branch direction of the moduli space opens up, while the other two are $A_0$ singularities. At the former, one has the Minahan--Nemeschansky $E_8$ SCFT \cite{Minahan:1996cj}, while the low-energy theory at the latter two points is a $U(1)$ theory with a massless charge-1 hypermultiplet with non-trivial KK-momenta. The full moduli space is depicted in Figure~\ref{fig:Moduli}.

Once another direction is compactified, the effective 3d theory will have two more compact scalars, whose vevs parametrize a torus fiber over a generic point on the 4d Coulomb branch. They can be thought of as the holonomy of the gauge field and the dual photon, and, when particles with either electric or magnetic charge become massless near the singularities of the 4d Coulomb branch, the torus fiber will become singular. This fibration is often known as the Seiberg--Witten geometry of the theory \cite{Seiberg:1994rs,Seiberg:1994aj}. The total space, which can be identified with the moduli space of an effective 3d $\mathcal{N}=4$ theory, is hyper-Kähler, with the fibration being holomorphic in one of its complex structures and the fibers being generically abelian varieties. In nice cases, they are Jacobians of Riemann surfaces, which are commonly referred to as Seiberg--Witten curves, and the ``Seiberg--Witten solution'' of a theory is usually given as a family of Seiberg--Witten curves parametrized by the 4d Coulomb  branch.

In the rank-1 case, meaning the 4d Coulomb branch is of complex dimension one, it is conventionally parametrized by a complex parameter $u$ and sometimes referred to as the ``$u$-plane.'' The fibers are elliptic curves dual to the Seiberg--Witten curves, which are also generically elliptic curves. The massless points on the 4d Coulomb branch are often labeled by the singular type of the fiber via Kodaira classification. For the case of the ($T^2$-compactified) E-string theory, the two types that appear are II$^*$ and I$_1$. 

When some of the six directions are compactified, one can also turn on fluxes and holonomies of the $E_8$ global symmetry along the compact directions. While the compactification with flux leads to interesting theories in lower dimensions---and the reader is referred to \cite{Kim:2017toz,Gukov:2018iiq,Pasquetti:2019hxf,Bah:2017gph,Razamat:2020bix,Razamat:2022gpm} and references therein for more details on the study of the physics as well as partition functions---in this paper we only utilize holonomies of the $E_8$ symmetry. It turns out that they are sufficient to make the partition function finite and well-defined in all cases that we are interested in.

The conjugacy class of an $E_8$ holonomy on $\mathcal{C}\simeq T^2$ can be described by eight Jacobi variables, acted upon by the Weyl group of $E_8$ and the SL$(2,\mathbb{Z})$ mapping class group, the latter of which also acts on the complex moduli $\sigma$ of $\CC$.\footnote{Later, we will also use the letter $\sigma$ for the signature of the four-manifold. Although what the symbol is referring to should be clear from the context, we occasionally use $\sigma(M_4)$ for the signature, while $\sigma(\CC)$ for the modulus parameter.} When their values are small, in the 4d theory at low energy, they are scalars in a background $E_8$ vector multiplet and can be viewed as complex masses of the theory. Therefore, we will denote them as $\{m_i\}$ with $i=1,\ldots,8$ and frequently refer to them as ``masses,'' even though each $m_i$ is actually valued in the torus $\mathbb{C}/\langle 2\pi ,2\pi \sigma\rangle$ 
before the quotient by the Weyl group.

The Seiberg--Witten solution of the ($T^2$-compactified) E-string theory will then be given by a $(u,\sigma,\{m_i\})$-family of elliptic curves, which captures much information about the 4d low-energy effective theory and serves as the starting point of later analysis.

\subsection{Seiberg--Witten geometry}
Various aspects of the Seiberg--Witten geometry for the E-string theory have been studied \cite{Ganor:1996pc,Minahan:1998vr,Eguchi:2002fc,Eguchi:2002nx,Mohri:2001zz,Sakai:2011xg,Sakai:2012ik,Chen:2021ivd}. The Seiberg--Witten curve can be expressed in the following Weierstrass form,
\begin{equation}
\label{eq:ellipticCurveES}
    \Sigma\,:\qquad y^2=4x^3-f(u;\sigma,\{m_i\})x-g(u;\sigma,\{m_i\})\, .
\end{equation}
The coefficients can be written as
\begin{align}
\label{eq:f}
    f(u;\sigma,\{m_i\})&=\sum_{i=0}^4 a_{i}(\sigma,\{m_i\})u^{4-i}\, ,\\
\label{eq:g}
    g(u;\sigma,\{m_i\})&=\sum_{i=0}^6b_i(\sigma,\{m_i\})u^{6-i}\, ,
\end{align}
where $\sigma$, living in the upper-half plane $\mathbb{H}$, is the complex modulus of the spacetime torus $\CC$, $u\in\mathbb{C}$ parameterizes the Coulomb branch, and $m_i$'s collectively describe the $E_8$ holonomy on $\mathcal{C}$. The total space of this fibration is a rational elliptic surface known as a ``$\frac{1}{2}$K3''---an almost del Pezzo surface that can be constructed from blowing up $9$ points in $\mathbb{P}^2$ (see Appendix~\ref{sec:E-string constructions} for more details and the relation to the F-theory construction of the E-string theory). In this setup, $u$ can then be interpreted as parameterizing the base $\mathbb{P}^1$ and $\sigma$ is the modulus at $u=\infty$, while $m_i$'s represent the positions of 8 of the blow-up points. 
The coefficients $\{a_i\}$ and $\{b_i\}$ are given explicitly using $E_8$-Jacobi forms in \cite[App.~A]{Sakai:2011xg}. 
The discriminant of the Seiberg--Witten curve is 
\begin{equation}
\label{eq:relative discriminant}
    \Delta_{\Sigma}=f(u;\sigma,\{m_i\})^3-27g(u;\sigma,\{m_i\})^2\, .
\end{equation}
Furthermore, the Seiberg--Witten curve is equipped with the holomorphic differential
\begin{equation}
    \omega=\frac{\dd x}{y}\,,
\end{equation}
which is the derivative of the meromorphic SW differential $\lambda_{\SW}$,
\begin{equation}
    \frac{\partial \lambda_{\SW}}{\partial u}=\omega.
\end{equation}
The special coordinates 
\begin{equation}
    a=\int_{A}\lambda_{\SW},\qquad a_D=\int_{B}\lambda_{\SW}\,,
\end{equation}
 are periods of $\lambda_{\SW}$ along $A$- and $B$-cycles of $\Sigma$, and their derivatives are related to the periods of the torus $\omega_{1,2}$ via 
\begin{equation}
\label{eq:Seiberg--Witten coordinates}
    \frac{\dd a}{\dd u}=\omega_1=\int_{A}\omega,\qquad \frac{\dd a_D}{\dd u}=\omega_2=\int_{B}\omega\,.
\end{equation}
Away from singularities on the Coulomb branch, the periods $\omega_{1,2}$ are finite, but they can diverge when approaching a singularity. The coupling constant of the abelian gauge theory in the low-energy effective theory is
\begin{equation}
    \tau=\frac{\omega_2}{\omega_1}=\frac{\partial a_D}{\partial a}\,.
\end{equation}
This quantity $\tau\in\mathbb{H}$ can also be viewed as the modulus of a torus, which is the dual of the Seiberg--Witten curve and the fiber in the Seiberg--Witten geometry, and it is important to not confuse it with $\mathcal{C}$ whose modulus is $\sigma$. Lastly, using these ingredients, we can compute the holomorphic prepotential via 
\begin{equation}
    \frac{\partial \mathcal{F}}{\partial a}=a_D\,.
\end{equation}

To see that the Seiberg--Witten curve is a generically smooth torus, one can solve for $y$ appearing in the $\Sigma$ as
\begin{equation}
    y_{\pm}=\pm \sqrt{4x^3-f(u;\sigma,\{m_i\})x-g(u;\sigma,\{m_i\})}\,.
\end{equation}
This then describes the Seiberg--Witten curve as a two-sheet cover of a $\mathbb{CP}^1$ with four ramification points,
\begin{equation}
\label{eq:x singularities}
    \begin{cases}
    x_1=\frac{\left(\sqrt{-3\Delta}+9 g\right)^{2/3}+3^{1/3} f}{2\cdot 3^{2/3}\left(\sqrt{-3\Delta}+9 g\right)^{1/3}}\,,\\
    x_2=\frac{-2\cdot({-3})^{1/3} f+i \left(\sqrt{3}+i\right) \left(\sqrt{-3\Delta}+9g\right)^{2/3}}{4\ 3^{2/3} \left({\sqrt{-3\Delta}+9 g}\right)^{1/3}}\,,\\
    x_3=\frac{(-3)^{2/3} f-\left({-3}\right)^{1/3} \left(\sqrt{-3\Delta}+9 g\right)^{2/3}}{6\left({\sqrt{-3\Delta}+9 g}\right)^{1/3}}\,,\\
    x_4=\infty\,,
    \end{cases}
\end{equation}
with the branched covering illustrated below.
\begin{equation*}
    \centering
\begin{tikzpicture}
    \def\centeryp{1}
    \def\centerym{-1}
    \draw (-1.5,0.75+\centeryp)--(1.5,0.75+\centeryp);
    \draw (-1.5,-0.75+\centeryp)--(1.5,-0.75+\centeryp);
    \draw (-1.5,0.75+\centeryp) arc(90:270:0.75) -- (-1.5,-0.75+\centeryp);
    \draw (1.5,-0.75+\centeryp) arc(270:450:0.75) -- (1.5,0.75+\centeryp);
    \draw (-1.5,0.75+\centerym)--(1.5,0.75+\centerym);
    \draw (-1.5,-0.75+\centerym)--(1.5,-0.75+\centerym);
    \draw (-1.5,0.75+\centerym) arc(90:270:0.75) -- (-1.5,-0.75+\centerym);
    \draw (1.5,-0.75+\centerym) arc(270:450:0.75) -- (1.5,0.75+\centerym);
    \node(yp) at (5,\centerym)[right]{$y=y_+$};
    \node(ym) at (5,\centeryp)[right]{$y=y_-$};
    \node[circle,fill=black,inner sep=0pt,minimum size=3pt, label=left:{$x_1$}](x1p) at (-1.25,-0.35+\centeryp) {};
    \node[circle,fill=black,inner sep=0pt,minimum size=3pt, label=left:{$x_2$}](x2p) at (-1,0.35+\centeryp) {};
    \node[circle,fill=black,inner sep=0pt,minimum size=3pt, label=right:{$x_3$}](x3p) at (1,0.35+\centeryp) {};
    \node[circle,fill=black,inner sep=0pt,minimum size=3pt, label=right:{$x_4$}](x4p) at (1.25,-0.35+\centeryp) {};
    \node[circle,fill=black,inner sep=0pt,minimum size=3pt, label=left:{$x_1$}](x1m) at (-1.25,-0.35+\centerym) {};
    \node[circle,fill=black,inner sep=0pt,minimum size=3pt, label=left:{$x_2$}](x2m) at (-1,0.35+\centerym) {};
    \node[circle,fill=black,inner sep=0pt,minimum size=3pt, label=right:{$x_3$}](x3m) at (1,0.35+\centerym) {};
    \node[circle,fill=black,inner sep=0pt,minimum size=3pt, label=right:{$x_4$}](x4m) at (1.25,-0.35+\centerym) {};
    \path[draw=red,fill=blue!50,line join=round,line cap=round,miter limit=10.00,line width=1.152pt] (x1p)--(x2p)--(x2m)--(x1m)--(x1p)-- cycle;
    \path[draw=red,fill=blue!50,line join=round,line cap=round,miter limit=10.00,line width=1.152pt] (x3p)--(x4p)--(x4m)--(x3m)--(x3p)-- cycle;
    \draw[color=blue, thick] (-1.15,\centeryp) -- (1.15,\centeryp);
    \draw[color=blue, thick] (1.15,\centerym) -- (-1.15,\centerym);
    \draw[color=blue, thick,dotted] (1.15,\centeryp) -- (1.15,\centerym);
    \draw[color=blue, thick,dotted] (-1.15,\centeryp) -- (-1.15,\centerym);
    \node[color=blue](b) at (0,\centeryp+0.25){$B$};
    \draw[color=purple, thick] (0,\centerym+0.75) to[out=225,in=135] (0,\centerym-0.75);
    \draw[color=purple, thick, dotted] (0,\centerym-0.75) to[out=45,in=-45] (0,\centerym+0.75);
    \node[color=purple](a) at (0.5,\centerym-0.35){$A$};
\end{tikzpicture}
\end{equation*} 
 
\begin{figure}
    \centering
\tikzset{every picture/.style={line width=0.75pt}} 

\tikzset{every picture/.style={line width=0.75pt}} 

\begin{tikzpicture}[x=0.75pt,y=0.75pt,yscale=-.8,xscale=.9]

\draw   (508.05,194) -- (734.5,194) -- (637.45,261) -- (411,261) -- cycle ;
\draw  [color={rgb, 255:red, 0; green, 0; blue, 0 }  ,draw opacity=1 ][fill={rgb, 255:red, 0; green, 0; blue, 0 }  ,fill opacity=1 ] (444,248.5) .. controls (444,246.57) and (445.57,245) .. (447.5,245) .. controls (449.43,245) and (451,246.57) .. (451,248.5) .. controls (451,250.43) and (449.43,252) .. (447.5,252) .. controls (445.57,252) and (444,250.43) .. (444,248.5) -- cycle ;
\draw  [color={rgb, 255:red, 0; green, 0; blue, 0 }  ,draw opacity=1 ][fill={rgb, 255:red, 0; green, 0; blue, 0 }  ,fill opacity=1 ] (646,220.5) .. controls (646,218.57) and (647.57,217) .. (649.5,217) .. controls (651.43,217) and (653,218.57) .. (653,220.5) .. controls (653,222.43) and (651.43,224) .. (649.5,224) .. controls (647.57,224) and (646,222.43) .. (646,220.5) -- cycle ;
\draw  [color={rgb, 255:red, 0; green, 0; blue, 0 }  ,draw opacity=1 ][fill={rgb, 255:red, 0; green, 0; blue, 0 }  ,fill opacity=1 ] (576,206.5) .. controls (576,204.57) and (577.57,203) .. (579.5,203) .. controls (581.43,203) and (583,204.57) .. (583,206.5) .. controls (583,208.43) and (581.43,210) .. (579.5,210) .. controls (577.57,210) and (576,208.43) .. (576,206.5) -- cycle ;
\draw  [color={rgb, 255:red, 0; green, 0; blue, 0 }  ,draw opacity=1 ][fill={rgb, 255:red, 0; green, 0; blue, 0 }  ,fill opacity=1 ] (554,248.5) .. controls (554,246.57) and (555.57,245) .. (557.5,245) .. controls (559.43,245) and (561,246.57) .. (561,248.5) .. controls (561,250.43) and (559.43,252) .. (557.5,252) .. controls (555.57,252) and (554,250.43) .. (554,248.5) -- cycle ;
\draw  [color={rgb, 255:red, 0; green, 0; blue, 0 }  ,draw opacity=1 ][fill={rgb, 255:red, 0; green, 0; blue, 0 }  ,fill opacity=1 ] (587,246.5) .. controls (587,244.57) and (588.57,243) .. (590.5,243) .. controls (592.43,243) and (594,244.57) .. (594,246.5) .. controls (594,248.43) and (592.43,250) .. (590.5,250) .. controls (588.57,250) and (587,248.43) .. (587,246.5) -- cycle ;
\draw  [color={rgb, 255:red, 0; green, 0; blue, 0 }  ,draw opacity=1 ][fill={rgb, 255:red, 0; green, 0; blue, 0 }  ,fill opacity=1 ] (628,249.5) .. controls (628,247.57) and (629.57,246) .. (631.5,246) .. controls (633.43,246) and (635,247.57) .. (635,249.5) .. controls (635,251.43) and (633.43,253) .. (631.5,253) .. controls (629.57,253) and (628,251.43) .. (628,249.5) -- cycle ;
\draw  [color={rgb, 255:red, 0; green, 0; blue, 0 }  ,draw opacity=1 ][fill={rgb, 255:red, 0; green, 0; blue, 0 }  ,fill opacity=1 ] (516,208.5) .. controls (516,206.57) and (517.57,205) .. (519.5,205) .. controls (521.43,205) and (523,206.57) .. (523,208.5) .. controls (523,210.43) and (521.43,212) .. (519.5,212) .. controls (517.57,212) and (516,210.43) .. (516,208.5) -- cycle ;
\draw  [color={rgb, 255:red, 0; green, 0; blue, 0 }  ,draw opacity=1 ][fill={rgb, 255:red, 0; green, 0; blue, 0 }  ,fill opacity=1 ] (693,201.5) .. controls (693,199.57) and (694.57,198) .. (696.5,198) .. controls (698.43,198) and (700,199.57) .. (700,201.5) .. controls (700,203.43) and (698.43,205) .. (696.5,205) .. controls (694.57,205) and (693,203.43) .. (693,201.5) -- cycle ;
\draw  [color={rgb, 255:red, 0; green, 0; blue, 0 }  ,draw opacity=1 ][fill={rgb, 255:red, 0; green, 0; blue, 0 }  ,fill opacity=1 ] (533,225.5) .. controls (533,223.57) and (534.57,222) .. (536.5,222) .. controls (538.43,222) and (540,223.57) .. (540,225.5) .. controls (540,227.43) and (538.43,229) .. (536.5,229) .. controls (534.57,229) and (533,227.43) .. (533,225.5) -- cycle ;
\draw  [color={rgb, 255:red, 0; green, 0; blue, 0 }  ,draw opacity=1 ][fill={rgb, 255:red, 0; green, 0; blue, 0 }  ,fill opacity=1 ] (485,219.5) .. controls (485,217.57) and (486.57,216) .. (488.5,216) .. controls (490.43,216) and (492,217.57) .. (492,219.5) .. controls (492,221.43) and (490.43,223) .. (488.5,223) .. controls (486.57,223) and (485,221.43) .. (485,219.5) -- cycle ;
\draw  [color={rgb, 255:red, 0; green, 0; blue, 0 }  ,draw opacity=1 ][fill={rgb, 255:red, 0; green, 0; blue, 0 }  ,fill opacity=1 ] (489,244.5) .. controls (489,242.57) and (490.57,241) .. (492.5,241) .. controls (494.43,241) and (496,242.57) .. (496,244.5) .. controls (496,246.43) and (494.43,248) .. (492.5,248) .. controls (490.57,248) and (489,246.43) .. (489,244.5) -- cycle ;
\draw  [color={rgb, 255:red, 0; green, 0; blue, 0 }  ,draw opacity=1 ][fill={rgb, 255:red, 0; green, 0; blue, 0 }  ,fill opacity=1 ] (613,209.5) .. controls (613,207.57) and (614.57,206) .. (616.5,206) .. controls (618.43,206) and (620,207.57) .. (620,209.5) .. controls (620,211.43) and (618.43,213) .. (616.5,213) .. controls (614.57,213) and (613,211.43) .. (613,209.5) -- cycle ;
\draw  [dash pattern={on 4.5pt off 4.5pt}]  (447.5,142) -- (447.5,248.5) ;
\draw  [dash pattern={on 4.5pt off 4.5pt}]  (696.5,96) -- (696.5,202.5) ;
\draw   (125.5,194) -- (351.95,194) -- (254.9,261) -- (28.45,261) -- cycle ;
\draw  [dash pattern={on 4.5pt off 4.5pt}]  (488.5,112) -- (488.5,218.5) ;
\draw    (345.5,231.95) -- (408.5,231) ;
\draw [shift={(342.5,232)}, rotate = 359.13] [fill={rgb, 255:red, 0; green, 0; blue, 0 }  ][line width=0.08]  [draw opacity=0] (10.72,-5.15) -- (0,0) -- (10.72,5.15) -- (7.12,0) -- cycle    ;
\draw  [color={rgb, 255:red, 189; green, 16; blue, 224 }  ,draw opacity=1 ][fill={rgb, 255:red, 189; green, 16; blue, 224 }  ,fill opacity=1 ] (186.7,227.5) .. controls (186.7,225.57) and (188.27,224) .. (190.2,224) .. controls (192.13,224) and (193.7,225.57) .. (193.7,227.5) .. controls (193.7,229.43) and (192.13,231) .. (190.2,231) .. controls (188.27,231) and (186.7,229.43) .. (186.7,227.5) -- cycle ;
\draw  [color={rgb, 255:red, 0; green, 0; blue, 0 }  ,draw opacity=1 ][fill={rgb, 255:red, 0; green, 0; blue, 0 }  ,fill opacity=1 ] (98,238) .. controls (98,236.07) and (99.57,234.5) .. (101.5,234.5) .. controls (103.43,234.5) and (105,236.07) .. (105,238) .. controls (105,239.93) and (103.43,241.5) .. (101.5,241.5) .. controls (99.57,241.5) and (98,239.93) .. (98,238) -- cycle ;
\draw  [color={rgb, 255:red, 0; green, 0; blue, 0 }  ,draw opacity=1 ][fill={rgb, 255:red, 0; green, 0; blue, 0 }  ,fill opacity=1 ] (301,206.5) .. controls (301,204.57) and (302.57,203) .. (304.5,203) .. controls (306.43,203) and (308,204.57) .. (308,206.5) .. controls (308,208.43) and (306.43,210) .. (304.5,210) .. controls (302.57,210) and (301,208.43) .. (301,206.5) -- cycle ;
\draw  [dash pattern={on 4.5pt off 4.5pt}]  (304.5,100) -- (304.5,206.5) ;
\draw  [dash pattern={on 4.5pt off 4.5pt}]  (101.5,135) -- (101.5,241.5) ;
\draw   (78.64,122.51) .. controls (78.64,115.62) and (88.88,110.03) .. (101.5,110.03) .. controls (114.12,110.03) and (124.36,115.62) .. (124.36,122.51) .. controls (124.36,129.41) and (114.12,135) .. (101.5,135) .. controls (88.88,135) and (78.64,129.41) .. (78.64,122.51) -- cycle ;
\draw    (90.5,118) .. controls (98.5,121) and (100.5,155) .. (111.5,116) ;
\draw    (94.5,123) .. controls (103.01,118.66) and (101.61,118.18) .. (109.05,120.89) ;
\draw  [fill={rgb, 255:red, 0; green, 0; blue, 0 }  ,fill opacity=1 ] (152.86,156.04) .. controls (152.86,153.82) and (154.41,152.02) .. (156.31,152.02) .. controls (158.22,152.02) and (159.77,153.82) .. (159.77,156.04) .. controls (159.77,158.26) and (158.22,160.06) .. (156.31,160.06) .. controls (154.41,160.06) and (152.86,158.26) .. (152.86,156.04) -- cycle ;
\draw  [fill={rgb, 255:red, 0; green, 0; blue, 0 }  ,fill opacity=1 ] (145.95,156.04) .. controls (145.95,153.82) and (147.49,152.02) .. (149.4,152.02) .. controls (151.31,152.02) and (152.86,153.82) .. (152.86,156.04) .. controls (152.86,158.26) and (151.31,160.06) .. (149.4,160.06) .. controls (147.49,160.06) and (145.95,158.26) .. (145.95,156.04) -- cycle ;
\draw  [fill={rgb, 255:red, 0; green, 0; blue, 0 }  ,fill opacity=1 ] (159.77,156.04) .. controls (159.77,153.82) and (161.32,152.02) .. (163.22,152.02) .. controls (165.13,152.02) and (166.68,153.82) .. (166.68,156.04) .. controls (166.68,158.26) and (165.13,160.06) .. (163.22,160.06) .. controls (161.32,160.06) and (159.77,158.26) .. (159.77,156.04) -- cycle ;
\draw  [fill={rgb, 255:red, 0; green, 0; blue, 0 }  ,fill opacity=1 ] (166.68,156.04) .. controls (166.68,153.82) and (168.23,152.02) .. (170.14,152.02) .. controls (172.04,152.02) and (173.59,153.82) .. (173.59,156.04) .. controls (173.59,158.26) and (172.04,160.06) .. (170.14,160.06) .. controls (168.23,160.06) and (166.68,158.26) .. (166.68,156.04) -- cycle ;
\draw  [fill={rgb, 255:red, 0; green, 0; blue, 0 }  ,fill opacity=1 ] (180.5,156.04) .. controls (180.5,153.82) and (182.05,152.02) .. (183.96,152.02) .. controls (185.87,152.02) and (187.41,153.82) .. (187.41,156.04) .. controls (187.41,158.26) and (185.87,160.06) .. (183.96,160.06) .. controls (182.05,160.06) and (180.5,158.26) .. (180.5,156.04) -- cycle ;
\draw  [fill={rgb, 255:red, 0; green, 0; blue, 0 }  ,fill opacity=1 ] (173.59,156.04) .. controls (173.59,153.82) and (175.14,152.02) .. (177.05,152.02) .. controls (178.96,152.02) and (180.5,153.82) .. (180.5,156.04) .. controls (180.5,158.26) and (178.96,160.06) .. (177.05,160.06) .. controls (175.14,160.06) and (173.59,158.26) .. (173.59,156.04) -- cycle ;
\draw  [fill={rgb, 255:red, 0; green, 0; blue, 0 }  ,fill opacity=1 ] (187.41,156.04) .. controls (187.41,153.82) and (188.96,152.02) .. (190.87,152.02) .. controls (192.78,152.02) and (194.32,153.82) .. (194.32,156.04) .. controls (194.32,158.26) and (192.78,160.06) .. (190.87,160.06) .. controls (188.96,160.06) and (187.41,158.26) .. (187.41,156.04) -- cycle ;
\draw  [fill={rgb, 255:red, 0; green, 0; blue, 0 }  ,fill opacity=1 ] (152.86,148.01) .. controls (152.86,145.79) and (154.41,143.99) .. (156.31,143.99) .. controls (158.22,143.99) and (159.77,145.79) .. (159.77,148.01) .. controls (159.77,150.23) and (158.22,152.02) .. (156.31,152.02) .. controls (154.41,152.02) and (152.86,150.23) .. (152.86,148.01) -- cycle ;
\draw  [dash pattern={on 4.5pt off 4.5pt}]  (190.87,160.06) -- (190.19,224.29) ;
\draw  [fill={rgb, 255:red, 0; green, 0; blue, 0 }  ,fill opacity=1 ] (139.04,156.04) .. controls (139.04,153.82) and (140.58,152.02) .. (142.49,152.02) .. controls (144.4,152.02) and (145.95,153.82) .. (145.95,156.04) .. controls (145.95,158.26) and (144.4,160.06) .. (142.49,160.06) .. controls (140.58,160.06) and (139.04,158.26) .. (139.04,156.04) -- cycle ;
\draw   (280.64,88.51) .. controls (280.64,81.62) and (290.88,76.03) .. (303.5,76.03) .. controls (316.12,76.03) and (326.36,81.62) .. (326.36,88.51) .. controls (326.36,95.41) and (316.12,101) .. (303.5,101) .. controls (290.88,101) and (280.64,95.41) .. (280.64,88.51) -- cycle ;
\draw    (292.5,84) .. controls (300.5,87) and (302.5,121) .. (313.5,82) ;
\draw    (296.5,89) .. controls (305.01,84.66) and (303.61,84.18) .. (311.05,86.89) ;
\draw   (423.64,128.51) .. controls (423.64,121.62) and (433.88,116.03) .. (446.5,116.03) .. controls (459.12,116.03) and (469.36,121.62) .. (469.36,128.51) .. controls (469.36,135.41) and (459.12,141) .. (446.5,141) .. controls (433.88,141) and (423.64,135.41) .. (423.64,128.51) -- cycle ;
\draw    (435.5,124) .. controls (443.5,127) and (445.5,161) .. (456.5,122) ;
\draw    (439.5,129) .. controls (448.01,124.66) and (446.61,124.18) .. (454.05,126.89) ;
\draw   (464.64,98.51) .. controls (464.64,91.62) and (474.88,86.03) .. (487.5,86.03) .. controls (500.12,86.03) and (510.36,91.62) .. (510.36,98.51) .. controls (510.36,105.41) and (500.12,111) .. (487.5,111) .. controls (474.88,111) and (464.64,105.41) .. (464.64,98.51) -- cycle ;
\draw    (476.5,94) .. controls (484.5,97) and (486.5,131) .. (497.5,92) ;
\draw    (480.5,99) .. controls (489.01,94.66) and (487.61,94.18) .. (495.05,96.89) ;
\draw   (672.64,82.51) .. controls (672.64,75.62) and (682.88,70.03) .. (695.5,70.03) .. controls (708.12,70.03) and (718.36,75.62) .. (718.36,82.51) .. controls (718.36,89.41) and (708.12,95) .. (695.5,95) .. controls (682.88,95) and (672.64,89.41) .. (672.64,82.51) -- cycle ;
\draw    (684.5,78) .. controls (692.5,81) and (694.5,115) .. (705.5,76) ;
\draw    (688.5,83) .. controls (697.01,78.66) and (695.61,78.18) .. (703.05,80.89) ;

\draw (581,128.4) node [anchor=north west][inner sep=0.75pt]    {$\dotsc $};
\draw (344,207.4) node [anchor=north west][inner sep=0.75pt]    {$\{m_i\}\rightarrow 0$};
\draw (166.31,127.93) node [anchor=north west][inner sep=0.75pt]    {$E_{8}$};

\draw (92,88.4)node [anchor=north west][inner sep=0.75pt] {I$_1$} ;

\draw (295, 53) node [anchor=north west][inner sep=0.75pt] {I$_1$};

\draw (477, 62) node [anchor=north west][inner sep=0.75pt] {I$_1$};

\draw (437, 92) node [anchor=north west][inner sep=0.75pt] {I$_1$};

\draw (657, 53) node [anchor=north west][inner sep=0.75pt] {I$_1$};

\end{tikzpicture}
    \caption{\textbf{Seiberg--Witten geometry of the E-string theory.} On the left is the massless limit where black points indicate type I$_1$ singularities while the purple point indicates the type II$^*$ singularities. On the right is the schematic geometry with 12 I$_1$ singularities for generic masses.}
    \label{fig:SW elliptic fibration}
\end{figure}
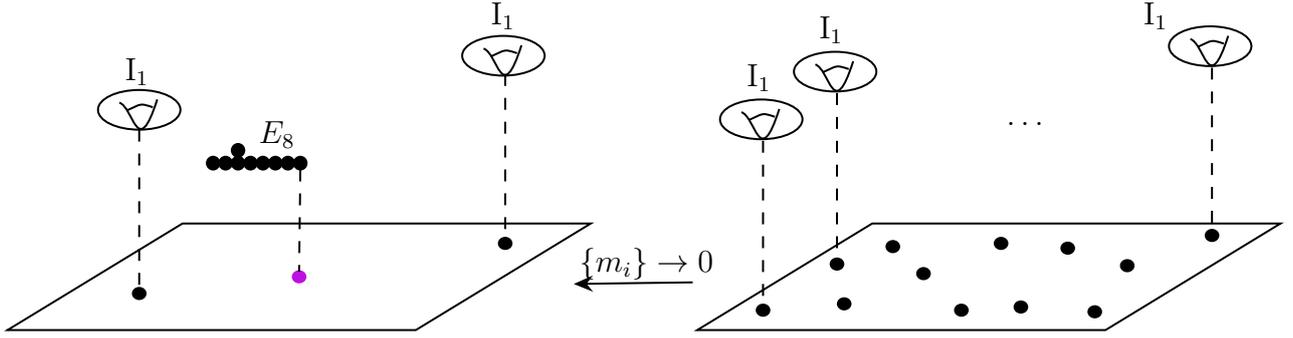

 Once the ``UV parameters'' $\sigma$ and $\{m_i\}$ are fixed, this describes a family of (possibly singular) tori over the Coulomb branch parametrized by $u$, with singularities at special values of $u$ given by the vanishing of the discriminant $\Delta_\Sigma(u)=0$. Since $\Delta_\Sigma$ is a degree-12 polynomial in $u$, generically, there are 12 solutions, giving rise to 12 $A_0$ (I$_1$ in Kodaira's classification) singularities. This is illustrated in Figure~\ref{fig:SW elliptic fibration}.

Near a root of $\Delta_{\Sigma}$, some BPS particles become massless. We can associate a monodromy to each singularity on the Coulomb branch by how charges of particles transform when we ``move around'' the singularity. In the low-energy 4d $\mathcal{N}=2$ $U(1)$ gauge theories, we have BPS states with quantum numbers $(n_e,n_m)$ denoting the electric and magnetic charges. Then, a monodromy $M$ corresponds to how these quantum numbers change when going around a singularity,
\begin{equation}
    \begin{pmatrix}
    n_e'\\
    n_m'
    \end{pmatrix}
    =
    M
    \begin{pmatrix}
    n_e\\
    n_m
    \end{pmatrix}\,.
\end{equation}
For type I$_1$ and II$^*$ singular fibers, the monodromies are respectively conjugate to $T$ and $ST$, where $S,T$ are the generators of $\SL(2,\mathbb{Z})$.

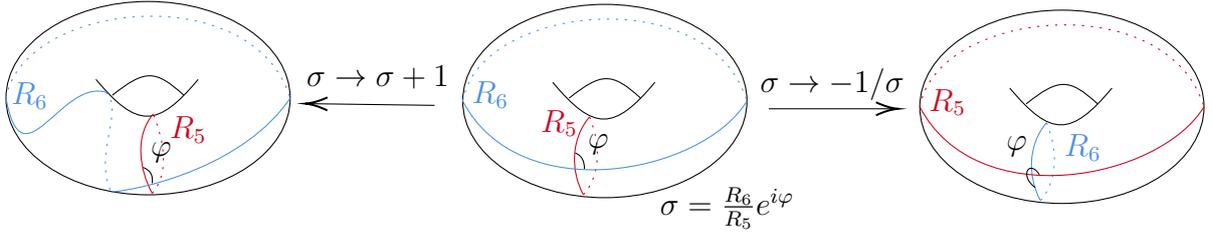
\begin{figure}
    \centering

\begin{tikzpicture}[x=0.75pt,y=0.75pt,yscale=-1,xscale=1]

\draw   (305,315.48) .. controls (305,288.63) and (336.75,266.87) .. (375.93,266.87) .. controls (415.1,266.87) and (446.86,288.63) .. (446.86,315.48) .. controls (446.86,342.33) and (415.1,364.1) .. (375.93,364.1) .. controls (336.75,364.1) and (305,342.33) .. (305,315.48) -- cycle ;
\draw    (349.9,306.05) .. controls (370.31,330.72) and (380.52,330.72) .. (400.93,306.05) ;
\draw    (358.07,314.03) .. controls (374.4,300.97) and (379.5,300.97) .. (393.79,314.03) ;
\draw [color={rgb, 255:red, 208; green, 2; blue, 27 }  ,draw opacity=1 ]   (365.21,362.65) .. controls (358.58,350.31) and (359.09,334.35) .. (368.27,323.46) ;
\draw [color={rgb, 255:red, 208; green, 2; blue, 27 }  ,draw opacity=1 ] [dash pattern={on 0.84pt off 2.51pt}]  (365.21,362.65) .. controls (373.38,353.21) and (374.4,338.7) .. (368.27,323.46) ;
\draw [color={rgb, 255:red, 74; green, 144; blue, 226 }  ,draw opacity=1 ]   (305,315.48) .. controls (333.06,369.9) and (424.4,352.49) .. (446.86,315.48) ;
\draw [color={rgb, 255:red, 74; green, 144; blue, 226 }  ,draw opacity=1 ] [dash pattern={on 0.84pt off 2.51pt}]  (305,315.48) .. controls (313.16,266.87) and (428.49,254.53) .. (446.86,315.48) ;
\draw    (361.13,341.6) .. controls (364.7,343.06) and (365.72,345.96) .. (365.72,349.59) ;
\draw    (290.93,317.66) -- (228.5,316.82) ;
\draw [shift={(226.5,316.79)}, rotate = 0.77] [color={rgb, 255:red, 0; green, 0; blue, 0 }  ][line width=0.75]    (10.93,-3.29) .. controls (6.95,-1.4) and (3.31,-0.3) .. (0,0) .. controls (3.31,0.3) and (6.95,1.4) .. (10.93,3.29)   ;
\draw    (520.2,319.39) -- (457.77,319.39) ;
\draw [shift={(522.2,319.39)}, rotate = 180] [color={rgb, 255:red, 0; green, 0; blue, 0 }  ][line width=0.75]    (10.93,-3.29) .. controls (6.95,-1.4) and (3.31,-0.3) .. (0,0) .. controls (3.31,0.3) and (6.95,1.4) .. (10.93,3.29)   ;
\draw   (533.14,318.39) .. controls (533.14,291.54) and (564.9,269.77) .. (604.07,269.77) .. controls (643.24,269.77) and (675,291.54) .. (675,318.39) .. controls (675,345.23) and (643.24,367) .. (604.07,367) .. controls (564.9,367) and (533.14,345.23) .. (533.14,318.39) -- cycle ;
\draw    (578.05,308.95) .. controls (598.46,333.62) and (608.66,333.62) .. (629.07,308.95) ;
\draw    (586.21,316.93) .. controls (602.54,303.87) and (607.64,303.87) .. (621.93,316.93) ;
\draw [color={rgb, 255:red, 74; green, 144; blue, 226 }  ,draw opacity=1 ]   (593.35,365.55) .. controls (586.72,353.21) and (587.23,337.25) .. (596.42,326.37) ;
\draw [color={rgb, 255:red, 74; green, 144; blue, 226 }  ,draw opacity=1 ] [dash pattern={on 0.84pt off 2.51pt}]  (593.35,365.55) .. controls (601.52,356.12) and (602.54,341.6) .. (596.42,326.37) ;
\draw [color={rgb, 255:red, 208; green, 2; blue, 27 }  ,draw opacity=1 ]   (533.14,318.39) .. controls (561.21,372.8) and (652.55,355.39) .. (675,318.39) ;
\draw [color={rgb, 255:red, 208; green, 2; blue, 27 }  ,draw opacity=1 ] [dash pattern={on 0.84pt off 2.51pt}]  (533.14,318.39) .. controls (541.31,269.77) and (656.63,257.44) .. (675,318.39) ;
\draw    (590.8,359.02) .. controls (582.64,353.21) and (587.74,344.51) .. (592.84,352.49) ;
\draw   (77,314.03) .. controls (77,287.18) and (108.76,265.42) .. (147.93,265.42) .. controls (187.1,265.42) and (218.86,287.18) .. (218.86,314.03) .. controls (218.86,340.88) and (187.1,362.65) .. (147.93,362.65) .. controls (108.76,362.65) and (77,340.88) .. (77,314.03) -- cycle ;
\draw    (121.9,304.6) .. controls (142.32,329.27) and (152.52,329.27) .. (172.93,304.6) ;
\draw    (130.07,312.58) .. controls (146.4,299.52) and (151.5,299.52) .. (165.79,312.58) ;
\draw [color={rgb, 255:red, 208; green, 2; blue, 27 }  ,draw opacity=1 ]   (149.97,361.92) .. controls (143.34,349.59) and (141.3,332.9) .. (150.48,322.01) ;
\draw [color={rgb, 255:red, 208; green, 2; blue, 27 }  ,draw opacity=1 ] [dash pattern={on 0.84pt off 2.51pt}]  (149.97,361.92) .. controls (158.14,352.49) and (156.6,337.25) .. (150.48,322.01) ;
\draw [color={rgb, 255:red, 74; green, 144; blue, 226 }  ,draw opacity=1 ]   (77,314.03) .. controls (85.16,359.02) and (104.56,303.87) .. (127.01,311.13) ;
\draw [color={rgb, 255:red, 74; green, 144; blue, 226 }  ,draw opacity=1 ] [dash pattern={on 0.84pt off 2.51pt}]  (77,314.03) .. controls (85.16,265.42) and (200.49,253.08) .. (218.86,314.03) ;
\draw    (145.38,348.86) .. controls (148.95,350.31) and (150.48,352.49) .. (149.97,356.84) ;
\draw [color={rgb, 255:red, 74; green, 144; blue, 226 }  ,draw opacity=1 ] [dash pattern={on 0.84pt off 2.51pt}]  (127.01,311.13) .. controls (136.19,320.56) and (120.37,342.33) .. (129.56,361.2) ;
\draw [color={rgb, 255:red, 74; green, 144; blue, 226 }  ,draw opacity=1 ]   (129.56,361.2) .. controls (147.42,359.74) and (196.41,347.41) .. (218.86,314.03) ;

\draw (342.78,318.67) node [anchor=north west][inner sep=0.75pt]  [font=\normalsize,color={rgb, 255:red, 208; green, 2; blue, 27 }  ,opacity=1 ]  {$R_{5}$};
\draw (308.45,304.16) node [anchor=north west][inner sep=0.75pt]  [color={rgb, 255:red, 74; green, 144; blue, 226 }  ,opacity=1 ]  {$R_{6}$};
\draw (402.23,357.72) node [anchor=north west][inner sep=0.75pt]    {$\sigma =\frac{R_{6}}{R_{5}} e^{i\varphi }$};
\draw (366.27,330.93) node [anchor=north west][inner sep=0.75pt]    {$\varphi $};
\draw (225.89,296.27) node [anchor=north west][inner sep=0.75pt]    {$\sigma \rightarrow \sigma +1$};
\draw (452.23,297.72) node [anchor=north west][inner sep=0.75pt]    {$\sigma \rightarrow -1/\sigma $};
\draw (604.17,331.1) node [anchor=north west][inner sep=0.75pt]  [font=\normalsize,color={rgb, 255:red, 208; green, 2; blue, 27 }  ,opacity=1 ]  {$\textcolor[rgb]{0.29,0.56,0.89}{R_{6}}$};
\draw (536.59,308.06) node [anchor=north west][inner sep=0.75pt]  [color={rgb, 255:red, 74; green, 144; blue, 226 }  ,opacity=1 ]  {$\textcolor[rgb]{0.82,0.01,0.11}{R_{5}}$};
\draw (575.17,331.73) node [anchor=north west][inner sep=0.75pt]    {$\varphi $};
\draw (157.54,321.94) node [anchor=north west][inner sep=0.75pt]  [font=\normalsize,color={rgb, 255:red, 208; green, 2; blue, 27 }  ,opacity=1 ]  {$R_{5}$};
\draw (78.45,304.71) node [anchor=north west][inner sep=0.75pt]  [color={rgb, 255:red, 74; green, 144; blue, 226 }  ,opacity=1 ]  {$R_{6}$};
\draw (148.01,335) node [anchor=north west][inner sep=0.75pt]    {$\varphi $};

\end{tikzpicture}
\caption{\textbf{Duality from the spacetime torus.} $T^2_\sigma$ can also be parameterized by a parallelogram with lengths $R_5,R_6$ and angle $\varphi$. Under a $T$-transformation, the $S^1_{R_{6}}$ winds over $S^1_{R_5}$ once, while under an $S$-transformation, the two radii exchange values leaving the shape of the torus unchanged.}
\label{fig:torus}
\end{figure}

One important aspect of the E-string theory compactified on $T^2$ is that it has an SL$(2,\mathbb{Z})$ duality from the mapping class group of the $T^2$. This is analogous to the duality of 4d $\mathcal{N}=4$ super-Yang--Mills theory and in fact better, as the latter example depends on the choice of ``polarization on $T^2$'' which is only invariant under the $\Gamma_0(4)$ subgroup (see e.g.~\cite{Gukov:2020btk,GHP2} for a more detailed exposition of this phenomenon). For the E-string theory, no such choice is involved, and the entire SL$(2,\mathbb{Z})$ group serves as a duality group for the theory. The action of the two generators $S$ and $T$ is illustrated in Figure~\ref{fig:torus}.

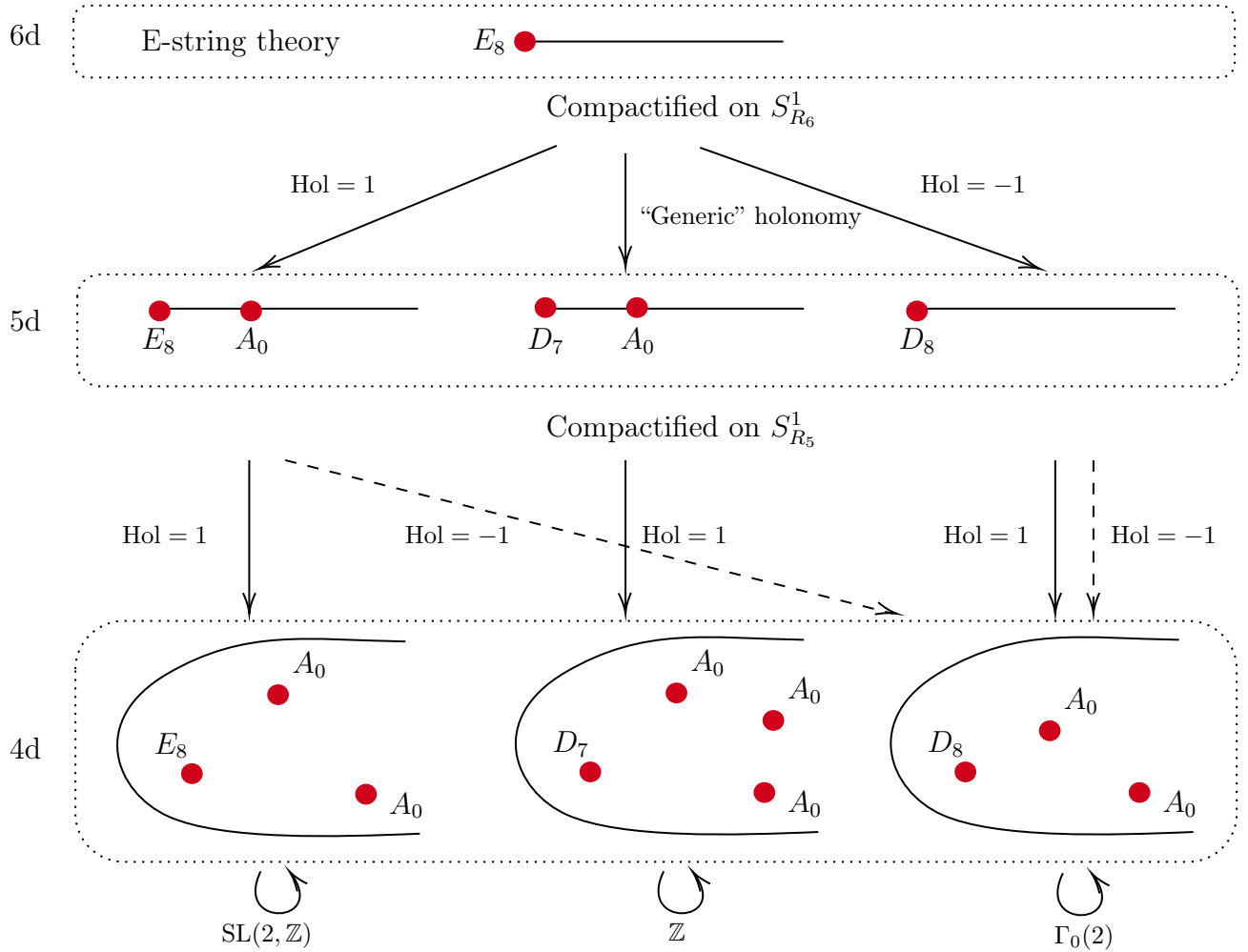
\begin{figure}
    \centering

\tikzset{every picture/.style={line width=0.75pt}} 

\begin{tikzpicture}[x=0.75pt,y=0.75pt,yscale=-1,xscale=1]

\draw    (91.07,167) -- (226.38,167) ;
\draw  [color={rgb, 255:red, 208; green, 2; blue, 27 }  ,draw opacity=1 ][fill={rgb, 255:red, 208; green, 2; blue, 27 }  ,fill opacity=1 ] (85.9,168.24) .. controls (85.9,165.49) and (88.21,163.26) .. (91.07,163.26) .. controls (93.92,163.26) and (96.24,165.49) .. (96.24,168.24) .. controls (96.24,170.99) and (93.92,173.22) .. (91.07,173.22) .. controls (88.21,173.22) and (85.9,170.99) .. (85.9,168.24) -- cycle ;
\draw  [color={rgb, 255:red, 208; green, 2; blue, 27 }  ,draw opacity=1 ][fill={rgb, 255:red, 208; green, 2; blue, 27 }  ,fill opacity=1 ] (133.82,168.24) .. controls (133.82,165.49) and (136.14,163.26) .. (138.99,163.26) .. controls (141.84,163.26) and (144.16,165.49) .. (144.16,168.24) .. controls (144.16,170.99) and (141.84,173.22) .. (138.99,173.22) .. controls (136.14,173.22) and (133.82,170.99) .. (133.82,168.24) -- cycle ;
\draw    (293.09,167) -- (428.4,167) ;
\draw  [color={rgb, 255:red, 208; green, 2; blue, 27 }  ,draw opacity=1 ][fill={rgb, 255:red, 208; green, 2; blue, 27 }  ,fill opacity=1 ] (287.92,166.43) .. controls (287.92,163.68) and (290.24,161.45) .. (293.09,161.45) .. controls (295.95,161.45) and (298.26,163.68) .. (298.26,166.43) .. controls (298.26,169.18) and (295.95,171.41) .. (293.09,171.41) .. controls (290.24,171.41) and (287.92,169.18) .. (287.92,166.43) -- cycle ;
\draw  [color={rgb, 255:red, 208; green, 2; blue, 27 }  ,draw opacity=1 ][fill={rgb, 255:red, 208; green, 2; blue, 27 }  ,fill opacity=1 ] (335.85,166.43) .. controls (335.85,163.68) and (338.16,161.45) .. (341.01,161.45) .. controls (343.87,161.45) and (346.18,163.68) .. (346.18,166.43) .. controls (346.18,169.18) and (343.87,171.41) .. (341.01,171.41) .. controls (338.16,171.41) and (335.85,169.18) .. (335.85,166.43) -- cycle ;
\draw    (487.6,167) -- (622.91,167) ;
\draw  [color={rgb, 255:red, 208; green, 2; blue, 27 }  ,draw opacity=1 ][fill={rgb, 255:red, 208; green, 2; blue, 27 }  ,fill opacity=1 ] (482.43,168.24) .. controls (482.43,165.49) and (484.74,163.26) .. (487.6,163.26) .. controls (490.45,163.26) and (492.77,165.49) .. (492.77,168.24) .. controls (492.77,170.99) and (490.45,173.22) .. (487.6,173.22) .. controls (484.74,173.22) and (482.43,170.99) .. (482.43,168.24) -- cycle ;
\draw  (138,247) -- (138,328) ;
\draw [shift={(138,328)}, rotate = 270] [color={rgb, 255:red, 0; green, 0; blue, 0 }  ][line width=0.75]    (10.93,-3.29) .. controls (6.95,-1.4) and (3.31,-0.3) .. (0,0) .. controls (3.31,0.3) and (6.95,1.4) .. (10.93,3.29)   ;
\draw    (227.32,444.13) .. controls (183.15,445.94) and (123.96,447.76) .. (95.77,435.08) .. controls (67.58,422.4) and (53.48,384.36) .. (94.83,359.91) .. controls (136.17,335.46) and (169.06,341.8) .. (219.8,342.7) ;
\draw  [color={rgb, 255:red, 208; green, 2; blue, 27 }  ,draw opacity=1 ][fill={rgb, 255:red, 208; green, 2; blue, 27 }  ,fill opacity=1 ] (102.81,412.44) .. controls (102.81,409.69) and (105.13,407.46) .. (107.98,407.46) .. controls (110.84,407.46) and (113.15,409.69) .. (113.15,412.44) .. controls (113.15,415.19) and (110.84,417.42) .. (107.98,417.42) .. controls (105.13,417.42) and (102.81,415.19) .. (102.81,412.44) -- cycle ;
\draw  [color={rgb, 255:red, 208; green, 2; blue, 27 }  ,draw opacity=1 ][fill={rgb, 255:red, 208; green, 2; blue, 27 }  ,fill opacity=1 ] (193.96,423.3) .. controls (193.96,420.55) and (196.27,418.32) .. (199.13,418.32) .. controls (201.98,418.32) and (204.3,420.55) .. (204.3,423.3) .. controls (204.3,426.06) and (201.98,428.29) .. (199.13,428.29) .. controls (196.27,428.29) and (193.96,426.06) .. (193.96,423.3) -- cycle ;
\draw  [color={rgb, 255:red, 208; green, 2; blue, 27 }  ,draw opacity=1 ][fill={rgb, 255:red, 208; green, 2; blue, 27 }  ,fill opacity=1 ] (147.92,370.78) .. controls (147.92,368.03) and (150.23,365.8) .. (153.08,365.8) .. controls (155.94,365.8) and (158.25,368.03) .. (158.25,370.78) .. controls (158.25,373.53) and (155.94,375.76) .. (153.08,375.76) .. controls (150.23,375.76) and (147.92,373.53) .. (147.92,370.78) -- cycle ;
\draw    (632.3,443.23) .. controls (588.14,445.04) and (528.94,446.85) .. (500.75,434.17) .. controls (472.56,421.49) and (458.47,383.46) .. (499.81,359.01) .. controls (541.16,334.55) and (574.04,340.89) .. (624.78,341.8) ;
\draw  [color={rgb, 255:red, 208; green, 2; blue, 27 }  ,draw opacity=1 ][fill={rgb, 255:red, 208; green, 2; blue, 27 }  ,fill opacity=1 ] (507.8,411.53) .. controls (507.8,408.78) and (510.11,406.55) .. (512.97,406.55) .. controls (515.82,406.55) and (518.14,408.78) .. (518.14,411.53) .. controls (518.14,414.28) and (515.82,416.51) .. (512.97,416.51) .. controls (510.11,416.51) and (507.8,414.28) .. (507.8,411.53) -- cycle ;
\draw  [color={rgb, 255:red, 208; green, 2; blue, 27 }  ,draw opacity=1 ][fill={rgb, 255:red, 208; green, 2; blue, 27 }  ,fill opacity=1 ] (598.94,422.4) .. controls (598.94,419.65) and (601.26,417.42) .. (604.11,417.42) .. controls (606.97,417.42) and (609.28,419.65) .. (609.28,422.4) .. controls (609.28,425.15) and (606.97,427.38) .. (604.11,427.38) .. controls (601.26,427.38) and (598.94,425.15) .. (598.94,422.4) -- cycle ;
\draw  [color={rgb, 255:red, 208; green, 2; blue, 27 }  ,draw opacity=1 ][fill={rgb, 255:red, 208; green, 2; blue, 27 }  ,fill opacity=1 ] (551.96,389.8) .. controls (551.96,387.05) and (554.28,384.82) .. (557.13,384.82) .. controls (559.98,384.82) and (562.3,387.05) .. (562.3,389.8) .. controls (562.3,392.55) and (559.98,394.78) .. (557.13,394.78) .. controls (554.28,394.78) and (551.96,392.55) .. (551.96,389.8) -- cycle ;
\draw    [dash pattern={on 4.5pt off 4.5pt}] (156.84,247) -- (475.07,328) ;
\draw [shift={(477,328)}, rotate = 194.9] [color={rgb, 255:red, 0; green, 0; blue, 0 }  ][line width=0.75]    (10.93,-3.29) .. controls (6.95,-1.4) and (3.31,-0.3) .. (0,0) .. controls (3.31,0.3) and (6.95,1.4) .. (10.93,3.29)   ;
\draw  (560,247) -- (560,328) ;
\draw [shift={(560,328)}, rotate = 269.35] [color={rgb, 255:red, 0; green, 0; blue, 0 }  ][line width=0.75]    (10.93,-3.29) .. controls (6.95,-1.4) and (3.31,-0.3) .. (0,0) .. controls (3.31,0.3) and (6.95,1.4) .. (10.93,3.29)   ;
\draw    (435.92,443.23) .. controls (391.75,445.04) and (332.56,446.85) .. (304.37,434.17) .. controls (276.18,421.49) and (262.08,383.46) .. (303.43,359.01) .. controls (344.77,334.55) and (377.66,340.89) .. (428.4,341.8) ;
\draw  [color={rgb, 255:red, 208; green, 2; blue, 27 }  ,draw opacity=1 ][fill={rgb, 255:red, 208; green, 2; blue, 27 }  ,fill opacity=1 ] (311.41,411.53) .. controls (311.41,408.78) and (313.73,406.55) .. (316.58,406.55) .. controls (319.44,406.55) and (321.75,408.78) .. (321.75,411.53) .. controls (321.75,414.28) and (319.44,416.51) .. (316.58,416.51) .. controls (313.73,416.51) and (311.41,414.28) .. (311.41,411.53) -- cycle ;
\draw  [color={rgb, 255:red, 208; green, 2; blue, 27 }  ,draw opacity=1 ][fill={rgb, 255:red, 208; green, 2; blue, 27 }  ,fill opacity=1 ] (402.56,422.4) .. controls (402.56,419.65) and (404.87,417.42) .. (407.73,417.42) .. controls (410.58,417.42) and (412.9,419.65) .. (412.9,422.4) .. controls (412.9,425.15) and (410.58,427.38) .. (407.73,427.38) .. controls (404.87,427.38) and (402.56,425.15) .. (402.56,422.4) -- cycle ;
\draw  [color={rgb, 255:red, 208; green, 2; blue, 27 }  ,draw opacity=1 ][fill={rgb, 255:red, 208; green, 2; blue, 27 }  ,fill opacity=1 ] (356.52,369.87) .. controls (356.52,367.12) and (358.83,364.89) .. (361.69,364.89) .. controls (364.54,364.89) and (366.85,367.12) .. (366.85,369.87) .. controls (366.85,372.62) and (364.54,374.85) .. (361.69,374.85) .. controls (358.83,374.85) and (356.52,372.62) .. (356.52,369.87) -- cycle ;
\draw  [color={rgb, 255:red, 208; green, 2; blue, 27 }  ,draw opacity=1 ][fill={rgb, 255:red, 208; green, 2; blue, 27 }  ,fill opacity=1 ] (407.26,384.36) .. controls (407.26,381.61) and (409.57,379.38) .. (412.43,379.38) .. controls (415.28,379.38) and (417.59,381.61) .. (417.59,384.36) .. controls (417.59,387.11) and (415.28,389.34) .. (412.43,389.34) .. controls (409.57,389.34) and (407.26,387.11) .. (407.26,384.36) -- cycle ;
\draw  [dash pattern={on 4.5pt off 4.5pt}]   (580,247) -- (580,328) ;
\draw [shift={(580,328)}, rotate = 269.35] [color={rgb, 255:red, 0; green, 0; blue, 0 }  ][line width=0.75]    (10.93,-3.29) .. controls (6.95,-1.4) and (3.31,-0.3) .. (0,0) .. controls (3.31,0.3) and (6.95,1.4) .. (10.93,3.29)   ;
\draw    (335,247) -- (335,328) ;
\draw [shift={(335,328)}, rotate = 267.93] [color={rgb, 255:red, 0; green, 0; blue, 0 }  ][line width=0.75]    (10.93,-3.29) .. controls (6.95,-1.4) and (3.31,-0.3) .. (0,0) .. controls (3.31,0.3) and (6.95,1.4) .. (10.93,3.29)   ;
\draw  [dash pattern={on 0.84pt off 2.51pt}] (46,14.6) .. controls (46,10.4) and (49.4,7) .. (53.6,7) -- (646.4,7) .. controls (650.6,7) and (654,10.4) .. (654,14.6) -- (654,37.4) .. controls (654,41.6) and (650.6,45) .. (646.4,45) -- (53.6,45) .. controls (49.4,45) and (46,41.6) .. (46,37.4) -- cycle ;
\draw  [dash pattern={on 0.84pt off 2.51pt}] (48,160.8) .. controls (48,154.28) and (53.28,149) .. (59.8,149) -- (644.2,149) .. controls (650.72,149) and (656,154.28) .. (656,160.8) -- (656,196.2) .. controls (656,202.72) and (650.72,208) .. (644.2,208) -- (59.8,208) .. controls (53.28,208) and (48,202.72) .. (48,196.2) -- cycle ;
\draw  [dash pattern={on 0.84pt off 2.51pt}] (47,357.27) .. controls (47,343.22) and (58.39,331.84) .. (72.43,331.84) -- (629.57,331.84) .. controls (643.61,331.84) and (655,343.22) .. (655,357.27) -- (655,433.57) .. controls (655,447.61) and (643.61,459) .. (629.57,459) -- (72.43,459) .. controls (58.39,459) and (47,447.61) .. (47,433.57) -- cycle ;
\draw    (299,81) -- (144.85,145.23) ;
\draw [shift={(143,146)}, rotate = 337.38] [color={rgb, 255:red, 0; green, 0; blue, 0 }  ][line width=0.75]    (10.93,-3.29) .. controls (6.95,-1.4) and (3.31,-0.3) .. (0,0) .. controls (3.31,0.3) and (6.95,1.4) .. (10.93,3.29)   ;
\draw    (335,85) -- (335,145) ;
\draw [shift={(335,145)}, rotate = 269.05] [color={rgb, 255:red, 0; green, 0; blue, 0 }  ][line width=0.75]    (10.93,-3.29) .. controls (6.95,-1.4) and (3.31,-0.3) .. (0,0) .. controls (3.31,0.3) and (6.95,1.4) .. (10.93,3.29)   ;
\draw    (374,82) -- (550.12,145.32) ;
\draw [shift={(552,146)}, rotate = 199.78] [color={rgb, 255:red, 0; green, 0; blue, 0 }  ][line width=0.75]    (10.93,-3.29) .. controls (6.95,-1.4) and (3.31,-0.3) .. (0,0) .. controls (3.31,0.3) and (6.95,1.4) .. (10.93,3.29)   ;
\draw    (144,464) .. controls (129.23,494.54) and (179.46,492.08) .. (159.94,465.24) ;
\draw [shift={(159,464)}, rotate = 51.84] [color={rgb, 255:red, 0; green, 0; blue, 0 }  ][line width=0.75]    (10.93,-3.29) .. controls (6.95,-1.4) and (3.31,-0.3) .. (0,0) .. controls (3.31,0.3) and (6.95,1.4) .. (10.93,3.29)   ;
\draw    (564,465) .. controls (549.23,495.54) and (599.46,493.08) .. (579.94,466.24) ;
\draw [shift={(579,465)}, rotate = 51.84] [color={rgb, 255:red, 0; green, 0; blue, 0 }  ][line width=0.75]    (10.93,-3.29) .. controls (6.95,-1.4) and (3.31,-0.3) .. (0,0) .. controls (3.31,0.3) and (6.95,1.4) .. (10.93,3.29)   ;
\draw    (355,464) .. controls (340.23,494.54) and (390.46,492.08) .. (370.94,465.24) ;
\draw [shift={(370,464)}, rotate = 51.84] [color={rgb, 255:red, 0; green, 0; blue, 0 }  ][line width=0.75]    (10.93,-3.29) .. controls (6.95,-1.4) and (3.31,-0.3) .. (0,0) .. controls (3.31,0.3) and (6.95,1.4) .. (10.93,3.29)   ;
\draw    (282.35,26) -- (417.65,26) ;
\draw  [color={rgb, 255:red, 208; green, 2; blue, 27 }  ,draw opacity=1 ][fill={rgb, 255:red, 208; green, 2; blue, 27 }  ,fill opacity=1 ] (277.18,26) .. controls (277.18,23.25) and (279.49,21.02) .. (282.35,21.02) .. controls (285.2,21.02) and (287.51,23.25) .. (287.51,26) .. controls (287.51,28.75) and (285.2,30.98) .. (282.35,30.98) .. controls (279.49,30.98) and (277.18,28.75) .. (277.18,26) -- cycle ;

\draw (80,18) node [anchor=north west][inner sep=0.75pt]   [align=left] {E-string theory};
\draw (292,220.05) node [anchor=north west][inner sep=0.75pt]  [rotate=-360] [align=left] {Compactified on $S^1_{R_5}$};
\draw (80.13,176) node [anchor=north west][inner sep=0.75pt]    {$E_{8}$};
\draw (128.93,176) node [anchor=north west][inner sep=0.75pt]    {$A_{0}$};
\draw (158.11,95) node [anchor=north west][inner sep=0.75pt]   [align=left] {\footnotesize $\mathrm{Hol}=1$};
\draw (282.06,176) node [anchor=north west][inner sep=0.75pt]    {$D_{7}$};
\draw (330.95,176) node [anchor=north west][inner sep=0.75pt]    {$A_{0}$};
\draw (342,111.76) node [anchor=north west][inner sep=0.75pt]   [align=left] {\footnotesize ``Generic'' holonomy};
\draw (476.57,176) node [anchor=north west][inner sep=0.75pt]    {$D_{8}$};
\draw (70,280) node [anchor=north west][inner sep=0.75pt]   [align=left] {\footnotesize $\mathrm{Hol}=1$};
\draw (86.71,389.35) node [anchor=north west][inner sep=0.75pt]    {$E_{8}$};
\draw (159,346.78) node [anchor=north west][inner sep=0.75pt]    {$A_{0}$};
\draw (209.74,421.04) node [anchor=north west][inner sep=0.75pt]    {$A_{0}$};
\draw (491.6,388.44) node [anchor=north west][inner sep=0.75pt]    {$D_{8}$};
\draw (563.04,365.8) node [anchor=north west][inner sep=0.75pt]    {$A_{0}$};
\draw (614.72,420.14) node [anchor=north west][inner sep=0.75pt]    {$A_{0}$};
\draw (218,280) node [anchor=north west][inner sep=0.75pt]   [align=left] {\footnotesize $\mathrm{Hol}=-1$};
\draw (500,280) node [anchor=north west][inner sep=0.75pt]   [align=left] {\footnotesize $\mathrm{Hol}=1$};
\draw (295.22,388.44) node [anchor=north west][inner sep=0.75pt]    {$D_{7}$};
\draw (367.6,345.88) node [anchor=north west][inner sep=0.75pt]    {$A_{0}$};
\draw (418.34,420.14) node [anchor=north west][inner sep=0.75pt]    {$A_{0}$};
\draw (418.34,360.37) node [anchor=north west][inner sep=0.75pt]    {$A_{0}$};
\draw (587.66,280) node [anchor=north west][inner sep=0.75pt]   [align=left] {\footnotesize $\mathrm{Hol}=-1$};
\draw (11,15) node [anchor=north west][inner sep=0.75pt]   [align=left] {6d};
\draw (11,166) node [anchor=north west][inner sep=0.75pt]   [align=left] {5d};
\draw (11,392) node [anchor=north west][inner sep=0.75pt]   [align=left] {4d};
\draw (292,51.05) node [anchor=north west][inner sep=0.75pt]  [rotate=-360] [align=left] {Compactified on $S^1_{R_6}$};
\draw (487.88,95) node [anchor=north west][inner sep=0.75pt]   [align=left] {\footnotesize $\mathrm{Hol}=-1$};
\draw (122,489.4) node [anchor=north west][inner sep=0.75pt]    {\footnotesize $\SL( 2,\mathbb{Z})$};
\draw (558,490.4) node [anchor=north west][inner sep=0.75pt]    {\footnotesize $\Gamma _{0}( 2)$};
\draw (362,490) node [anchor=north ][inner sep=0.75pt]    {\footnotesize $\Z$};
\draw (342,280) node [anchor=north west][inner sep=0.75pt]   [align=left] {\footnotesize $\mathrm{Hol}=1$};
\draw (253.13,18) node [anchor=north west][inner sep=0.75pt]    {$E_{8}$};

\end{tikzpicture}
    \caption{\textbf{Coulomb branch of the KK-compactified E-string theory with special holonomies.} We illustrate a one-parameter family of holonomies preserving a $D_7\subset E_8$ subgroup. It starts with the trivial holonomy (Hol$=1$) and ends with ``Hol$=-1$'' which refers to the unique element of $E_8$ with $D_8\simeq Spin(16) /\mathbb{Z}_2$ stabilizer. The ``$-1$'' holonomy breaks $\SL(2,\Z)$ into $\Gamma_0(2)$, which is the subgroup that leaves the configuration invariant. The other two configurations in the SL$(2,\Z)$ orbit can be reached by turning on the holonomy in different ways, which are also illustrated with dashed lines.  
    When the holonomy is ``generic'' (still required to be $D_7$-preserving), only a $\Z$ subgroup generated by $T$ is expected to be preserved.}
    \label{fig:KK compactification of E-string}
\end{figure}

With masses turned on, one also needs to transform the mass parameters as Jacobi variables. For special masses, they can be invariant under a subgroup of SL$(2,\mathbb{Z})$, leading to theories with some modular invariance. One example is illustrated in Figure~\ref{fig:KK compactification of E-string}.

The physics of the families of KK-compactified theories parametrized by $\sigma$ and $m_i$'s is remarkably rich. 
In the next subsection, we will focus on the massless limit, deferring the discussion of some other limits to Appendix~\ref{sec:specialMass}.

\subsection{Singularities and modularity}
\label{sec:modularity}

In the massless limit, the Seiberg--Witten curve of the E-string theory simplifies to \cite{Eguchi:2002fc,Ganor:1996pc},
\begin{equation}
\label{eq:E8curve}
        \Sigma\,:\qquad y^2=4x^3-\frac{1}{12}E_4u^4x-\left(\frac{1}{216}E_6u^6-4u^5\right)\,,
\end{equation}
where $E_4$ and $E_6$ are Eisenstein series in $q:=e^{2\pi i \sigma}$ (see Appendix~\ref{sec:modular forms} for a review).
Its discriminant is given by
\begin{equation}
        \Delta_\Sigma(u,\sigma)=f^3-27g^2=\frac{1}{12^3}E_4^3u^{12}-27\left(\frac{1}{216}E_6u^6-4u^5\right)^2\,,
\end{equation}
from which we find that the curve becomes singular at the following loci,
    \begin{equation}
    \label{eq:E8singularities}
    \begin{cases}
    u_{s,0}=0\, ,\\
    u_{s,+}=-\frac{864(E_6+E_4^{3/2})}{E_4^3-E_6^2}\, ,\\
    u_{s,-}=-\frac{864(E_6-E_4^{3/2})}{E_4^3-E_6^2}\, .
    \end{cases}
    \end{equation}
The vanishing orders for $f$, $g$ and $\Delta$ are summarized in Table~\ref{table:masslessSingularities}. Thus, we can identify $u_{s,\pm}$ as the two I$_1$ singularities and $u_{s,0}$ as the II$^*$ singularity. 

\begin{table}[h!]
    \centering
    \begin{tabular}[t]{lcccc}
    \toprule
         &  $\text{ord}_{\Sigma}(f)$ & $\text{ord}_{\Sigma}(g)$ & $\text{ord}_{\Sigma}(\Delta)$ & Kodaira type\\
    \midrule
         $u_s=0$& 4 & 5 & 10& II${}^*$\\
         $u_s=u_{s,\pm}$& 0& 0& 1& I${}_1$\\
    \bottomrule
    \end{tabular}
    \caption{\textbf{Singularities on the Coulomb branch}. The three singularities in the massless limit, $u_{s,0},u_{s,\pm}$, can be classified according to the behavior of the coefficients and the discriminant of the Seiberg--Witten curve as they approach 0. For the full list of Kodaira--Tate classifications of singularities, see e.g.,~\cite{husemoller2013elliptic}. 
    }
    \label{table:masslessSingularities}
    \end{table}

When approaching an I$_1$ singularity, one cycle of the elliptic curve degenerates, which corresponds to collisions of the branch points. Indeed, when $u=u_{s,+}$, one has $x_1=x_3$, where $x_{1,2,3,4}$ are the massless limits of~\eqref{eq:x singularities}, while for $u=u_{s,-}$, one has $x_2=x_3$ instead.

\subsubsection*{$T$-transformation}

As can be observed from~\eqref{eq:E8curve}, the coefficients in the massless limit are simply Eisenstein series, which are modular forms under $\SL(2,\mathbb{Z})$. (See Appendix~\ref{sec:modular forms} for a brief review on modular forms.) In what follows, we will aim to establish a better understanding of how the configuration of singularities varies when $\sigma$ is varied and in what sense the configuration is modular.

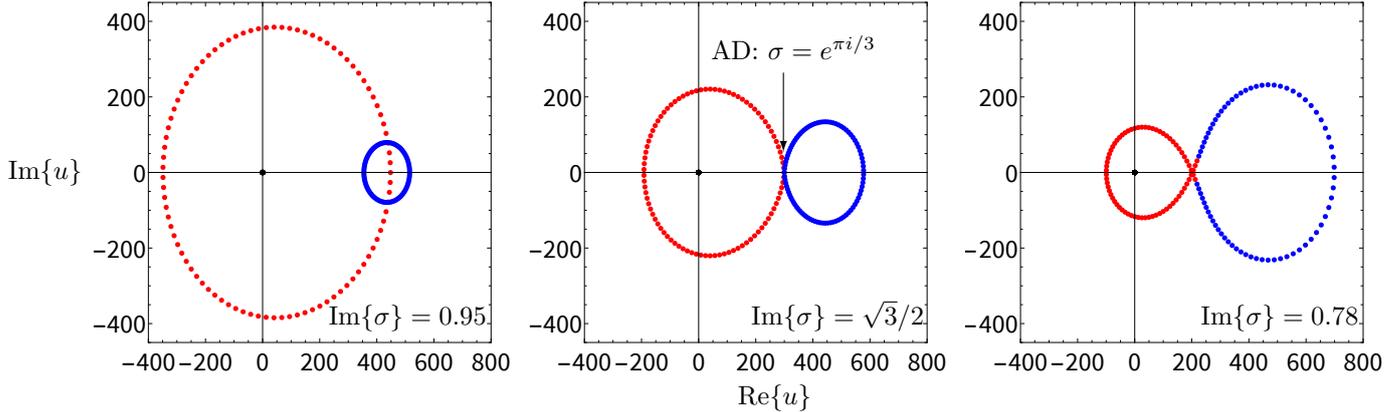
\begin{figure}
    \centering
    \vskip 0pt
    \begin{subfigure}{0.325\textwidth}
    \centering
    \includegraphics[width=\linewidth]{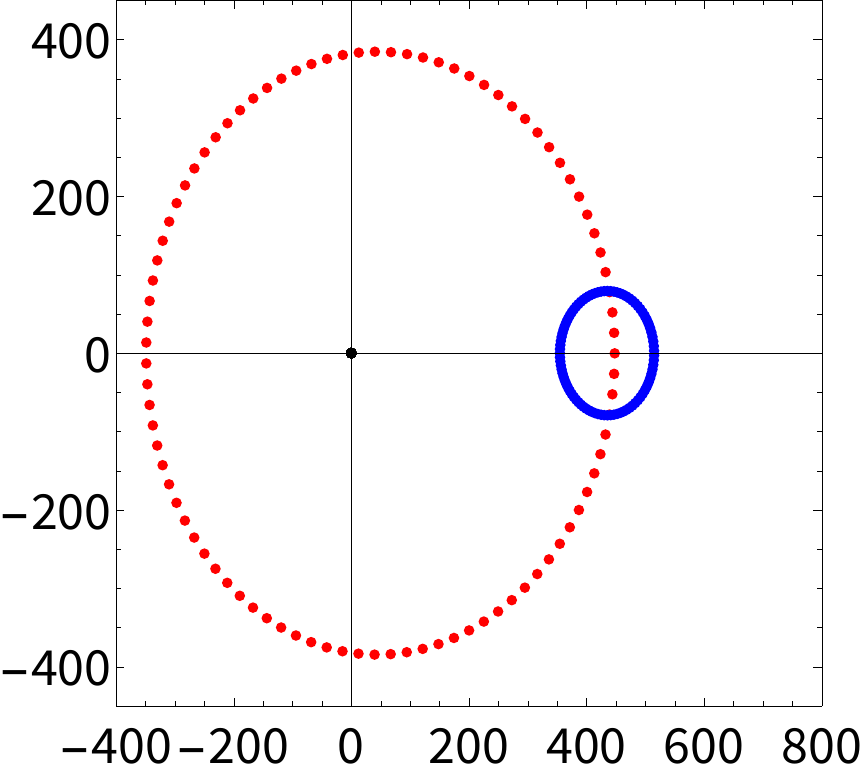}
    \begin{picture}(0,0)\vspace*{-1.2cm}
    \put(-110,90){\footnotesize $\Im{u}$}
    \put(10,35){\footnotesize $\Im{\sigma}=0.95$}
    \end{picture}
    \end{subfigure}
    \hfill
    \begin{subfigure}{0.325\textwidth}
    \centering
    \includegraphics[width=\linewidth]{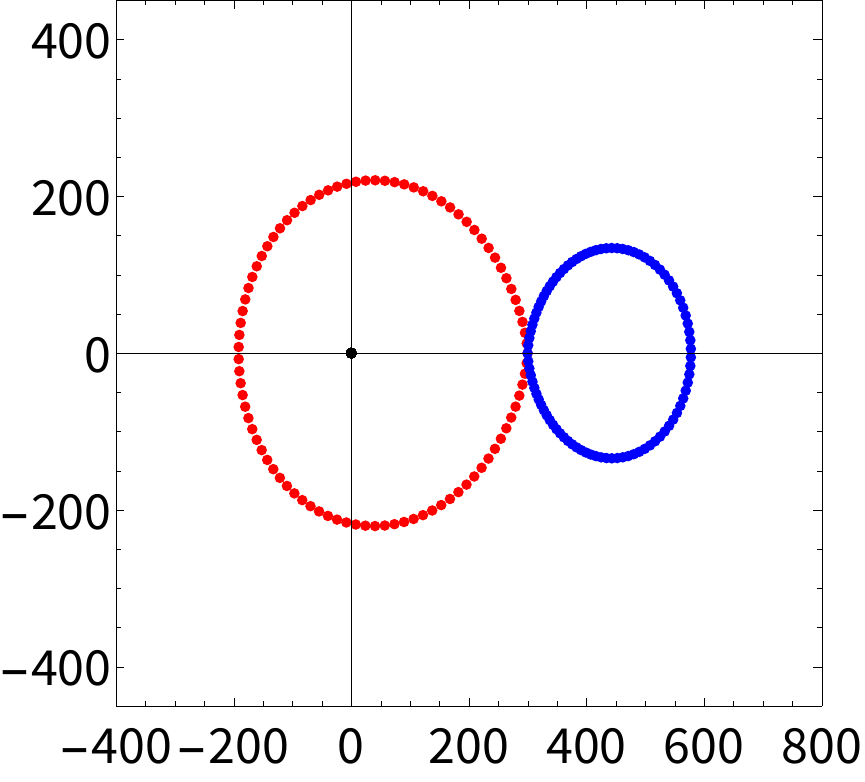}
    \begin{picture}(0,0)\vspace*{-1.2cm}
    \put(0,5){\footnotesize $\Re{u}$}
    \put(5,35){\footnotesize $\Im{\sigma}=\sqrt{3}/2$}
    \put(17,130){\vector(0,-1){30}}
    \put(-10,135){\footnotesize AD: $\sigma=e^{\pi i/3}$}
    \end{picture}
    \end{subfigure}
    \hfill
    \begin{subfigure}{0.325\textwidth}
    \centering
    \includegraphics[width=\linewidth]{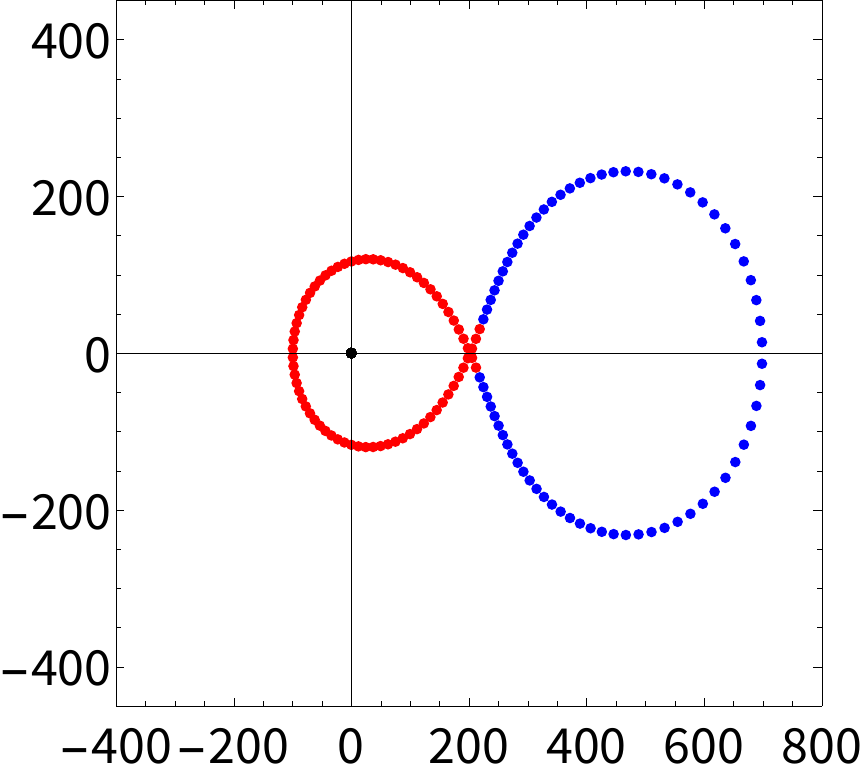}
    \begin{picture}(0,0)\vspace*{-1.2cm}
    \put(10,35){\footnotesize $\Im{\sigma}=0.78$}
    \end{picture}
    \end{subfigure}\vspace*{0.0cm}
    \caption{\textbf{``Phases'' of $T$-transformations of massless singularities:} The trajectories of the singularities in the massless limit of the E-string theory at various fixed $\Im{\sigma}$ as $\Re{\sigma}$ slowly varies from $-0.5$ to $0.5$ with each discrete step set to $\delta \sigma=1/91$. Black, blue, and red dots each indicate the trajectories of $u_{s,0}$, $u_{s,+}$, and $u_{s,-}$, respectively.
    }
    \label{fig:Tsingularity}
\end{figure}

To begin with, we have the $T$-transformation associated to $\sigma\to \sigma+1$. When we continuously perform this transformation in the moduli space of $\mathcal{C}$, the singularities exhibit a ``phase transition'' as we also vary $\Im{\sigma}$ as can be seen in Figure~\ref{fig:Tsingularity}. In particular, the transition occurs at $\Im{\sigma}=\sqrt{3}/2$ which is precisely the line where the special value $\sigma=e^{\pi i/3}$ lies. For this value of $\sigma$,  $E_4(\sigma=e^{\pi i/3})=0$ and the two I$_1$ singularities collide into a type-II singularity (also known as $H_0$ where the SCFT is the original Argyres--Douglas theory~\cite{Argyres:1995jj}).\footnote{The monodromy around this type-II singularity is $(ST)^{-1}$, which is the inverse of the monodromy around the type-II$^*$ (i.e.,~$E_8$) singularity. This is consistent with the monodromy at the cylindrical end $u=\infty$ being trivial, a generic feature of 6d SCFTs on $T^2$. The fact that this singularity arises when $E_4(\sigma)=0$ can be understood using the 8-dimensional F-theory model with $E_8\times E_8$ symmetry~\cite{Morrison:1996pp},
\begin{equation}
    y^2=x^3+\alpha x z^4+ z^5(1+\beta z+z^2)\, ,
\end{equation}
where the moduli of a D3 probe near one of the $E_8$ singularities at either $z=0$ or $z=\infty$ is the Coulomb branch of the E-string theory on $T^2$~\cite{Ganor:1996pc}. To have another higher order pole of the discriminant  necessarily requires that $\alpha=0$, and therefore $E_4=0$ for this family of elliptic curves. Then, with large $\beta$, we have an $H_0$ singularity at around $z\approx -1/\beta$ from two coalescing D7-branes. } 

When $\Im{\sigma}<\sqrt{3}/2$ but above $\frac{\sqrt 3}{6}$ (where an SL$(2,\Z)$ image of $e^{\pi i/3}$ lives), the singularities are swapped as we perform a $T$-transformation following the loop with fixed $\Im{\sigma}$. Below $\Im\sigma=\sqrt3/6$ they are not swapped at any fixed height, since the further images of $e^{\pi i/3}$ come in pairs. 

It is a general feature that after traversing along a generic loop in the parameter space (given by $\sigma$ and $m_i$'s in the present case), in general the singularities are non-trivially braided. Such braiding is an invariant of the homotopy class of the loop upon small deformations, but there can be ``wall-crossings'' once the loop sweeps past points in the parameter space that give rise to enhanced singularities. In the massless case, such a point is $\sigma=e^{\pi i/3}$, which causes the two I$_1$ singularities to collide.  

What is the significance of such a phase transition? One thing it reveals is that the contribution of a single singularity cannot be fully modular under the entire SL$(2,\Z)$ but instead has branched cuts. Therefore, full modularity of the partition function is expected to be a highly non-trivial property that requires delicate cooperation and cancellation among its various ingredients.

\subsubsection*{$S$-transformation}

Now, let us examine how these singularities transform under an $S$-transformation, namely~$\sigma\mapsto-1/\sigma$. As the Eisenstein series $E_{2n}$ each have modular weight $2n$, to keep the Seiberg--Witten curve in the massless limit invariant under an $S$-transformation, we need to rescale $(u,x,y)\mapsto (\sigma^{-6}u,\sigma^{-10}x,\sigma^{-15}y)$.\footnote{The necessity of re-scaling $u$ upon an $S$-transformation can be seen in $g$. Namely, consider an $S$-transformation which sends $\sigma\to-1/\sigma$. Then, we have
\begin{equation*}
        \frac{1}{216}E_6u^6-4u^5\to \frac{1}{216}\sigma^6E_6\tilde{u}^6-4\tilde{u}^5\, .
\end{equation*}
To remain invariant, $u$ must transform under the $S$-transformation as $u\to\tilde{u}=\sigma^{-6}u$ giving $u$ a nontrivial $-6$ modular weight.} This implies that $(u,x,y)$ have modular weights $(-6,-10,-15)$. As the spacetime torus remains unchanged under an $\SL(2,\Z)$-transformation, it is desirable to not have to scale $u$, but instead at most multiply it by a phase. This is because distances on the Coulomb branch are physical quantities and there should be a way of fixing the relative position of the singularities, while an overall rotation of the Coulomb branch is ``unphysical'' (although we will see later that this is actually related to the gravitational anomaly of the 6d theory). 

One method to rectify this re-scaling of physical quantities in the 4d theory under the $S$-transformation of the spacetime torus is to ensure that $\text{Area}(T^2)$ remains constant after the $\SL(2,\mathbb{Z})$ action by incorporating dimensionful quantities into the Seiberg--Witten curve. This alone will not be enough as one also needs factors that transform under $S$ with powers of $\sigma$ and are invariant under $T$, both up to a phase. Such invariance was realized in \cite{Ganor:1996pc} with coefficients that are non-holomorphic. We will achieve this in a slightly different way with the Dedekind eta function $\eta(\sigma)$, which indeed transforms as
\begin{equation}
    \eta(\sigma+1)=e^{2\pi i/24}\cdot\eta(\sigma)\,,\quad \eta(-1/\sigma)=\sqrt{-i\sigma}\cdot\eta(\sigma)\,.
\end{equation}
Using powers of $\eta$ to compensate for the modular weight of $E_4$ and $E_6$ with the scaling $(x,y,u)\mapsto (x/\eta^{20},y/\eta^{30},u/\eta^{12})$, we have
    \begin{equation}
    \label{eq:E8masslessDimensionful}
        y^2=4x^3-\frac{A_{T^2}^2}{12}\frac{E_4}{\eta^8}u^4x-\left(\frac{A_{T^2}^3}{216}\frac{E_6}{\eta^{12}}u^6-4A_{T^2}^{5/2}{}u^{5}\right)\, ,
    \end{equation}
    where we also restored the dependence on the area of the torus.
Under $T$, the coefficients will pick up phases
\begin{equation}
    \frac{E_4}{\eta^8}\mapsto e^{-2\pi i/3}\frac{E_4}{\eta^8}\,,\quad \frac{E_6}{\eta^{12}}\mapsto -\frac{E_6}{\eta^{12}}\,,
\end{equation}
while under $S$ the former is invariant but the latter will pick up a minus sign.
Therefore, with the normalization \eqref{eq:E8masslessDimensionful}, the Coulomb branch geometry becomes $\SL(2,\Z)$ invariant, up to an overall rotation. 

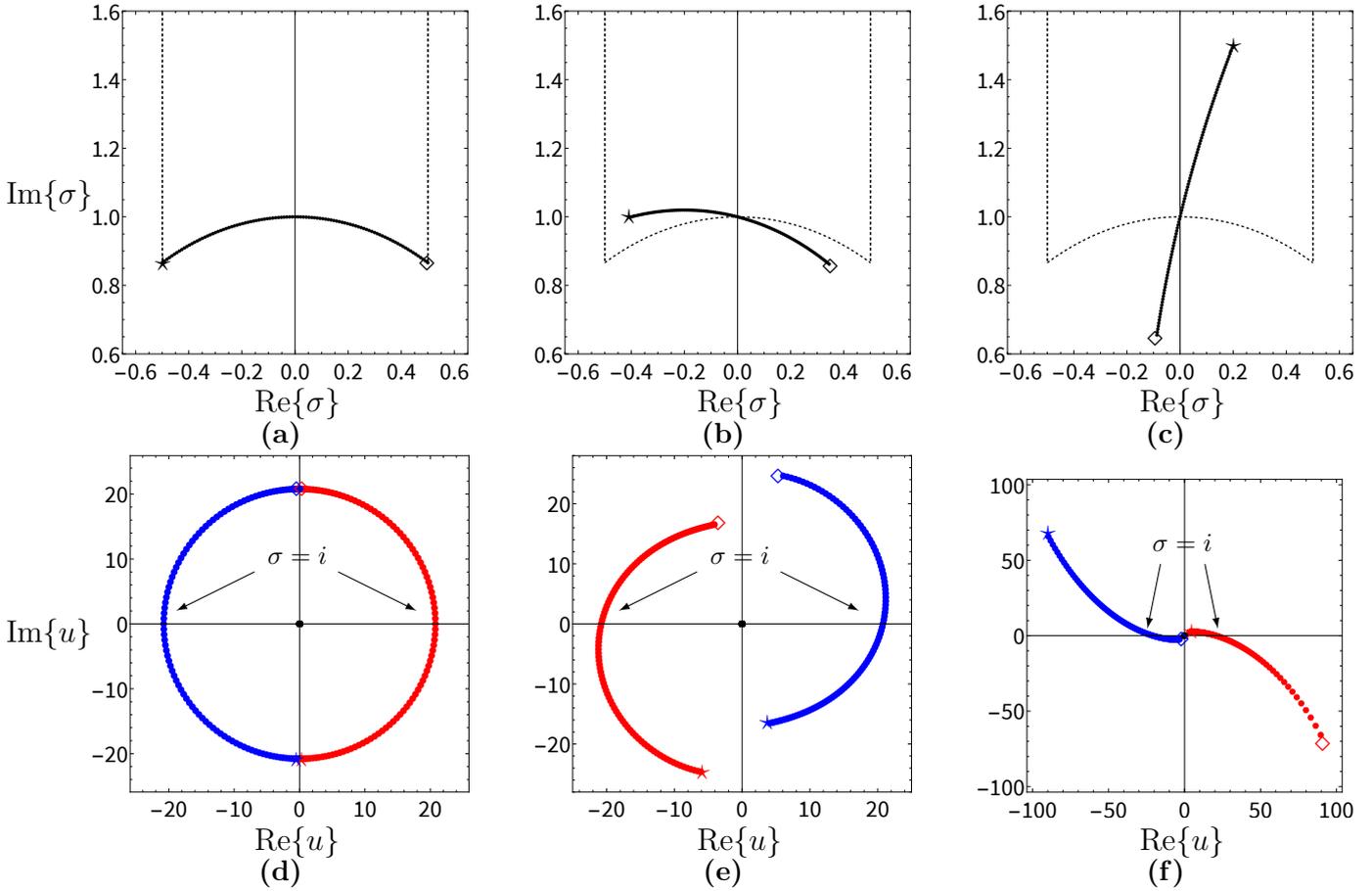
\begin{figure}[htb!]
    \centering
    \vskip 0pt
    \begin{subfigure}{0.3\textwidth}
    \centering
    \includegraphics[width=\linewidth]{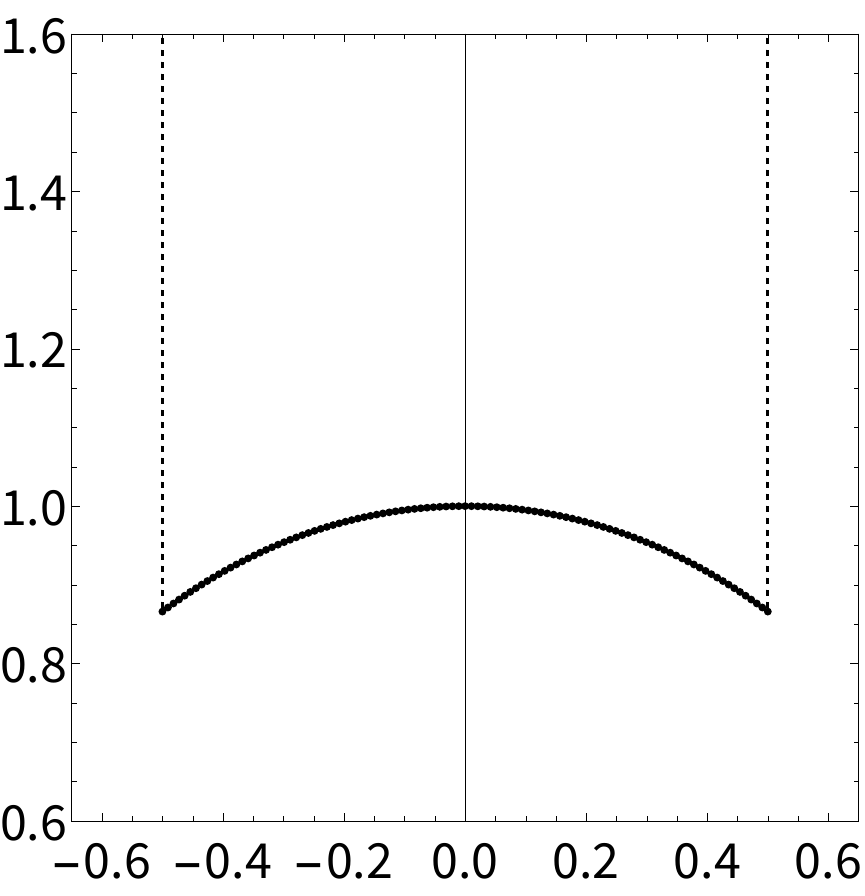}
    \begin{picture}(0,0)\vspace*{-1.2cm}
    \put(-10,5){$\Re{\sigma}$}
    \put(-105,85){$\Im{\sigma}$}
    \put(-48,59){$\star$}
    \put(53,59){$\diamond$}
    \end{picture}\vspace*{-0.3cm}
    \caption{ }
    \label{fig:S-transformation geodesic}
    \end{subfigure}
    \hfill
    \begin{subfigure}{0.3\textwidth}
    \centering
    \includegraphics[width=\linewidth]{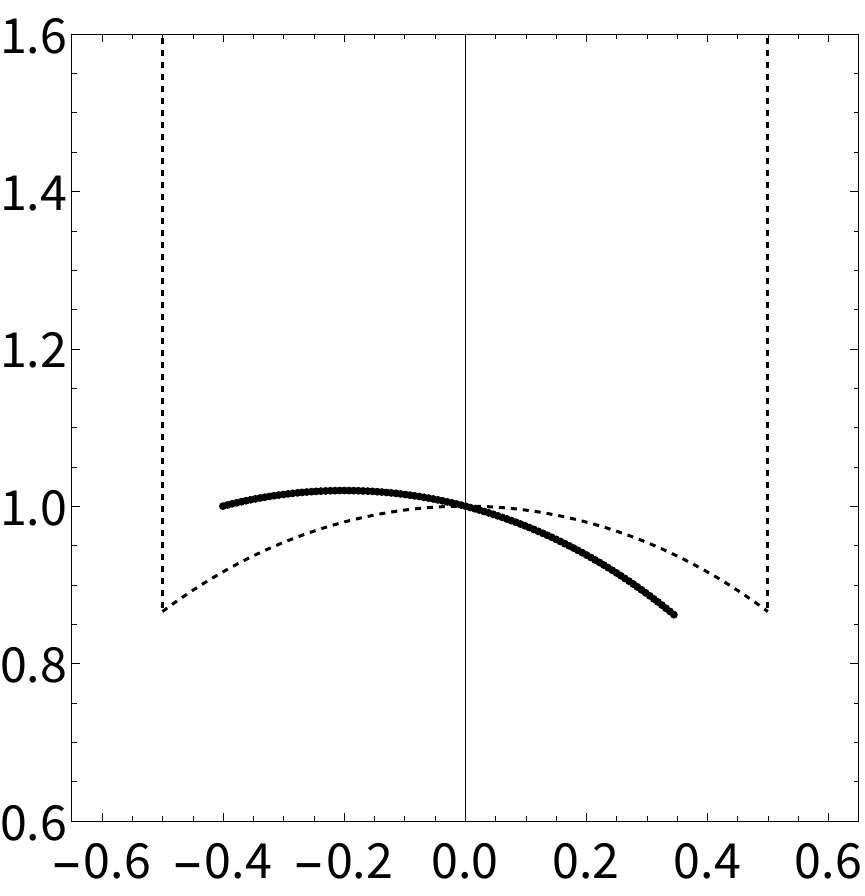}
    \begin{picture}(0,0)\vspace*{-1.2cm}
    \put(-10,5){$\Re{\sigma}$}
    \put(-39,77){$\star$}
    \put(38,58){$\diamond$}
    \end{picture}\vspace*{-0.3cm}
    \caption{ }
    \label{fig:S-transformation geodesic 2}
    \end{subfigure}
    \hfill
    \begin{subfigure}{0.3\textwidth}
    \centering
    \includegraphics[width=\linewidth]{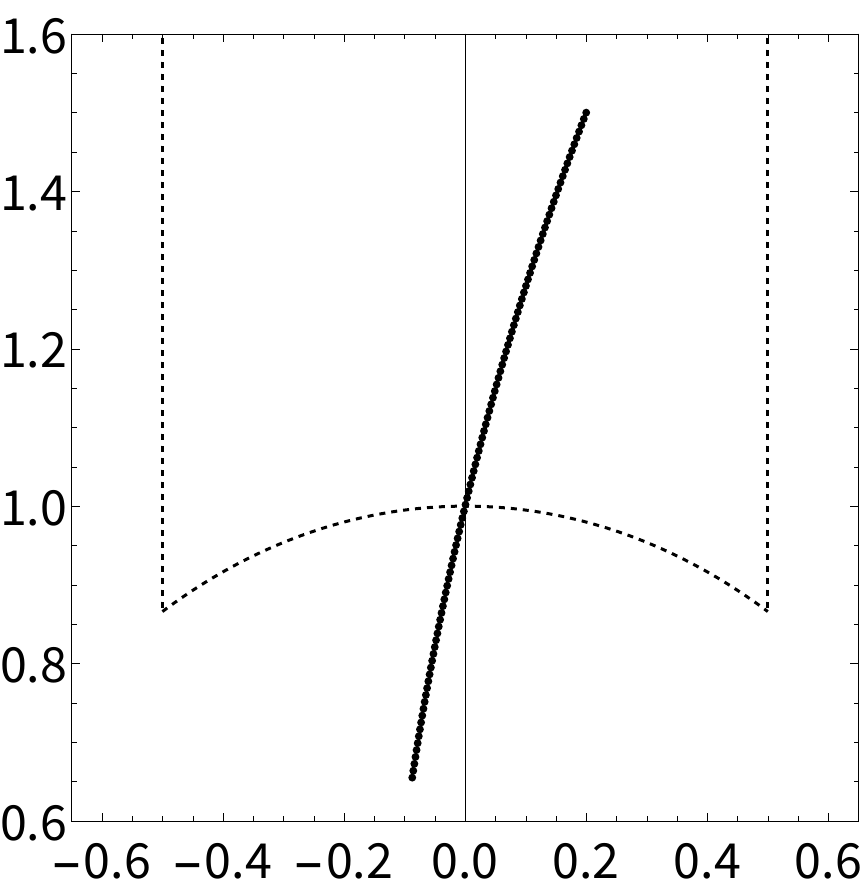}
    \begin{picture}(0,0)\vspace*{-1.2cm}
    \put(-10,5){$\Re{\sigma}$}
    \put(23,143){$\star$}
    \put(-7,30){$\diamond$}
    \end{picture}\vspace*{-0.3cm}
    \caption{ }
    \label{fig:S-transformation geodesic 3}
    \end{subfigure}
    \hfill
    \begin{subfigure}{0.3\textwidth}
    \centering
    \includegraphics[width=\linewidth]{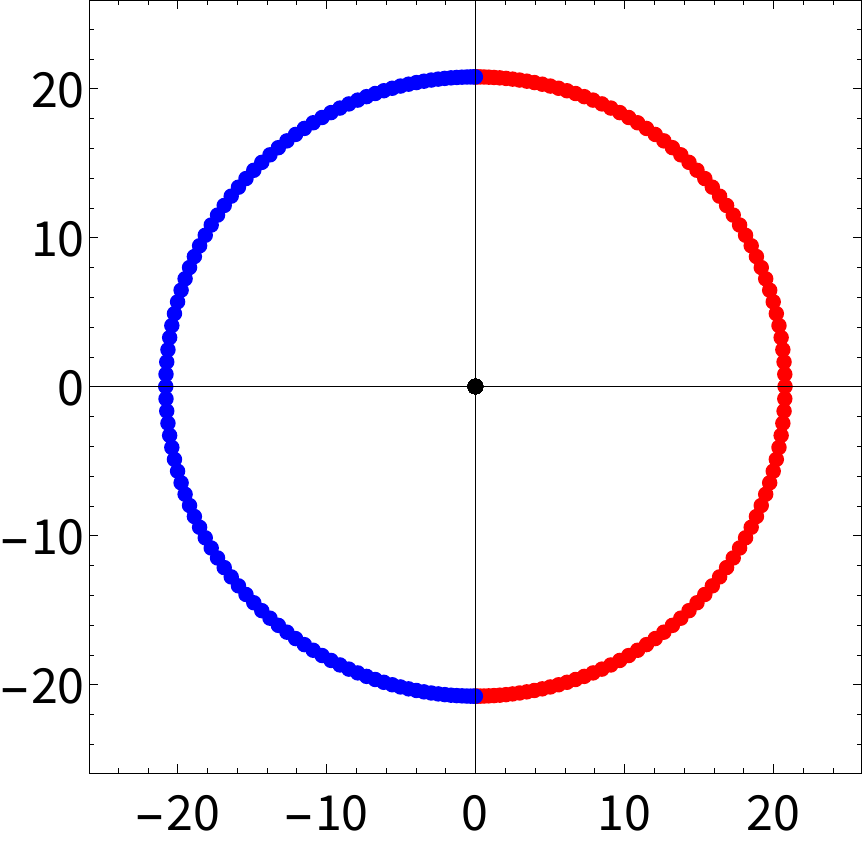}
    \begin{picture}(0,0)\vspace*{-1.2cm}
    \put(-105,85){$\Im{u}$}
    \put(-10,5){$\Re{u}$}
    \put(5,37){\textcolor{red}{$\star$}}
    \put(3,37){\textcolor{blue}{$\star$}}
    \put(5,141){\textcolor{red}{$\diamond$}}
    \put(3,141){\textcolor{blue}{$\diamond$}}
    \put(22,112){\vector(2,-1){30}}
    \put(-10,112){\vector(-2,-1){30}}
    \put(-5,115){\footnotesize $\sigma=i$}
    \end{picture}\vspace*{-0.3cm}
    \caption{ }
    \label{fig:S-transformation singularities trajectory}
    \end{subfigure}
    \hfill
    \begin{subfigure}{0.3\textwidth}
    \centering
    \includegraphics[width=\linewidth]{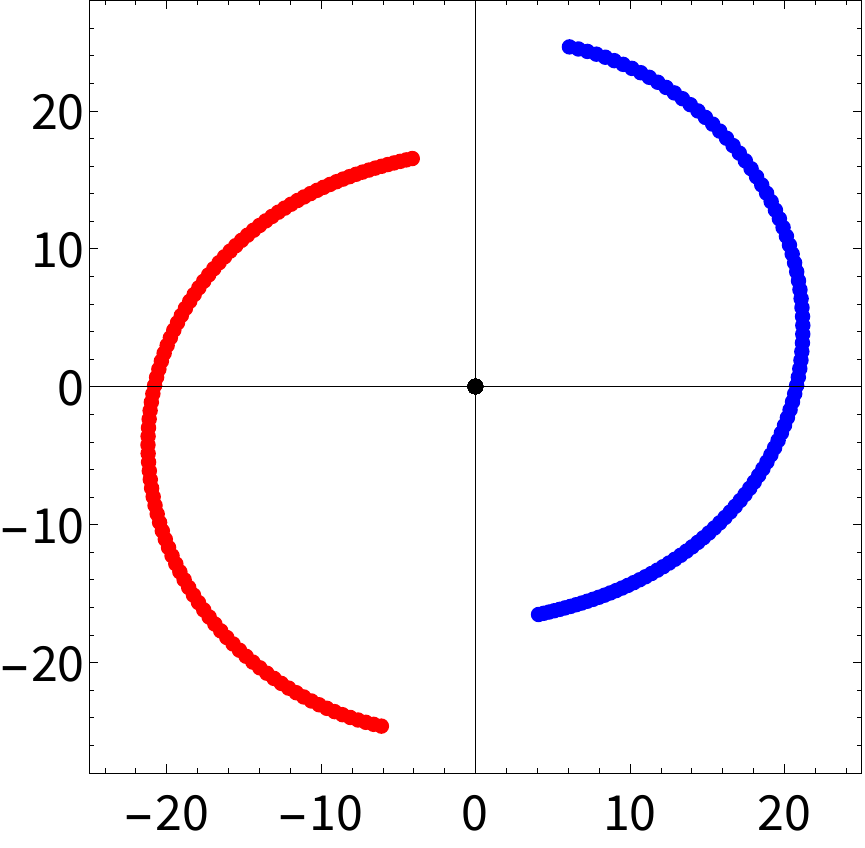}
    \begin{picture}(0,0)\vspace*{-1.2cm}
    \put(-10,5){$\Re{u}$}
    \put(14,51){\textcolor{blue}{$\star$}}
    \put(-5,128){\textcolor{red}{$\diamond$}}
    \put(18,146){\textcolor{blue}{$\diamond$}}
    \put(-11,32){\textcolor{red}{$\star$}}
    \put(22,112){\vector(2,-1){30}}
    \put(-10,112){\vector(-2,-1){30}}
    \put(-5,115){\footnotesize $\sigma=i$}
    \end{picture}\vspace*{-0.3cm}
    \caption{ }
    \label{fig:S-transformation singularities trajectory 2}
    \end{subfigure}
    \hfill
    \begin{subfigure}{0.3\textwidth}
    \centering
    \includegraphics[width=\linewidth]{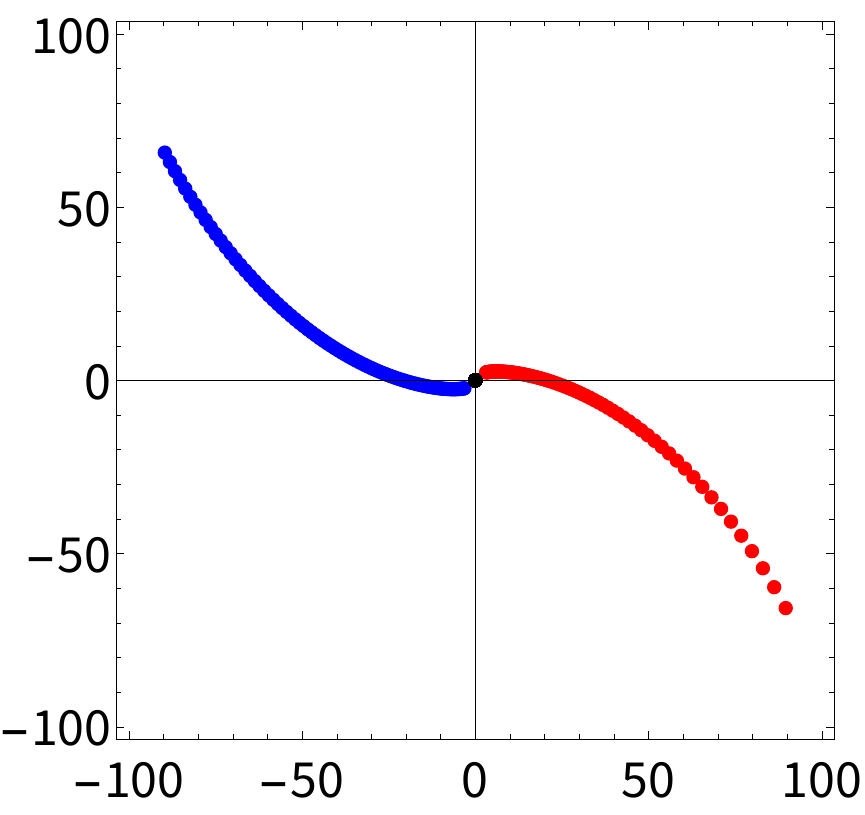}
    \begin{picture}(0,0)\vspace*{-1.2cm}
    \put(-10,5){$\Re{u}$}
    \put(-48,124){\textcolor{blue}{$\star$}}
    \put(57,43){\textcolor{red}{$\diamond$}}
    \put(7,86){\textcolor{red}{$\star$}}
    \put(3,83){\textcolor{blue}{$\diamond$}}
    \put(15,115){\vector(1,-5){5}}
    \put(-2,115){\vector(-1,-5){5}}
    \put(-5,120){\footnotesize $\sigma=i$}
    \end{picture}\vspace*{-0.3cm}
    \caption{ }
    \label{fig:S-transformation singularities trajectory 3}
    \end{subfigure}\vspace*{-0.3cm}
    \caption{\textbf{$S$-transformations of singularities:} The trajectories of the singularities in the massless limit of the E-string theory under an $S$-transformation as we slowly vary along the geodesic connecting the two $S$-dual end points. In the top figures~\ref{fig:S-transformation geodesic},~\ref{fig:S-transformation geodesic 2}, and~\ref{fig:S-transformation geodesic 3}, the $\star$ and $\diamond$ indicate the beginning and end points of the trajectory, respectively. In particular, from top left to top right, we have $\sigma_{\star}=-0.5+\sqrt{3}i/2,-0.4+i,0.2+1.5i$ and $\sigma_{\diamond}=-1/\sigma_{\star}\approx 0.5+\sqrt{3}i/2,0.34+0.86i,-0.087 + 0.66i$, respectively. The solid black line indicates the geodesic path connecting the two points and the dotted lines indicate the fundamental domain for $\SL(2,\mathbb{Z})$. In Figures~\ref{fig:S-transformation singularities trajectory},~\ref{fig:S-transformation singularities trajectory 2}, and~\ref{fig:S-transformation singularities trajectory 3}, the black, blue, and red dots each indicate the trajectories of $u_{s,0}$ (which in fact stays at the origin), $u_{s,+}$, and $u_{s,-}$ where the end points of the trajectory are indicated by the corresponding shape and color. In particular, the two I$_1$ singularities collide to form a type II singularity at $\sigma_{\star}$ and $\sigma_{\diamond}$ in~\ref{fig:S-transformation singularities trajectory}, thus overlapping the blue and red dotted lines in this figure at the end points. In each of the figures in the second row, we marked the locations of the two I$_1$ singularities for the special value of $\sigma=i$. 
    }
    \label{fig:Ssingularity}
\end{figure}

The locations of the singularities are
    \begin{equation}
    \label{eq:modular function singularities}
        \begin{cases}
            u'_{s,0}=0\, ,&\text{Type II}^*\, ,\\
            u'_{s,\pm}=A_{T^2}^{-1/2}\eta^{12}u_{s,\pm}\, ,&\text{Type I}_1\, ,
        \end{cases}
    \end{equation}
where $u_{s,\pm}=-\frac{864(E_6\pm E_4^{3/2})}{E_4^3-E_6^2}$ is the value before the normalization. As can be observed in Figure~\ref{fig:Ssingularity}, the above singularities are, up to a $\pi$-rotation, invariant under an $S$-transformation. Indeed, using \eqref{eq:modular function singularities}, we have
\begin{equation}
    u'_{s,\pm}(-1/\sigma)=-u'_{s,\mp}(\sigma)\, .
\end{equation}
Therefore, one expects the two I$_1$ singularities to exchange values after the $S$-action (up to a sign). Note, however, this is again a path-dependent statement like what we have seen for the $T$-transformation. For a different homotopy class of trajectories on $\mathbb{H}$ with the zeroes of $E_4$ removed, the statement can be different.

The fact that there is no normalization for the Seiberg--Witten curve such that it is fully modular invariant might seem puzzling. However, this is actually a manifestation of the gravitational anomaly of the theory. On a geometry of the kind $ M_4\times T^2$, the anomaly can be seen in lower dimensions in multiple ways. One is by first compactifying on $T^2$, and instead of having a 4d theory that is strictly SL$(2,\Z)$ invariant, non-trivial background topological terms of the form 
\begin{equation}
    a(\sigma)\int \mathcal{R}\^ \mathcal{R} + b(\sigma) \int \tr F_R \^ F_R\,,
\end{equation}
involving the Ricci curvature and the curvature for the background $R$-symmetry can be generated. This is related to the fermions in the KK-tower, whose masses depend on the shape of the torus in a chiral fashion and can be shifted or permuted upon the action of SL$(2,\Z)$, leading to a background term. (See~\cite{GHP2} for more details.) Therefore, the partition function is not expected to be fully modular (i.e.~a modular function), but instead turns out to be a modular form. 

To see that it should be a modular form, it is convenient to first compactify on the four-manifold $M_4$ to obtain a 2d $(0,1)$ theory, $\CT[M_4]$. This 2d theory will have a gravitational anomaly valued in the dual of $\Omega_4^{\text{Spin}}\simeq\Z$, which can be understood as the mismatch of the left- and right-moving central charges $2(c_R-c_L)$ after flowing to an IR SCFT point. Such an anomaly causes the partition function to acquire a phase under the $T$-action. This phase can be normalized by a Dedekind eta function, which makes the elliptic genus a modular form so that it is invariant under $T$ but picks up a pre-factor under the $S$-transformation. We will discuss this and related issues in more detail in Section~\ref{sec:Anomalies}. Before that, we first briefly discuss the topological twist of the E-string theory, with the emphasis on explaining why the theory still makes sense on a non-spin manifold.

\subsection{Topological twist}\label{sec:topological twist}

To preserve supersymmetry on $ M_4 \times T^2$ for a general smooth, oriented and closed $M_4$, one can choose a non-trivial background of R-symmetry on $M_4$ via a process known as (partial) ``topological twist''~\cite{Witten:1988ze}.

In the M-theory construction (cf.~Table~\ref{table:The E-string theory's M-theory's brane configuration}), the surviving supercharges on the M5-brane satisfy
\begin{equation}
    \text{M9}\,:\, -i\Gamma^{0123456789}\epsilon=\Gamma^{(10)}\epsilon=\epsilon\,,\qquad \text{M5}\,:\,i\Gamma^{012345}\epsilon=\epsilon\,.
\end{equation}
Here, $\Gamma^{\mu}$ denotes the 32-by-32 $\Gamma$ matrices in eleven dimensions, $\Gamma^{\mu_1\dots \mu_k}=\Gamma^{[\mu_1}\Gamma^{\mu_2}\dots\Gamma^{\mu_k]}$, and $\epsilon$ denotes the 32-component spinor.\footnote{We are implicitly working in Euclidean signature, where a convenient way to construct eleven-dimensional $\Gamma$ matrices is to use the 2-by-2 Pauli matrices
\begin{align*}
    \Gamma^0=\sigma_1\otimes\sigma_0\otimes\sigma_0\otimes\sigma_0\otimes\sigma_0&\,,\qquad 
    \Gamma^1=\sigma_2\otimes\sigma_0\otimes\sigma_0\otimes\sigma_0\otimes\sigma_0\,,\\
    \Gamma^2=\sigma_3\otimes\sigma_1\otimes\sigma_0\otimes\sigma_0\otimes\sigma_0&\,,\qquad 
    \Gamma^3=\sigma_3\otimes\sigma_2\otimes\sigma_0\otimes\sigma_0\otimes\sigma_0\,,\\
    \Gamma^4=\sigma_3\otimes\sigma_3\otimes\sigma_1\otimes\sigma_0\otimes\sigma_0&\,,\qquad 
    \Gamma^5=\sigma_3\otimes\sigma_3\otimes\sigma_2\otimes\sigma_0\otimes\sigma_0\,,\\
    \Gamma^6=\sigma_3\otimes\sigma_3\otimes\sigma_3\otimes\sigma_1\otimes\sigma_0&\,,\qquad 
    \Gamma^7=\sigma_3\otimes\sigma_3\otimes\sigma_3\otimes\sigma_2\otimes\sigma_0\,,\\
    \Gamma^8=\sigma_3\otimes\sigma_3\otimes\sigma_3\otimes\sigma_3\otimes\sigma_1&\,,\qquad 
    \Gamma^9=\sigma_3\otimes\sigma_3\otimes\sigma_3\otimes\sigma_3\otimes\sigma_2\,,\\
    \Gamma^{(10)}=\sigma_3\otimes\sigma_3\otimes\sigma_3\otimes\sigma_3\otimes\sigma_3&=-i\Gamma^{0123456789}\,,
\end{align*} and $[\mu_1\mu_2\dots\mu_n]$ denotes the anti-symmetrization of the indices. In particular, we will have
\begin{equation}
    \Gamma^{\mu_1\mu_2\dots\mu_n}=\frac{1}{n!}\delta_{\nu_1\nu_2\dots\nu_n}^{\mu_1\mu_2\dots\mu_n}\Gamma^{\nu_1\nu_2\dots\nu_n}\,.
\end{equation}} 
Thus, the M5/M9-brane configuration breaks $\frac{3}{4}$ of the supersymmetry present in M-theory and preserves 8 supercharges on the worldvolume theory of the M5-brane. Examining the symmetry present in this configuration, we also have a Spin$(4)\cong SU(2)_R\times SU(2)_F$ symmetry that rotates $(x^6,x^7,x^8,x^9)$. The $SU(2)_R$ is generated by $\Sigma^{(+)}_a=\frac14\sum_{i<j}(\eta^{(+)}_{ij})_a\Gamma^{ij}$ while $SU(2)_F$ is generated by $\Sigma^{(-)}_a=\frac14\sum_{i<j}(\eta^{(-)}_{ij})_a\Gamma^{ij}$ where $i,j=6,7,8,9$ and $a=1,2,3$. Here, we have used the standard 't Hooft symbol $(\eta^{(\pm)}_{ij})_a=\epsilon_{aij9}\pm \delta_{ai}\delta_{j9}\mp\delta_{aj}\delta_{i9}$ such that $\frac12\sum_{i<j}(\eta_{ij}^{(\pm)})_a(\eta_{ij}^{(\pm)})_b=\delta_{ab}$ and $[(\Sigma^{\pm})_a,(\Sigma^{\pm})_b]=-\epsilon_{abc}(\Sigma^{\pm})_c$ with $[(\Sigma^{+})_a,(\Sigma^{-})_b]=0$. Now we can identify the positive chirality subspace, where the supercharges of interest live, with the image of the projector $P_+=\frac14\big(1+\Gamma^{(10)}\big)\big(1+i\Gamma^{012345}\big)$, where $P_+\Sigma^{(+)}P_+=\frac{i}{2}\tau_a$ and $P_+\Sigma_a^{(-)}P_+=0$. With this, we can identify $SU(2)_R$ as the $R$-symmetry of the superconformal algebra that rotates the charges $\Sigma^{(+)}_a\epsilon=\frac{i}{2}\tau_a\epsilon$ while $SU(2)_F$ as the flavor symmetry present at higher-rank E-string theories acting trivially on the supercharge $\Sigma^{(-)}_a\epsilon=0$.
In the rank-1 case, $SU(2)_F$ only acts on a ``center-of-mass hypermultiplet'' that is decoupled from the E-string theory.\footnote{One can choose to restore this hypermultiplet, enlarging the Higgs branch into $\C^2\times \mathbb{O}_{\text{min}}^{E_8}$ where $\C^2$ parametrizes the location of the instanton. The $SU(2)_R\times SU(2)_F$ symmetry acts on this $\C^2$ factor via the quotient map to $SO(4)$.
} 

Splitting the world-volume $\R^6$ into $\R^4\times \R^2$, the system has the following global symmetries
\begin{equation}
    \text{Spin}(4)_E\times U(1)_E\times SU(2)_R\times E_8\cong SU(2)_+\times SU(2)_-\times U(1)_E \times SU(2)_R\times E_8\,,
\end{equation}
where $\text{Spin}(4)_E\times U(1)_E\subset \text{Spin}(6)_E$ consists of spacetime symmetries. Upon compactifying $\R^2$ to $T^2$, the system still preserves all 8 supercharges, while $U(1)_E$ will be replaced by the (orientation-preserving) isometry group of $T^2$, which, for generic $\sigma$, is the product of two $U(1)$ symmetries. From the point of view of the effective 4d theory $\CT[T^2]$, they act on the tower of KK-modes.

On $T^2$, one can consider the following different types of backgrounds:
\begin{itemize}
    \item Size $A_{T^2}$ of the torus. The size $A_{T^2}$ is the only scale in the system, and all dimensionful quantities in the 4d theory, such as masses of BPS particles, geodesic distances between points on the moduli space, etc., all scale with $A_{T^2}$. However, the supersymmetric partition function $Z_{\CT}[M_4\times T^2]$ is an elliptic genus from the 2d point of view, which cannot depend on $A_{T^2}$.
    \item The modulus $\sigma$. The ratios and phases of the masses of the KK tower will depend on $\sigma$. In the limit $u\rightarrow \infty$, the 4d theory at low energy approaches a $U(1)$ gauge theory with coupling constant $\sigma$. From the principle of holomorphy, one expects that the Seiberg--Witten geometry, as well as the partition function, depends on $\sigma$ in a holomorphic fashion. The mapping class group $\SL(2,\Z)$ acts on $\sigma$, leading to dualities of the compactified theory. Due to the gravitational anomaly, the partition function actually picks up phases under the $\SL(2,\Z)$, which we will analyze in Section~\ref{sec:Anomalies}.
    \item A flux of $E_8$. This background is labeled by a dominant coweight of $E_8$. This breaks $E_8$ into a subgroup, which is classified by subdiagrams of the Dynkin diagram of $E_8$. Although this provides an interesting discrete family where each member is $\SL(2,\Z)$ invariant, we will not consider these in the present work. 
    \item Holonomies of $E_8$. This also breaks $E_8$ into a subgroup. In addition, this breaks the $\SL(2,\Z)$ symmetry for $\sigma$, which can be restored by combining the holonomies along the two cycles into a Jacobi variable that also transforms under SL$(2,\Z)$.  
    \item Spin structures. Naively, only the periodic-periodic spin structure can preserve supersymmetry and should be the one that we use. However, for any 6d $(1,0)$ theory, one can compensate for an anti-periodic boundary condition with a central holonomy in $SU(2)_R$ so that the supercharges are still periodic. This combination of $(-1)^F$ and the center $\Z_2$ of $SU(2)_R$ is referred to as $\Z_2^U$ in \cite{GHP2}, whose action on E-string theory is expected to be trivial for $Q=1$ but non-trivial for $Q>1$. Therefore, there should be four different 4d theories labeled by the spin structures in general, but they are expected to be identical for $Q=1$. We will see later that the E-string theory can be made independent of the spin structure on the entire $ M_4\times T^2$. 
\end{itemize}

The 8 supercharges $Q^i_{\alpha}$ and $\widetilde{Q}_{\dot{\alpha}j}$ transform under the symmetry $SU(2)_+\times SU(2)_-\times SU(2)_R\times U(1)_E$ as $(\mathbf{2},\mathbf{1},\mathbf{2})^{+1/2}$ and $(\mathbf{1},\mathbf{2},\mathbf{2})^{-1/2}$ respectively.\footnote{The numbers in the parentheses denote the $SU(2)_+\times SU(2)_-\times SU(2)_R$ representation, whose subspaces are indexed by $\alpha,\dot{\alpha},i=1,2$, while the superscript denotes (one half of) the $U(1)_E$ charge.} The supersymmetry algebra is then
\begin{equation}
    \{Q^i_{\alpha},\widetilde{Q}_{\dot{\alpha}j}\}=2\delta^{i}_{~j}\sigma^{\mu}_{\alpha\dot{\alpha}}P_{\mu}\,,\quad \{Q^i_{\alpha},Q^j_{\beta}\}=0\,,\quad \{\widetilde{Q}_{\dot{\alpha}i},\widetilde{Q}_{\dot{\beta}j}\}=0\,,
\end{equation}
where $\sigma^{\mu}=(1,\vec{\sigma})$ and $P_{\mu}$ is the energy-momentum operator.

In the Donaldson--Witten twist \cite{Witten:1988ze}, one replaces $SU(2)_+\subset \text{Spin}(4)_E$ with the diagonal subgroup $SU(2)_{\rm diag}:=\text{diag}(SU(2)_+\times SU(2)_R)\subset SU(2)_+\times SU(2)_R$, and the resulting group $\text{Spin}(4)'_E:=SU(2)_{\rm diag}\times SU(2)_-$ now consists of twisted spacetime symmetries. Upon such a twist, a linear combination of supercharges, 
\begin{equation}
    \mathcal{Q}=\delta^{\alpha j}Q_{\alpha j}\,,
\end{equation}
becomes a 4d scalar that transforms under $SU(2)_{\rm diag}\times SU(2)_-\times U(1)_E$ as $(\mathbf{1},\mathbf{1})^{+1/2}$. 

This procedure can be performed at the level of the 6d theory as a partial twist, then $\mathcal{Q}$ will be a spinor in the remaining two directions, generating a $(0,1)$ supersymmetry. The reader is referred to \cite{Gukov:2018iiq} for more details. From the point of view of M-theory, the partial twist can be understood as wrapping the M5-brane on $M_4$ as a Cayley submanifold  of a local Spin(7) space (which can be taken to live in the $x^{2-9}$ directions in Table~\ref{table:The E-string theory's M-theory's brane configuration}). The normal bundle, occupying the $x^{6,7,8,9}$ directions, is identified with a spinor bundle on $M_4$. This raises the question of how to deal with non-spin 4-manifolds, and when $M_4$ is spin, what the dependence on the spin structure is. 

\subsubsection{Twisting on non-spin manifolds}

In a Lagrangian theory, to make sure that the topological twist makes sense on general $M_4$, one only has to find a way to make the fields well-defined, even without a spin structure. As the E-string theory is non-Lagrangian, this is not the right procedure, but it nonetheless serves as a starting point of our discussion. 

A non-spin manifold can be viewed as one with ``non-trivial 't Hooft flux'' for both $SU(2)_\pm$ symmetries described by a $\Z_2$-valued 2-cocycle, which obstructs the existence of the spinor bundles $S^+$ and $S^-$. This is not a problem for the supercharges, as they also transform under $SU(2)_R$, and, upon identifying $SU(2)_+$ and $SU(2)_R$, all of the supercharges become well defined, giving rise to a scalar, a self-dual 2-form and a vector. 

The same is true for the 6d tensor and vector multiplet, whose fermions transform in the same way as the supercharges. On the other hand, one should be more concerned with the hypermultiplet, whose fermions won't transform under $SU(2)_R$ and whose bosons transform under $SU(2)_R$ but not under $SU(2)_+$. This might lead one to conclude that any theory with a Higgs branch cannot be twisted unless one cancels the obstruction in some other ways (e.g.,~with flux in global or gauge symmetries).

However, this is not necessarily true for strongly coupled theories. In fact, for the E-string theory, the center of $SU(2)_R$ acts trivially on the Higgs branch \cite{kronheimer1990instantons}, allowing it to fiber consistently over $M_4$. The well-definedness of the massless bosons in this way ensures the well-definedness of the fermions due to the existence of the scalar supercharge. Encouraged by this, one can conjecture that the action of the $\Z_2$ center of $SU(2)_R$ is always the same as the diagonal $\Z_2$ of the center of $SU(2)_+\times SU(2)_-$ on the full theory, which ensures that the theory is well defined on a non-spin $M_4$. This will be tested later when we show that the partition function on non-spin manifolds also possesses the expected properties, such as being integral modular forms.

One can also have a similar statement for general rank $Q>1$, namely, the diagonal $\Z_2\subset SU(2)_+\times SU(2)_-\times SU(2)_R\times SU(2)_F$ acts trivially on the theory. In other words, the true symmetry of the theory is $(SU(2)_+\times SU(2)_-\times SU(2)_R\times SU(2)_F)/\Z_2$. 

This can be justified by observing that, in the M-theory construction for the E-string theory, this diagonal $\Z_2$ is given by a 2$\pi$-rotation on the worldvolume of M5-branes composed with one in the orthogonal directions. Since a $2\pi$-rotation just amounts to $(-1)^F$, they should just cancel each other. In fact, if one also includes $\Z_2^{\mathcal{F}}$ generated by $(-1)^F$, the true symmetry is expected to be $(SU(2)_+\times SU(2)_-\times SU(2)_R\times SU(2)_F\times \Z_2^{\mathcal{F}})/(\Z_2\times \Z_2)$, where the two $\Z_2$'s are respectively the diagonal $\Z_2$ of $SU(2)_+\times SU(2)_-\times \Z_2^{\mathcal{F}}$ and of $SU(2)_R\times SU(2)_F\times \Z_2^{\mathcal{F}}$.

As a slightly trivial check of the proposal, if one restores the $SU(2)_F$ global symmetry in the $Q=1$ case by adding back the decoupled free hypermultiplet, the free hypermultiplet is indeed compatible with this conjecture, as bosons transform as $(++--+)$ and fermions transform as $(-++--)$ or $(+-+--)$ under the five $\Z_2^+\times \Z_2^-\times \Z_2^R\times\Z_2^F\times \Z_2^{\mathcal{F}}$, with both diagonal combinations acting trivially. 

For the 6d theory, the $SU(2)_+\times SU(2)_-\subset \text{Spin}(6)_E$ is a subgroup and the 2$\pi$-rotation is the subgroup $\Z_2\subset\Z_4$ of the center. Then the statement is that the true symmetry is $(\text{Spin}(6)_E\times SU(2)_R\times SU(2)_F\times \Z_2^{\mathcal{F}})/(\Z_2\times \Z_2)$, which becomes $(\text{Spin}(6)_E \times SU(2)_R\times \Z_2^{\mathcal{F}})/(\Z_2\times\Z_2)$ in the $Q=1$ case.

\subsubsection{Spin-$SU(2)$ theories from the E-string theory}

One potential confusion of the statement that the E-string theory can be defined on general 4-manifolds is about what happens for some of the low-energy effective theories that have weak-coupling descriptions. It is necessary that they are also well-defined, although the mechanism can be different for different low-energy theories. 

For a family of theories that can be obtained from the 5d $D_8$ theory, which admit a Lagrangian description as a $Sp(1)=SU(2)$ gauge theory with 8 quark multiplets, it is expected that, on a general background allowing $w_2$ of the tangent bundle, one actually has a spin-$SU(2)$ theory.\footnote{Even on the flat space, there is a physical difference between this theory and the $SU(2)$ theory, which can be detected via the spin and statistics of extended operators. For example, the fundamental Wilson line should be fermionic in the Spin-$SU(2)$ theory but bosonic in the usual $SU(2)$ theory. This property is protected by RG and can be checked against the UV computation of the E-string elliptic genus. For example, the minus sign in \cite[eq.~3.19]{Kim:2014dza} is compatible with the expectation that, for winding number being 1 and arbitrary KK-momentum, the E-string produces a fermionic line operator in the 5d theory.} In other words, the flux for $SU(2)_\pm$ is canceled by an 't Hooft flux of the gauge $SU(2)$. If one then deforms its Seiberg--Witten geometry by a generic mass, all the 8 quark singularities are naturally expected to have spin$^c$ gauge groups. As the two monopole and two dyon singularities also have spin$^c$ gauge groups, all the 12 I$_1$ singularities have spin$^c$ gauge groups satisfying $2c_1(\text{gauge})\equiv w_2(T)\pmod2$.

On the other hand, if one deforms the theory onto the $E_8$ Higgs branch, at low energy one has 29 free hypermultiplets. The emergent $SU(2)'_R\times SU(2)_F'\times E_7$ should be such that their fluxes cancel, making the hypermultiplet well defined on general $M_4$. 

One expects a similar phenomenon also for higher-rank E-string theories, whose compactification on an $S^1$ with a ``$-1$-holonomy'' should lead to, at low energy,\footnote{Notice that this is a genuine low-energy limit, and the 6d theory is expected to actually have KK-modes that are not captured by the 5d gauge theory. To see this, one can move on to the Higgs branch of the 6d theory (along a direction not lifted by the ``$-1$'' holonomy) and one sees that some KK modes of the 6d free hypermultiplet cannot be present in the 5d gauge theory. This is in contrast with the case of the 6d $(2,0)$ theory, all of whose KK-modes on $S^1$ are expected to be present in the 5d theory as modes with non-trivial instanton backgrounds (see \cite{Kim:2011mv} and references therein).} a Spin-$Sp(Q)$ gauge theory. This can be understood from the ADHM construction of the moduli space of $Q$ $D_8$ instantons as the Higgs branch of an $\CN=4$ supersymmetric quantum mechanics with an $Sp(Q)$ vector multiplet, a $Sp(Q)-SO(16)$ bifundamental hypermultiplet $\phi$, and a hypermultiplet $M$ in the antisymmetric representation of $Sp(Q)$. Notice that the diagonal $\Z_2$ of the center of $SU(2)_R\times SU(2)_F$ acts indeed as $(-1)^F$ after the quotient (i.e.~it acts trivially on the bosonic zero modes), thanks to the action of the center of $Sp(Q)$ on $\phi$. Since the Higgs branch can be consistently fibered over $M_4$ with some flux for the $SU(2)_F$, one must be able to cancel the 't Hooft flux of the $SU(2)_R$, which is in turn determined by that of the $SU(2)_+$ via the topological twist. For $M$, this is possible by turning on a flux background for the $SU(2)_F$ part, but for $\phi$, it has to be an 't Hooft flux of the gauge $Sp(Q)$ given by the $w_2$ of the tangent bundle. Therefore, the theory is a Spin-$Sp(Q)$ theory.

\subsubsection{Spin structures}

We have seen that one can make the KK-compactified theory independent of the spin structure on $T^2$. This turns out to be also possible for the entire $M_6= M_4 \times T^2$, which can be viewed as another consequence of the true symmetry of the E-string theory being $(\text{Spin}(6)_E \times SU(2)_R\times \Z_2^{\mathcal{F}})/(\Z_2\times\Z_2)$. Namely, changing the spin structure can be viewed as turning on holonomies in the $\Z_2\subset\Z_4$ part of the center of Spin$(6)_E$, which, for the E-string theory, can always be canceled by holonomies in the center of $SU(2)_R$. Therefore, the E-string theory can be made independent of the spin structure on the entire $M_6$.

Without topological twist, this is really a choice, since one can choose not to turn on the holonomies of $SU(2)_R$ or choose a different set of holonomies. However, when the topological twist is performed on $M_6=M_4 \times T^2$, it is no longer a choice, but one has to turn on the holonomies of $SU(2)_R$ in exactly the same way as $SU(2)_+$. Therefore, the twisted theory is automatically independent of the spin structure.

The story is slightly more interesting in the higher-rank case, where one has to turn on holonomies of $SU(2)_F$ as well to make it independent of the spin structure. But such a set of holonomies is a genuine choice, and one can choose to not perform such an action. Therefore, the higher-rank E-string theories are sensitive to the spin structure of the underlying manifold, but in a quite predictable way (i.e.~can be compensated by $SU(2)_R$).

One consequence of having a non-trivial $M_4$ with a non-trivial background (in particular, it is ``chiral'' in the sense that it treats $SU(2)_+$ and $SU(2)_-$ differently) is that the 2d theory obtained from the compactification on $M_4$ will have anomalies. These are non-perturbative and robust features of the theory and provide powerful checks on the computation for the partition function. We will discuss a few aspects of anomalies next.

\subsection{Anomalies}\label{sec:Anomalies}

We start with the anomaly polynomial of the E-string theory given as a characteristic class in eight dimensions \cite{Ohmori-Shimizu-Tachikawa-2014,Kim:2017toz},
\begin{equation}
\label{eq:e-string anomaly polynomial}
    I_8^{\rm E-string}=\frac{1}{5760}\left(3120 c_2(R)^2+203p_1(T)^2-116 p_2(T)-1320c_2(R)p_1(T)\right)+I_8^{(\rm flavor)}\, ,
\end{equation}
where $c_2(R)$ is the second Chern class of the $SU(2)_R$ R-symmetry bundle, $p_1(T),p_2(T)$ denote the first and second Pontryagin classes of the tangent bundle, and 
\begin{equation}
    I_8^{(\rm flavor)}=\frac{1}{32}(\Tr F_{E_8}^2)^2-\frac14\Tr F_{E_8}^2c_2(R)+\frac{1}{16}\Tr F_{E_8}^2p_1(T)\,,
\end{equation}
captures the (mixed) anomalies of the $E_8$ global symmetry.\footnote{For general rank-$Q$ E-string theories, there will also be an additional global symmetry present, namely $SU(2)_F$, which the anomaly polynomial will capture via $c_2(F)$ similar to $c_2(R)$.} This degree-8 characteristic class describes the anomaly of the 6d theory via the standard descent process,
\begin{equation*}
    I_8=\dd I_7\, ,\qquad \delta I_7=\dd I_6\, .
\end{equation*}

Similarly, after compactification, the anomalies of the 2d theory $\CT[M_4]$ can be described by a degree-4 anomaly polynomial,
\begin{equation}
    I_4=\frac{d}{48}\cdot p_1(T)+\frac{k}{4}\Tr F^2\,,
\end{equation}
where $d\in\Z$ measures the gravitational anomaly and equals $2( c_R-c_L)$ of the IR SCFT while $k$ measures the anomaly for the global $E_8$ symmetry.

\subsubsection{The gravitational anomaly}

Upon integrating the 8-form anomaly polynomial \eqref{eq:e-string anomaly polynomial} for the E-string theory over $M_4$, we obtain the gravitational anomaly of $\CT[M_4]$~\cite{Benini:2013cda,Gukov:2018iiq},
\begin{equation}
    d=36\cdot\left(-\frac{1320}{5760}-8\frac{203}{5760}+4\frac{116}{5760}\right)\sigma(M_4)-24\cdot\frac{1320}{5760}\chi(M_4)=-\frac{31}{2}\sigma(M_4)-\frac{11}{2}\chi(M_4)\in\mathbb{Z}\, .
\end{equation}
Here, $\chi(M_4)$ is the Euler characteristic of $M_4$ and $\sigma(M_4)$ is the signature of $M_4$ (not to be confused with the complex structure of $\CC$). The expression above is an integer since $\chi\equiv \sigma \pmod2.$ 

One can think of a 2d theory with gravitational anomaly $d$ as living on the boundary of a 3d gravitational Chern--Simons theory with level $d$. As a consequence, the $T$-transformation leads to a phase for the partition function of the E-string theory on $M_4\times T^2$,\footnote{In our convention, the SL$(2,\Z)$ action on the Hilbert space of the level-$d$ gravitational Chern--Simons theory on $T^2$ with periodic-periodic spin structure is given by $S=e^{2\pi i d/8}$ and $T=e^{-2\pi i d/24}$.}
\begin{equation}
    Z[M_4\times T^2;\sigma+1]=e^{-2\pi id/24}Z[M_4\times T^2;\sigma]\, ,
\end{equation}
which implies that the partition function as a $q$-series actually has fractional powers,
\begin{equation}
    Z[M_4\times T^2]\in q^{(31\sigma(M_4)+11\chi(M_4))/48}\mathbb{Z}(\!(q)\!)\, .
\end{equation}

Similarly, assuming no backgrounds for $E_8$ on $M_4$, one obtains the flavor anomaly of the 2d theory, 
\begin{equation}
\label{eq:flavor anomaly}
    k=\frac{\chi+3\sigma}{2}=2\chi_h+\sigma\,\in \Z\,,
\end{equation}
where the holomorphic Euler characteristic is given by $\chi_h:=\frac{\chi+\sigma}{4}$. 

Notice that the partition function $Z$ transforms under both $S$ and $T$ with phases. To apply the theory of modular (and Jacobi) forms, which are conventionally chosen to be invariant under $T$ but transform under $S$, one can normalize the elliptic genus by a factor given by a $d$-th power of the Dedekind eta function,\footnote{For positive $d$, this can be thought of physically as adding $d$ copies of left-moving Majorana fermions to cancel the anomaly and computing the correlation function with insertions of the fermion fields (say, at the origin of $T^2$) which absorb the zero modes. 
} 
\begin{equation}\label{Normalize}
    Z_{\rm norm}(\sigma):=\eta^{d}\cdot Z(\sigma)\,.
\end{equation}
Since $\eta\in q^{1/24}\cdot \mathbb{Z}[\![q]\!]$ and the phase of $\eta$ under $S$ or $T$ exactly cancels that of the partition function, leaving only a $\sigma^{d/2}$ factor under the $S$-action, $Z_{\rm norm}$ is now expected to be a weight-$\frac{d}{2}$ modular form (and, furthermore, an $E_8$ Jacobi form of index $k$ in the presence of mass parameters) with an integral $q$-series.\footnote{Although the phase for $Z$ is a 24-th root of unity and one can choose to normalize the theory with $Z'(\sigma)=\eta^{m}Z(\sigma)$ as long as $m\equiv d \pmod{24}$, it is natural to actually take $m=d$ as the 7d and 3d TQFTs capturing the anomaly of the E-string theory and its compactification ``know'' about the full value of $d$, not just $d \bmod 24$. This will be important in connection with TMF, as different choices of $m$ differ by a power of $\Delta(\sigma):=\eta(\sigma)^{24}$, which is not a topological modular form by itself.}

This normalization procedure \eqref{Normalize} is reminiscent of the scaling of the expression for $\Sigma$, keeping the $j$-invariant unchanged but making it behave differently under the $\SL(2,\Z)$ action on the UV parameter $\sigma$. There, we have used a similar convention that the coefficients of the elliptic curve $\Sigma$ will remain invariant under a $T$-transformation, but require a re-scaling of $(u,x,y)$ under an $S$-transformation to stay invariant. On the other hand, the unnormalized $Z$ is analogous to the other choice~\eqref{eq:E8masslessDimensionful} for $\Sigma$, where $S$ and $T$ both leave the curve invariant up to a rotation of the $u$-plane. 

In the next section, we will perform computation with the curve~\eqref{eq:E8curve}, and the partition function will be implicitly normalized by copies of $\eta$ and carry a modular weight. Therefore, the partition function in the massless limit would transform under $S$ as
\begin{equation}
    Z_{\text{norm}}[M_4\times T^2;-1/\sigma]=\sigma^{d/2}\cdot Z_{\text{norm}}[M_4\times T^2;\sigma]\, .
\end{equation}
To avoid clutter, we will omit the subscript and simply refer to it as $Z$.

\subsubsection{Relation with the R-symmetry anomaly of 4d $E_8$ SCFT}\label{sec:E8Anomaly}

The particular combination of $11\chi+31\sigma$ might seem at first arbitrary but becomes intriguing once one notices that the very same combination would appear in the anomaly of the $U(1)_R$ symmetry of the twisted $E_8$ SCFT. Indeed, the $E_8$ theory has conformal central charges $a=\frac{95}{24}$ and $c=\frac{31}{6}$~\cite{Argyres:2007tq}, and the $U(1)_R$ ``ghost number'' of the partition function is~\cite{Shapere:2008zf}
\begin{equation}
    \Delta_R = 2(2a-c)\chi +3c\sigma = \frac{11\chi+31\sigma}{2}\,.
\end{equation}

It is not a mere coincidence that the gravitational anomaly of the $\CT[M_4]$ theory obtained from the E-string and the $U(1)_R$ anomaly of the vacuum of the $E_8$ theory are proportional. To obtain the $E_8$ SCFT from the rank-1 E-string theory, one can make the area of the $T^2$ small. Then the KK modes become very massive, and the $U(1)_R$ emerges in the IR as, in a sense, a rotation of the $x_5$ and $x_6$ directions. Since the $S$-transformation is a $-\pi/2$-rotation for the square torus, one has
\begin{equation}
    e^{2\pi i d/8}=e^{-i\frac\pi 2 \cdot \frac{\Delta_R}{2}}\,,
\end{equation}
where the $\frac12$ factor on the right is due to the fact that $U(1)_R$ is in fact the spin double cover of the Euclidean rotation in the $(x_5,x_6)$ plane. Indeed, the relation above is consistent with $d=-\Delta_R$, but only requires the equality mod 8.

To see the full agreement, one observes that $d$ measures the difference in the degrees of freedom between the right- and left-moving sectors of $\CT[M_4]$. But such a difference---in particular, of their zero modes---is exactly what causes the $U(1)_R$ anomaly of the vacuum of the 4d theory in the zero-area limit. The minus sign in $d=-\Delta_R$ is an artifact of the conventions used. Namely, a left-moving boson or fermion will contribute $-2$ or $-1$ to $d$, but their zero modes from the 4d point of view have positive $U(1)_R$-charge $2$ and $1$.

A correlation function in the 4d theory can be, at least conceptually, thought of as an integral over the moduli space of the theory, which is given by a system of partial differential equations on $M_4$ when the 4d theory has a Lagrangian description. The quantity $-\Delta_R$ then measures the virtual dimension of the moduli space when the 4d theory is superconformal. Only when it is zero is the partition function without insertion expected to be non-vanishing. From the 2d point of view, this moduli space is the target of the sigma model at low energy, and the (equivariant) TMF invariant associated with the $M_4$ is roughly the (equivariant) topological elliptic genus of this moduli space. Then, indeed, its virtual dimension determines the gravitational anomaly $d$ of the sigma model. Of course, even when the space is singular or non-existent (due to the non-Lagrangian nature of the 6d theory), the 2d theory $\CT[M_4]$ still makes sense and one still expects a TMF invariant for $M_4$, which can be non-vanishing even when $\Delta_R\neq 0$. This is supposed to be one of the advantages of the approach advocated in~\cite{Gukov:2018iiq} compared to traditional ones.

From the point of view of the partition function, it is less straightforward to talk about the zero-area limit, as $Z[M_4\times T^2]$ doesn't depend on the relative size of $M_4$ and $T^2$. However, guided by the discussion above, we expect that the partition function for the $E_8$ SCFT is only defined and non-zero when $d=\Delta_R=0$. Then the partition function of the E-string theory is expected to become a modular function, given by a polynomial in $j:=\frac{E_4^3}{\Delta}$. As the contribution of the KK-modes should be modular and the partition function of the $E_8$ SCFT should have no dependence on the modulus of $T^2$, it is natural to speculate that the latter is then given by the constant term in this polynomial. However, it turns out that the partition function of the $E_8$ SCFT in general doesn't actually have a well-defined massless limit, and, to have something well defined, we will have to subtract the $E_8$ part and work with the rest of the E-string partition function.\footnote{\label{foot:masslesslimit} The value of the partition function can be made finite but depends on the direction of the limit. This is because the contribution of the II$^*$ singularity is now a $W(E_8)$-invariant rational function of degree zero in the masses (cf.~behavior of $f(m_1,m_2)=m_1/m_2+m_2/m_1$ in the limit $m_{1,2}\to 0$). } Therefore, the expectation is that the E-string partition function, after removing the ambiguous contribution of the $E_8$ SCFT and written as a polynomial in $j$, should not have a constant term.\footnote{
Arguably, this is a rather wild speculation, as the rest of the theory can still contribute to the constant term. While this assumption leads to the right prediction that we confirm when we evaluate the partition function in Section~\ref{sec:zeroweight}, it would be interesting to revisit and substantiate this claim. }

The story will be more interesting and clearer if one either considers the correlation function with insertions of operators or turns on mass deformations. While we will not discuss the former topic in this paper, deferring the systematic study of such correlation functions to future work, we do make use of the mass deformation occasionally. These are given by a flat connection of the $E_8$ flavor symmetry on $T^2$, whose anomaly we discuss next.

\subsubsection{Anomaly of the $E_8$ global symmetry}

In the rest of this section, we will briefly remark on the effect of turning on the masses in the theory. Using the $E_8$ symmetry of the theory might sound intimidating at first, but it can provide valuable insight into the theory and its partition function, even if one is only interested in the massless limit.

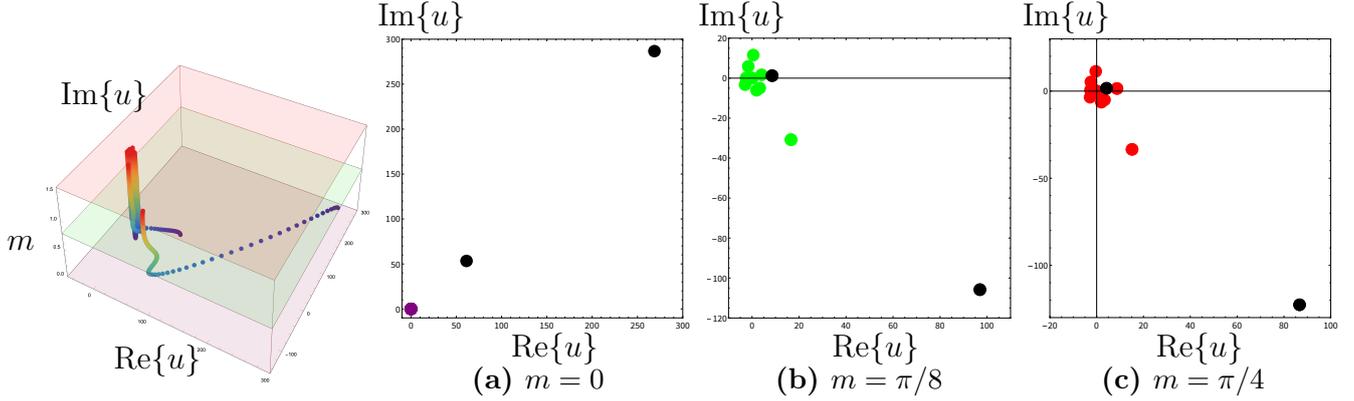
\begin{figure}
    \centering
    \begin{subfigure}{0.25\textwidth}
    \centering
    \includegraphics[width=\linewidth]{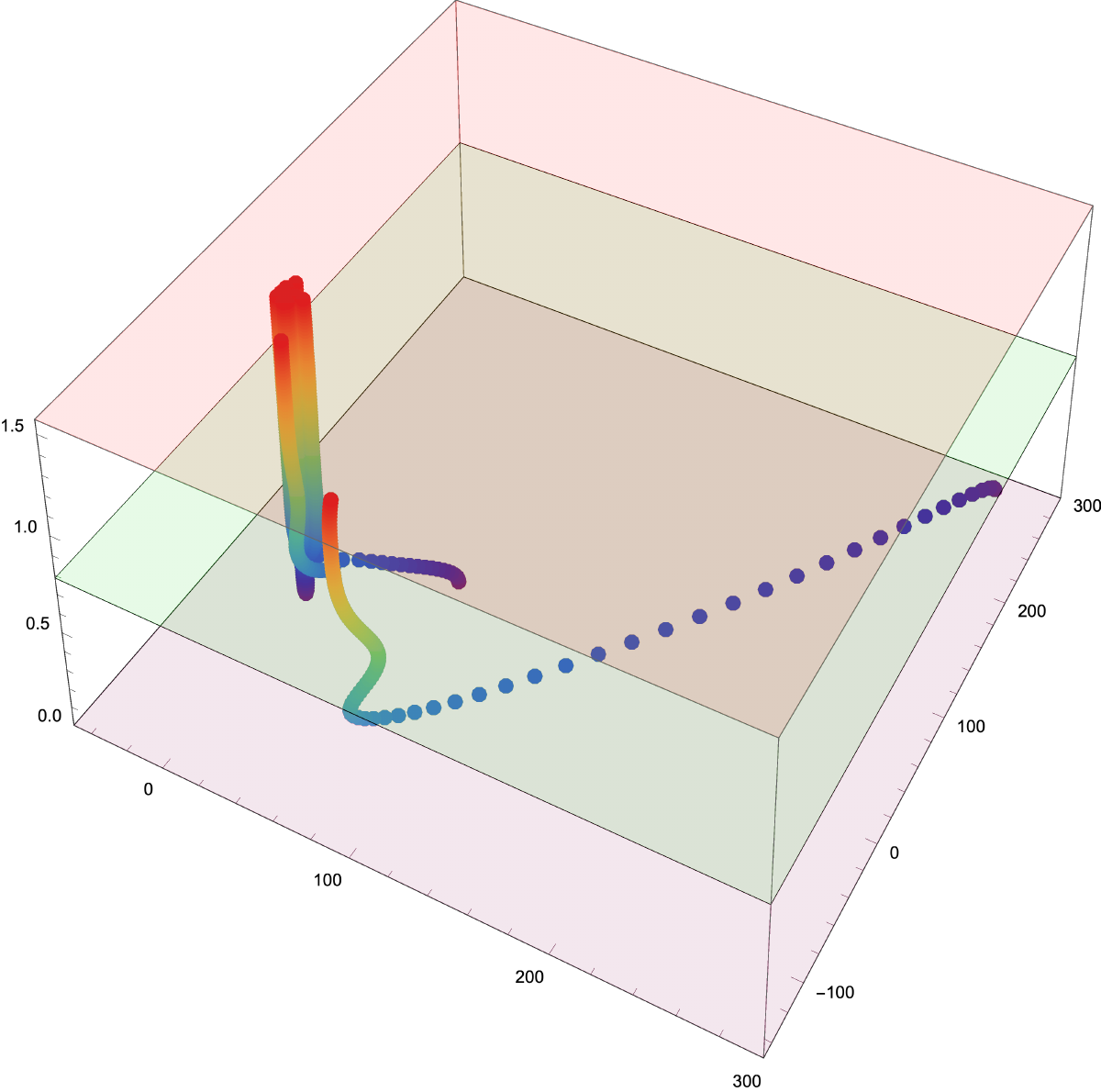}
    \begin{picture}(0,0)\vspace*{-1.2cm}
    \put(-75,65){$m$}
    \put(-35,20){$\Re{u}$}
    \put(-55,120){$\Im{u}$}
    \end{picture}\vspace*{-0.3cm}
    \end{subfigure}
    \hfill
    \begin{subfigure}{0.235\textwidth}
    \centering
    \includegraphics[width=\linewidth]{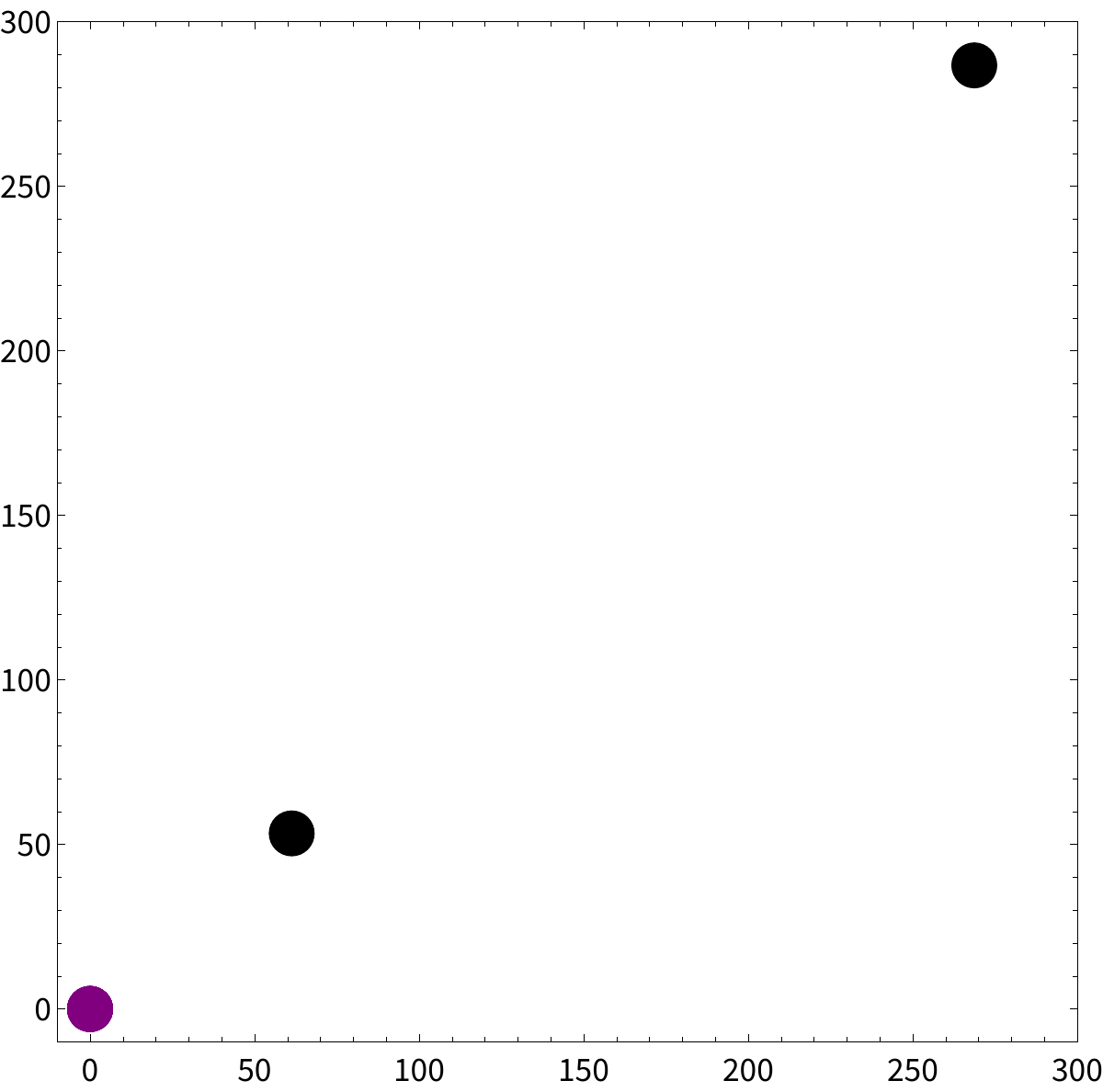}
    \begin{picture}(0,0)\vspace*{-1.2cm}
    \put(-60,130){$\Im{u}$}
    \put(-10,5){$\Re{u}$}
    \end{picture}\vspace*{-0.3cm}
    \caption{$m=0$}
    \label{fig:massive singularities initial}
    \end{subfigure}
    \hfill
    \begin{subfigure}{0.235\textwidth}
    \centering
    \includegraphics[width=\linewidth]{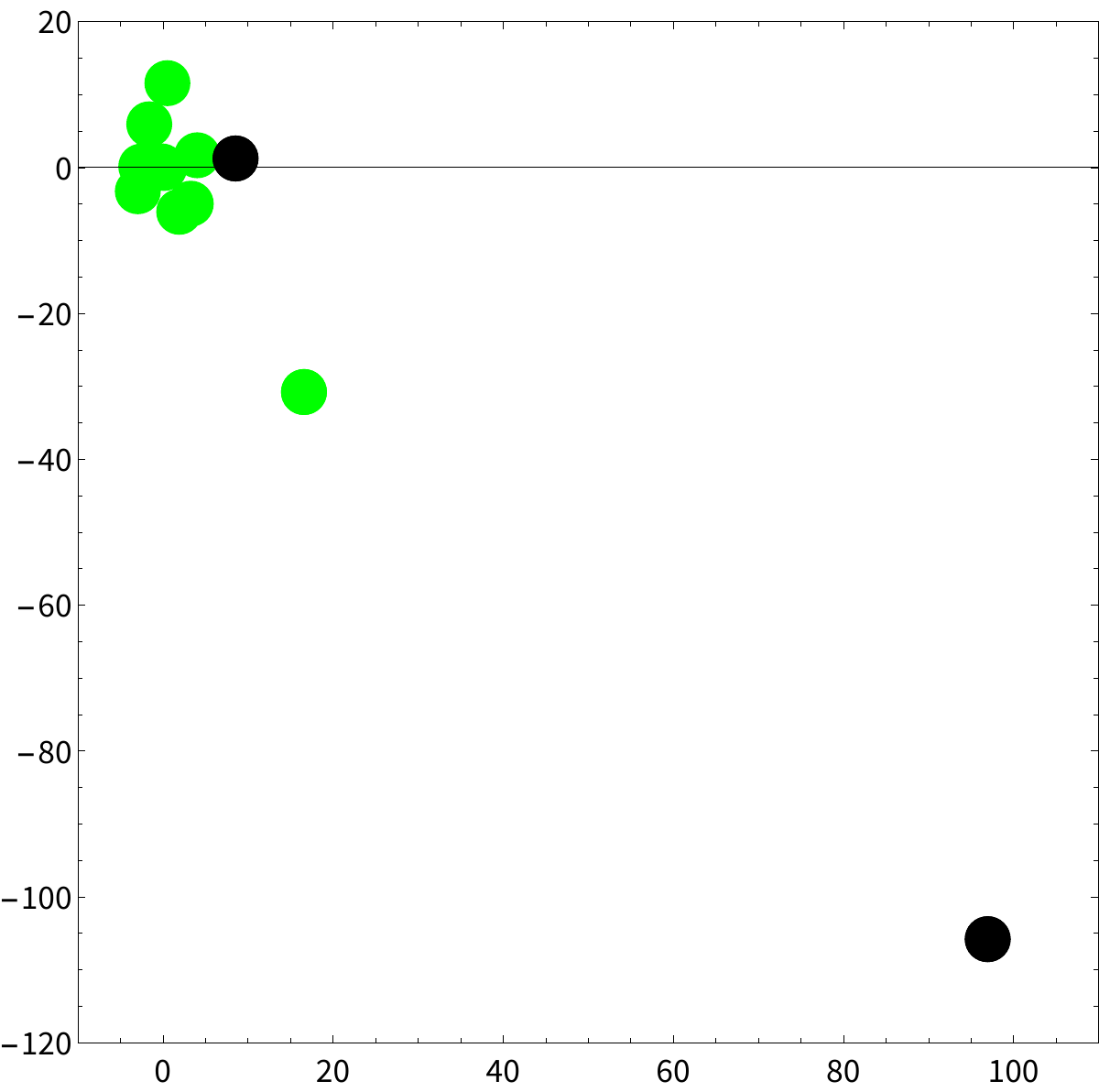}
    \begin{picture}(0,0)\vspace*{-1.2cm}
    \put(-60,130){$\Im{u}$}
    \put(-10,5){$\Re{u}$}
    \end{picture}\vspace*{-0.3cm}
    \caption{$m=\pi/8$}
    \label{fig:massive singularities half}
    \end{subfigure}
    \hfill
    \begin{subfigure}{0.24\textwidth}
    \centering
    \includegraphics[width=\linewidth]{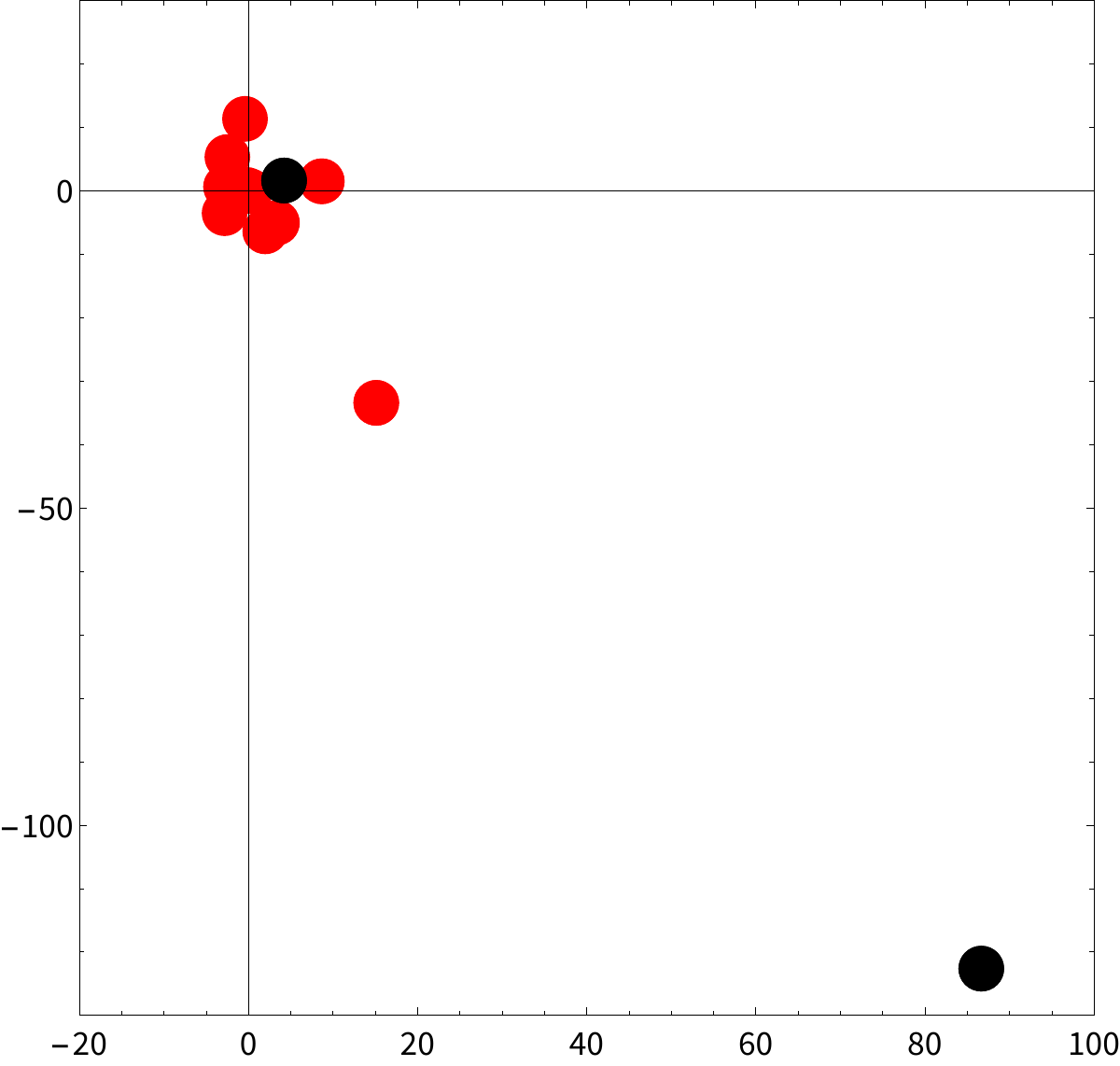}
    \begin{picture}(0,0)\vspace*{-1.2cm}
    \put(-60,130){$\Im{u}$}
    \put(-10,5){$\Re{u}$}
    \end{picture}\vspace*{-0.3cm}
    \caption{$m=\pi/4$}
    \label{fig:massive singularities final}
    \end{subfigure}\vspace*{-0.3cm}
    \caption{\textbf{Singularities on the Coulomb branch of the E-string theory on $T^2$ with generic $E_8$ holonomies.} In the leftmost figure, we plot the positions of singularities as we vary the mass parameters of the E-string theory. The eight mass parameters are parameterized as $\{m_i\}\approx m(0.299, 0.824, 1.200, 2.475, 3.022, 1.532, 0.916, 2.422)$ with $m\in[0,\pi/4]$ multiplying a set of randomly generated numbers, while we have set $\sigma=0.3+0.7i$. 
    In the following three figures (\ref{fig:massive singularities initial},\ref{fig:massive singularities half},\ref{fig:massive singularities final}), we take horizontal slices of the leftmost figure, i.e.~configurations of singularities at a fixed $m$. We indeed observe explicitly that the type $\mathrm{II}^*$ singularity (indicated by \tikzcircle[purple,fill=purple]{3pt}) appearing in Fig.~\ref{fig:massive singularities initial} gets deformed into 10 $\mathrm{I}_1$ singularities in both Fig.~\ref{fig:massive singularities half} (indicated by \tikzcircle[green,fill=green]{3pt}) and \ref{fig:massive singularities final} (indicated by \tikzcircle[red,fill=red]{3pt}). The two black dots \tikzcircle[black,fill=black]{3pt} appearing in Fig.~\ref{fig:massive singularities initial},~\ref{fig:massive singularities half},~\ref{fig:massive singularities final} indicate the value of the original two $\mathrm{I}_1$ singularities existing prior to any mass deformations.}
    \label{fig:singularities}
\end{figure}

In general, it is difficult to analytically compute the roots of an order-12 polynomial, but there are various special points in the parameter space that have been understood previously (see e.g.,~\cite{Ganor:1996pc, Eguchi:2002fc,Eguchi:2002nx}). However, at a generic point, the Coulomb branch consists of 12 type I$_1$ singularities, depicted in Figure~\ref{fig:singularities}.

The monodromy around each I$_1$ singularity can be described by $M=ATA^{-1}$ with the 2-by-2 matrix $A$  specifying a weakly coupled local duality frame wherein the low-energy effective theory is given by a $U(1)$ vector multiplet coupled to a charge-1 hypermultiplet whose mass becomes zero at the singularity. Therefore, the contribution of these singularities to the partition function on $ M_4 \times T^2$ will be proportional to the Seiberg--Witten invariants of $M_4$~\cite{Witten:1994cg}. Since the partition function, as a BPS-protected quantity, depends meromorphically on the masses, one can always obtain the result for a singular configuration (e.g.~the massless limit where the $E_8$ SCFT emerges) by performing the computation for generic masses and then taking the limit, whenever it exists. This will be one of the computational tools in Section~{\ref{sec:3}}, although our use of it remains limited. Given the richness of limits in the parameter space of the compactified E-string theory, it would be particularly rewarding to better understand the partition function with generic masses, as the partition functions of many other interesting theories in 5d and 4d are encoded in that of the E-string theory. 

We now study the behavior of these singularities by analyzing the modularity of the coefficients in $f$ and $g$. To streamline this process, we use the following $u$-expansion of the discriminant,
\begin{equation*}
    \Delta_\Sigma=\sum_{i=0}^{12}D_iu^{12-i}=D_0\prod_{i=1}^{12}(u-u_{s,i})=f^3-27g^2.
\end{equation*}
 We list the first two of the coefficients $D_i$ below, with the rest given explicitly in Appendix~\ref{sec:modularity of coefficients}.
\begin{itemize}
    \item $i=0$:
    \begin{equation}
        D_0=\frac{E_4^3-E_6^2}{1728}=\eta(\sigma)^{24}\,
    \end{equation}
    equals the modular discriminant, a modular form of weight $12$.\footnote{We opt to write it as $\eta^{24}$ rather than the more customary $\Delta$ to avoid potential confusion with $\Delta_\Sigma$ and the appearance of fractional powers of $\Delta$.}
    \item $i=1$:
    \begin{equation}
    \label{D1}
        D_1=\frac{54\cdot 4}{216}\frac{E_6P(\sigma,\{m_i\})}{E_4}\,,
    \end{equation}
    which is an $E_8$ Jacobi form of weight $6$ and index $1$. Here $P$ is given in~\eqref{eq:P}. See Appendix~\ref{sec:modular forms} for a brief review on the properties of Jacobi forms.
\end{itemize}
Conveniently, all coefficients, $D_i$, are meromorphic $E_8$ Jacobi forms under the full group $\SL(2,\mathbb{Z})$.
Therefore, the collection of all the singularities is also an $\SL(2,\mathbb{Z})$ Jacobi form in the following sense. Under the action of 
\begin{equation}
\label{eq:ellipticTransformations}
    T:\ \sigma\to\sigma+1,\qquad S:\ \sigma\to -\frac{1}{\sigma},\ \{m_i\}\to \frac{\{m_i\}}{\sigma}\,,
\end{equation}
the Seiberg--Witten curve of the E-string theory $\Sigma$ remains invariant under a $T$-transformation while the invariance under an $S$-transformation is up to a rescaling of the curve itself  \cite{Eguchi:2002nx},
\begin{equation*}
    (u,x,y)\to (\sigma^{-6}e^{im^2/4\pi\sigma}u,\sigma^{-10}e^{im^2/2\pi\sigma}x,\sigma^{-15}e^{i3m^2/4\pi\sigma}y),
\end{equation*} 
where $m^2=\sum_i (m_i)^2$. This re-scaling reveals that the singularities of the mass-deformed E-string theory ``behave as Jacobi forms'' of weight $-6$ and index $1$. Notice that, just like in the massless case that we analyzed before, positions of individual singularities are not SL$(2,\Z)$ Jacobi forms but instead have branch cuts, while the entire collection is a Jacobi form in the sense explained above. 

We expect that, from this modular weight and index of singularities, the partition function of the E-string theory will also be a meromorphic Jacobi form with the weight and index expected from the gravitational anomaly and the flavor anomaly, which we will verify in Section~\ref{sec:3}. Before that, we will first consider an extension/completion of the E-string theory, which provides interesting insights into certain properties of the E-string partition function.

\subsection{The $E_8\times E_8$ theory}\label{sec:E8E8}

In the M-theory setup for the E-string theory, one can add back the other end-of-the-world M9-brane to get the entire  Ho\v{r}ava--Witten picture of the $E_8\times E_8$ heterotic string. An M5-brane probe in this geometry will then have the full $E_8\times E_8$ symmetry in its 6d world-volume theory, but at the expense of introducing a scale $l$ that measures the distance between the two M9-branes.\footnote{We thank Edward Witten for numerous comments and suggestions which motivated us to look further into this topic and the issue of full topological invariance.}

The 4d effective theory---which can be understood as that on a D3-brane in an 8-dimensional geometry given by F-theory on an elliptic K3 (or, equivalently, IIB on a two-manifold $B\simeq \mathbb{P}^1$ with 24 D7-branes)---only depends on the dimensionless quantity $\log\rho \sim l\cdot{A_{T^2}}\cdot M_{\text{Planck}}^3$,\footnote{In the 4d theory, $\rho$ is in fact a complex parameter (e.g.,~can be viewed as the exponentiated complexified Kähler moduli of $T^2$ in the heterotic string theory). However, the phase will not play a role in our analysis and we will keep $\rho$ real.} along with the moduli $\sigma$ of the internal $T^2$ and the $E_8\times E_8$ holonomies, and one can perform the computation in the large-$\rho$ limit and expect that the partition function decomposes as two copies of that of the E-string theory.

Although this partition function---if it indeed decomposes in this way---cannot detect additional information about the 4-manifold, its existence might lead to consequences for the E-string partition function.
Indeed, as the moduli of the $E_8\times E_8$ model is compact (with a generic mass deforming away the Higgs branch), its partition function is expected to be fully topological. This leads to the natural question whether the partition function of the E-string theory is also a full topological invariant, even in the $b_2^+\le 1$ case.\footnote{The reason why the full topological invariance of the E-string theory doesn't directly follow from that of the $E_8\times E_8$ is because the $\Z_2$ exchange symmetry, acting as a charge conjugation in the 4d theory, can multiply the $u$-plane integrand by a sign $(-1)^{\chi_h}$ \cite{Witten:1994cg}. Also, it is not true that the E-string theory on an arbitrary $M_4$ can always be embedded into the $E_8\times E_8$ model, as the latter satisfies the condition $\int_{M_4}dH=0$ for the 3-form flux, leading to the constraint $n_1+n_2=-\chi-3\sigma$ for the instanton numbers $(n_1,n_2)$ of the two $E_8$ factors. This becomes the familiar $n_1+n_2=24$ when $M_4$ is a K3 surface. Supersymmetry in our twisted background then forces $\chi+3\sigma\le 0$.\label{foot:E8E8}
} This would open up interesting possibilities such as detecting exotic smooth structures when $b_2^+=0$.

We will revisit this topological invariance after laying out more details of the partition function computation.

\section{Partition function}
\label{sec:3}

In this section, we will compute and study the partition function of the E-string theory on $M_4 \times T^2$, where $M_4$ is closed and oriented. From the 4d point of view, such a partition function for a topologically twisted theory is often known as the ``Donaldson--Witten partition function'' \cite{Witten:1988ze,Witten:1994cg,Witten:1994ev}. What sets the E-string theory apart from a garden variety 4d theory is its 6d nature, which leads to the following list of expectations for its partition function $Z$:
\begin{itemize}
    \item It is a weight-$\frac d2$ and index-$k$ $E_8$ meromorphic Jacobi form with integer coefficients under the full modular group $\SL(2,\mathbb{Z})$;
    \item It can be lifted to an $E_8$-equivariant topological modular form, $\CZ\in \widetilde{\mathrm{TMF}}^{E_8}_{d,k}$ (cf.~Footnote~\ref{foot:tilde}); 
    \item In the massless limit (when it exists), $Z\in\mathbb{Z}(\!(q)\!)$ is an integral $q$-series and, furthermore, as a modular form, $Z\in \mathrm{MF}^{\mathbb{Z}}_{d/2}$, it can be expressed as a $\Z$-linear combination of $E_4^iE_6^j\Delta^{k}$, with the coefficients compatible with a lift of it to a topological modular form, $\CZ\in\pi_d(\TMF)$.
\end{itemize}

With the above expectations, let us now pivot to explain how the partition function is computed by integrating over the Coulomb branch of the theory. Without insertion of any topological operators, the integral can be expressed as \cite{Moore:1997pc} (see also \cite{Marino:1998tb,Korpas:2019cwg,Moore:2017cmm})
\begin{equation}
\label{eq:donaldson-invariants}
    Z_\CT[M_4\times T^2;\sigma,\{m_i\}]=\int_{\mathcal{M}_C}\frac{\dd u\dd\bar{u}}{(\Im\tau)^{1/2}}\cdot \mu(u;\sigma,\{m_i\})\Psi(u;\sigma,\{m_i\})\,.
\end{equation}
Here $\Psi(u)\cdot (\Im\tau)^{-1/2}$ is the partition function of the low-energy effective theory at $u$---generically a $U(1)$ gauge theory with a $u$-dependent coupling constant $\tau\in\mathbb{H}$---while
$\mu$, together with ${\dd u\dd \bar{u}}$, can be viewed as the ``$u$-plane measure'' which also takes into account couplings of the effective theory to gravitational backgrounds. Both of them depend on $\sigma$ and $\{m_i\}$, which are UV parameters from the point of view of the 4d theory $\CT[T^2]$.

In the simply-connected case---which we assume from now on---with $b_2^+>1$,\footnote{Recall that $b_2(M_4)=\dim(H^2(M_4,\R))$ is the second Betti number of $M_4$, which can be decomposed as $b_2(M_4)=b_2^+(M_4)+b_2^-(M_4)$ where $b_2^+(M_4)$ and $b_2^-(M_4)$ respectively denote the number of independent self-dual and anti-self-dual harmonic two-forms on $M_4$.} due to fermionic zero modes coming from the (topologically twisted) abelian vector multiplet, the integrand vanishes except at singular points on the Coulomb branch, where the integral picks up a delta-function contribution. In the low-energy effective theory, these are places where some BPS particles become massless, invalidating the low-energy description as a pure gauge theory. As these new degrees of freedom are associated with the emergence of a Higgs branch at the singularity, we will refer to the contribution of these delta functions as ``Higgs branch contributions.'' We denote them as $Z_{{\rm Higgs},s}$ where $s$ labels the singularities on the Coulomb branch. 

On the other hand, when $b_2^+\leq1$, the integrand does not vanish, and there is a non-trivial Coulomb branch contribution $Z_{\rm Coulomb}$---often known as the ``$u$-plane integral.'' In the $b_2^+=1$ case, it is given by \eqref{eq:donaldson-invariants} with 
\begin{align*}
    \mu(u;\sigma,\{m_i\})&=\alpha^{\chi(M_4)}\beta^{\sigma(M_4)}\frac{\dd\bar{\tau}}{\dd\bar{u}}\left(\frac{\dd a}{\dd u}\right)^{1-\chi(M_4)/2}\left(\frac{\Delta_\Sigma}{D_0}\right)^{\sigma(M_4)/8}\,,\\
    \Psi(u;\sigma,\{m_i\})&=\sum_{\lambda\in\text{spin}^c}(-1)^{\lambda\cdot w_2} \cdot (\lambda,\omega)\cdot \exp[-i\pi\bar{\tau}(\lambda_+)^2-i\pi\tau(\lambda_-)^2]\, .
\end{align*}
Here, $\alpha$ and $\beta$ are constants coming from background topological terms in the action, $\omega\in H^{2}_+(M_4)$ is a choice of a normalized self-dual 2-form with $(\omega,\omega)=1$, the spin$^c$ structure $\lambda$ labels the topological type of the $U(1)$ gauge bundle as explained in Section~\ref{sec:topological twist},\footnote{One way to think about $\lambda$ is as an element in $H^2(M_4,\Z)$ but shifted by $\frac{1}{2}w_2$. Then $\frac{1}{2}w_2^2$ is understood as the Pontryagin square of $w_2$ valued in $H^4(M_4,\Z_4)$. In particular, $(-1)^{\lambda\cdot w_2}$ can be $\pm i$ for non-spin $M_4$. Another convention is to use $s=2\lambda\in H^2(M_4;\mathbb{Z})$ to label spin$^c$ structures. Physically, $\lambda$ is the flux of the twisted $U(1)$ bundle, which could have half-integer fluxes, while $2\lambda$ can be viewed as the Chern class of the line bundle associated with a charge-2 particle, which is globally well defined. } and $w_2$ is the second Stiefel--Whitney class of $M_4$. Also, we have not included the delta functions supported at the singularities. Once we do, we have that the partition function decomposes into
\begin{equation}
\label{eq:DW decomposition}
    Z=\sum_sZ_{\text{Higgs},s}+Z_{\rm Coulomb}\, .
\end{equation}
As the partition function depends meromorphically on the mass parameters $\{m_i\}$, it is sufficient to perform the computation for generic masses, which deform the type II$^*$ singularity into 10 type I${}_1$ singularities (cf.~Figure~\ref{fig:singularities}). The partition function for special values of the masses (reviewed in Appendix~\ref{sec:specialMass}) can be obtained as limits, which can diverge due to non-compact Higgs branches opening up. 

For generic masses, the only type of singularities that appears is of type I$_1$, where the Higgs branch is a point and $Z_{\text{Higgs},s}$ is related to the Seiberg--Witten invariants.\footnote{Although it is still technically correct to refer to it as the ``Higgs branch contribution'' in this case, it is customary and perhaps better to use the term ``monopole contribution'' or ``Seiberg--Witten contribution'' to set it apart from cases with non-compact Higgs branches.} Therefore, the contribution of type I${}_1$ singularities to the integral~\eqref{eq:donaldson-invariants} is
\begin{equation}
\label{eq:estringZ}
    Z_{\text{Higgs},s}=\sum_{\lambda}C_{s,\lambda}\cdot \SW(\lambda)\, ,
\end{equation}
where $C_{s,\lambda}$ depends on the choice of singularity $s$ and a spin$^c$ structure $\lambda$, and $\SW(\lambda)$ denotes the Seiberg--Witten invariant of $M_4$ with the spin$^c$ structure $\lambda$. The appearance of spin$^c$ structures (as opposed to the untwisted $U(1)$ that usually appears at quark singularities) for each I$_1$ has been anticipated from the discussion in  Section~\ref{sec:topological twist}.

There are also 4-manifolds for which $Z_{\text{Higgs}}=0$. For example, if $M_4$ admits a metric with positive scalar curvature, the SW invariants vanish for any $\lambda$ \cite{Witten:1994cg}, and $Z_{\text{Higgs}}$, which is a linear combination of SW invariants, also vanishes. This, e.g.,~will be the case for $\mathbb{CP}^2$ (see Appendix~\ref{appendix:SW} for more details as well as some other conditions on $M_4$ that lead to vanishing SW invariants). Arguably, such 4-manifolds are natural venues for studying the partition functions of 6d theories as only $Z_{\rm Coulomb}$ contributes.

However, in the present work, our focus is on the opposite type of 4-manifolds with $b_2^+>1$ and hence $Z_{\rm Coulomb}=0$. For manifolds with $b_2^+\le 1$, we only discuss $Z_{\rm Coulomb}$ in the context of topological invariance in Section~\ref{sec:FullTop}, and will leave the evaluation of the $u$-plane integral and a more detailed study of four-manifolds with $b_2^+(M_4)\leq 1$ to future work. 

\subsection{The Higgs branch contribution}
\label{sec:Seiberg-Witten-Partition-Function}

For generic masses, the Higgs branches are only points located at the I$_1$ singularities, where a BPS particle (often referred to as the ``monopole'' in the context of the SW equations) becomes massless in the low-energy description. Their contribution is then given by the SW invariant multiplied by a ``$u$-plane measure.'' The latter can be computed via quantities in the weakly-coupled local duality frame (such that the massless particle carries only an electric charge) \cite{Moore:1997pc}. 

In such a duality frame, the elliptic curve can be expressed as
\begin{equation}
\label{I1curve}
    y^2=4x^3-\frac{4\pi^4E_4(\tau_s)}{3\omega_1^4}x-\frac{8\pi^6E_6(\tau_s)}{27\omega_1^6}\,.
\end{equation}
At the singularity, the coupling constant $\tau_s$ tends to $i\infty$, while $\omega_1$ is the period of the elliptic curve that remains finite in the limit $q_s:=e^{2\pi i\tau_s}\to 0$.

Comparing this with the SW curve $\Sigma$ allows one to fix $\tau_s$ and $\omega_1$ as functions of $u$. However, it is more convenient to use $q_s$---which is a well-behaved local coordinate near a type I$_1$ singularity---to express the other quantities, 
\begin{subequations}
\begin{align}
\label{eq:uexpansions}
    u&=u_s+\mu_1 q_s+\mu_2 q_s^2+\dots\, ,\\
\label{eq:aexpansions}
    a_s&=\kappa_1q_s+\kappa_2q_s^2+\dots\, .
\end{align}
\end{subequations}
Here $a_s$ is the special coordinate near the singularity $u_s$, given by the period of the SW differential along the vanishing cycle. 

Switching to the local duality frame won't affect the $j$-invariant, leading to
\begin{equation}
\begin{split}
\label{eq:useries}
    27\cdot\frac{g(u;\sigma,\{m_i\})^2}{f(u;\sigma,\{m_i\})^3}=\frac{E_6(\tau_s)^2}{E_4(\tau_s)^3}\, .
\end{split}
\end{equation}
Using this relation, we can then derive the coefficients appearing in the $q_s$-expansion of $u$ by plugging \eqref{eq:uexpansions} into \eqref{eq:useries}. As expected, the coefficients are also functions of $\sigma$ and $\{m_i\}$. 
Furthermore, identifying the differential $dx/y$ between $\Sigma$ and the local curve gives
\begin{equation}
\label{eq:omega1 expression}
    \omega_1^2=2\left(\frac{\pi}{3}\right)^2\frac{E_6(\tau_s)}{E_4(\tau_s)}\cdot \frac{f(u;\sigma,\{m_i\})}{g(u;\sigma,\{m_i\})}\, .
\end{equation}
Then, by using
\begin{equation}
\label{eq:dasdu relation to period}
    \frac{\dd a_s}{\dd u}
    =\frac{i}{2\pi}\omega_1\, ,
\end{equation}
we can express $\dd a_s/\dd u$ using~\eqref{eq:omega1 expression}.
More details, such as the $q_s$-expansion of $u$ and $a_s$ near I$_1$ singularities, can be found in Appendix~\ref{sec:details on the partition function}.

Recall that $\Sigma$ is invariant under the transformations~\eqref{eq:ellipticTransformations} (upon suitable rescalings of $(x,y,u)$ for the $S$-transformations). Therefore, $f(u;\sigma,\{m_i\}),g(u;\sigma,\{m_i\})$ are Jacobi forms with modular weights dependent on the re-scalings. Now, recall that 
from our previous calculations, we have shown that the locations of singularities, as a combination of $a_i,b_i$'s appearing in $f(u;\sigma,\{m_i\}),g(u;\sigma,\{m_i\})$, are modular forms (possibly with branch cuts related to non-trivial braiding of the singularities after a modular transformation). This directly implies that $\mu_{n}(\sigma,\{m_i\})$ and $\kappa_n(\sigma,\{m_i\})$ are both modular forms. See Appendix~\ref{sec:modularity of coefficients} for the analysis on the modularity of the coefficients in~\eqref{eq:ellipticCurveES}.

Near an I$_1$ singularity, it is customary to use a different normalization of the elliptic curve, e.g.,~\eqref{eq:tildetauellipticcurve}, whose discriminant $\Delta_{0}$ is related to the previous one via a $\sigma$-dependent factor as $\Delta_\Sigma/D_0=\eta(\sigma)^{-24}\Delta_{0}(\tau_s)$~\cite{Shapere:2008zf}. It can be expressed in terms of the Dedekind $\eta$-function and its period $\omega_1$.
Then, near an I${}_1$ singularity in the Coulomb branch, we can apply~\eqref{eq:dasdu relation to period} to express $\Delta_{0}(\tau_s)$ as
\begin{equation}
\label{eq:local discriminant}
    \Delta_0(\tau_s)=(2\pi)^{12}\omega_1^{-12}\eta(\tau_s)^{24}=\left(\frac{\dd a_s}{\dd u}\right)^{-12}\eta(\tau_s)^{24}\,.
\end{equation}

In this paper, we only consider the partition function of the E-string theory without insertions of operators except for some brief remarks towards the end in Section~\ref{sec:5}. In the absence of these, the integral~\eqref{eq:donaldson-invariants} at an I${}_1$ singularity receives the following Higgs branch contribution \cite{Witten:1994cg,Moore:1997pc,Moore:2017cmm},
\begin{align}
\label{eq:Zsw}
\begin{split}
    Z_{\text{Higgs},s}&=\alpha^{\chi}\beta^{\sigma}\iota_s\,i^{3\sigma/2}\eta^{-3\sigma}\\
    &\quad \times\sum_{\lambda\in \text{spin}^c} (-1)^{n(\lambda)+\lambda\cdot w_2}\cdot \rm{SW}(\lambda)\\
    &\qquad \qquad \times\left[\left(\frac{a_s}{q_s}\right)^{\chi_h-1}\frac{\dd u}{\dd q_s}\left(\frac{\eta(\tau_s)^{24}}{q_s}\right)^{\sigma/8}\left(\frac{\dd a_s}{\dd u}\right)^{1-2\chi_h-\sigma}q_s^{-n(\lambda)}\right]_{q_s^0}\,.\\
\end{split}
\end{align}
Here $[\cdot]_{q_s^0}$ denotes the constant term of the Laurent expansion in powers of $q_s$, and $\iota_s$, a root of unity,  defines a multiplier system that renders the function single-valued by fixing the choice of the branch. Furthermore, 
\begin{equation}
\chi_h=\frac{\chi(M_4)+\sigma(M_4)}{4}\, ,
\end{equation}
is the holomorphic Euler characteristic, which is in general a half-integer,
and $n(\lambda)$ is half the virtual dimension of the Seiberg--Witten moduli space for a given spin$^c$ structure $\lambda$ on the four-manifold given by
\begin{equation}
    n(\lambda)=\delta-\chi_h\,,
\end{equation}
where $\delta:=\frac{1}{2}(\lambda^2-\sigma/4)$ is the index of the spin$^c$ Dirac operator. We focus on $M_4$ that are simply connected, for which we have $\chi=2+b_2^++b_2^-$ and $\sigma=b_2^+-b_2^-$.\footnote{Another widely used pair consists of the following combinations 
\begin{equation}
\label{eq:M4data}    \int_{M_4}c_1^2=2\chi(M_4)+3\sigma(M_4), \qquad \int_{M_4} c_2=\chi(M_4),
\end{equation}
which are identified with the first and top Chern classes when $M_4$ is a complex surface. Later, we often use the pair $(\chi_h,\sigma)$.
}

In addition, for a given $M_4$, we have the following relationship regarding the Seiberg--Witten invariants \cite{Witten:1994cg}
\begin{equation}
\label{eq:SW symmetries}
    \text{SW}(-\lambda)=(-1)^{\chi_h}\text{SW}(\lambda)\, .
\end{equation}
Using this relation, we can simplify the sum over spin$^c$ structures by first averaging between $\lambda$ and $-\lambda$, as the coefficients for them are almost identical given that only $\lambda^2$ appears, except for a possible sign from $(-1)^{\lambda\cdot w_2}$ in the non-spin case.

Now, for the E-string theory, we can express the Higgs branch contribution to the partition function shown in \eqref{eq:Zsw} in the following explicit form
\begin{multline}
\label{eq:finalSWpartition}
    Z_{\rm Higgs}[M_4\times T^2; \sigma,\{m_i\}]=\alpha^{\chi}\beta^{\sigma}2^{(\chi+5\sigma)/8}3^{(\chi+5\sigma)/4}\,i^{(\sigma-\chi)/4}\eta^{-3\sigma}\\
    \times \sum_s\iota_s\cdot\sum_{\substack{\lambda\in \text{spin}^c(M_4)}}(-1)^{n(\lambda)+\lambda\cdot w_2}\cdot\left[\left(\frac{f(u_s)}{g(u_s)}\right)^{\frac{\chi_h-1}{2}}\left(\frac{f}{g}\right)^{\frac{1-2\chi_h-\sigma}{2}}\left(\mu_1\check{E}_1(q_s)\right)^{\chi_h-1}\right.\\
    \times \left(\frac{\eta(\tau_s)^{24}}{q_s}\right)^{\sigma/8}\left.\left(\sum_{l\geq 1}l\mu_l q_s^{l-1}\right)
    \left(\frac{E_6(\tau_s)}{E_4(\tau_s)}\right)^{(1-2\chi_h-\sigma)/2}\right]_{q_s^{n(\lambda)}}\SW(\lambda)\, ,
\end{multline}
where the function $\check{E}_1(\tau_s)$ is given explicitly in \eqref{eq:checkE1} as a power series in $q_s$ and $f/g$ is evaluated at $u(q_s)$. 
The above expression depends on the underlying four-manifold via the summation over spin$^c$ structures $\lambda$, the Seiberg--Witten invariants SW$(\lambda)$, and the topological data $\sigma(M_4)$ and $\chi(M_4)$.

\subsubsection{The $(-1)^{\lambda\cdot w_2}$ phase factor and homological orientation}\label{sec:Pontryagin}

We have now seen the phase factor $(-1)^{\lambda\cdot w_2}$ enough times to justify a slightly more detailed discussion than it has so far received. 

First recall that it is $\Z_4$-valued because $\lambda\cdot w_2\in\frac12 \Z$. Viewing $\lambda$ as $\frac{1}2 w_2 +\tilde \lambda$ with $\tilde\lambda\in H^2(M_4,\Z)$, the phase $(-1)^{\lambda\cdot w_2}=(-1)^{w_2^2/2}\cdot (-1)^{w_2\cdot\tilde\lambda}$ consists of two parts. While the second factor is unambiguous, the first should be interpreted as $e^{\frac{\pi i}2  \CP(w_2)}$, where $\CP(w_2)$ is the Pontryagin square of $w_2$.\footnote{Notice that the other option is $e^{-\frac{\pi i}2  \CP(w_2)}$. However, this is not a choice to be made. Instead, once we fix all the relevant conventions, this should be determined by performing a computation via the path integral of the topologically twisted theory similar to \cite[Sec.~3]{Moore:1997pc}.} This first factor can be further replaced by $i^{\sigma}$ due to the relation
\begin{equation}
    \CP(w_2)\equiv p_1+2w_4 \equiv \sigma \pmod 4\,.
\end{equation}
This is analogous to the Spin-$SU(2)$ version of Donaldson theory which we explain first in a bit more detail. 

In the $SU(2)$ Donaldson theory, such a phase is trivial, while in the $SO(3)$ version, one will need to make a choice to fix the value of such a phase when $M_4$ is non-spin. This is part of the ``homological orientation'' and is needed to give an orientation of the instanton moduli space in the non-spin case (for a detailed discussion, see \cite[Sec.~7.1.6]{donaldson1997geometry}). It corresponds to an integral lift of $w_2(SO(3)_{\text{gauge}})$. For the Spin-$SU(2)$ theory, there is a similar phase, except one doesn't need to choose a lift, as it would necessarily agree with the integral lift of $w_2(M_4)$ and the phase is just determined by the Pontryagin square of $w_2(M_4)$ together with an unambiguous part. 

The partition function of the E-string theory, in this aspect, is analogous to the Spin-$SU(2)$ version of the Donaldson theory---and in fact expected to reduce to it in a certain limit. However, there is still a choice of homological orientation on $H^0\oplus H^1\oplus H^2_+$, which is needed to fix the sign of the Seiberg--Witten invariants. Changing such an orientation would flip the overall sign of the E-string theory partition function. From the 2d $(0,1)$ theory side, we expect that such an orientation arises when we want to fix the ``vacuum parity'' of $\CT[M_4]$, with different choices differing by an ``Arf theory,'' whose effect is to flip bosonic states to fermionic and vice versa.  Although it would be interesting to understand the physics interpretation of this choice better (e.g.~as a result of flipping masses of fermions in the 2d theory), in this paper, we will pay less attention to the overall minus sign. Instead, we will make use of the following two important properties of the phase $(-1)^{\lambda\cdot w_2}$:
\begin{itemize}
    \item When $\sigma$ is odd, the phase factor $\sim \pm i$ is imaginary;
    \item There can be a relative phase between $\lambda$ and $-\lambda$, given by $(-1)^{2\lambda\cdot w_2}=(-1)^{\sigma}$. 
\end{itemize}
     They will play interesting roles later in our analysis in Section~\ref{sec:examples}, where they respectively ensure the reality and modularity of the partition function when the signature $\sigma$ is odd.

\subsection{Full topological invariance}\label{sec:FullTop}

As discussed previously, the fact that the E-string theory is half of the $E_8\times E_8$ theory suggests that its partition function might be fully topological, but there are at least two subtleties pointed out in Footnote~\ref{foot:E8E8}.

In this section, we will analyze this issue more directly and see whether/when the contribution of $u=\infty$ to the metric dependence of the E-string partition function vanishes. When this happens, one can have a metric-independent combination,
\begin{equation}
    Z=Z_{\rm Coulomb}+\sum_\lambda C_\lambda \cdot \SW(\lambda),
\end{equation}
even though the individual terms on the right-hand side have metric dependence.\footnote{In the $b_2^+=0$ case, the usual Seiberg--Witten invariant cannot be defined as there will be reducible solutions for any metric and perturbation. While the right way to deal with the reducible solutions should be encoded in the behavior of the $u$-plane integral near the corresponding I$_1$ singularity, one possibility is to still count only irreducible solutions similar to the approach taken in~\cite{OkonekTeleman+2000+347+358}. } We now analyze the contribution at $u=\infty$ in more detail, focusing on the $b_2^+=1$ case. 

The analysis can be carried out either within the E-string theory at $u\to \infty$, or by embedding it into the $E_8\times E_8$ theory in the large $\rho$ limit, where the Seiberg--Witten geometry for the 4d effective theory with unbroken $E_8\times E_8$ is given by \cite{Morrison:1996pp}
\begin{equation}
    \Sigma_{\rm full}:\qquad y^2=x^3-3\rho^4 u^4E_4(\sigma)x+u^5-2\rho^6 E_6(\sigma) u^6+u^7.
\end{equation}
The two II$^*$ singularities are at $u=0$ and $\infty$, and two of the four I$_1$'s are at distance $\sim \rho^{-6}$ from 0, while the other two are far away at distance $\sim \rho^6$. In the intermediate region $\rho^{-6}\ll|u|\ll\rho^6$, the geometry looks like a long cylinder, and cutting it open along the $|u|=1$ circle gives two copies of the E-string theory. 

A $\Z_2$ global symmetry of the 6d theory from exchanging the two M9-branes now becomes the action $u\mapsto 1/u$ on the moduli space, which also scales $x$ and $y$ as
\begin{equation}
    x\mapsto u^{-4} x,\quad y\mapsto u^{-6} y
\end{equation}
to leave the SW curve invariant. The differential $\dd x/y$---as well as its period $\frac{\dd a}{\dd u}$---then scales by a $u^2$ factor. This ensures that  $a$ (up to a sign), the Seiberg--Witten differential $\lambda_{\SW}$, the prepotential $\CF$ and physical quantities computed from them remain unchanged.

However, as the embedding of two copies might not always be supersymmetric, we will instead work with a single copy of the E-string theory, for which the metric dependence can be analyzed directly on the $u$-plane. 

At $b_2^+=1$ the integrand of~\eqref{eq:donaldson-invariants} is a total derivative, $(\Im\tau)^{-1/2}\Psi=\partial_{\bar\tau}\widehat{G}$, with $\widehat{G}(\tau,\bar\tau)$ the modular completion of a mock modular form $G(\tau)$ (see~\cite{Moore:1997pc,Korpas:2019cwg} for its construction). Stokes' theorem reduces the $u$-plane integral to boundary terms,
\begin{equation}
\label{eq:uplaneStokes}
    Z_{\rm Coulomb}=-\pi\mathcal{N}\Big[\operatorname*{Res}_{u=\infty}\big(\nu\,\widehat{G}\,\dd u\big)+\sum_s\operatorname*{Res}_{u=u_s}\big(\nu\,G\,\dd u\big)\Big]\,,\qquad \nu=\left(\frac{\dd a}{\dd u}\right)^{1-\frac{\chi}{2}}\left(\frac{\Delta_\Sigma}{D_0}\right)^{\frac{\sigma}{8}}\,,
\end{equation}
where $\mathcal{N}$ collects the constants of the measure. The wall crossing of terms associated to the I$_1$ singularities  cancels that of $\sum_\lambda C_\lambda\cdot\SW(\lambda)$ as for other $\CN=2$ theories~\cite{Moore:1997pc}. 

At the cylindrical end, as $u\to\infty$, $\tau(u)\to\sigma$ stays in the interior of the upper half-plane, and $\widehat{G}$ does not reduce to its holomorphic part. On a large circle $|u|=R\to\infty$ only the terms that are holomorphic in $u$ at fixed $\bar\tau=\bar\sigma$ survive. With $v=1/u$, $\frac{\dd a}{\dd u}=\frac{i}{2\pi}\,\omega$ and $u\,\omega=2\pi+\CO(v)$, one finds that
\begin{equation}
\label{eq:Binf}
    B_\infty:=\operatorname*{Res}_{u=\infty}\big(\nu\,\widehat{G}\,\dd u\big)\;\propto\;\Big[v^{(\chi+3\sigma)/2}\Big]\left\{(u\,\omega)^{1-\frac{\chi}{2}-\frac{3\sigma}{2}}\,\eta\big(\tau(v)\big)^{3\sigma}\,\widehat{G}\big(\tau(v),\bar\sigma\big)\right\}\,
\end{equation}
is given by the coefficient of $v^{(\chi+3\sigma)/2}$ of the power series in $v$. 

Thus the end of the cylinder contributes through the first $(\chi+3\sigma)/2$ orders of the expansion around $u=\infty$, and $B_\infty=0$ whenever $\chi+3\sigma<0$. This is because the fiber at $u=\infty$ is smooth, and the expression in braces is a power series in $v$ without negative powers.

For $b_1=0$ and $b_2^+=1$ one has $\chi+3\sigma=4+2\sigma$. When $\sigma$ is even, the phase $(-1)^{\lambda\cdot w_2}$ takes the same value on $\lambda$ and $-\lambda$, so $\Psi$ is odd under $\lambda\to-\lambda$ and $Z_{\rm Coulomb}=0$ identically. Hence the contribution of $u=\infty$ can only be non-zero for $\sigma=1$ (e.g.~$\mathbb{CP}^2$) and for $\sigma=-1$ (e.g.~$\mathbb{CP}^2\#2\overline{\mathbb{CP}}{}^2$). In both cases it is found to be non-zero (and for $\sigma=-1$ with continuous metric dependence via the period point) but will not be discussed in detail here.

This is consistent with the topological invariance of the $E_8\times E_8$ theory, due to the two subtleties of Footnote~\ref{foot:E8E8}. The two cases with $B_\infty\neq0$ have $\chi+3\sigma=6$ and $2$, so they admit no supersymmetric $E_8\times E_8$ background, which needs $\chi+3\sigma\le0$. Moreover, at $b_2^+=1$ and $b_1=0$ one has $\chi_h=1$, so the $\Z_2$ symmetry of the $E_8\times E_8$ theory introduces an additional minus sign for the $u$-plane integrand, causing the boundary terms of the two halves along $|u|=1$ to cancel, whatever the value of $B_\infty$. 

Conversely, whenever the embedding exists, we indeed have $B_\infty=0$: for $\chi+3\sigma<0$ by the counting above, and for $\chi+3\sigma=0$, i.e.~$\sigma=-2$, because $\sigma$ is even. In summary, for $b_1=0$, the full partition function is metric independent except possibly for the two special cases above.

It would be interesting to perform this computation more systematically and extend it to the following cases.
\begin{itemize}
    \item $b_2^+=0.$ This would be the starting point for constructing a potential new invariant obtained by canceling the pathologies of the Seiberg--Witten invariants with the $u$-plane integral to obtain a genuine smooth invariant for such 4-manifolds. This also requires extending the analysis around the I$_1$ singularities.
        \item Beyond the leading order in $1/\rho$. This would serve as a check of the metric-independence of the partition function of the full $E_8\times E_8$ theory. In particular, there should be no metric dependence to all orders of $1/\rho$, while for the E-string theory one still expects it to be only non-zero for $\sigma=\pm 1$ when $b_1=0$.
    \item Other 6d theories. It would be interesting to understand the $u$-plane integral for other 6d theories whose 4d KK theories have a Coulomb branch with a cylindrical end (or, in other words, a ``pure tensor branch'' in 6d) and examine whether some of our findings above are universal features of such theories.
\item New perspectives on 4d theories. As a mass deformation is unlikely to affect the analysis at $u=\infty$ of the E-string theory, there can be interesting implications for various 4d theories at different special values of masses. One of the most intriguing is perhaps the $D_4\oplus D_4$ point, whose existence might suggest that two (or four) copies of the 4d $SU(2)$ $N_f=4$ theory can be combined in a non-trivial way, together with a $u$-plane integral over the region between them, to achieve topological invariance.  
\end{itemize}
We will leave these interesting questions for future work.

\subsection{Modularity}

As explained in Section~\ref{sec:modularity}, the partition function is expected to be a modular form, transforming as $Z(-1/\sigma)=\sigma^{d/2}\cdot Z(\sigma)$, due to gravitational anomalies of the E-string theory. In what follows, we will examine the modularity of the partition function, with the aim of demonstrating that it possesses the expected modular weight.

Let us take this opportunity to stress again the difference between the two modular parameters: $\sigma$ and $\tau_{s}$. Note that $\tau_s$ is a local parameter around a singularity, therefore $Z_{\text{Higgs},s}$ is independent of $\tau_s$, which is not well-defined globally, as is evident from the operation $[\cdot ]_{q_s^0}$. On the other hand, as $\sigma$ is the complex structure of the spacetime torus---a UV parameter from the point of view of the 4d KK theory---we do expect the partition function to be modular with respect to $\sigma$, and, as a prerequisite, we have seen the location of singularities exhibit particular modular behaviors in Section~\ref{sec:modularity}. 

To see the full modularity of the partition function with respect to $\sigma$, we need to compute the entire partition function by summing up different contributions. On the other hand, although in general the individual contribution of a particular singularity cannot be itself an $\SL(2,\Z)$ modular form, it should be one with branch cuts, which can perhaps be alternatively viewed as a component of a vector-valued modular form. As a consequence, one can still talk about its modular weight, and compare it with the expectation from gravitational anomalies. This is what we will do in the remainder of this subsection.

For a single singularity $s$, we will focus on a single term at a time in the summation over spin$^c$ classes $\lambda$. If the coefficients $C_{s,\lambda}$ are all modular with the same weight, the full partition function $Z[M_4\times T^2]$ is guaranteed to be modular---up to branch cuts---in situations where the Coulomb branch itself doesn't contribute. As $C_{s,\lambda}$ is given in \eqref{eq:finalSWpartition} as a product, we need only to examine the $\sigma$-dependence of individual components.

As they are all expressed using the coefficients in the SW curve $\Sigma$, it is straightforward to verify that $C_{s,\lambda}(\sigma,\{m_i\})$ is indeed modular. To check the modular weight, one can work in the massless limit, where the $\sigma$-dependent terms are\footnote{Naively, the pre-factor $\alpha^\chi\beta^\sigma\eta^{-3\sigma}$ can also depend on the modulus parameter $\sigma$, but this turns out to be a constant independent of $\sigma$ and, therefore, doesn't affect the modular weight. See Section~\ref{sec:effective gravitational couplings} for the determination of $\alpha$ and $\beta$.}
\begin{multline}
\label{eq:masslessPartitionFunction}
    C_{s,\lambda}(\sigma,\mathbf{0})\sim
     \left[\left(\frac{f(u_s)}{g(u_s)}\right)^{\frac{\chi_h-1}{2}}\left(\frac f g\right)^{\frac{1-2\chi_h-\sigma}{2}}\left(\mu_1\check{E}_1(q_s)\right)^{\chi_h-1}\left(\frac{\eta(\tau_s)^{24}}{q_s}\right)^{\sigma/8}\right.\\
    \times \left(\sum_l l\mu_lq_s^{l-1}\right)\left.\left(\frac{E_6(\tau_s)}{E_4(\tau_s)}\right)^{(1-2\chi_h-\sigma)/2}\right]_{q_s^{n(\lambda)}}\,.
\end{multline}
 The modular weights (and indices, which will be relevant momentarily) of various ingredients are listed below (see Appendix~\ref{sec:modularity of coefficients} for more details):
\begin{center}
\begin{tabular}{c|c|c|c|c|c}
    function & $E_4$ & $E_6$ & $\check{E}_1$ & $f/g$ & $\mu_l$\\
    \hline
    weight & $4$ & $6$ & 0 & $10$ & $-6$\\
    \hline
    index & 0 & 0 & 0 & $-2$ & $1$
\end{tabular}
\end{center}
Notice that $\tau_s$-dependent factors such as $E_{4,6}(\tau_s)$, $q_s$ and $\eta(\tau_s)$ have no $\sigma$ dependence, and won't contribute to the modular weight of $C_{s,\lambda}$. In the end, we find the modular weight of $C_{s,\lambda}$ to be $-(11\chi+31\sigma)/4$, which is indeed consistent with the prediction from gravitational anomalies. 

As mentioned above, there is a caveat in sending all the mass parameters to 0, when 10 of the 12 I$_1$ singularities collide to form the II$^*$ singularity. In this limit, depending on the four-manifold, the partition function may diverge due to the emergence of a Higgs branch. We will return to this point shortly in Section~\ref{sec:massless}, but, nevertheless, the modular weight found is still correct, which can be confirmed with a computation using generic masses.

We will now restore the mass parameters so that we can also check that the partition function has the correct index compatible with the anomaly of the global $E_8$ symmetry. The partition function is now expected to be an integral meromorphic $E_8$ Jacobi form of weight $\frac d2$ and index $k$,
\begin{equation*}
    Z[M_4\times T^2;\sigma,\{m_i\}]\in  \mathrm{MF}^{E_8,\Z}_{\frac d2,k}.
\end{equation*}
See \cite{Sakai:2011xg,Sakai:2024vby} for a description of the ring structure of the $E_8$ Jacobi forms with a list of generators. The partition function is also expected to be an ``integral Jacobi form'' with an integral $q$-series expansion, in the $E_8$-equivariant sense (i.e.~coefficients are $\Z$-linear combinations of irreducible $E_8$ characters).

Consider again the coefficients $C_{s,\lambda}$ studied above. With masses turned on, the formula for them takes exactly the same form, with the only meaningful difference being that various factors are now Jacobi forms. Among them, only $f/g$ and $\mu_l$ have non-trivial indices. In the end, we find that $C_{s,\lambda}$ is an $E_8$ Jacobi form---again with branch cuts---of weight $\frac d2=-\frac{11\chi+31\sigma}4$ and index $k=2\chi_h+\sigma$, which is in perfect agreement with anomalies of $\CT[M_4]$. 

 The prediction from the 6d theory is that, once we sum over $s$ and $\lambda$ (and including also the $u$-plane integral $Z_{\rm Coulomb}$ if non-vanishing), the full partition function is modular under the entire $\SL(2,\Z)$. This amounts to requiring that all the branch cuts cancel and the different $C_{s,\lambda}$, individually being components of a certain vector-valued Jacobi form, are assembled into a genuine Jacobi form. And under the action of any element $\begin{pmatrix}a&b\\c&d\end{pmatrix}\in \SL(2,\mathbb{Z})$ or $p,q\in\Gamma_{E_8}$, the following two equalities hold,
\begin{align}
    Z\left(\frac{a\sigma+b}{c\sigma+d},\frac{\{m_i\}}{c\sigma+d}\right)&=(c\sigma+d)^{d/2}\exp\left(\frac{ik}{4\pi }\cdot\frac{c\sum m_i^2}{c\sigma+d}\right)Z(\sigma,\{m_i\})\, ,\\
    Z\left(\sigma,\{m_i\}+2\pi\{p_i\}+2\pi\sigma\{q_i\}\right)&=\exp\left[-ik\left(\pi\sigma\sum q_i^2+\sum m_iq_i\right)\right]Z(\sigma,\{m_i\})\,.
\end{align}

It turns out that this ``full modularity'' requires a collection of properties of the Seiberg--Witten invariants, which would be very interesting to better understand. We will not carry out a systematic investigation about the condition for the full modularity in the present work, but will discuss this in examples in Section~\ref{sec:examples} in the massless limit.

\subsection{Convergence in the massless limit}
\label{sec:massless}

We have seen previously that the type II${}^*$ singularity can be smoothly deformed into 10 type I${}_1$ singularities. However, the partition function can diverge in the massless limit (or in limits approaching other special values of $\{m_i\}$). We now analyze the condition on the four-manifold such that this limit is finite. 

From the point of view of the Coulomb branch, the potential divergence is related to the collision of singularities that allows different species of BPS particles to become massless simultaneously. Therefore, one way of studying the divergence is to first work with a set of generic holonomy parameters $\{m_i\}$, and then consider the limit $m_i\rightarrow 0$ to extract the leading behavior.\footnote{This is similar to how the contribution of the Argyres--Douglas point is analyzed in \cite{Marino:1998tb}. See also \cite{Dedushenko:2017tdw,Aspman:2022sfj,Aspman:2023ate} for more recent studies of the behavior of the Donaldson--Witten partition function when singularities collide.} This would be a natural starting point for the investigation of the full $E_8$-equivariant partition function, which we will leave for future work. 

Another approach, which is arguably simpler and what we will pursue here, is to count the number of zero modes that can either cause or attenuate the divergence. For the limit where all masses tend to zero, there are two relevant sets of the zero modes, one from the $E_8$ SCFT, another from the Argyres--Douglas theory at the special value of $\sigma(\CC)=e^{\pi i/3}$. The former is captured by the vacuum R-charge of the $E_8$ theory, given by
\begin{equation}
    \Delta_R=\frac{11\chi+31\sigma}{2}.
\end{equation}
With a minus sign, $-\Delta_R$ counts the effective number of chiral multiplets minus fermi multiplets in $\CT[M_4]$, which can be thought of as the real dimension of the moduli space of the $E_8$ SCFT on $M_4$. Therefore, its partition function is expected to behave as 
\begin{equation}
    Z_{E_8}\sim m^{\frac{\Delta_R}{2}}
\end{equation}
to the leading order in $m$, with a zero (or pole if the power is negative) of order $\frac{\Delta_R}{2}$. Here, the $\frac{1}{2}$ factor is to take into account the fact that each \emph{complex} non-compact direction of the moduli space contributes a factor $\sim m^{-1}$ from the point of view of equivariant localization.

On the other hand, when $\sigma(\CC)=e^{i\pi/3}$, the two I$_1$ singularities collide, forming an Argyres--Douglas point. The number of zero modes is proportional to $7\chi+11\sigma$, giving a potential divergence for the partition function. As this is a compact theory with the Higgs branch being a point, the partition function should not actually diverge. However, this is implemented by the sum rule of the Seiberg--Witten invariants \cite{Marino:1998tb}, and we still expect $C_\lambda$ to diverge.

Therefore, for the massless limit of the E-string partition function, the $11\chi+31\sigma=0$ and $7\chi+11\sigma=0$ lines divide the geography of simply-connected 4-manifolds (hence positive $\chi$ and $\chi_h$) into three regions:
\begin{itemize}
    \item The ``convergent region'' with $11\chi+31\sigma>0$. There, the partition function of the E-string theory will have a well-defined massless limit,
\begin{equation}
    \lim_{\{m_i\}\to 0}Z[M_4\times T^2;\sigma,\{m_i\}]=Z_{\rm Higgs,II^*}+Z_{\rm Higgs,I_1,+}+Z_{\rm Higgs,I_1,-}+Z_{\rm Coulomb}\,.
\end{equation}
In particular, when furthermore $b_2^+>1$, only the middle two terms are non-vanishing, greatly simplifying our analysis.
\item The ``strongly divergent'' region with $7\chi+11\sigma<0$.\footnote{It is a slight misnomer to call it ``strongly divergent,'' as in fact the full partition functions with $E_8$ holonomies are expected to be zero in most of this region. This is expected physically because the full partition function should not have a pole at $\sigma(\CC)=e^{\pi i /3}$, and when $C_\lambda$ (with generic $E_8$ holonomies turned on) does have such a pole, the sum of the Seiberg--Witten invariants has to be zero. This turns out to happen for $7\chi+11\sigma\le -16$---exactly when the sum rule of \cite{Marino:1998tb,Feehan:2014gea} holds. Above the sum-rule region, there is in fact not much difference with the weakly divergent region. 
} In this region, $C_{\lambda}$ has two types of divergences, and one cannot take the massless limit. Although the contribution from the two I$_1$ singularities $Z_{\rm Higgs,I_1,+}+Z_{\rm Higgs,I_1,-}$ stays finite, one might not expect them to be either integral or modular. However, we observe in all examples tested that the contribution of the I$_1$ singularities to $C_\lambda$, though generically non-integral, is still modular, but possibly with negative powers of $E_4$, consistent with the predicted divergence at $\sigma=e^{\pi i/3 }$. One conceptual explanation of this modularity is the following. Under the action of SL$(2,\Z)$ the II$^*$ singularity is invariant by itself while the two I$_1$ singularities are braided (cf.~Figures~\ref{fig:Tsingularity} and~\ref{fig:Ssingularity}).
Therefore, the two orbits are expected to be separately modular.
\item The ``weakly divergent'' (or ``weakly convergent,'' depending on the point of view) region with $-\frac{7}{11}\chi\le \sigma \le -\frac{11}{31}\chi$. This is the transition zone between the first two regions. There, the contribution of the $E_8$ singularity still diverges in the massless limit,\footnote{On the $11\chi+31\sigma=0$ line, the situation is slightly different, as the value of the partition function can be made finite but is ambiguous (cf.~Footnote~\ref{foot:masslesslimit}). } but the contributions of the two I$_1$ singularities are expected to be modular but not necessarily integral, and now have non-negative powers of $E_4$, with the divergence at $\sigma=e^{\pi i/3}$ gone.

\end{itemize}
The analysis above might seem somewhat crude, but we will support it by demonstrating in Section~\ref{sec:examples} that there indeed exist three qualitatively different regions.

\begin{figure}[tb!]
    \centering
    \includegraphics[width=0.75\linewidth]{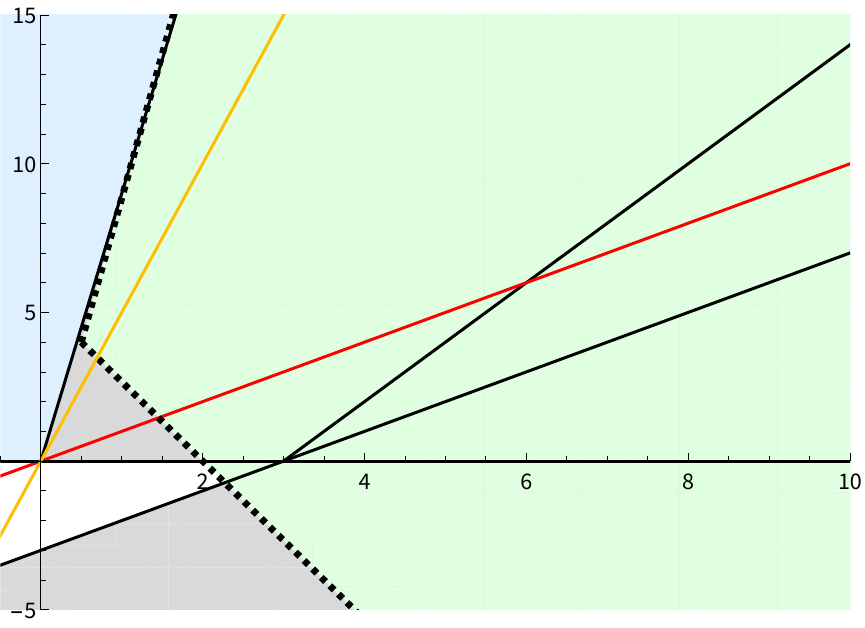}
    \begin{picture}(0,0)\vspace*{-1.2cm}
    \put(-355,280){$c_1^2$}
    \put(10,70){$\chi_h$}
    \put(-295,280){$9\chi_h$}
    \put(-264,280){\textcolor{orange}{$5.8\chi_h$}}
    \put(10,200){\textcolor{red}{$c_1^2=\chi_h$}}
    \put(10,250){$2\chi_h-6$}
    \put(10,160){$\chi_h-3$}
    \put(-215,-10){$11/8$}
    \put(-230,200){\footnotesize minimal surfaces}
    \put(-225,190){\footnotesize of general type}
    \put(-100,190){\rotatebox{38}{\footnotesize \!\!\!Noether inequality}}
    \put(-350,100){\rotatebox{73}{\footnotesize Bogomolov--Miyaoka--Yau}}
    \put(-330,20){\footnotesize MMP sum rules}
    \put(-284,67){$\bullet$}
    \put(-280,75){\footnotesize K3}
    \put(-335,117){$\bullet$}
    \put(-327,124){\footnotesize $S^4$}
    \end{picture}\vspace*{0.3cm}
    \caption{\textbf{Comparing the critical lines of the E-string theory to the geography of 4-manifolds.}  The orange line indicates the bound $c_1^2=5.8\chi_h$, above which the contribution of the $E_8$ SCFT to the partition function of the E-string theory has a well-defined massless limit. Below the orange line is the region where such a limit is expected to diverge.  The red line $c_1^2=\chi_h$ (or equivalently $7\chi+11\sigma=0$) is parallel to the boundary line of the region of SW sum rules related to the Argyres--Douglas theory studied in \cite{Marino:1998tb}, but with a different intercept. The 3/2-conjecture for irreducible manifolds is $c_1^2\ge0$, which is represented in the figure by the $\chi_h$ axis and the four-manifolds that satisfy this inequality are within the blue shaded region. Additionally the $11/8$ conjecture for simply-connected spin four-manifolds is shown where the allowed $(\chi_h,c_1^2)$'s live in the green shaded region between the two dashed lines.
    }
    \label{fig:geography}
\end{figure}
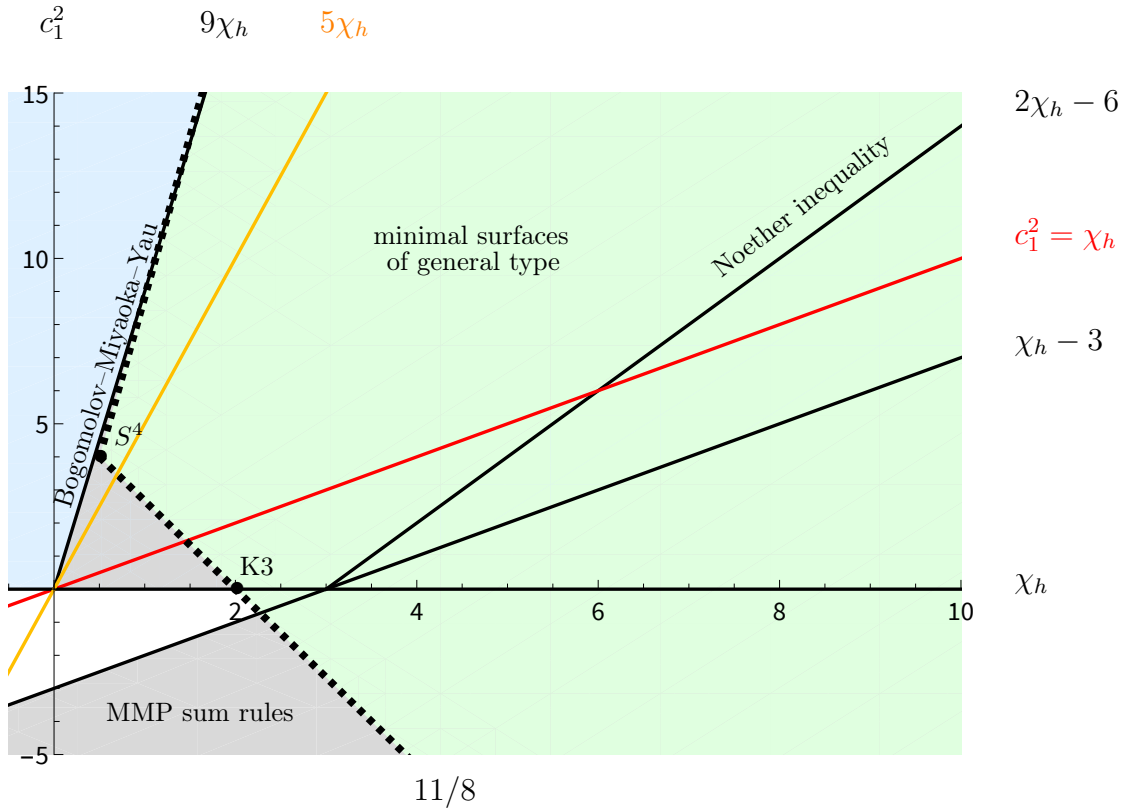

It is often customary to use $c_1^2=2\chi+3\sigma$ and $\chi_h$ when discussing the geography of 4-manifolds. Then the two bounds $7\chi+11\sigma\geq0$ and $11\chi+31\sigma\geq0$
correspond, respectively, to the following inequalities,
\begin{equation}
\label{eq:new topological bound}
    c_1^2\geq \chi_h\,\quad\text{and} \quad c_1^2\geq 5.8\,\chi_h,
\end{equation}
which are illustrated in Figure~\ref{fig:geography} where we compare them to other special regions in the geography of 4-manifolds. In fact, most minimal surfaces of general type are in the weakly convergent region. The bound $\chi\ge -\frac{11}{7}\sigma$ might appear to be close to the bound given by either the 11/8-conjecture for spin 4-manifolds~\cite{Matsumoto:1982} or the 3/2-conjecture for irreducible 4-manifolds. However, in the $c_1^2$--$\chi_h$ plane, their slopes are positive, negative, and zero, respectively.\footnote{Here, the $11/8$ conjecture $b_2(M_4)\geq\frac{11}{8}|\sigma(M_4)|$ can be re-expressed as $12\chi_h-c_1^2-2+2b_1\geq \frac{11}{8}|c_1^2-8\chi_h|$. The lines corresponding to both signs are drawn in Figure~\ref{fig:geography}. One branch appears to be close to the Bogomolov--Miyaoka--Yau (BMY) bound~\cite{Miyaoka1977Chern,Yau1977Calabi,Yau1978Ricci}, but in fact they have different slopes ($\frac{8\cdot 23}{19}\approx9.68$ vs 9) and intercepts.}

\subsection{Effective gravitational couplings}
\label{sec:effective gravitational couplings}

To completely fix the normalization of the partition function, one needs to fix topological terms such as $\frac{b(u)}{32\pi^2}\cdot\tr R\wedge \widetilde{R}$ and $\frac{c(u)} {24\pi^2}\cdot\tr R\wedge R$ in the action of the low-energy effective theory, whose contributions to the partition function are incorporated by the $u$-plane measure $\mu(u;\sigma,\{m_i\})$. They are highly constrained and, explicitly, contribute a factor $e^{\chi\cdot b(u)}\cdot e^{\sigma\cdot c(u)}$ with~\cite{Witten:1995gf,Moore:1997pc}\footnote{Recall the following formulas for the Euler characteristic and the signature for a four-manifold $M_4$
\begin{equation*}
    \chi(M_4)=\int_{M_4} e(M_4)=\frac{1}{32\pi^2}\int_{M_4}\tr R\wedge \widetilde{R}\,,\quad \sigma(M_4)=\frac{1}{3}\int_{M_4}p_1(M_4)=\frac{1}{24\pi^2}\int_{M_4}\tr R\wedge R\,,
\end{equation*}
where $e(M_4)$ denotes the Euler class of $M_4$.} 
\begin{equation*}
    e^{b(u)}\sim \left(\frac{\dd u}{\dd a_s}\right)^{1/2}\,,\qquad e^{c(u)}\sim \Delta^{1/8}\, .
\end{equation*}
The proportional constants in the two relations above are meaningful, as they lead to an overall factor $\alpha^{\chi}\beta^\sigma$ for the partition function. In general, they could depend on UV parameters (such as $\sigma$ and $m_i$'s in our case of the E-string theory), and fixing them is the prerequisite for determining the integrality and (topological) modularity of the partition function.

For a 4d theory, the coefficients have some intrinsic ambiguities as one can always add certain local topological terms in the action. However, for a 6d theory, such ambiguities are absent. In particular, for the E-string theory, the partition function on $ M_4 \times T^2$ counts BPS states in string theory or M-theory with certain backgrounds and, as a consequence, has a canonical normalization. 

Therefore, one strategy for fixing $\alpha$ and $\beta$ is to compare our 4d computation with genuinely six-dimensional computations in some cases. One such case, which is arguably the best understood, is when $M_4$ is not a compact 4-manifold but instead $\R^4$ with an omega background \cite{Nekrasov:2002qd}.
In other words, we use the Nekrasov partition function whose $\varepsilon$-expansion gives~\cite{Manschot:2019pog}\footnote{We have omitted the linear term in $\varepsilon_1,\varepsilon_2$ on the RHS in~\eqref{eq:nekrasov expansion} as it vanishes for the E-string theory similar to most cases studied in~\cite{Manschot:2019pog}. See also \cite{Closset:2021lhd} for similar analysis for the KK-reductions of 5d $\CN=1$ SCFTs.}
\begin{equation}
\label{eq:nekrasov expansion}
    \varepsilon_1\varepsilon_2\log Z(u)=-\mathcal{F}(u)+\varepsilon_1\varepsilon_2 b(u)+\frac{\varepsilon_1^2+\varepsilon_2^2}{3}c(u)+\mathcal{O}(\varepsilon^3)\,.
\end{equation}
 Here $\varepsilon_{1,2}$ are the deformation parameters characterizing the $\Omega$-background, with $\varepsilon_1\varepsilon_2$ and $\frac{\varepsilon_1^2+\varepsilon_2^2}{3}$ being the ``$\chi$ and $\sigma$'' of $\R^4_{\varepsilon_1,\varepsilon_2}$, and $\mathcal{F}$ is the prepotential.  For the E-string theory, the Nekrasov partition function has been computed in~\cite{Minahan:1998vr,Kim:2014dza,Sakai:2014hsa} and has the following structure,
\begin{equation}
    \varepsilon_1\varepsilon_2\log Z=\sum_{n,g\geq 0}(\varepsilon_1+\varepsilon_2)^{2n}(\varepsilon_1\varepsilon_2)^gF^{(n,g)}(\sigma,\{m_i\};u)\,,
\end{equation}
where $F^{(n,g)}$ denotes the topological string amplitudes on $\frac{1}{2}$K3. Then, we can make the following identification 
\begin{align}
\mathcal{F}&=-F^{(0,0)}\,,\\
    b(u)&=F^{(0,1)}+2F^{(1,0)}\,,\\
    c(u)&=3F^{(1,0)}\,.
\end{align}
Before carrying out this computation, we first address a related issue that naively leads to an ambiguity in this matching. 

\subsubsection*{Scale and normalization independence}

As the E-string theory is superconformal, the partition function is expected to be independent of the overall size of $M_4\times T^2$. The independence on the relative size between the two factors is evident from the point of view of the elliptic genus of $\CT[M_4]$, which can only depend on the shape of the $T^2$ but not on either the area of $T^2$ or a mass scale in the theory determined by the size of $M_4$. 

This scale independence might seem non-trivial from the point of view of the 4d low-energy effective theory described by its SW geometry, as the computation of its partition function is somewhat convoluted. However, as $A_{T^2}$ is the only dimensionful UV parameter, all the dimensionful quantities (such as $a$, $a_D$, $u$, $\lambda_{\SW}$, and $\CF$) in the theory scale with $A_{T^2}$, and one can work with dimensionless combinations formed out of them and powers of $A_{T^2}$. This can effectively be achieved by setting $A_{T^2}=1$, while scaling behaviors of various quantities in the theory can be readily restored by looking at their mass dimensions. In particular, the partition function is expected to be dimensionless and hence independent of $A_{T^2}$. Due to the topological twist, there is also no dependence on the metric on $M_4$, and we arrive at the same conclusion that the partition function of the 6d theory on $M_4\times T^2$ only depends on the modulus $q$ of $T^2$ (as well as backgrounds for $E_8$ if turned on).

The reader might worry about another issue which is closely related: the partition function, e.g.,~given by \eqref{eq:finalSWpartition}, apparently depends on a normalization of the SW curve as it is computed from $f$ and $g$. Namely, consider the following rescaling
\[
f\to \tilde{f}= c^4\cdot f \,,\qquad  g\to \tilde g= {c^6}\cdot  g\,.
\]
This leaves the family of elliptic curves invariant as the $j$-invariant is unchanged under such a re-scaling. However, to keep the physics unchanged, one will need to ensure that the SW differential $\lambda_{\SW}$ stays constant. One can try to keep the Weierstrass form of the SW curve the same by switching to new variables $x' =c^{2}\cdot x$ and $y'=c^3\cdot y$. This tells us that the differential that one should use is actually
\begin{equation}
    \tilde{\frac{d x}{y}}=\frac1c\frac{dx}{y}.
\end{equation}
Taking this into account, we can work out how various quantities in the low-energy effective theory get modified. As expected, all physical quantities, such as the periods $\omega_{1,2}$ and the special coordinates $a$ and $a_D$ are unchanged. From this perspective, there is a canonical normalization for $f$ and $g$, in the sense that the low-energy dynamics is given by the differential $dx/y$ as opposed to a scalar multiple of it. This normalization can be fixed by matching $a(u)$ in the $u\to\infty$ limit. Namely, at the cylindrical end of the moduli space, one wants the limiting behavior to be
\begin{equation}
    \frac{\dd a}{\dd u}\sim \frac{i}{u}
\end{equation}
with the right coefficient compatible with the mass of the KK modes. We use the curve of \cite{Eguchi:2002fc} with this normalization (see Appendix~\ref{app:normalization}).

\subsubsection*{Matching with refined topological strings on local $\frac12$K3}

Now, we are in a position to identify $\alpha$ and $\beta$ and fix the normalization for the partition function of the E-string theory. For this purpose, it is enough to make the comparison in the large-$u$ limit. As the behavior in this limit is independent of the mass parameters, one can set an arbitrary value for them. 

In the large-$u$ limit, we have the following expansions for $b(u),c(u)$,
\begin{align}
\begin{split}
\label{b(u)}
b(u)&=\log\alpha-\frac12\log i+\frac12\log u+\CO(u^{-1})\,,
\end{split}\\
\begin{split}
c(u)&=\log\beta+\frac32\log u+\CO(u^{-1})\,,
\end{split}
\end{align}
where $D_{0}=a_0^3-27b_0^2=\eta^{24}(\sigma)$ is the modular discriminant.

On the side of topological strings, $F^{(1,0)}$ and $F^{(0,1)}$ are given by (see, e.g.,~\cite{Iqbal:2007ii,Klemm:1996hh,Minahan:1997ct,Sakai:2014hsa,Huang:2013yta,Chen:2021ivd,Mohri:2001zz})
\begin{align}
    F^{(0,1)}&=-2\log \eta-\frac{1}{2}\log u+\CO(u^{-1})\,,\\
    F^{(1,0)}&=\log\eta+\frac{1}{2}\log u+\CO(u^{-1})\,.
\end{align}
Combining these leads to the following predictions for the large-$u$ behavior of the gravitational couplings,\footnote{To determine $\alpha$ and $\beta$, one only needs the perturbative part of the Nekrasov partition function. One can include the instanton corrections, and also match the higher order terms. For example, the next-to-leading order terms of $b(u)$ and $c(u)$ can be computed from both the topological string amplitudes and the Seiberg--Witten geometry to be 
\[
b(u)\,:\,\frac{(E_2E_4-E_6)b_1}{96D_0}\frac{1}{u}\,,\qquad c(u)\,:\,-\frac{E_6b_1}{32D_0}\frac{1}{u}\,,
\]
where $b_1$ is the coefficient of $u^5$ in $g$ ($b_1=-4$ in the massless limit~\eqref{eq:E8curve}),
with a perfect agreement between the two approaches.
}
\begin{align}
    b(u)&=\frac12\log u+\CO(u^{-1})\,,\qquad c(u)=3\log\eta+\frac32\log u+\CO(u^{-1})\,.
\end{align}
Thus, by comparing with the expressions for $b(u)$ and $c(u)$ obtained from the Seiberg--Witten geometry, we can fix the coefficients, for the normalization~\eqref{eq:dasdu relation to period}, to be
\begin{equation}
    \alpha=i^{1/2}\,,\qquad \beta=D_0^{1/8}\,.
\end{equation}

However, the matching above determines 
\begin{equation}
    \tilde c = 
    K_0\,\alpha^{\chi}\beta^{\sigma}\eta^{-3\sigma}\,(2\pi\varrho)^{3\sigma/2-d_T/2}\,18^{d_T/4}=K_0\cdot18^{d_T/4}\cdot(2\pi\varrho)^{\chi_h}\,,
\end{equation}
where
\begin{equation}
    d_T:=2(\chi_h+\sigma)\,,
\end{equation}
and the choice of its notation stems from the fact that it equals the gravitational anomaly of the free tensor multiplet when reduced on $M_4$. 

What remains to be determined are the two other constants $K_0$ and $\varrho$. This is carried out in detail in Appendix~\ref{app:normalization}, where we use the central charges of BPS states with non-zero KK momentum to derive $\varrho=i/2\pi$, and the Witten index of the E-string theory to fix $K_0=1$.  

Therefore, in the end we have
\begin{equation}
\label{eq:overall coefficient}
    \tilde c(\chi,\sigma)=i^{\chi_h}\cdot 2^{(\chi+5\sigma)/8}3^{(\chi+5\sigma)/4}=i^{\chi_h}\cdot 18^{(\chi_h+\sigma)/2}\,.
\end{equation}
Notice that the $\eta^{-3\sigma}$ factor cancels the $\eta$-powers in $\beta^\sigma$, making $\tilde c$ a genuine constant with no dependence on the modulus parameter $\sigma(\CC)$. It might appear concerning that the phase factor $i^{\chi_h}$ can be imaginary when $\chi_h$ is odd. However, this phase, when combined with the $(-1)^{\lambda\cdot w_2}$ factor, turns out to be crucial for the reality of the partition function.

The powers of 2 and 3 are also important, and we will see in Section~\ref{sec:examples} that, with this overall factor, the partition function is indeed always an integral $q$-series in cases that we can check, assuming the so-called ``simple-type conjecture'' of 4-manifolds, which we will explain in detail later.

\subsection{Integrality}\label{sec:integrality}

The (normalized) partition function $Z[M_4\times T^2]$ of the 6d theory can be identified with the (normalized) elliptic genus of the 2d theory $\CT[M_4]$, given by~\cite{witten1987elliptic}
\begin{equation}
    Z_{\CT[M_4]}[T^2_{\sigma}]=\eta^{d}\cdot\Tr_R(-1)^Fq^{L_0}\,,
\end{equation}
where the trace is taken over the Ramond sector of the Hilbert space on a circle. The $q$-series of powers of $\eta$ is always integral, and therefore the entire partition function is expected to be an integral $q$-series. In particular, the coefficients in the series carry physical meaning as counting BPS states, either of the 2d theory on $S^1$ or the 6d theory on $M_4\times S^1$. 

This seemingly simple fact turns out to be highly constraining, as it has to be satisfied for all 4-manifolds. Assuming $b_2^+>1$, we have
\begin{equation}
    Z[M_4\times T^2]=\sum_\lambda C_\lambda \cdot\SW(\lambda).
\end{equation}
As $C_\lambda$ contains the factor $18^{(\chi_h+\sigma)/2}$ whose exponent can be chosen to be arbitrarily positive or negative, it seems impossible to have integrality in all cases. The situation becomes even worse once one notices that fractional coefficients of $C_\lambda$'s can't always be cured by having many $\lambda$'s or having really large $\SW(\lambda)$, as, for many $\chi$ and $\sigma$, there are 4-manifolds that have only a pair of $\{\lambda,-\lambda\}$ with non-vanishing SW$=\pm 1$.\footnote{One class of such examples is minimal surfaces of general type, which we will also assume to be simply-connected to apply our formulas. They are in the region above the $\chi_h$ axis between the Noether inequality and the BMY bound illustrated in Figure~\ref{fig:geography}. One can also fill in the region between the Noether inequality and the $7\chi+11\sigma$ line and further extend to the ``half-Noether line'' $c_1^2=\chi_h-3$---also the upper part of the MMP sum-rule line---via the rational blowdown constructions of \cite{fintushel2002rational}, which result in simply-connected symplectic manifolds with one ``basic class.'' And by a theorem of Taubes \cite{Taubes-1994}, the values of the SW invariant for them are again $\pm1$.} Hence it is required that $C_\lambda$ is integral or at least half of an integer, $2C_\lambda\in \Z$, for these values of $\chi$ and $\sigma$. 

In Table~\ref{tab:original grav coef}, we list the leading coefficients of the $q$-series of $C_\lambda$, up to a phase, computed using
\begin{equation}
\label{eq:Ztest}
C_\lambda=\tilde{c}(\chi,\sigma(M_4))\sum_{s}\mu_1^{\chi_h}\left(\frac{f}{g}\right)^{-(\chi_h+\sigma(M_4))/2}\,,
\end{equation}
where we assumed $n(\lambda)=0$ to simplify the expression. This condition is related to the simple-type conjecture, which we will discuss further in Section~\ref{sec:examples}. In the table, we restrict ourselves to $\chi$ and $\sigma$ satisfying
\begin{alignat*}{2}
\text{simply-connected with $b_2^+\ge1$: }&&\quad \chi+\sigma-2&\ge2\,,\\
\text{even modular weight: }&&\chi-3\sigma&\equiv0\pmod8\,,\\\text{non-trivial SW: }&&\chi+\sigma&\equiv0\pmod 4\,.
\end{alignat*}
The ``$x$ and $y$ coordinates'' for the table are chosen to be $\chi_h$ and $c_1^2-5\chi_h=\frac14(3\chi+7\sigma)$. Then the three conditions require that $\chi_h\in \Z_+$ is a positive integer and $c_1^2-5\chi_h$ is even.  
In the region $c_1^2\le 5.8\chi_h$, as we don't yet have access to the contribution of the $E_8$ singularity, the sum is only over the two I$_1$'s. We denote such a sum as $C'_\lambda$, which agrees with $C_\lambda$ when $c_1^2>5.8\chi_h$.

\begin{table}[htb!]
    \centering
    \centering
    \begin{tblr}{hlines, vlines, colsep=4pt,
             colspec = {Q[l, wd=1.8cm] |*{8}{Q[l, wd=1.55cm]}},
             column{1} = {c},
             column{2} = {c}, 
             cells = {c}
             }
    {\begin{tikzpicture}[baseline={(0,0.475cm-0.5ex)}]
        \useasboundingbox (0,0) rectangle (1.8cm,0.95cm);
        \draw[overlay] (-4pt,0.95cm+2pt) -- (1.8cm+4pt,-2pt);
        \node[anchor=north east,inner sep=0pt] at (1.8cm,0.95cm) {$\chi_h$};
        \node[anchor=base west,inner sep=0pt] at (-3pt,1pt) {\small $c_1^2-5\chi_h$};
    \end{tikzpicture}}
    & 1 & 2 & 3 & 4 & 5 & 6 & 7 & 8\\
    20 & \SetCell[]{bg=blue!20}$1$ & \SetCell[]{bg=blue!20}$1$ & \SetCell[]{bg=blue!20}$1$ & \SetCell[]{bg=blue!20}$1$ & \SetCell[]{bg=green!10}$1$ & \SetCell[]{bg=green!10}$1$ & \SetCell[]{bg=green!10}$1$ & \SetCell[]{bg=green!10}$1$ \\
    18 & \SetCell[]{bg=blue!20}$1$ & \SetCell[]{bg=blue!20}$1$ & \SetCell[]{bg=blue!20}$1$ & \SetCell[]{bg=green!10}$1$ & \SetCell[]{bg=green!10}$1$ & \SetCell[]{bg=green!10}$1$ & \SetCell[]{bg=green!10}$1$ & \SetCell[]{bg=green!10}$1$\\
    16 & \SetCell[]{bg=blue!20}$1$ & \SetCell[]{bg=blue!20}$1$ & \SetCell[]{bg=blue!20}$1$ & \SetCell[]{bg=green!10}$1$ & \SetCell[]{bg=green!10}$1$ & \SetCell[]{bg=green!10}$1$ & \SetCell[]{bg=green!10}$1$ & \SetCell[]{bg=green!10}$1$\\
    14 & \SetCell[]{bg=blue!20}$1$ & \SetCell[]{bg=blue!20}$1$ & \SetCell[]{bg=green!10}$1$ & \SetCell[]{bg=green!10}$1$ & \SetCell[]{bg=green!10}$1$ & \SetCell[]{bg=green!10}$1$ & \SetCell[]{bg=green!10}$1$ & \SetCell[]{bg=green!10}$1$\\
    12 & \SetCell[]{bg=blue!20}$1$ & \SetCell[]{bg=blue!20}$1$ & \SetCell[]{bg=green!10}$1$ & \SetCell[]{bg=green!10}$1$ & \SetCell[]{bg=green!10}$1$ & \SetCell[]{bg=green!10}$1$ & \SetCell[]{bg=green!10}$1$ & \SetCell[]{bg=green!10}$1$\\
    10 & \SetCell[]{bg=blue!20}$1$ & \SetCell[]{bg=blue!20}$1$ & \SetCell[]{bg=green!10}$1$ & \SetCell[]{bg=green!10}$1$ & \SetCell[]{bg=green!10}$1$ & \SetCell[]{bg=green!10}$1$ & \SetCell[]{bg=green!10}$1$ & \SetCell[]{bg=green!10}$1$\\
    8 & \SetCell[]{bg=blue!20}$1$ & \SetCell[]{bg=green!10}$1$ & \SetCell[]{bg=green!10}$1$ & \SetCell[]{bg=green!10}$1$ & \SetCell[]{bg=green!10}$1$ & \SetCell[]{bg=green!10}$1$ & \SetCell[]{bg=green!10}$1$ & \SetCell[]{bg=green!10}$1$\\
    6 & \SetCell[]{bg=blue!20}$1$ & \SetCell[]{bg=green!10}$1$ & \SetCell[]{bg=green!10}$1$ & \SetCell[]{bg=green!10}$1$ & \SetCell[]{bg=green!10}$1$ & \SetCell[]{bg=green!10}$1$ & \SetCell[]{bg=green!10}$1$ & \SetCell[]{bg=cyan!20}$1$\\
    4 & \SetCell[]{bg=green!10}$1$ & \SetCell[]{bg=green!10}$1$ & \SetCell[]{bg=green!10}$1$ &\SetCell[]{bg=green!10}$1$ & \SetCell[]{bg=cyan!20}$1$ & \SetCell[]{bg=cyan!20}$1$ & \SetCell[]{bg=cyan!20}$1$ & \SetCell[]{bg=cyan!20}$1$\\
    2 & \SetCell[]{bg=green!10}$1$ & \SetCell[]{bg=green!10}$1$ & \SetCell[]{bg=cyan!20}$1$ & \SetCell[]{bg=cyan!20}$1$& \SetCell[]{bg=cyan!20}$1$& \SetCell[]{bg=cyan!20}$1$& \SetCell[]{bg=cyan!20}$1$& \SetCell[]{bg=cyan!20}$1$\\
    0 & \SetCell[]{bg=cyan!20}$2$ & \SetCell[]{bg=cyan!20}$2$ & \SetCell[]{bg=cyan!20}$2$& \SetCell[]{bg=cyan!20}$2$& \SetCell[]{bg=cyan!20}$2$& \SetCell[]{bg=cyan!20}$2$& \SetCell[]{bg=cyan!20}$2$& \SetCell[]{bg=cyan!20}$2$\\
    $-2$ & \SetCell[]{bg=yellow!30}$2^{-8}3^{-6}$ & \SetCell[]{bg=yellow!30}$2^{-8}3^{-6}$ & \SetCell[]{bg=yellow!30}$2^{-8}3^{-6}$ & \SetCell[]{bg=yellow!30}$2^{-8}3^{-6}$ & \SetCell[]{bg=yellow!30}$2^{-8}3^{-6}$ & \SetCell[]{bg=yellow!30}$2^{-8}3^{-6}$ & \SetCell[]{bg=yellow!30}$2^{-8}3^{-6}$ & \SetCell[]{bg=yellow!30}$2^{-8}3^{-6}$\\
    $-4$ & \SetCell[]{bg=yellow!30}$2^{-16}3^{-12}$ & \SetCell[]{bg=yellow!30}$2^{-16}3^{-12}$ & \SetCell[]{bg=yellow!30}$2^{-16}3^{-12}$ & \SetCell[]{bg=yellow!30}$2^{-16}3^{-12}$ & \SetCell[]{bg=yellow!30}$2^{-16}3^{-12}$ & \SetCell[]{bg=yellow!30}$2^{-16}3^{-12}$ & \SetCell[]{bg=yellow!30}$2^{-16}3^{-12}$ & \SetCell[]{bg=yellow!30}$2^{-16}3^{-12}$\\
    $-6$ & \SetCell[]{bg=red!30}$2^{-24}3^{-18}$ & \SetCell[]{bg=yellow!30}$2^{-24}3^{-18}$ & \SetCell[]{bg=yellow!30}$2^{-24}3^{-18}$ & \SetCell[]{bg=yellow!30}$2^{-24}3^{-18}$ & \SetCell[]{bg=yellow!30}$2^{-24}3^{-18}$ & \SetCell[]{bg=yellow!30}$2^{-24}3^{-18}$ & \SetCell[]{bg=yellow!30}$2^{-24}3^{-18}$ & \SetCell[]{bg=yellow!30}$2^{-24}3^{-18}$\\
    $-8$ & \SetCell[]{bg=red!30}$2^{-32}3^{-24}$ & \SetCell[]{bg=yellow!30}$2^{-32}3^{-24}$ & \SetCell[]{bg=yellow!30}$2^{-32}3^{-24}$ & \SetCell[]{bg=yellow!30}$2^{-32}3^{-24}$ & \SetCell[]{bg=yellow!30}$2^{-32}3^{-24}$ & \SetCell[]{bg=yellow!30}$2^{-32}3^{-24}$ & \SetCell[]{bg=yellow!30}$2^{-32}3^{-24}$ & \SetCell[]{bg=yellow!30}$2^{-32}3^{-24}$ \\
    $-10$ & \SetCell[]{bg=red!30}$2^{-40}3^{-30}$ & \SetCell[]{bg=red!30}$2^{-40}3^{-30}$ & \SetCell[]{bg=yellow!30}$2^{-40}3^{-30}$ & \SetCell[]{bg=yellow!30}$2^{-40}3^{-30}$ & \SetCell[]{bg=yellow!30}$2^{-40}3^{-30}$ & \SetCell[]{bg=yellow!30}$2^{-40}3^{-30}$ & \SetCell[]{bg=yellow!30}$2^{-40}3^{-30}$ & \SetCell[]{bg=yellow!30}$2^{-40}3^{-30}$\\
    $-12$ & \SetCell[]{bg=red!30}$2^{-48}3^{-36}$ & \SetCell[]{bg=red!30}$2^{-48}3^{-36}$ & \SetCell[]{bg=yellow!30}$2^{-48}3^{-36}$ & \SetCell[]{bg=yellow!30}$2^{-48}3^{-36}$ & \SetCell[]{bg=yellow!30}$2^{-48}3^{-36}$ & \SetCell[]{bg=yellow!30}$2^{-48}3^{-36}$ & \SetCell[]{bg=yellow!30}$2^{-48}3^{-36}$ & \SetCell[]{bg=yellow!30}$2^{-48}3^{-36}$ \\
    $-14$ & \SetCell[]{bg=red!30}$2^{-56}3^{-42}$ & \SetCell[]{bg=red!30}$2^{-56}3^{-42}$ & \SetCell[]{bg=red!30}$2^{-56}3^{-42}$ & \SetCell[]{bg=yellow!30}$2^{-56}3^{-42}$ & \SetCell[]{bg=yellow!30}$2^{-56}3^{-42}$ & \SetCell[]{bg=yellow!30}$2^{-56}3^{-42}$ & \SetCell[]{bg=yellow!30}$2^{-56}3^{-42}$ & \SetCell[]{bg=yellow!30}$2^{-56}3^{-42}$\\
    $-16$ & \SetCell[]{bg=red!30}$2^{-64}3^{-48}$ & \SetCell[]{bg=red!30}$2^{-64}3^{-48}$ & \SetCell[]{bg=red!30}$2^{-64}3^{-48}$ & \SetCell[]{bg=yellow!30}$2^{-64}3^{-48}$ & \SetCell[]{bg=yellow!30}$2^{-64}3^{-48}$ & \SetCell[]{bg=yellow!30}$2^{-64}3^{-48}$ & \SetCell[]{bg=yellow!30}$2^{-64}3^{-48}$ & \SetCell[]{bg=yellow!30}$2^{-64}3^{-48}$\\
    $-18$ & \SetCell[]{bg=red!30}$2^{-72}3^{-54}$ & \SetCell[]{bg=red!30}$2^{-72}3^{-54}$ & \SetCell[]{bg=red!30}$2^{-72}3^{-54}$ & \SetCell[]{bg=red!30}$2^{-72}3^{-54}$ & \SetCell[]{bg=yellow!30}$2^{-72}3^{-54}$ & \SetCell[]{bg=yellow!30}$2^{-72}3^{-54}$ & \SetCell[]{bg=yellow!30}$2^{-72}3^{-54}$ & \SetCell[]{bg=yellow!30}$2^{-72}3^{-54}$
    \end{tblr}
    \caption{\textbf{Leading coefficients in the $q$-series of $C'_\lambda$ up to a phase.} The coefficients are integral (green, cyan and blue) above the $c_1^2-5\chi_h=0$ (or equivalently $3\chi+7\sigma=0$) line, a region that contains the convergent region $c_1^2>5.8\chi_h$ (green) and the upper part of the weakly divergent region, $5\chi_h\le c_1^2\le5.8\chi_h$ (cyan). Below this line (yellow), the coefficients of the $q$-series begin to become fractional. This pattern continues, with more negative powers, into the ``strongly divergent region'' (red). The blue region is given by $\sigma+2>\chi$ (or equivalently $c_1^2-10\chi_h>-1$), where simply-connected 4-manifolds can no longer exist. 
    }
    \label{tab:original grav coef}
\end{table}

It is actually quite a ``boring'' table as the values turn out to be constant along rows. Nevertheless, we find that the coefficients are integral for $c_1^2-5\chi_h\ge 0$, i.e.~in the convergent region and in the upper part $5\chi_h\le c_1^2\le 5.8\chi_h$ of the weakly divergent region, but start to become fractional immediately below this line. 

\subsection{TMF constraints}
\label{sec:TMF}
As highlighted in the introduction, 6d $(1,0)$ SCFTs, four-manifolds, and topological modular forms (TMF) are interconnected through a combination of 2d/4d compactifications, a topological twist, and the Segal--Stolz--Teichner conjecture. One consequence is that the partition function of a 6d theory on $M_4\times T^2$, when finite, cannot be a general modular form, but one that is liftable to a topological modular form along the map
\begin{align}
\begin{split}
     \pi_d(\TMF)&\rightarrow \MF_{d/2}^{\mathbb{Z}}\,,\\
    \mathcal{Z}_{\CT[M_4]} &\mapsto Z,
\end{split}
\end{align}
where $Z$ is the partition function of the 6d theory on $ M_4\times T^2$ (with trivial holonomy along $T^2$ for flavor symmetries) while
$\mathcal{Z}_{\CT[M_4]}$ denotes the hypothetical ``{topological Witten genus}'' of the 2d $(0,1)$ theory $\CT[M_4]$.\footnote{When we allow general $E_8$ holonomies on $T^2$, the partition function is expected to be in the image of 
\begin{equation}
     \widetilde\TMF_{d,k}^{E_8}\rightarrow \mathrm{MF}^{E_8,\Z}_{d/2,k}\,,
\end{equation}
where $ \widetilde\TMF_{d,k}^{E_8}$ denotes the space of degree-$d$ index-$k$ $E_8$-equivariant meromorphic topological modular forms and $\mathrm{MF}^{E_8,\Z}_{d/2,k}$ denotes the ring of meromorphic integral $E_8$ Jacobi forms with weight $d/2$ and index $k$. Note that neither $\widetilde{\TMF}^{E_8}$ nor this arrow has been constructed mathematically---the arrow expresses a physical expectation, modeled on the map between their holomorphic versions constructed in \cite{Gepner_Meier_2023}. Also, note that while there are different versions of topological modular forms with the most used variants being tmf, Tmf and TMF, what has a direct connection to 2d $(0,1)$ theories is $\TMF$, the 576-periodic version of topological modular forms.} See \cite{Gukov:2018iiq} for more details including a review of the connection between TMF and quantum field theories. 

In the present work, we don't need the full details of the ring $\pi_*(\TMF)$, but only its image in modular forms. We use the presentation $\MF_{*}^{\mathbb{Z}}=\mathbb{Z}[E_4,E_6,\Delta^{\pm}]/(E_4^3-E_6^2-1728\Delta)$, and the partition function $Z[M_4\times T^2]$, when well defined with a trivial background for flavor symmetries, can be expressed as a polynomial in $E_4$, $E_6$, and $\Delta^{\pm}$, as is true for any integral modular forms. However, for it to reside in the image of $\pi_*(\TMF)$, the coefficients $a_{ijk}$  in front of $E_4^iE_6^j\Delta^k$, where $i\in \Z_{\ge0}$, $j=0$ or 1, $k\in\Z$, and $4i+6j+12k=d/2$, are subject to a divisibility constraint \cite{Hopkins:2002math.....12397H}
\begin{equation}
\label{eq:tmf divisibility}
    \text{$a_{ijk}$ is divisible by}
    \begin{cases}
        \frac{24}{\text{gcd}(24,k)}\,,&i=j=0\,,\\
        2\,,&j=1\,,\\
        1\,,&\text{otherwise}\,.
    \end{cases}
\end{equation}
See e.g.~\cite{Gaiotto:2018ypj,Lin:2021bcp} for discussions about the relevance of the constraints to physics as well as some tests and applications. For any given 6d theory $\CT$, it would be interesting to test whether the partition function $Z_\CT[M_4\times T^2]$ actually satisfies these constraints and their equivariant refinements, which is non-trivial if $Z$ is computed from the 4d perspective using the low-energy effective theory. In the present case with $\CT$ being the E-string theory, the constraint \eqref{eq:tmf divisibility} is indeed always satisfied in all ``simple-type'' examples---often very ``tightly''---which is a non-trivial check on our results. It would be interesting to understand and utilize the $E_8$-equivariant version of the constraints which might lead to genuinely new predictions about the Seiberg--Witten invariants.

\section{Examples}
\label{sec:examples}

We will now combine all the ingredients to compute the partition function for some example four-manifolds to test whether our listed expectations are all (non-trivially) satisfied. This turns out to rely on some properties of the Seiberg--Witten invariants. Some of them are standard and well known, such as the facts that they are all integers and that SW$(-\lambda)=(-1)^{\chi_h}\SW(\lambda)$. Some are not yet well established, such as the SW simple-type conjecture, which we explain below before delving into examples.

\paragraph{The simple-type conjecture.} A four-manifold is said to be of Seiberg--Witten simple type (SWST) when it satisfies the following condition: If a {spin}$^c$ structure $\lambda$ is such that $\lambda^2$ is not equal to $\frac14 c_1^2=\frac14\left(2\chi+3\sigma\right)$, then $\text{SW}(\lambda)=0$. An alternative way to state this is that all basic classes\footnote{The basic classes are defined as spin$^c$ structures such that their SW invariants are non-zero. Due to the relation between SW$(\lambda)$ and SW$(-\lambda)$, such a pair is usually considered collectively as a single basic class.} $\lambda$ must have the virtual dimension $n(\lambda)$ equal to zero. It is conjectured that any smooth, closed, oriented and simply-connected 4-manifolds with $b_2^+>1$ have SW simple type \cite{Witten:1994cg,Kronheimer_Mrowka_2007}. The SWST conjecture has been proved in large classes of 4-manifolds (e.g.~all symplectic manifolds with $b_2^+>1$ \cite{taubes1999rm}) and no simply-connected counter-example is currently known.\footnote{The simple-type conjecture can be understood from the physics side as certain properties of the 0-observables in the 4d theory (see \cite{Witten:1994cg,Moore:1997pc,Moore:2017cmm} for more discussions from this perspective). Results in this section suggest that 6d theories can potentially provide a physics explanation of the SWST conjecture. It is therefore desirable to first interpret the conjecture from the 6d perspective. One such potential interpretation involves a one-parameter deformation of the 6d partition function ``with two $q$'s'' \cite{GHP2}. Namely, for a 6d theory $\CT$ on $T^2$, if the low-energy theory for $\CT[T^2]$ also has a gauge-theory description, one can deform the theory by changing its low-energy gauge couplings, away from the ``natural values'' given by $\sigma$. It is conceivable that one combination of the two $q$'s exactly keeps track of $n(\lambda)$ in the decomposition of the partition function as a sum over SW contributions from I$_1$ singularities when $M_4$ is simply connected with $b_2^+>1$. }

In what follows, we will look at seven classes of four-manifolds which illustrate different unique features of the partition function, with one persistent highlight being a surprising connection between its topological modularity and the simple-type conjecture of 4-manifolds. Throughout this section, we will use $Z_\lambda:=\sum_sZ_{\lambda,s}$ to denote the contribution of a given spin$^c$ class $\lambda$ to the partition function, which can be written as $Z_\lambda=C_\lambda\cdot \SW(\lambda)$, where we emphasize the dependence on $\lambda$ while suppressing the UV parameter $\sigma$ (as well as $m_i$'s but they are set to zero for most of the time in this section). 

\subsection{$(\chi,\sigma)=(8,0)$: TMF and the simple-type conjecture}

Let us first consider the case of simply-connected 4-manifolds with $\chi=8$ and $\sigma=0$.\footnote{Note that throughout the present and the previous section, we want the manifold $M_4$ to be simply connected. Otherwise, the form of the partition function would be modified, significantly altering the analysis. If $M_4$ were allowed to be non-simply-connected, one could construct a family with $\sigma=0$ as the product of two Riemann surfaces with genus $g_1$ and $g_2$, $M_4=\Sigma_{g_1}\times \Sigma_{g_2}$, with $\chi(M_4)=(2-2g_1)(2-2g_2)$. This class of examples is more closely related to the cases studied in \cite{Gukov:2018iiq}.} 
From Freedman's theorem, these manifolds are homeomorphic to $\#^3(S^2\times S^2)$ or $\#^3(\cp^2\#\overline{\cp}^2)$ depending on whether they are spin. See e.g.~\cite{Gompf1995,park2002geography,fintushel2007reverse,akhmedov2010exotic} for constructions of 4-manifolds on the ``zero-signature line'' which are relevant for the discussion here and in Section~\ref{sec:8nplus4}.
Such 4-manifolds have $b_2^+=3>1$, allowing us to obtain the partition function as a sum over Seiberg--Witten invariants. In this case, we have
\[
\chi_h=2\,,\qquad c_1^2=16\,,\qquad n(\lambda)=\frac12\lambda^2-2\,,
\]
and the modular weight of the partition function is
\begin{equation}
    -\frac{11}{4}\chi-\frac{31}{4}\sigma=-22\,.
\end{equation}
If $M_4$ is a minimal complex surface, it is then of general type, since $c_1^2=16>0$ and $\chi_h=2$ exclude rational and ruled surfaces.
Now, the constant appearing in front of the partition function becomes
\begin{equation}
    \tilde{c}(\chi,\sigma)=-2\cdot3^{2}\,.
\end{equation}
We have the following expansion within the sum over spin$^c$ structures
\begin{align*}
\left(\frac{E_6}{E_4}\right)^{-3/2}&=1+1116q_s+798228q_s^2+510427584q_s^3+\dots
\end{align*}

We first consider basic classes $\lambda$ satisfying $n(\lambda)=0$, for which
\begin{equation}
C_\lambda\sim \sum_s\left(\frac{f(u_s)}{g(u_s)}\right)^{-1}\mu_1^2=\sum_s \frac{E_4^5u_s^7}{18(-864+E_6u_s)}\,
\end{equation}
has no $\lambda$ dependence except for the overall $\Z_4$ phase $(-1)^{\lambda\cdot w_2}$. For each singularity, we will then have the following $q$-series 
\begin{align*}
    u_{s,0}&\,: \quad 0\\
    u_{s,+}&\,: \quad -\left(\frac{1}{q^6}+\frac{552}{q^5}+\frac{68400}{q^4}-\frac{4709280}{q^3}+\dots\right)\,,\\
    u_{s,-}&\,: \quad 2^{24}3^{18}\left(1+1272q+706680q^2+227946720q^3+\dots\right)\,.
\end{align*}
Therefore, altogether, the partition function is 
\begin{equation}
    C_\lambda=-\left(\frac{1}{q^6}+\frac{552}{q^5}+\frac{68400}{q^4}-\frac{4709280}{q^3}-\frac{291759180}{q^2}+\frac{71469296208}{q}+\dots\right)\,.
\end{equation}
Rewriting the above into Eisenstein series, we have the general ansatz,
\begin{equation}
    C_\lambda=\sum_{k=3}^6 a_{ijk}E_4^iE_6^j\Delta^{-k}\,,
\end{equation}
where $4i+6j-12k=-22$. We can further simplify the expression via restricting to $j=0,1$ due to the relation $E_6^2=E_4^3-1728\Delta$. 
Then, solving for the coefficients, we have the partition function
\begin{equation}
    C_\lambda=-\left(E_4^{11}E_6\Delta^{-6}-1728E_4^8E_6\Delta^{-5}+559872E_4^5E_6\Delta^{-4}\right)\,.
\end{equation}

Notice that while we have $a_{ijk}\in\Z$ and the above is an integral modular form, it does not lift to TMF due to the first coefficient. However, $\lambda$ and $-\lambda$ contribute together $2C_\lambda\cdot\SW(\lambda)$, so the partition function satisfies the TMF divisibility constraint~\eqref{eq:tmf divisibility} in a non-trivial fashion.

\subsubsection*{Contribution from classes with $n(\lambda)= 1$}

From the analysis above, we see that if $M_4$ is of SWST, then the partition function is liftable to TMF. Now, let us assume to the contrary that there exists a simply-connected four-manifold with $(\chi,\sigma)=(8,0)$ which is not of SWST. In particular, let us first suppose that there exists a basic class $\lambda$ with $n(\lambda)=1$.\footnote{In general, $n(\lambda)$ can be half-integral. However, in such cases all SW invariants vanish if $M_4$ is simply connected, as the integrand over the SW moduli space will be of the ``wrong degree.''} Then, its contribution to the partition function is
\begin{equation}
\label{eq:nonSWST}
    Z_\lambda=\tilde{c}\cdot \SW(\lambda)\cdot \sum_s\left[\left(\frac{f(u_s)}{g(u_s)}\right)^{-1}\mu_1\left(2\mu_2+\frac{\kappa_2}{\kappa_1}\mu_1\right)+\frac{(E_6(\sigma)u_s-432)}{6E_4(\sigma)}\mu_1^3+\frac{62E_4^5u_s^7}{-864+E_6u_s}\right]\,.
\end{equation}
Then, plugging in the values of $\mu_2$ and $\kappa_2/\kappa_1$ derived in~\eqref{eq:mu2} and~\eqref{eq:kappa2}, we obtain the following $q$-series associated to each singularity
\begin{align*}
    u_{s,0}&\,:\quad 0\,,\\
    u_{s,+}&\,:\quad -\frac{1}{2}\left(
    \frac{11}{q^7}+\frac{7332}{q^6}+\frac{1560780}{q^5}+\frac{86768320}{q^4}-\frac{5886577710}{q^3}+\dots
    \right)\,,\\
    u_{s,-}&\,:\quad -2^{25}3^{19}\left(
    -215-444552q-390068424q^2-196722373920q^3+\dots
    \right)\,.
\end{align*}
The ansatz for $C_\lambda$ is given by
\begin{equation}
C_\lambda=\sum_{k=3}^{11}c_{k}E_4^{-7+3k}E_6\Delta^{-k}\,,
\end{equation}
which actually has the unique solution
\begin{equation}
\label{eq:nonSWSTmodular}
C_\lambda=-\frac{1}{2}\left(11E_4^{14}E_6\Delta^{-7}-25932\,E_4^{11}E_6\Delta^{-6}+16070400\,E_4^{8}E_6\Delta^{-5}-2102879232\,E_4^{5}E_6\Delta^{-4}\right)\,,
\end{equation}
which gives an integral modular form once $\lambda$ and $-\lambda$ are combined, since $\tilde{c}/36=-\frac12$. The TMF divisibility constraint~\eqref{eq:tmf divisibility} then further requires the number of such $\{\lambda,-\lambda\}$ pairs---assuming SW$(\lambda)$=SW$(-\lambda)=\pm1$---to be even. 

\subsubsection*{Classes with $n(\lambda)> 1$}

Similarly, let us consider another hypothetical basic class $\lambda$ with $n(\lambda)=2$
. Its contribution is proportional to
\begin{multline}    Z_{\lambda}=\tilde{c}\cdot\SW(\lambda)\cdot\sum_s\left[\left(\frac{f}{g}\right)^{-1}\mu_1\left(3\mu_3+2\frac{\kappa_2}{\kappa_1}\mu_2+\frac{\kappa_3}{\kappa_1}\mu_1\right)+\frac{(E_6(\sigma)u_s-432)}{6E_4(\sigma)}\mu_1^2\left(2\mu_2+\frac{\kappa_2}{\kappa_1}\mu_1\right)\right.\\
    +\frac{E_6(\sigma)\mu_1^2-864\mu_2+2E_6(\sigma)\mu_2u_s}{12E_4(\sigma)}\mu_1^2+\frac{(E_6(\sigma)u_s-432)^2\mu_1^4}{12E_4(\sigma)u_s(E_6(\sigma)u_s-864)}\\
    +1116\left(\left(\frac{f(u_s)}{g(u_s)}\right)^{-1}\mu_1\left(2\mu_2+\frac{\kappa_2}{\kappa_1}\mu_1\right)+\frac{(E_6(\sigma)u_s-432)}{6E_4(\sigma)}\mu_1^3\right)\\
    \left.+\frac{44346E_4^5u_s^7}{-864+E_6u_s}\right]\,,
\end{multline}
with the various contributions arising from each singularity being
\begin{align*}
    u_{s,0}&\,:\quad 0\,,\\
    u_{s,+}&\,:\quad -\frac{1}{6}\left(
    \frac{107}{q^8}+\frac{79728}{q^7}+\frac{21209112}{q^6}+\frac{2299746368}{q^5}+\frac{64205189400}{q^4}+\dots
    \right)\,,\\
    u_{s,-}&\,:\quad -2^{27}3^{19}\left(
    -29921-98121336q-130630154616q^2-97811215632096q^3+\dots
    \right)\,.
\end{align*}
Then, again, by modular weight, we can assume the following ansatz,
\begin{equation}
\label{eq:ansatznonSWST}
C_\lambda=\sum_{k=3}^{12}\tilde{c}_{k}E_4^{-7+3k}E_6\Delta^{-k}\,,
\end{equation}
which has the unique solution
\begin{multline}
C_\lambda=-\frac{1}{6}\left(107E_4^{17}E_6\Delta^{-8}-323448\,E_4^{14}E_6\Delta^{-7}+303529104\,E_4^{11}E_6\Delta^{-6}\right.\\
\left.-93486431232\,E_4^{8}E_6\Delta^{-5}+6100230942720\,E_4^{5}E_6\Delta^{-4}\right)\,.
\end{multline}
This is a modular form, but not an integral one unless one multiplies it by 6.
In the same way one finds, for a class with $n(\lambda)=3$,
\begin{multline}
C_\lambda=-\frac{1}{12}\left(533E_4^{20}E_6\Delta^{-9}-1977372\,E_4^{17}E_6\Delta^{-8}+2514486780\,E_4^{14}E_6\Delta^{-7}\right.\\
-1277980814640\,E_4^{11}E_6\Delta^{-6}+227770651081728\,E_4^{8}E_6\Delta^{-5}\\
\left.-8465961882193920\,E_4^{5}E_6\Delta^{-4}\right)\,,
\end{multline}
again modular and now with a denominator 12, while for $n(\lambda)=4$ the denominator becomes $20$. When $n(\lambda)$ increases,  we find that the denominators keep growing (2, 6, 12, 20 and 60 for $n(\lambda)=1,\dots,5$), with the first few further contributions being
\begin{itemize}
    \item $n(\lambda)=4$:
    \begin{multline}
        C_\lambda=-\frac{1}{20}\left(1879\,E_4^{23}E_6\Delta^{-10}-8287944\,E_4^{20}E_6\Delta^{-9}+13379905836\,E_4^{17}E_6\Delta^{-8}\right.\\
        -9659673444000\,E_4^{14}E_6\Delta^{-7}+3045105386348400\,E_4^{11}E_6\Delta^{-6}\\
        \left.-344598453331522560\,E_4^{8}E_6\Delta^{-5}+7985881103358873600\,E_4^{5}E_6\Delta^{-4}\right)\,,
    \end{multline}
    \item $n(\lambda)=5$:
    \begin{multline}
        C_\lambda=-\frac{1}{60}\left(10655\,E_4^{26}E_6\Delta^{-11}-54566808\,E_4^{23}E_6\Delta^{-10}+107150666148\,E_4^{20}E_6\Delta^{-9}\right.\\
        -101172141275872\,E_4^{17}E_6\Delta^{-8}+47160155187916500\,E_4^{14}E_6\Delta^{-7}\\
        -9988135546134978000\,E_4^{11}E_6\Delta^{-6}+765094083458276090880\,E_4^{8}E_6\Delta^{-5}\\
        \left.-11800279116430888366080\,E_4^{5}E_6\Delta^{-4}\right)\,.
    \end{multline}
\end{itemize}

While the partition function is modular whether or not $M_4$ is of SW simple type, integrality and the TMF condition do put constraints on the non-simple-type basic class. For $(\chi,\sigma)=(8,0)$, it requires the sum of $(-1)^{\lambda\cdot w_2}\SW(\lambda)$ over basic classes with $n(\lambda)=2,3,4$ to be divisible by 12, $24$, and $40$, if no classes with $n(\lambda)\ge 5$ occur. 

For the present choice of $(\chi,\sigma)$, such conditions are compatible with a result of Bauer and Furuta \cite[Cor.~3.6,~Thm.~3.7]{bauer2004stable}, which in the present case guarantees the liftability of the partition function to TMF, and it would be interesting to see whether considering refinement by $E_8$ holonomies can lead to stronger conditions.\footnote{For example, we are never able to use the stronger divisibility involving $\Delta$ in the massless limit, as the coefficient of the pure power of $\Delta$ (present only when 12 divides the weight $w$) equals $C_\lambda/\Delta^{w/12}$ at $\sigma=e^{\pi i/3}$, where every other monomial with $E_4$ vanishes. However, in the convergent region where $C_\lambda$ can be defined in the massless limit, it simply vanishes at this point due to the zero modes of the Argyres--Douglas theory.
}

It would also be interesting to see whether considering more general correlation functions can lead to more constraints on how the simple-type conjecture can be violated. For example, to get a stronger statement using modularity (as opposed to integrality), one can consider insertion of surface defects in the 4d theory, which will preserve modularity if they are engineered with codimension-two defects in the E-string theory wrapping $T^2$. 

\subsection{$(\chi,\sigma)=(10,-2)$: non-spin manifolds}

As explained in Section~\ref{sec:topological twist}, the E-string theory on $M_4\times T^2$ remains canonically well defined even when $M_4$ is non-spin. Here, we verify this claim by demonstrating that there is indeed a well-behaved partition function. With $\chi=10$ and $\sigma=-2$, we have $b_2^+=3$ and $\chi_h=2$ as $M_4$ is simply connected, and the fact that $16\nmid \sigma$ indicates that the underlying manifold has to be non-spin.

The coefficient of $\SW(\lambda)$ for a basic class $\lambda$ with $n(\lambda)=0$ (i.e.~compatible with SWST) in the partition function reduces to 
\[
C_\lambda=\tilde{c}\cdot\sum_s\mu_1^2\,,
\]
where $\tilde{c}=-1$.
Each singularity contributes
\begin{align*}
    u_{s,0}&\,:\quad 0\,,\\
    u_{s,+}&\,:\quad -\left(\frac{1}{q^4}+\frac{528}{q^3}+\frac{65664}{q^2}-\frac{2598208}{q}-165147624+\dots\right)\,,\\
    u_{s,-}&\,:\quad -2^{16}3^{12}\left(1+1008q+422064q^2+96823872q^3+\dots\right)\,.
\end{align*}
Collecting all of these together yields
\begin{equation}
    C_\lambda=-\left(\frac{1}{q^4}+\frac{528}{q^3}+\frac{65664}{q^2}-\frac{2598208}{q}+34663369752 + 35141075728992 q+\dots\right)\,.
\end{equation}
The modular weight of the partition function is expected to be
\begin{equation}
    -\frac{11}{4}\chi-\frac{31}{4}\sigma=-12\,.
\end{equation}
With this modular weight, we can re-express the partition function with $C_\lambda=a_{9,0,-4}E_4^{9}\Delta^{-4}+a_{6,0,-3}E_4^6\Delta^{-3}+a_{3,0,-2}E_4^{3}\Delta^{-2}$.
Plugging this ansatz back into the $q$-series and multiplying by the appropriate normalization factor, we obtain 
\begin{equation}
C_\lambda=-\left(E_4^9\Delta^{-4}-1728E_4^6\Delta^{-3}+373248E_4^3\Delta^{-2}\right)\,,
\end{equation}
which is an integral modular form. Therefore, in this non-spin four-manifold example, $Z_\lambda$, and hence the entire partition function $Z$, assuming $M_4$ is of SWST, is integral, modular and liftable to TMF as expected.

\subsubsection*{Classes with $n(\lambda)>0$}

Now, in this non-spin case, let us examine the contribution of a non-simple-type class. Similar to the previous example, we can consider a spin$^c$ structure $\lambda$  with $n(\lambda)>0$ violating the SWST conjecture. With $(\chi,\sigma)=(10,-2)$, the next relevant non-negative $n(\lambda)$ equals $1$ with $(2\lambda)^2=22$. 
In this case, we also have to account for the following $q_s$-series
\begin{align*}
    \left(\frac{\eta(\tau_s)^{24}}{q_s}\right)^{-1/4}&=1+6q_s+27q_s^2+98q_s^3+315q_s^4+\dots\,,\\
    \left(\frac{E_6(\tau_s)}{E_4(\tau_s)}\right)^{-1/2}&=1+372q_s+127692q_s^2+57980064 q_s^3 + 26762059596 q_s^4+\dots\,,\\
    \left(\frac{f(u_s)}{g(u_s)}\right)^{1/2}\left(\frac{f}{g}\right)^{-1/2}&=1+\rho_1q_s+\rho_2q_s^2+\dots\,,
\end{align*}
where $\rho_1=D-432$, $\rho_2=D^2-744D+41472$, and $D=864-E_6u_s$ as in Appendix~\ref{sec:details on the partition function}.
Therefore, the contribution to the partition function for a spin$^c$ structure with $n(\lambda)=1$ is
\begin{equation}
\label{eq:Zchi8sigma-2nlambda1}
    Z_\lambda=\tilde{c}\cdot\SW(\lambda)\cdot \sum_s\mu_1\left(\left(\frac{\kappa_2}{\kappa_1}+6+372+\rho_1\right)\mu_1+2\mu_2\right)\,.
\end{equation}
 Then, the coefficient of SW$(\lambda)$ in $Z_\lambda$ is the sum of the following contributions from the singularities:
\begin{align*}
    u_{s,0}&\,:\quad 0\,,\\
    u_{s,+}&\,:\quad -\left(\frac{7}{2q^5}+\frac{1980}{q^4}+\frac{335430}{q^3}+\frac{15380480}{q^2}+\frac{92612475}{q}+\dots\right)\,,\\
    u_{s,-}&\,:\quad -2^{19}3^{13}\left(23+50400q+40407120q^2+17118564480q^3+\dots\right)\,.
\end{align*}
Then, we can suppose the partition function has the following ansatz
\[
Z_\lambda=\sum_{k=1}^{6}c_kE_4^{-3+3k}\Delta^{-k}\,,
\]
with the unique solution
\begin{equation}
Z_\lambda=-\SW(\lambda)\left(\frac72E_4^{12}\Delta^{-5}-8520\,E_4^{9}\Delta^{-4}+4924800\,E_4^{6}\Delta^{-3}-358318080\,E_4^{3}\Delta^{-2}\right)\,,
\end{equation}
giving an integral modular form of weight $-12$ once $\lambda$ and $-\lambda$ are combined.

Similarly, a class with $n(\lambda)=2$ contributes
\begin{multline}
\label{eq:Zchi10sigma-2nlambda2}
    Z_\lambda=\tilde{c}\cdot\SW(\lambda)\cdot \sum_s\mu_1\left[\left(\frac{\kappa_3}{\kappa_1}+(378+\rho_1)\frac{\kappa_2}{\kappa_1}+129951+378\rho_1+\rho_2\right)\mu_1\right.\\
    \left.+2\left(\frac{\kappa_2}{\kappa_1}+378+\rho_1\right)\mu_2+3\mu_3\right]\,,
\end{multline}
where $378$ and $129951$ are the coefficients of $q_s$ and $q_s^2$ in the product of the first two series above.
The two singularities contribute
\begin{align*}
    u_{s,0}&\,:\quad 0\,,\\
    u_{s,+}&\,:\quad -\left(\frac{47}{6q^6}+\frac{4577}{q^5}+\frac{850275}{q^4}+\frac{53427460}{q^3}+\frac{1126225070}{q^2}+\frac{4508770266}{q}+\dots\right)\,,\\
    u_{s,-}&\,:\quad -2^{16}3^{13}\left(36013+180685872q+269006937840q^2+195983593922880q^3+\dots\right)\,,
\end{align*}
and their sum is the modular form
\begin{multline}
Z_\lambda=-\SW(\lambda)\left(\frac{47}{6}E_4^{15}\Delta^{-6}-24751\,E_4^{12}\Delta^{-5}+23329767\,E_4^{9}\Delta^{-4}\right.\\
\left.-6392872512\,E_4^{6}\Delta^{-3}+211726794240\,E_4^{3}\Delta^{-2}\right)\,,
\end{multline}
which is not integral with the $-\frac{47}{6}$ coefficient, similar to the spin case.

\subsection{$(\chi,\sigma)=(8n+4,0)$: SW$(\lambda)$ vs.~SW$(-\lambda)$}\label{sec:8nplus4}

We now consider a family of examples that have $\chi=8n+4$ and $\sigma=0$ with $n\ge0$. If such an $M_4$ is spin---which we won't assume---then it is homeomorphic to $\#^{4n+1}(S^2\times S^2)$ from Freedman's classification theorem. For $n\ge 1$, such 4-manifolds satisfy $b_2^+>1$, and the formula \eqref{eq:masslessPartitionFunction} can be applied, with the normalization constant being $\tilde{c}=i(-1)^n2^{\frac{1}{2}(1+2n)}3^{1+2n}$ that satisfies $\tilde{c}/(i\sqrt{2})\in\mathbb{Z}$ for all $n$.

We will start with $n=0$, which corresponds to $\chi=4$, $\sigma=0$, and $b_2^+=1$, and one should also take into account the contribution of the $u$-plane. However, it was found to vanish identically in this case in Section~\ref{sec:FullTop}. Therefore, we just need to focus on the Seiberg--Witten contributions.

The contribution of a basic class $\lambda$ with $n(\lambda)=0$ is proportional to
\begin{equation}
C_{\lambda}\sim \sum_s\mu_1\left(\frac{f}{g}\right)^{-1/2}\,.
\end{equation}
Now, each singularity contributes the following $q$-series
\begin{align*}
    u_{s,0}&\,:\quad 0\,,\\
    u_{s,+}&\,:\quad -i\left(\frac{1}{q^3}+\frac{276}{q^2}-\frac{3888}{q}-1281552+\dots\right)\,,\\
    u_{s,-}&\,:\quad -i\left(80621568+51275317248q+12181273952256q^2+\dots\right)\,.
\end{align*}
Combining them, we have
\begin{equation}
    C_\lambda=-i\left(\frac{1}{q^3}+\frac{276}{q^2}-\frac{3888}{q}+79340016+51475587738q+\dots\right)\,.
\end{equation}
Can this expression be modular? Notice that the expected modular weight of the partition function is now an odd number, 
\begin{equation}
    -\frac{31}{4}\sigma(M_4)-\frac{11}{4}\chi=-11\,.
\end{equation}
As there are no full SL$(2,\Z)$ modular forms with odd weight, the expression $C_\lambda$ cannot be fully modular. This phenomenon persists for larger $n$, as the modular weight $-11(2n+1)$ is always odd while $C_\lambda$ is a non-trivial $q$-series. For example, when $n=2$, we have 
\begin{equation}
    C_{\lambda}=i\left(-\frac{1}{q^{15}}-\frac{1380}{q^{14}}-\frac{742320}{q^{13}}-\frac{182376240}{q^{12}}-\ldots\right)\,.
\end{equation}
How can the partition function be modular then? This paradox is resolved once one notices that there is a cancellation between $\lambda$ and $-\lambda$. Indeed, we have $\SW(\lambda)=(-1)^{\chi_h}\SW(-\lambda)$ and, furthermore, 
\begin{equation}
    C_\lambda=(-1)^\sigma C_{-\lambda}.
\end{equation}
In the present case with $\chi_h=2n+1$ being odd, the sum
\begin{equation}
    C_\lambda\cdot\SW(\lambda)+ C_{-\lambda}\cdot\SW(-\lambda)=0
\end{equation}
is simply zero. Therefore, the partition function is identically zero, and there is no contradiction with having an odd modular weight. This family of examples also highlights the value of having operator insertions, which can shift the modular weight, potentially making the partition function non-zero.

Notice that, more generally, for $b_2^+=1$ and even $\sigma$, $Z_{\rm Coulomb}$ vanishes identically (cf.~Section~\ref{sec:FullTop}), consistent with the partition function having an odd modular weight $-11-5\sigma$ and SW contributions canceling each other.

\subsection{$(\chi,\sigma)=(24,-16)$: K3 and the divergent region}\label{sec:divergence}

We now consider simply-connected four-manifolds with $\chi=24$ and $\sigma=-16$, of which one example is the K3 surface. Such $\chi$ and $\sigma$ violate the bound $7\chi+11\sigma\ge 0$, and we expect the partition function to diverge in the massless limit with a possible pole in $C'_\lambda$ at $\sigma(\CC)=e^{\pi i/3}$ (which doesn't actually happen in the present example), due to the contribution of the Higgs branch. However, we will use this as an example to illustrate that the partition function should only form an integral modular form when all three singularities are considered. Namely, if one only keeps the contribution of the two I$_1$ singularities, which is still finite in the massless limit, the partition function is still modular but not integral. In this aspect, it is completely analogous to the weakly divergent region that we will discuss later.

For this $\chi$ and $\sigma$, assuming $M_4$ is of SW simple type,\footnote{The K3 surface, for example, is of SW simple type, with the only basic class given by $\lambda=0\in H^2$ and SW$(0)=1$. See, e.g.,~\cite{MorgenBook} for a nice review of the Seiberg--Witten invariants for complex surfaces.} one has $n(\lambda)=\frac{1}{2}\lambda^2=0$ for any basic class $\lambda$. We then have
\begin{equation}
C_{\lambda}=\tilde{c}\cdot\sum_s\mu_1^2\left(\frac{f}{g}\right)^{7}\,,
\end{equation}
where $\tilde{c}=-2^{-7}3^{-14}$.
In the massless limit, the contribution of the type II$^*$ singularity diverges, while the $q$-series associated to the two type I$_1$ singularities are
\begin{align*}
    u_{s,+}&\,: \quad -\left(q^{10}+360q^{11}+62640 q^{12} + 6179680 q^{13}+\dots \right)\,,\\
    u_{s,-}&\,: \quad -\frac{1}{2^{40}3^{30}}\left(-1 + 840 q - 381240 q^2 + 125951520 q^3-34424679480 q^4 +\dots\right)\,.
\end{align*}
Thus, when only incorporating the contribution from $u_{s,\pm}$, we have
\begin{equation}
    C'_{\lambda}=-\frac{1}{2^{40}3^{30}}\left(-1+840q-381240q^2+125951520q^3-34424679480 q^4+\dots \right)\,,
\end{equation}
where the prime is a reminder that it doesn't include all singularities. We can observe that $C'_\lambda$ is not an integral $q$-series but rather one with fractional coefficients. Barring the possibility of having a very large number of basic classes or a very large value for SW$(\lambda)$ (neither of which happens for the standard K3 and many known exotic K3's), the partition function needs completion, namely by the contribution of the $E_8$ singularity that diverges in this particular limit. 

Now we look at modularity, which is heuristically expected as the two I$_1$ singularities form a closed SL$(2,\Z)$ orbit by themselves. The modular weight for the partition function is
\begin{equation}
    -\frac{11}{4}\chi-\frac{31}{4}\sigma=58\,,
\end{equation}
and one can attempt to express the combined contributions from the two I$_1$ singularities as a modular form as
\[
C'_\lambda\sim (a_{13,1,0}E_4^{13}+a_{10,1,1}\Delta E_4^{10}+a_{7,1,2}\Delta^2E_4^{7}+a_{4,1,3}\Delta^3E_4^4+a_{1,1,4}\Delta^4E_4)E_6\,.
\]
This linear system has the unique solution
\begin{equation}
C'_\lambda=\frac{1}{2^{40}3^{30}}\,E_4E_6\left(E_4^6-1296E_4^3\Delta+186624\Delta^2\right)\left(E_4^6-2160E_4^3\Delta+933120\Delta^2\right)\,,
\end{equation}
so $C'_\lambda$ is a holomorphic modular form of weight 58, but with non-integral coefficients. Notice that $C'_\lambda$ is regular at the Argyres--Douglas point with $\sigma=e^{\pi i/3}$, which makes sure that the AD point doesn't give a divergent contribution for the K3 partition function. Negative powers of $E_4$ will in general appear once we enter the region of the MMP sum rules (cf.~Figure~\ref{fig:geography}), which will make the full partition function with general $E_8$ holonomies vanish.

\subsection{$\chi_h\leq c_1^2\le5.8\chi_h$: the weakly divergent region}

Now let us consider four-manifolds that lie within the region between the two critical lines, in a family given by $\chi_h=n$ and $c_1^2=5n-4$ where $n\in\mathbb{Z}_+$. In this case, we have $c_1^2=5\chi_h-4$ which naturally satisfies $\chi_h\le c_1^2<5\chi_h$. We have
\begin{equation}
    \chi=4+7n\,,\qquad \sigma=-4-3n\,.
\end{equation}
The resulting partition function is expected to have modular weight $4(5+n)$. 

By direct computation, one can verify that the $C'_\lambda$ coefficients (again assuming $n(\lambda)=0$) for 4-manifolds in this family---and more generally in this entire region for all cases we checked---can always be expressed as modular forms, but in general they have fractional coefficients.

As an example, when $n=2$, the $q$-series of $C'_{\lambda}$ is given by
\begin{equation}
    C'_{\lambda}= -\frac{1}{2^{16}3^{12}}\left(1-48q-19440q^2+4574400q^3+34188692496 q^4+\dots \right)\,,
\end{equation}
where we used $\tilde{c}=-2^{-4}3^{-8}$.
As the expected modular weight is $28$, one can use the following ansatz for $C'_{\lambda}$,
\[
    C'_{\lambda}=\sum_{k=0}^2c_kE_4^{7-3k}\Delta^k\,.
\]
There is actually a solution for the coefficients $c_k$,
\begin{equation}
    C'_{\lambda}=-2^{-16}\cdot 3^{-12}\cdot\left(E_4^7-1728E_4^4\Delta+373248E_4\Delta^2\right)\,,
\end{equation}
which confirms that $C'_{\lambda}$ is indeed still modular even though it doesn't incorporate the contribution of the $E_8$ SCFT. 

This is a general feature in this region, which we further illustrate with examples saturating the inequality $7\chi+11\sigma\ge0$.

\subsubsection*{On the border line $c_1^2=\chi_h$}

Let us consider $\chi=176$ and $\sigma=-112$. This leads to $c_1^2=\chi_h=16$ and $b_2^+=31>1$. We will consider a basic class $\lambda$ with $n(\lambda)=0$.

For this choice of $\chi$ and $\sigma$, one has
\begin{equation}
    C'_\lambda=\tilde{c}\cdot\sum_{s=\pm}\mu_1^{16}\left(\frac{f}{g}\right)^{48}=\frac{1}{2^{320}3^{192}}\cdot\left[(E_4^{3/2}-E_6)^{64}+(E_4^{3/2}+E_6)^{64}\right]\,,
\end{equation}
which is fully modular, as terms with fractional powers of $E_4$ cancel between the two singularities, but has fractional coefficients. 

To see the latter fact more clearly, we can re-express the above in the following standard form
\begin{equation}
    C'_{\lambda}=
    \sum_{k=0}^{32}c_kE_4^{96-3k}\Delta^k\,,
\end{equation}
where the coefficients $c_k$ are fractional for all $k$. Concretely, the first six values are
\begin{alignat*}{3}
    c_0&=2^{-256}3^{-192}\,,\qquad &&c_1=-2^{-246}3^{-189}\,, \qquad &&c_2=2^{-243}3^{-186}\cdot61\,,\\
    c_3&=-2^{-237}3^{-183}\cdot 295\,,\qquad &&c_4=2^{-236}3^{-180}\cdot 32509\,,
    \qquad &&c_5=-2^{-229}3^{-177}\cdot 42427\,,\\
    & &&\qquad \vdots &&
\end{alignat*}

Therefore, after subtracting the divergent $E_8$ contribution, the remaining part of the partition function,
\begin{equation}
    Z'=\sum_\lambda C'_\lambda\cdot \SW(\lambda),
\end{equation}
is still modular but no longer guaranteed to be integral.

\subsubsection*{Integrality at $c_1^2=5\chi_h$}

The $c_1^2=5.8\chi_h$ line divides the convergent and weakly divergent regions, but one observation is that, already at $c_1^2=5\chi_h$, the coefficient $C'_\lambda$ becomes integral. For 4-manifolds on this line, the modular weight of the partition function can be found to be 
\begin{equation}
    -\frac{11\chi+31\sigma}{4}=-5c_1^2+29\chi_h=4\chi_h.
\end{equation}

For instance, consider $\chi=14,\sigma=-6$ such that $c_1^2=10$ and $\chi_h=2$. We will again only consider basic classes $\lambda$ with $n(\lambda)=0$, and the Seiberg--Witten contributions from the two I$_1$ singularities are given with coefficient 
\begin{equation}
    C'_\lambda=-2 E_4^2\,.
\end{equation}
This behavior remains consistent along the $c_1^2=5\chi_h$ line---namely $C'_\lambda\propto 2E_4^{\chi_h}$ up to a phase for all 4-manifolds on this line. In fact, the phase can be imaginary when $\chi_h$ is odd, which we will also see in the next examples.

\subsection{Odd $\chi$ and $\sigma$}

For the previous examples, they typically have even $\chi$ and $\sigma$, but this is not a necessary requirement to have an interesting partition function. Let us now consider four-manifolds that have odd $\chi$ and $\sigma$. 

We want to consider four-manifolds that can lead to a partition function with even modular weight, so that it can be non-trivial in the massless limit. This amounts to requiring
\begin{equation}
    -\frac{11}{4}\chi-\frac{31}{4}\sigma\equiv 0\pmod{2}\,,
\end{equation}
in addition to $11\chi+31\sigma>0$. As we also want to have non-vanishing SW invariants, $\chi_h=(\chi+\sigma)/4$ should be integral. There are two possibilities, one with $\chi_h$ even, and one with $\chi_h$ odd. We discuss them separately below.

\subsubsection*{Odd $\chi_h$}
One can verify that, when the modular weight is even and $\chi_h$ is odd, we have 
\begin{equation}
    \chi-3\sigma\equiv\chi+\sigma+4\equiv 0\pmod{8}\,,
\end{equation}
which indeed forces both $\chi$ and $\sigma$ to be odd. 

A simple example among this class of four-manifolds satisfying the above constraint is one with $\chi=3,\sigma=1$ (which has $b_2^+=1$ and, therefore, a potentially non-trivial Coulomb branch contribution). Such a four-manifold gives an even modular weight of $-16$. We have
\begin{equation}
\label{eq:complete phase between spinc}
    C_{\lambda}\cdot \SW(\lambda)=(-1)^{\chi_h+\sigma}\cdot C_{-\lambda}\cdot \SW(-\lambda)=C_{-\lambda}\cdot \SW(-\lambda)\,,
\end{equation}
and there will be no cancellations among opposite spin$^c$ structures. The contribution to the partition function from each basic class $\lambda$ with $n(\lambda)=0$ is
\begin{equation}
\label{eq:odd chih even modular weight}
    C_{\lambda}= -i\left(\frac{1}{q^4}+\frac{288}{q^3}-\frac{5616}{q^2}-\frac{1879168}{q}+35124547752+26715275934912q+\dots\right)\,,
\end{equation}
where we used $\tilde{c}=i\,2\cdot 3^2$.
Now, one can re-express this $q$-series as a modular form,
\begin{equation}
C_{\lambda}=i\left(-E_4^8\Delta^{-4}+1728E_4^5\Delta^{-3}-373248E_4^2\Delta^{-2}\right)\,.
\end{equation}
 The modularity of $C_\lambda$ (and hence of the partition function when $b_2^+>1$) is tested in many other cases with odd $\chi_h$ and $\sigma$, which we will not list here.

Previously in Section~\ref{sec:8nplus4}, 4-manifolds with odd $\chi_h$ and even $\sigma$ lead to a trivial partition function due to cancellations between $\lambda$ and $-\lambda$ (this is the mechanism to avoid contradiction with having odd modular weights in that case). However, with odd $\sigma$, the situation is the polar opposite, with non-trivial partition function in the odd $\chi_h$ case, which we have seen above, and cancellations for the even $\chi_h$ case, which we discuss next.

\subsubsection*{Even $\chi_h$}

We now have 
\begin{equation}
\label{eq:odd chi sigma}
 \chi+\sigma\equiv 0\pmod{8}\,,
\end{equation}
which actually forces the modular weight to be odd. Therefore, the prefactors $C_\lambda$ individually cannot be full modular forms in the massless limit.

An example among this class is $\chi=7$ and $\sigma=1$. 
For this choice, we can compute the relevant $q$-series appearing in the partition function for a given $\lambda$
\begin{multline}
    C_{\lambda}=-\left(\frac{1}{q^{7}}+\frac{564}{q^{6}}+\frac{69984}{q^{5}}-\frac{5830480}{q^4}\right.\\
    -\frac{369601470}{q^3} + \frac{96337594944}{q^2}-\frac{9571149592768}{q}\\
    + (569331797433408 + 2807929681968365568i)\\
    \left.- 27\,(52230319437285 - 146012343462355009536 i)\,q+\dots\right)\,,
\end{multline}
where we used $\tilde{c}=-2^{3/2}3^3$.

This clearly does not vanish. But with $\chi_h=2$ and $\sigma$ odd, we have a cancellation between $\lambda$ and $-\lambda$ due to
\begin{equation}
       C_{\lambda}\cdot \SW(\lambda)=(-1)^{\chi_h+\sigma}C_{-\lambda}\cdot \SW(-\lambda)=-C_{-\lambda}\cdot \SW(-\lambda)\,. 
\end{equation}
Thus, the partition function vanishes
\begin{equation}    Z[M_4\times T^2]=\sum_{\lambda}C_{\lambda}\cdot\mathrm{SW}(\lambda)=0\,.
\end{equation}

For four-manifolds that satisfy~\eqref{eq:odd chi sigma}, the  $q$-series for $C_\lambda$ in general has complex coefficients.\footnote{The negative-power $q$-terms appearing in the partition function actually remain real. The source of the imaginary terms arises consistently from the $u_{s,-}$ singularity.}  This is, in a sense, a bigger pathology compared with non-modularity as the interpretation in terms of BPS state counting would fail if the whole partition function also behaves in this way. Fortunately, this problem is cured by the cancellation discussed above.

\subsection{$11\chi+31\sigma=0$: modular invariant partition functions} \label{sec:zeroweight}
\fixme{need edit}
The $11\chi+31\sigma=0$ line in the $\chi$--$\sigma$ plane divides the convergent and weakly divergent region. On this line, the partition function is expected to be a modular function (i.e.~modular form of weight zero) and strictly invariant under the SL$(2,\Z)$ action. 

However, as explained in Footnote~\ref{foot:masslesslimit}, 
on this borderline the ten I$_1$ singularities that coalesce to the II$^*$ point contribute, in the massless limit, a finite constant that depends on the direction in which the masses are sent to zero. So the full partition function doesn't really have a well-defined massless limit here. Therefore, we will again look at $C'_\lambda$, which is a modular function and can be written as a polynomial in the $j$-invariant. 

We can consider $M_4$ with $\chi=31$ and $\sigma=-11$. This corresponds to $\chi_h=5$, which allows the partition function to be non-trivial. In the massless limit, we have 
\begin{equation}
    C'_\lambda=i\left(-373248 j^2+1728j^3-j^4\right)\,, 
\end{equation}
where we used $\tilde{c}=2^{-3}3^{-6}\cdot i$. The ``two I$_1$'' partition function is then indeed (up to a phase) an integral modular function  with no constant term in $j$.

 This is in fact a general phenomenon observed for all such modular invariant cases, i.e., for all $(\chi,\sigma)=n\cdot (31,-11)$ with $n\in\mathbb{Z}_+$. For the $n=2$ case, we have
\begin{equation}
        C'_\lambda=-\left(69657034752\,j^4-1289945088 j^5+3732480 j^6-3456 j^7+j^8\right)\,,
\end{equation}
where we used $\tilde{c}=-2^{-6}3^{-12}$.
Similar to the case of $n=1$, this coefficient is an integral modular form but now starts with $j^4$. Also notice that the TMF constraint is automatically satisfied in these cases, as any integral modular forms of weight zero are also topological modular forms. 

For this family of examples---in fact for any other examples with $c_1^2\ge 5\chi_h$ we have studied---the powers of 2 and 3 always exactly cancel in the end (cf.~coefficients in green region of Table~\ref{tab:original grav coef}). Although the conceptual explanation for this fact remains unclear, it highlights how sharply the E-string theory captures the structure of 4-manifolds.

\section{Conclusions and future directions}
\label{sec:5}

In this paper, we studied the E-string theory and its compactifications, with a focus on the computation and analysis of the $M_4\times T^2$ partition function, primarily in the massless limit. The results illustrate several key advantages of employing 6d theories in the study of 4-manifolds, suggesting that this is an exciting avenue for future research. In this last section, we outline several directions that we believe are especially interesting to explore.

\paragraph{Inclusion of defects.} One of the most interesting features of 6d SCFTs is that they often have defects of codimension 2 and 4. After the compactification on $T^2$, they give rise to defects of all possible dimensions in the 4d effective theory. While all of them are interesting (the study of line defects, for example, is closely related to the computation of the partition function on non-simply-connected 4-manifolds) and incorporating them should vastly enhance the power of the 6d theory in the study of 4-manifolds, of special interest are point and surface defects, partially due to their connection with the Donaldson polynomial. In the E-string theory, one type of point operators is obtained from an M2-brane wrapping $T^2$. Its modular weight can be computed via anomaly inflow to be $-6$, suggesting that they are related to the ``$u$ operator'' in the low-energy description.\footnote{One often considers a generating function $e^{2pu}$ for this operator~\cite{Witten:1994cg,Moore:1997pc}, and $p$, for a 6d theory, will no longer be purely a number but carry certain modular weight.} For surface operators, there are two classes coming from M2- and M5-branes respectively. Those from M5-branes also carry modular weights while those of the other type are labeled by points on $T^2$ and have OPEs when two become closer on $T^2$ if they intersect in $M_4$. (See \cite{Alday:2009fs,Dedushenko:2017tdw} for related discussion in the context of 6d $(2,0)$ theories.)

\paragraph{Torsion-valued invariants.} Another exciting problem is---still in the framework of the 4d effective theory---to understand the TMF-valued invariant of $M_4$ when it is purely or partially torsion. This can in principle detect new information about 4-manifolds that is invisible to the SW invariants.\footnote{One such example is the effect of the ``nilpotent operations'' such as taking the connected sum with $S^2\times S^2$. (Recall that two homeomorphic but non-diffeomorphic 4-manifolds can become diffeomorphic after connected sum with $S^2\times S^2$ a certain number of times \cite{Wall}). Indeed, the modular weight for the partition function on $T^2\times S^2\times S^2$ is odd, forcing the non-torsion part to vanish. The torsion part, on the other hand, in general won't vanish, a phenomenon that can already be seen in one of the simplest examples where the 6d theory is taken to be a free tensor multiplet \cite{Gukov:2018iiq,Gukov:2025nmk}.} In fact, the computation of the partition function only utilizes a small fraction of the information contained in the 4d effective theory (i.e.~geometry in the neighborhood of I$_1$ singularities), and the modularity happens ``accidentally'' from the perspective of the 4d theory. To obtain the torsion-valued invariants, it seems necessary for one to tap further into the rich structures present in the family of Coulomb branches fibered over the moduli of elliptic curve $\mathcal{M}_{\text{ell}}$, such as the braiding of singularities, special behaviors at $\sigma=e^{\pi i/3}$ and $i$ (cf.~Figure~\ref{fig:Tsingularity} and~\ref{fig:Ssingularity}), and the BPS spectrum as well as its wall-crossing.  Among the simplest torsion-valued invariants are those in KO$(\!(q)\!)$. They correspond to lifting the 4d theory not directly to 6d but to the intermediate 5d, which is related to the next point. (See also \cite{Kim:2025fpz} for a study of the partition function of a 5d gauge theory on $S^1\times M_4$, and \cite{Gaiotto:2019asa,Gaiotto:2019gef,Tachikawa:2023nne,Tachikawa:2025flw} for more discussions on understanding the torsion-valued invariants of 2d $(0,1)$ theories.)

\paragraph{Categorification.} The E-string theory reduced on $M_4\times S^1$ leads to a 1d $\CN=1$ quantum mechanics, whose deformation classes for different KK-momentum sectors describe an element of KO$_{E_8}(\!(q)\!)$ (or, in general, the meromorphic version thereof). One can also attempt to ``categorify''  this invariant by considering the Hilbert space with the action of $E_8$ and time-reversal $\Z_2^T$ symmetry (related to flipping the internal $S^1$). This might seem inaccessible due to the strongly-coupled nature of the 5d $E_8$ theory. However, one can also turn on the holonomy of $E_8$ on the circle, and for the ``$-1$'' holonomy, the 5d theory admits a gauge theory description at low energy, whose Hilbert space can be computed as a Floer-type homology  \cite{Gukov:2018iiq}. 

\paragraph{Refinement by $E_8$ (topological) Jacobi forms.} In the present work, we only verify the full (topological) modularity of the partition functions in the massless limit. It is possible, although requiring significantly more effort, to express the partition functions as (meromorphic) $E_8$ Jacobi forms. This would allow one to go beyond the $11\chi+31\sigma$ line, while recovering integrality in the weakly divergent region. One can then ask about the lift to $\widetilde{\TMF}^{E_8}$. Currently, this subject is not sufficiently understood on the math side, but given the steady progress in equivariant TMF in recent years (see, e.g.~\cite{Gepner_Meier_2023,Lin:2024qqk,bauer2025topological}), this goal could be achievable in the not-so-distant future.

\paragraph{Dual 2d perspectives.} Another way to categorify the invariant is to consider the full 2d $(0,1)$ theory $\CT[M_4]$ obtained from compactifying the 6d theory on $M_4$. One sacrifice to make is that the theory can depend on the geometry and the presentation of $M_4$, with different choices related to each other via deformations and a web of dualities. For example, there can be dualities induced by Kirby moves, and the $(0,1)$ trialities of 2d theories discussed in~\cite{Gukov:2019lzi} might be some of the ``elementary transformations.''

\paragraph{TQFT structure.} The partition function $Z[M_4\times T^2]$ defines a 4d TQFT over $M_4$, and it would be interesting to understand its algebraic structures, including what the Hilbert spaces associated with 3-manifolds are and how cobordisms give maps between such Hilbert spaces. Such an understanding would, on one hand, allow the computation of the topological invariants via cutting-and-gluing methods such as trisection \cite{Gukov:2017zao} and, on the other hand, provide insights into the structural properties of the partition function such as those discovered in \cite{Gu:2019pqj}. Just like in the case of Donaldson and Seiberg--Witten invariants, this TQFT is not expected to satisfy the strict version of Atiyah--Segal axioms, as its Hilbert spaces on three-manifolds are, in general, infinite dimensional. One potential cure to this problem is lifting the 4d TQFT to one over (meromorphic $E_8$-equivariant) TMF, and the TMF-modules associated to 3-manifolds have a better chance of having finite rank.\footnote{This is the case for the 6d abelian tensor multiplet \cite{Gukov:2025nmk}, which is more ``non-compact''---in the sense of having two cylindrical ends in the 4d moduli space---compared with either the E-string theory or the $E_8\times E_8$ theory.}

\paragraph{Coulomb branch contribution.} Another direction that we did not adequately explore is the evaluation of the $u$-plane integral in the case of $b_2^+\leq 1$. In particular, for $b_2^+=0$, deriving the integrand and analyzing its behavior near the I$_1$ singularities would serve as the first step toward rigorously defining the partition function of the E-string theory as an invariant of 4-manifolds with vanishing $b_2^+$. Additional questions in this direction have also been suggested in Section~\ref{sec:FullTop}.

\paragraph{Beyond the E-string theory.} Another obvious generalization of the present work is to consider other 6d SCFTs, such as the $(2,0)$ M-string and the other rank-1 $\CO(-k)$ theories. Although we emphasized some of the advantages of the E-string theory over the others, it indeed possesses a Higgs branch of relatively high dimensions with a large global symmetry, which is a source of divergence in the massless limit. In the opposite regime, one has the ``non-Higgsable clusters'' that do not have any flavor symmetry, given by $k=3$, $4$, $6$, $8$, and $12$ \cite{Morrison:2012np}. (See \cite{Apruzzi:2016nfr} for a discussion on aspects of their compactification.) They have no Higgs branch, and one can expect that the partition function on $M_4\times T^2$ is always well defined. One problem is that they are in general relative theories living on the boundary of a non-trivial 7d TQFT, and, as a consequence, the partition function will be a level-$k$ modular form. In the special case of $k=4$, one can make the theory absolute (i.e.~the 7d TQFT admits a topological boundary condition; see \cite{Gukov:2020btk} for more discussion from this perspective), and the partition functions will be modular forms. The first step to perform such a computation is to better understand compactifications of the theory and, in particular, the details of its SW geometry. 

\smallskip
We hope the present work will inspire further research and exploration in the directions mentioned above and beyond.

\subsubsection*{Acknowledgements}
We would like to thank Sergei Gukov, Hee-Cheol Kim, Slava Krushkal, Jan Manschot, Greg Moore, Kaiwen Sun, Houri Tarazi, Cliff Taubes, Cumrun Vafa, Edward Witten, and Xinyu Zhang for interesting discussions and correspondences and two anonymous referees for their valuable comments and suggestions.  DW would also like to thank Junkai Dong for helpful discussions on Mathematica.  
The work of DP is supported by research grant 42125 from VILLUM FONDEN, ERC-SyG project No.~810573 ``Recursive and Exact New Quantum Theory,'' Simons Collaboration on ``New Structures in Low-dimensional Topology,'' and the National Science Foundation Grant No.~DMS-2424139, while he was in residence at the Simons Laufer Mathematical Sciences Institute (SLMath) during the Fall 2026 semester. The work of DW is supported by a grant from the Simons Foundation (602883, CV), the DellaPietra Foundation, and by the NSF grant PHY-2013858. We thank the organizers of the 2022 and 2023 Summer Workshops at the Simons Center for Geometry and Physics, the 29th Nordic Congress of Mathematicians, and the workshop ``Enumerative Invariants, Quantum Fields and String Theory'' at the Mittag-Leffler Institute for kind hospitality. DW thanks the Center for Quantum Mathematics at SDU for graciously hosting him during the final stage of this work.

\newpage
\appendix

\section{Modular and Jacobi forms}
\label{sec:modular forms}

In this appendix, we briefly review some definitions and basic properties of modular forms and Jacobi forms used in this paper. 
\subsection*{Modular forms}
\begin{definition}
    For a given holomorphic function $f:\mathbb{H}\to\mathbb{C}$ analytic in $\tau\in\mathbb{H}$, if for all elements $\gamma=
    \begin{pmatrix}
        a&b\\
        c&d
    \end{pmatrix} \in \SL(2,\mathbb{Z})$, one has
    \begin{equation}
    \label{eq:modularForms}
        f\left(\frac{a\tau+b}{c\tau+d}\right)=(c\tau+d)^{2k}f(\tau)\,,
    \end{equation}
    and $f$ can be expressed as $f=\sum_{n\geq 0}a_nq^n$ where $q:=\exp(2\pi i\tau)$, then $f$ is a (holomorphic) modular form of weight $2k$ where $k\in\mathbb{Z}$.
\end{definition}
There are various useful generalizations that relax part of the definition. 
\begin{itemize}
    \item 
A \emph{weakly holomorphic} modular form is an $f$ with the same transformation property under SL$(2,\Z)$ but some non-vanishing $a_n$ coefficient for $n<0$. These form a ring which we denote as MF$_{*}$. 
\item One can furthermore demand that the $a_n$'s are integers. The ring of such \emph{integral weakly holomorphic modular forms} will be denoted as MF$^\Z_*$.\footnote{This is sometimes denoted as wMF$_*$ to emphasize that they are only weakly holomorphic, e.g.~they are formal Laurent series of $q$ (valued in $\Z(\!(q)\!)$) as opposed to formal power series (in $\Z[\![q]\!]$).}
\item One can also relax the transformation property by demanding that it only has to hold for a subgroup $\Gamma\subset\SL(2,\Z)$. Then $f$ will be a modular form of level $\Gamma$. When $\Gamma=\Gamma_0(N)$ is a Hecke congruence subgroup given by requiring the matrix element $c\equiv 0 \pmod N$,\footnote{Recall that
\begin{equation*}
    \SL(2,\mathbb{Z}):=\left\{\left.\begin{pmatrix}a&b\\c&d\end{pmatrix}\right\vert a,b,c,d\in\mathbb{Z},ad-bc=1\right\}\,.
\end{equation*}
 A congruence subgroup contained in $\Gamma_0(N)$ is $\Gamma_1(N)$, which has $c\equiv0$ and $a\equiv d\equiv1\pmod N$.} such modular forms are often referred to as \emph{level-$N$ modular forms}.
\end{itemize}
\noindent In the following, we will mainly work with modular forms of the full $\SL(2,\mathbb{Z})$ group, unless noted otherwise. The generators of $\SL(2,\mathbb{Z})$ are 
\begin{equation}
\label{sl2z generators}
    S=
    \begin{pmatrix}
    0&-1\\
    1&0
    \end{pmatrix},\qquad
    T=
    \begin{pmatrix}
    1&1\\
    0&1
    \end{pmatrix}\,.
\end{equation}
Therefore, using the above definition of modular forms and generators of $\SL(2,\mathbb{Z})$, we can infer several properties of modular forms of $\SL(2,\mathbb{Z})$. In particular, if $f$ is a modular form of $\SL(2,\mathbb{Z})$, then $f(\tau+1)=f(\tau)$ is periodic, which makes the Fourier expansion in $q=e^{2\pi i\tau}$ convenient. Similarly, all modular forms of odd weight vanish as is evident by applying $S^2$ to $f$.\footnote{Note that this is a statement about SL$(2,\Z)$ modular forms and does not apply to modular forms under more general subgroups of $\SL(2,\mathbb{Z})$.} Furthermore, the set of all weight-$2k$ modular forms is a vector space. In the following, we will introduce two examples of modular forms that arise in the study of the E-string theory.

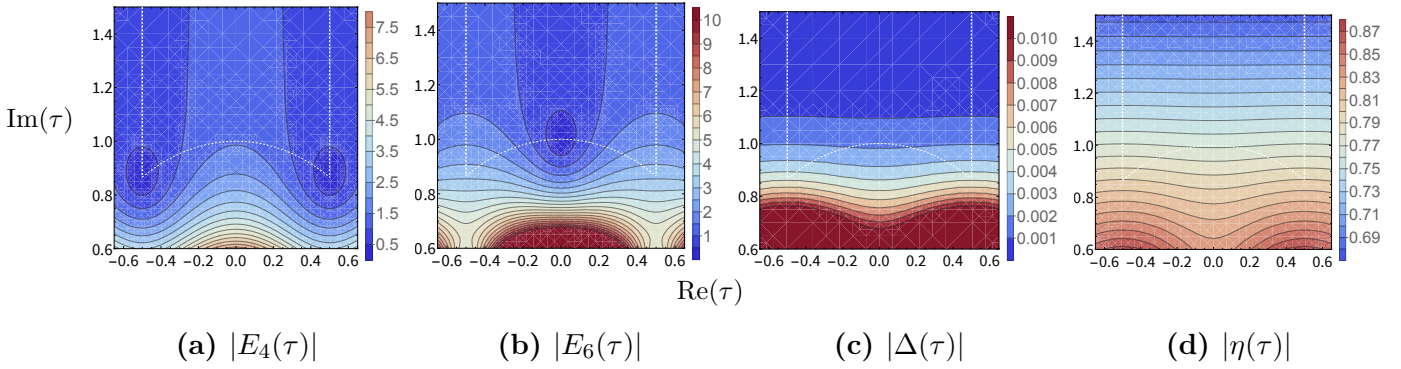
\begin{figure}
    \centering
    \vskip 0pt
    \begin{subfigure}{0.24\textwidth}
    \centering
    \includegraphics[width=\linewidth]{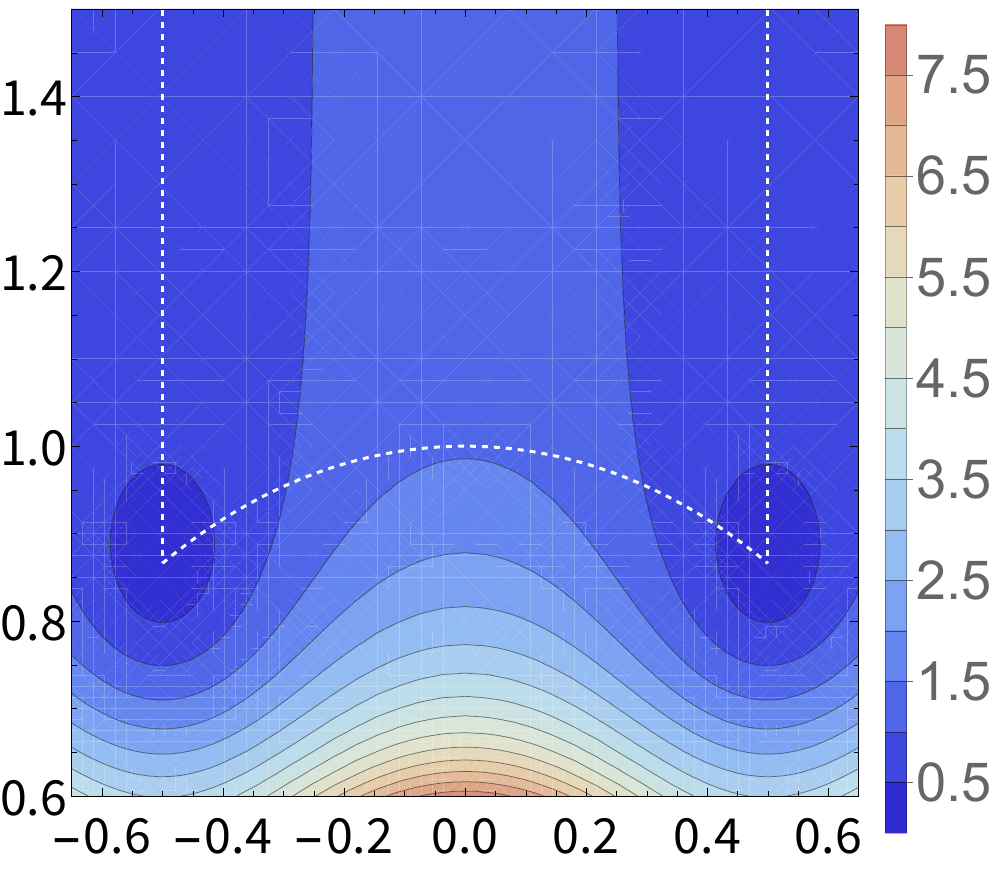}
    \begin{picture}(0,0)\vspace*{-1.2cm}
    \put(-90,70){\footnotesize $\Im(\tau)$}
    \end{picture}\vspace*{0.0cm}
    \caption{$|E_4(\tau)|$}
    \label{fig:modular forms: E4}
    \end{subfigure}
    \hfill
    \begin{subfigure}{0.24\textwidth}
    \centering
    \includegraphics[width=\linewidth]{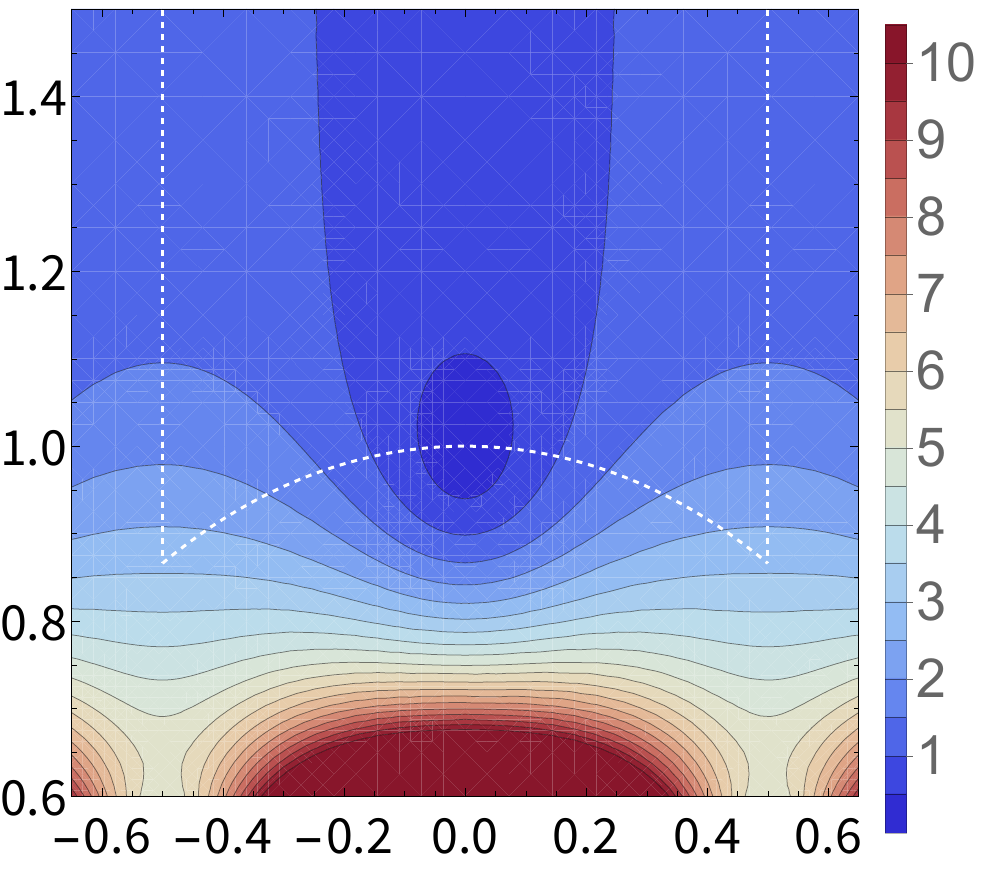}
    \begin{picture}(0,0)\vspace*{-1.2cm}
    \put(40,5){\footnotesize $\Re(\tau)$}
    \end{picture}\vspace*{0.0cm}
    \caption{$|E_6(\tau)|$}
    \label{fig:modular forms: E6}
    \end{subfigure}
    \hfill
    \begin{subfigure}{0.25\textwidth}
    \centering
    \includegraphics[width=\linewidth]{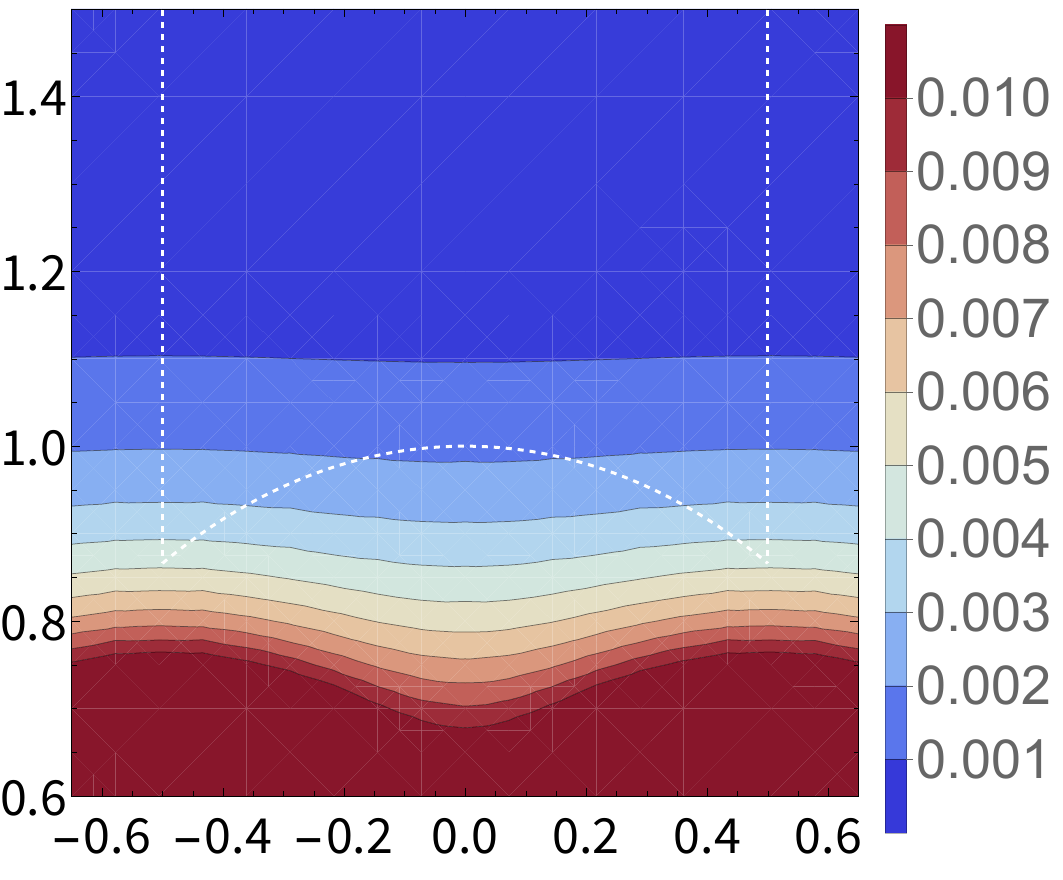}
    \begin{picture}(0,0)\vspace*{-1.2cm}
    \end{picture}\vspace*{0.0cm}
    \caption{$|\Delta(\tau)|$}
    \label{fig:modular forms: Delta}
    \end{subfigure}
    \hfill
    \begin{subfigure}{0.24\textwidth}
    \centering
    \includegraphics[width=\linewidth]{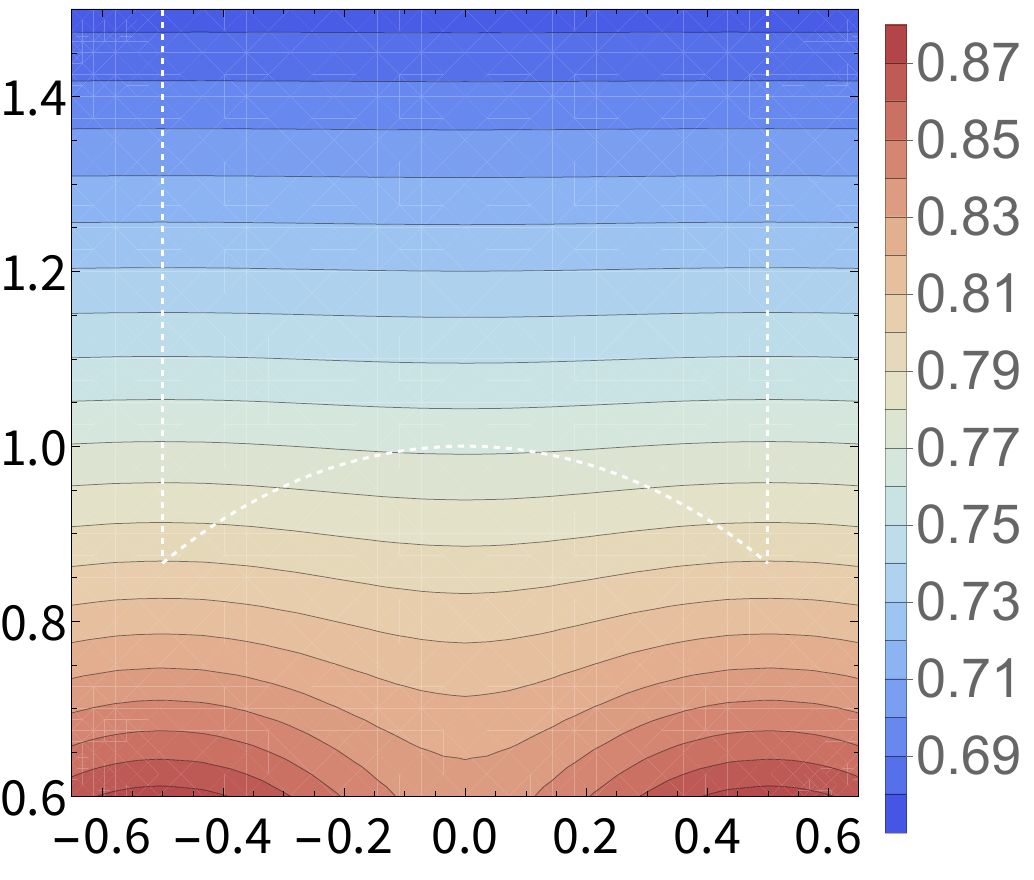}
    \begin{picture}(0,0)\vspace*{-1.2cm}
    \end{picture}\vspace*{0.0cm}
    \caption{$|\eta(\tau)|$}
    \label{fig:modular forms: Eta}
    \end{subfigure}\vspace*{-0.2cm}
    \caption{\textbf{Modular forms.} The absolute values of $E_4(\tau),E_6(\tau)$, $\Delta(\tau)$, and $\eta(\tau)$ on the upper half plane $\mathbb{H}$. The white dotted lines indicate the \textit{fundamental domain} of $\SL(2,\mathbb{Z})$.}
    \label{fig:modular forms}
\end{figure}

\paragraph{Eisenstein series.} This is really a family of modular forms, which are defined as $E_{2k}=G_{2k}(\tau)/(2\zeta(2k))$ with $k\in\Z_{\ge 2}$,
\begin{equation}
\label{eq:Eisenstein G}
    G_{2k}(\tau)=\sum_{(m,n)\in\mathbb{Z}^2\backslash \{0,0\}}\frac{1}{(m+n\tau)^{2k}}\,,
\end{equation}
and $\zeta$ being the Riemann zeta function. They are modular forms of weight $2k$. To see this, we observe that an $\SL(2,\mathbb{Z})$ transformation on $\tau$ will result in
\begin{align*}
    G_{2k}\left(\frac{a\tau+b}{c\tau+d}\right)&=\sum_{(m,n)\in\mathbb{Z}^2\backslash \{0,0\}}\left(\frac{c\tau+d}{m(c\tau+d)+n(a\tau+b)}\right)^{2k}=(c\tau+d)^{2k}\sum_{(m',n')\in\mathbb{Z}^2\backslash \{0,0\}}\frac{1}{(m'+n'\tau)^{2k}}\,.
\end{align*}
The Fourier expansion of the Eisenstein series is
\begin{equation*}
    E_{2k}=1-\frac{4k}{B_{2k}}\sum_{n\geq 1}\tilde{\sigma}_{2k-1}(n)q^n\,,
\end{equation*}
where $B_{2k}=\frac{(-1)^{k+1}2(2k)!}{(2\pi)^{2k}}\zeta(2k)$ are the Bernoulli numbers and $\tilde{\sigma}_{2k-1}(n):=\sum_{d\vert n}d^{2k-1}$ is the divisor sum function. For $E_4$ and $E_6$, their $q$-expansions are
\begin{align*}
    E_4&=1+240 q+2160 q^2+6720 q^3+17520 q^4+30240 q^{5}+\dots,\\
    E_6&=1-504 q-16632 q^2-122976 q^3-532728 q^4-1575504 q^{5}+\dots\,.
\end{align*}
The two functions are illustrated in Figure~\ref{fig:modular forms: E4} and \ref{fig:modular forms: E6}. The zeros of $E_4$ are located at $\tau=\pm 1/2+\sqrt{3}i/2$, while the zeros of $E_6$ are located at $\tau=i$. We can build more modular forms from polynomials in $E_4$ and $E_6$. In fact, over $\mathbb{Q}$, all modular forms can be written as such polynomials. In other words, the ring of holomorphic modular forms over $\mathbb{Q}$ is $\mathbb{Q}[E_4,E_6]$. To get a weakly holomorphic modular form, it turns out that one just needs to also include the inverse of the \textit{modular discriminant}.

\paragraph{Modular Discriminant.} It is defined as 
\begin{equation}
    \Delta=\frac{E_4^3-E_6^2}{1728}\,.
\end{equation} 
See  Figure~\ref{fig:modular forms: Delta} for a plot of this function. It is also an integral modular form with the following $q$-series expansion
\begin{equation*}
    \Delta=q-24 q^2+252 q^3-1472 q^4+4830 q^5+\dots
\end{equation*}
This is a modular form of weight $12$ that is nowhere vanishing in $\mathbb{H}$, and in particular can be inverted, with
\begin{equation}
    \Delta^{-1}=q^{-1}+24+...
\end{equation}
also an integral modular form. However, $\Delta^{-1}$ does have a pole at the cusp $\tau\rightarrow i \infty$, making it weakly holomorphic. The ring of weakly holomorphic modular forms is then MF$_{*}\simeq \C[E_4,E_6,\Delta^{-1}]$. To get the ring of integral modular forms, one also needs to add $\Delta$, leading to MF$^\Z_*\simeq\Z[E_4,E_6,\Delta^\pm]/(E_4^3-E_6^2-1728\Delta)$. One can in addition add the inverse of $E_4$ to get a larger ring, which is what we actually used in several cases.

Beyond these examples, there are also some commonly used functions that are not strictly modular forms, but possess some relaxed version of modularity. We mention two examples below.

\paragraph{The Eisenstein series $E_2$.} For the case of $k=1$, using \eqref{eq:Eisenstein G}, $E_2$ does not converge absolutely and is hence a \textit{quasi-modular form} which transforms as
\begin{equation}
\label{eq: E2 transformation}
    E_2\left(\frac{a\tau+b}{c\tau+d}\right)=(c\tau+d)^2E_2(\tau)-\frac{6ic}{\pi}(c\tau+d)\,.
\end{equation}
This can be derived by using ``Hecke's trick,'' see e.g., Chapter 4 of \cite{Zagier1992}. Finally, we state without proof the following Ramanujan identities:
\begin{subequations}
\begin{align}
    \label{eq:ramanujan identity E2}
    DE_2&=\frac{1}{12}(E_2^2-E_4)\,,\\
    \label{eq:ramanujan identity E4}
    DE_4&=\frac{1}{3}(E_4E_2-E_6)\,,\\
    \label{eq:ramanujan identity E6}
    DE_6&=\frac{1}{2}(E_6E_2-E_4^2)\,,
\end{align}
\end{subequations}
where we introduce the operator $D:=q\frac{d}{dq}$.

\paragraph{Dedekind eta function.} Unlike the above two examples, the eta function, defined as
\begin{equation}
    \eta(\tau)=q^{1/24}\prod_{n\geq 1}(1-q^n)=q^{1/24}\left(1-q-q^2+q^5+q^7+\dots \right),
\end{equation}
 is not a modular form in the strict sense, but a weight $1/2$ modular form with a ``multiplier system,'' i.e.~it carries a non-trivial complex phase when acted upon by elements in $\SL(2,\mathbb{Z})$.\footnote{Naively, having a fractional modular weight contradicts our definition of modular forms, i.e., \eqref{eq:modularForms}. However, functions that transform correctly with a multiplier system, $\epsilon(\gamma)$, that satisfies
\begin{equation*}
    f\left(\frac{a\tau+b}{c\tau+d}\right)=\epsilon(\gamma)(c\tau+d)^kf(\tau)\,,
\end{equation*}
where $k\in\mathbb{Q}$ and $|\epsilon(\gamma)|=1$ for all $\gamma\in \SL(2,\mathbb{Z})$ are also customarily referred to as modular forms. For the case of the Dedekind eta function, one should further lift $\gamma$ to Mp$(2,\Z)$---the metaplectic double cover of SL$(2,\Z)$---to make the function single valued.} For example, when we apply the generators of $\SL(2,\mathbb{Z})$ to $\eta$, we have
\begin{align*}
    T\,:&\qquad \eta(\tau+1)=e^{2\pi i(\tau+1)/24}\prod_{n\geq 1}(1-q^n)=e^{\pi i/12}\eta(\tau)\,,\\
    S\,:&\qquad \eta(-1/\tau)=\sqrt{-i\tau}\eta(\tau)\,.
\end{align*} 
Lastly, we have $\Delta(\tau)=\eta(\tau)^{24}$ which becomes a modular form in the strict sense, as the multiplier system of $\eta$ becomes trivialized.

\subsection*{$E_8$ Jacobi forms}

Another way to generalize modularity is by incorporating some Jacobi variables. In the context of the E-string theory, the relevant ones are $W(E_8)$-invariant Jacobi forms (or ``$E_8$ Jacobi forms'' for short), $g:\mathbb{H}\times \mathbb{C}^8\to\mathbb{C}$. Here we will only review some basic facts about them, and the reader is referred to~\cite[App.~A.1]{Sakai:2011xg} for more details and~\cite{Wang:2018fil,Sun:2021ije,Sakai:2024vby} for recent developments. 

The $W(E_8)$-invariant Jacobi forms are, as the name suggests, invariant under actions of the Weyl group of $E_8$ on $\{m_i\}\in\mathbb{C}^8$,
\begin{equation}
    g_{k,l}(\tau,w(\{m_i\}))=g_{k,l}(\tau,\{m_i\})\,,\qquad w\in W(E_8)\,.
\end{equation}
These Jacobi forms are also quasi-periodic under the following shift
\begin{equation}
    g_{k,l}(\tau,\{m_i\}+2\pi\alpha+2\pi\tau\beta)=\exp\left[-li\left(\pi\tau \beta^2+\langle \{m_i\},\beta\rangle\right)\right]g_{k,l}(\tau,\{m_i\})\, ,\qquad \alpha\,,\beta\in\Gamma_{E_8}\, ,
\end{equation}
where $\Gamma_{E_8}=\{\{m\}\in\mathbb{Z}^8\cup(\mathbb{Z}+1/2)^8\vert \sum_im_i=0\pmod{2}\}\subset \frak{t}_{E_8}\simeq \R^8$ is the (co-)root lattice of $E_8$ and $\langle\cdot\,,\cdot\rangle$ denotes the pairing with respect to the standard metric on $\R^8$. Here, the integer $l\in\mathbb{Z}$ is known as the ``index'' of the Jacobi form, which is often also referred to as the level  of the Jacobi form, as it is related to the level of the affine $E_8$ Lie algebra in the context of 2d CFTs. 

The lattice $\Gamma_{E_8}$ is generated by $\alpha_{\rm I}=\{\pm 1,\pm 1,0,\dots,0\}$ and $\alpha_{\rm II}=\{\pm 1,\pm 1,\dots,\pm 1\}/2$, where $\alpha_{\rm I}$ can take any permutations of the two $\pm 1$ with an arbitrary choice of signs, while components of $\alpha_{\rm II}$ have to sum to an even number.
The modularity of the $W(E_8)$-invariant Jacobi forms under an action of $\gamma=\begin{pmatrix}a&b\\c&d\end{pmatrix}\in \SL(2,\mathbb{Z})$ is given by the following transformation property, 
\begin{equation}
    g_{k,l}\left(\frac{a\tau+b}{c\tau+d},\frac{\{m_i\}}{c\tau+d}\right)=(c\tau+d)^{k}\exp\left[\frac{il}{4\pi}\frac{c\cdot m^2}{c\tau+d}
    \right]g_{k,l}(\tau,\{m_i\})\,
\end{equation}
which depends on both the weight $k$ and the index $l$. 

The building blocks of $W(E_8)$-invariant Jacobi forms are the theta functions defined as
\begin{subequations}
\begin{align}
\label{eq:theta1}
    \vartheta_1(\tau,z)&=i\sum_{n\in\mathbb{Z}}(-1)^ne^{iz(n-1/2)}q^{(n-1/2)^2/2}=2\sum_{n\geq 1}(-1)^{n-1}q^{(n-1/2)^2/2}\sin\left[\left(n-\frac12\right)z\right]\,,\\
    \label{eq:theta2}
    \vartheta_2(\tau,z)&=\sum_{n\in\mathbb{Z}}e^{iz(n-1/2)}q^{(n-1/2)^2/2}=2\sum_{n\geq 1}q^{(n-1/2)^2/2}\cos\left[\left(n-\frac12\right)z\right]\,,\\
    \label{eq:theta3}
    \vartheta_3(\tau,z)&=\sum_{n\in\mathbb{Z}}e^{izn}q^{n^2/2}=1+2\sum_{n\geq 1}q^{n^2/2}\cos[nz]\,,\\
    \label{eq:theta4}
    \vartheta_4(\tau,z)&=\sum_{n\in\mathbb{Z}}(-1)^ne^{izn}q^{n^2/2}=1+2\sum_{n\geq 1}(-1)^nq^{n^2/2}\cos[nz]\,.
\end{align}
\end{subequations}
(For a short-hand notation, we denote $\vartheta_l(\tau,0)$ as $ \vartheta_l(\tau)$ or simply $\vartheta_l$.) Up to linear order in the small-$z$ expansion, the elliptic theta functions have the following $q$-series,
\begin{alignat*}{2}\vartheta_1(\tau,z)&=z[\partial_z\vartheta_1(\tau)]+\mathcal{O}(z^2)\,,\qquad &&\vartheta_2(\tau,z)=\vartheta_2(\tau)+\mathcal{O}(z^2)\,,\\
    \vartheta_3(\tau,z)&=\vartheta_3(\tau)+\mathcal{O}(z^2)\,,\qquad &&\vartheta_4(\tau,z)=\vartheta_4(\tau)+\mathcal{O}(z^2)\,,
\end{alignat*}
where $\partial_z\vartheta_1(\tau):=\partial_z\vartheta_1(\tau,z)\vert_{z=0}$. We can also relate $\vartheta_1(\tau,z)$ to $\vartheta_{2,3,4}$ via the following transformations
\begin{equation}
\label{eq:elliptic theta 1 relations}
    \begin{gathered}\vartheta_1(\tau,z+\pi)=\vartheta_2(\tau,z)\, ,\quad \vartheta_1(\tau,z+\pi\tau)=i q^{-1/8}e^{-iz/2}\vartheta_4(\tau,z)\, ,\\ \vartheta_1(\tau,z+\pi+ \pi\tau)=q^{-1/8}e^{-iz/2}\vartheta_3(\tau,z)\, .\end{gathered}
\end{equation}
Furthermore, these theta functions transform under an action of $T$ as
\begin{align}
\label{eq:t to t+1}
\begin{split}
    \vartheta_1(\tau+1,z)=e^{\pi i/4}\vartheta_1(\tau,z)\, ,&\qquad \vartheta_2(\tau+1,z)=e^{\pi i /4}\vartheta_2(\tau,z)\, ,\\
    \vartheta_3(\tau+1,z)=\vartheta_4(\tau,z)\, ,&\qquad \vartheta_4(\tau+1,z)=\vartheta_3(\tau,z)\, ,
\end{split}
\end{align}
and when $z=0$, they transform under an action of $S$ as
\begin{equation}
    \vartheta_2(-1/\tau)=\sqrt{-i\tau}\vartheta_4(\tau)\, ,\qquad \vartheta_3(-1/\tau)=\sqrt{-i\tau}\vartheta_3(\tau)\, ,\qquad \vartheta_4(-1/\tau)=\sqrt{-i\tau}\vartheta_2(\tau)\, .
\end{equation}

From the above elliptic theta functions, we can construct the theta function associated to the $E_8$ lattice as 
\begin{equation}
\label{eq:P}
    P(\tau,\{m_i\})=\frac{1}{2}\left[\sum_{l=1}^4\prod_{i=1}^8\vartheta_l(\tau,m_i)\right].
\end{equation}
Thus, we have the following modular behavior under an action of $\gamma\in \SL(2,\mathbb{Z})$ for $P(\tau,\{m_i\})$
\begin{equation}
    P\left(\frac{a\tau+b}{c\tau+d},\frac{\{m_i\}}{c\tau+d}\right)=(c\tau+d)^4\exp(\frac{i}{4\pi}\frac{c}{c\tau+d}\sum_{j}m_jm^j)P(\tau,\{m_i\})\,.
\end{equation}
In particular, we observe that $P(\tau+1,\{m_i\})=P(\tau,\{m_i\})$ and $P(\tau,\{m_i\})$ is a $W(E_8)$-invariant Jacobi form of weight $4$ and index $1$. Using the elliptic theta functions and $P(\tau,\{m_i\})$, one can construct all the necessary $W(E_8)$-invariant Jacobi forms used in the Seiberg--Witten curve of the E-string theory as noted in~\cite{Sakai:2011xg}. For consistency with~\cite{Sakai:2011xg}, let us continue to denote these $W(E_8)$-invariant Jacobi forms as $A_l,B_k$ where the indices take values as $l=1,2,3,4,5$ and $k=2,3,4,6$.

Now, let us expand these Jacobi forms in the small $m$ limit (assuming all $m_i$'s are small),
\begin{align}
\label{eq:WE8 expansions}
    A_l(\tau;m)&=E_4-\frac{l}{24}\left(E_4E_2-E_6\right)\sum m_i^2+\dots\,,\\
    B_k(\tau;m)&=E_6-\frac{k}{24}(E_2E_6-E_4^2)\sum m_i^2+\dots\,,
\end{align}
where we have made use of the Ramanujan identities stated above. From these small-$m$ expansions, we can deduce how the coefficients appearing in the SW curve of the E-string scale with $m$. Namely, we have
\begin{equation}
\label{eq:small-m expansion of ai bi}
    a_i(\tau,\{m\})=a_i(\tau,0)+\tilde{a_i}(\tau)m^{2i}+\CO(m^{4i})\,,\quad b_j(\tau,\{m\})=b_j(\tau,0)+\tilde{b_j}(\tau)m^{2(j-1)}+\CO(m^{4(j-1)})\,,
\end{equation}
where $i,j>1$ and $\tilde{a_i},\tilde{b_j}$ denote the coefficients of the leading small-$m$ expansion of $a_i(\tau,\{m\})$ and $b_j(\tau,\{m\})$.

\section{More on the E-string theory}
\label{sec:More on the E-string theory}

In this appendix, we provide more background and furnish further technical details regarding the E-string theory.

\subsection{Stringy realizations}
\label{sec:E-string constructions}
In Section~\ref{sec:2}, when discussing the physics of the E-string theory, we primarily use its M-theory realization as the worldvolume theory of an M5-brane probing a Ho\v{r}ava--Witten wall. However, there also exist a few other realizations of the theory which, when combined, provide significant insights into its physics. Below, we briefly review several different approaches, mainly serving as a guide to the literature, with highlights on their relations with M-theory descriptions of the E-string theory and its compactifications, which are used more prominently in the main text. 

\subsubsection*{Heterotic string construction}
The E-string theory was originally discovered in the studies of small instantons in heterotic string theory. They were first studied in the K3-compactification of the $SO(32)$ heterotic string theory, which is a 6d $(1,0)$ supergravity theory on $\mathbb{R}^6$ \cite{Witten:1995gx}. In examining the low-energy worldsheet theory of the heterotic string with this geometry, the gauge bundle with instanton number 24 arises due to anomaly cancellation requirements. The size of the instanton is proportional to the length of the string, and the moduli space of this instanton develops a singularity at finite distance when the instanton shrinks to zero size, or, equivalently, when a string becomes tensionless. This is suggestive of an extra gauge symmetry at the core of the instanton, which upon constructing the moduli space of vacua via Higgsing free hypermultiplets, turns out to be $SU(2)$.

In a similar vein, the K3-compactification of the $E_8\times E_8$ heterotic string theory was later studied. In \cite{Ganor:1996mu}, the gauge fields of the instantons are assumed to be embedded in one of the $E_8$'s and, using T-duality, arguments regarding the $SO(32)$ heterotic string on $\mathbb{R}^5\times S^1_{r'}\times \mathrm{K3}$ were translated over to the $E_8\times E_8$ heterotic string with $r\leftrightarrow r'=1/r$. In particular, the 6d $(1,0)$ theory of a small $E_8$ instanton is the (rank-1) E-string theory. In the resulting 5d $\mathcal{N}=1$ theory, special holonomies along $S^1_r$ and $S^1_{r'}$ in both heterotic string theories were turned on, breaking both $E_8\times E_8$ and $SO(32)$ down to $SO(16)\times SO(16)$. A salient feature of these two special holonomies related by T-duality is that the quantum numbers of the BPS states created from wrapping small instantons around $S^1$ become invariant under $r\leftrightarrow r'$. The analysis regarding this special holonomy will be further elaborated in the following subsection from the perspective of the 4d Seiberg--Witten geometry. Meanwhile, the authors of \cite{Seiberg:1996vs} consider small instantons embedded in both of the $E_8$ factors contributing collectively towards the instanton number $24$. In both of these works, the connection to the worldvolume theory of M5-branes probing an M9-brane is made.

\subsubsection*{Type IIA construction}
Immediately following the discovery of this exotic interacting six-dimensional conformal field theory with minimal supersymmetry in the context of the heterotic string theories, efforts were driven to realize this with brane configurations. Initial attention was on discussions in the M-theory context where the Ho\v{r}ava--Witten wall was naturally utilized. However, soon afterwards, it was realized that the E-string theory can also be constructed in type IIA string theory, with the brane configuration \cite{Hanany:1997gh,Hanany:1997sa,Kim:2014dza}:
\begin{center}
    \begin{tabular}{c|c|c|c|c|c|c|c|c|c|c}
         & $x^0$ & $x^1$ & $x^2$ & $x^3$ & $x^4$ & $x^5$ & $x^6$ & $x^7$ & $x^8$ & $x^9$ \\
        \hline
        NS5 & $\times $ & $\times $& $\times $ & $\times $& $\times $& $\times $ & -- & -- & -- & $l$ \\
        \hline
        8 D8/O8$^{-}$ & $\times $& $\times $& $\times $& $\times $& $\times $& $\times $ & $\times$ & $\times$ & $\times$ & $0$  \\
        \hline
        D2 & $\times $ & $\times $& -- & -- & -- & -- & -- & -- & -- & $\times$ \\
    \end{tabular}
\end{center}

\noindent such that the geometry is $\mathbb{R}^9\times \mathbb{R}^+$ with the O8-plane serving as the boundary of spacetime. The $SU(2)_R$ is now identified with the Spin$(3)$ symmetry rotating $\mathbb{R}_{678}$. Furthermore, as we have a total of 16 D8-branes in the overall covering space before the $\Z_2$ orientifolding, the world-volume theory of these D8-branes then exhibits an $SO(16)$ gauge symmetry. The above configuration can be uplifted to M-theory on $\mathbb{R}^9\times \mathbb{R}^+\times S^1$ via duality, with the D8/O8$^-$ system uplifted to the M9-brane where the $SO(16)$ gauge symmetry enhances to an $E_8$ gauge symmetry on the M9-brane. The NS5-brane lifts to an M5-brane and the D2-branes get lifted to M2-branes.

\subsubsection*{Type IIB construction}
Similarly, the E-string theory can be constructed in type IIB string theory. From a series of studies \cite{Aharony:1997ju,Leung:1997tw,Aharony:1997bh}, it was found that a myriad of 5d $\mathcal{N}=1$ gauge theories can be engineered using $(p,q)$ branes. These ideas were further pursued in \cite{Kim:2015jba} to construct 5d $\mathcal{N}=1$ gauge theories via a $(p,q)$ 5-brane web system in the presence of a $[p,q]$ 7-brane configuration in the Type IIB setup. As $[p,q]$ 7-branes are moved across $(p,q)$ 5-branes, due to the Hanany--Witten effect, certain D5-branes are annihilated, leaving a brane-web system given by a loop of 5-branes encircling a 7-brane configuration. Therefore, to describe these theories, it is sufficient to specify the 7-branes within the 5-brane loop. With this, the E-string theory is then described as the UV fixed point of the following 7-brane configuration
\begin{equation}
    \mathbf{E}=\mathbf{ANCANCANCANC}\, ,
\end{equation}
encircled by a 5-brane loop where $\mathbf{A},\mathbf{N},\mathbf{C}$ denote $[1,0],[0,1],[1,1]$ 7-branes, respectively.

\subsubsection*{F-theory construction}
In more recent developments in this subject, F-theory often serves as one of the main tools for constructing and understanding 6d $(1,0)$ SCFTs. In particular, F-theory on an elliptic Calabi--Yau threefold with a non-compact base $\mathrm{ECY}_3 \longrightarrow B_2$ gives a powerful way to classify and study 6d theories \cite{Morrison:1996pp,Heckman:2015bfa}. For an excellent review on this topic, see \cite{Heckman:2018jxk}.

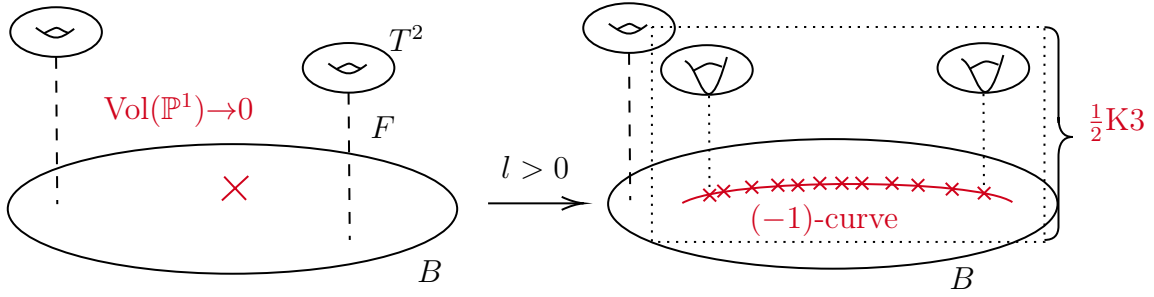
\begin{figure}
\centering

\tikzset{every picture/.style={line width=0.75pt}} 

\begin{tikzpicture}[x=0.75pt,y=0.75pt,yscale=-1,xscale=1]

\draw    (286.27,108.6) -- (331.95,108.6) ;
\draw [shift={(333.95,108.6)}, rotate = 180] [color={rgb, 255:red, 0; green, 0; blue, 0 }  ][line width=0.75]    (10.93,-3.29) .. controls (6.95,-1.4) and (3.31,-0.3) .. (0,0) .. controls (3.31,0.3) and (6.95,1.4) .. (10.93,3.29)   ;
\draw   (46.51,111.41) .. controls (46.51,93.41) and (96.73,78.81) .. (158.68,78.81) .. controls (220.63,78.81) and (270.85,93.41) .. (270.85,111.41) .. controls (270.85,129.41) and (220.63,144) .. (158.68,144) .. controls (96.73,144) and (46.51,129.41) .. (46.51,111.41) -- cycle ;
\draw  [dash pattern={on 0.84pt off 2.51pt}]  (396.56,54.59) -- (396.8,103.6) ;
\draw   (346.19,108.87) .. controls (346.19,90.87) and (396.41,76.28) .. (458.36,76.28) .. controls (520.31,76.28) and (570.53,90.87) .. (570.53,108.87) .. controls (570.53,126.87) and (520.31,141.46) .. (458.36,141.46) .. controls (396.41,141.46) and (346.19,126.87) .. (346.19,108.87) -- cycle ;
\draw [color={rgb, 255:red, 208; green, 2; blue, 27 }  ,draw opacity=1 ]   (383.22,108.44) .. controls (406.61,95.42) and (529.92,96.06) .. (548,108) ;
\draw  [color={rgb, 255:red, 208; green, 2; blue, 27 }  ,draw opacity=1 ] (393.7,101.29) -- (400.11,107.67)(399.99,101.41) -- (393.82,107.54) ;
\draw   (564,127) .. controls (568.67,127) and (571,124.67) .. (571,120) -- (571,85.2) .. controls (571,78.53) and (573.33,75.2) .. (578,75.2) .. controls (573.33,75.2) and (571,71.87) .. (571,65.2)(571,68.2) -- (571,27) .. controls (571,22.33) and (568.67,20) .. (564,20) ;
\draw  [dash pattern={on 0.84pt off 2.51pt}]  (533.56,53.59) -- (533.8,102.6) ;
\draw  [color={rgb, 255:red, 208; green, 2; blue, 27 }  ,draw opacity=1 ] (530.7,100.29) -- (537.11,106.67)(536.99,100.41) -- (530.82,106.54) ;
\draw  [color={rgb, 255:red, 208; green, 2; blue, 27 }  ,draw opacity=1 ] (414.7,97.29) -- (421.11,103.67)(420.99,97.41) -- (414.82,103.54) ;
\draw  [color={rgb, 255:red, 208; green, 2; blue, 27 }  ,draw opacity=1 ] (514.7,98.29) -- (521.11,104.67)(520.99,98.41) -- (514.82,104.54) ;
\draw  [color={rgb, 255:red, 208; green, 2; blue, 27 }  ,draw opacity=1 ] (484.7,95.29) -- (491.11,101.67)(490.99,95.41) -- (484.82,101.54) ;
\draw  [color={rgb, 255:red, 208; green, 2; blue, 27 }  ,draw opacity=1 ] (497.7,96.29) -- (504.11,102.67)(503.99,96.41) -- (497.82,102.54) ;
\draw  [color={rgb, 255:red, 208; green, 2; blue, 27 }  ,draw opacity=1 ] (459.7,95.29) -- (466.11,101.67)(465.99,95.41) -- (459.82,101.54) ;
\draw  [color={rgb, 255:red, 208; green, 2; blue, 27 }  ,draw opacity=1 ] (437.7,96.29) -- (444.11,102.67)(443.99,96.41) -- (437.82,102.54) ;
\draw  [color={rgb, 255:red, 208; green, 2; blue, 27 }  ,draw opacity=1 ][fill={rgb, 255:red, 208; green, 2; blue, 27 }  ,fill opacity=1 ] (154.54,95.15) -- (165.46,106.85)(166.58,94.86) -- (153.42,107.14) ;
\draw   (372.64,41.51) .. controls (372.64,34.62) and (382.88,29.03) .. (395.5,29.03) .. controls (408.12,29.03) and (418.36,34.62) .. (418.36,41.51) .. controls (418.36,48.41) and (408.12,54) .. (395.5,54) .. controls (382.88,54) and (372.64,48.41) .. (372.64,41.51) -- cycle ;
\draw    (384.5,37) .. controls (392.5,40) and (394.5,74) .. (405.5,35) ;
\draw    (388.5,42) .. controls (397.01,37.66) and (395.61,37.18) .. (403.05,39.89) ;
\draw   (510.64,40.51) .. controls (510.64,33.62) and (520.88,28.03) .. (533.5,28.03) .. controls (546.12,28.03) and (556.36,33.62) .. (556.36,40.51) .. controls (556.36,47.41) and (546.12,53) .. (533.5,53) .. controls (520.88,53) and (510.64,47.41) .. (510.64,40.51) -- cycle ;
\draw    (522.5,36) .. controls (530.5,39) and (532.5,73) .. (543.5,34) ;
\draw    (526.5,41) .. controls (535.01,36.66) and (533.61,36.18) .. (541.05,38.89) ;
\draw  [dash pattern={on 0.84pt off 2.51pt}] (368.05,19.89) -- (564,19.89) -- (564,128) -- (368.05,128) -- cycle ;
\draw  [color={rgb, 255:red, 208; green, 2; blue, 27 }  ,draw opacity=1 ] (400.7,99.29) -- (407.11,105.67)(406.99,99.41) -- (400.82,105.54) ;
\draw  [color={rgb, 255:red, 208; green, 2; blue, 27 }  ,draw opacity=1 ] (469.7,95.29) -- (476.11,101.67)(475.99,95.41) -- (469.82,101.54) ;
\draw  [color={rgb, 255:red, 208; green, 2; blue, 27 }  ,draw opacity=1 ] (448.7,95.29) -- (455.11,101.67)(454.99,95.41) -- (448.82,101.54) ;
\draw  [color={rgb, 255:red, 208; green, 2; blue, 27 }  ,draw opacity=1 ] (427.7,96.29) -- (434.11,102.67)(433.99,96.41) -- (427.82,102.54) ;
\draw   (193.78,40.44) .. controls (193.78,33.54) and (204.01,27.95) .. (216.63,27.95) .. controls (229.26,27.95) and (239.49,33.54) .. (239.49,40.44) .. controls (239.49,47.33) and (229.26,52.92) .. (216.63,52.92) .. controls (204.01,52.92) and (193.78,47.33) .. (193.78,40.44) -- cycle ;
\draw  [dash pattern={on 4.5pt off 4.5pt}]  (216.63,52.92) -- (217.23,127) ;
\draw    (206.53,38.27) .. controls (215.57,44.78) and (217.7,44.78) .. (225.67,38.81) ;
\draw    (210.79,40.98) .. controls (219.29,36.64) and (215.57,38.27) .. (223.01,40.98) ;
\draw   (47.78,22.44) .. controls (47.78,15.54) and (58.01,9.95) .. (70.63,9.95) .. controls (83.26,9.95) and (93.49,15.54) .. (93.49,22.44) .. controls (93.49,29.33) and (83.26,34.92) .. (70.63,34.92) .. controls (58.01,34.92) and (47.78,29.33) .. (47.78,22.44) -- cycle ;
\draw  [dash pattern={on 4.5pt off 4.5pt}]  (70.63,34.92) -- (71.23,109) ;
\draw    (60.53,20.27) .. controls (69.57,26.78) and (71.7,26.78) .. (79.67,20.81) ;
\draw    (64.79,22.98) .. controls (73.29,18.64) and (69.57,20.27) .. (77.01,22.98) ;
\draw   (333.78,20.44) .. controls (333.78,13.54) and (344.01,7.95) .. (356.63,7.95) .. controls (369.26,7.95) and (379.49,13.54) .. (379.49,20.44) .. controls (379.49,27.33) and (369.26,32.92) .. (356.63,32.92) .. controls (344.01,32.92) and (333.78,27.33) .. (333.78,20.44) -- cycle ;
\draw  [dash pattern={on 4.5pt off 4.5pt}]  (356.63,32.92) -- (357.23,107) ;
\draw    (346.53,18.27) .. controls (355.57,24.78) and (357.7,24.78) .. (365.67,18.81) ;
\draw    (350.79,20.98) .. controls (359.29,16.64) and (355.57,18.27) .. (363.01,20.98) ;

\draw (291.33,83.22) node [anchor=north west][inner sep=0.75pt]    {$l >0$};
\draw (249.35,135.75) node [anchor=north west][inner sep=0.75pt]    {$B$};
\draw (415.25,107.57) node [anchor=north west][inner sep=0.75pt]    {\textcolor[rgb]{0.82,0.01,0.11}{$(-1)$-curve}};
\draw (515.04,139.21) node [anchor=north west][inner sep=0.75pt]    {$B$};
\draw (583.93,57.16) node [anchor=north west][inner sep=0.75pt]    {\textcolor[rgb]{0.82,0.01,0.11}{$\frac{1}{2}$K3}};
\draw (226.11,63.38) node [anchor=north west][inner sep=0.75pt]    {$F$};
\draw (235.33,17.66) node [anchor=north west][inner sep=0.75pt]    {$T^{2}$};
\draw (93,52) node [anchor=north west][inner sep=0.75pt]   [align=left] {\textcolor[rgb]{0.82,0.01,0.11}{Vol($\mathbb{P}^1$)$\rightarrow$0}};

\end{tikzpicture}

\caption{\textbf{Elliptic Calabi--Yau threefold with $\mathcal{O}(-1)$ base.} The left figure represents the geometry of the singular elliptic Calabi--Yau threefold used in the F-theory construction of the E-string theory. In the right figure, the $(-1)$-curve acquires finite volume, which corresponds to moving on to the tensor branch. The total space of the elliptic fibration over the curve is an almost del Pezzo surface dP$_9$ (often referred to as ``$\frac12\mathrm{K3}$''). It can be obtained by blowing up 9 points on $\mathbb{P}^2$. As an elliptic surface, it generically has 12 singular fibers. }
\label{fig:the E-string theory from F-theory}
\end{figure}

For the E-string theory, its F-theory construction fits into a family of \textit{minimal} 6d $(1,0)$ SCFTs labeled by a positive integer $n$ up to 12, with the base $B_2$ being the total space of the $\mathcal{O}(-n)$-bundle over $\mathbb{P}^1$.\footnote{Minimal here means that the SCFT has a one-dimensional tensor branch, see e.g.,~\cite{Heckman:2015bfa,Haghighat:2014vxa}.} 
In this setup, the E-string theory is the ``$\mathcal{O}(-1)$-theory'' and is depicted in Figure~\ref{fig:the E-string theory from F-theory}.  We can describe the elliptic fibration over $B_2$ using the Weierstrass model,
\begin{equation}
    y^2=4x^3-fx-g\,,
\end{equation}
where $f$ and $g$ are sections of line bundles $-4K$ and $-6K$ of $B_2$. 

In the case of $n=1$, the total space of the elliptic fibration over the $(-1)$-curve is often known as a ``$\frac12$K3,'' which is an almost del Pezzo surface from blowing up 9 points in $\mathbb{P}^2$. The elliptic fibration can be understood from the Lefschetz pencil, $P(x,y)+u Q(x,y)$, of two cubics in $\mathbb{P}^2$ that intersect at 9 points. The fibration in Weierstrass form is given by
\begin{equation}
        y^2=4x^3-f(u,\sigma;m_i)x-g(u,\sigma;m_i)\,.
\end{equation}
Here $f$ and $g$ are of degree $4$ and $6$ in $u$ which parametrizes the base $\mathbb{P}^1$, $\sigma$ is the modulus of the elliptic fiber at $u=\infty$, and $m_i$'s are given by positions of the blow-up points. This fibration in general has 12 singular fibers with generic $\sigma$ and $m_i$'s.

In this construction of the 6d theory, the elliptic fiber is the ``F-theory torus,'' but in a different duality frame of the compactified E-string theory, the entire $\frac12$K3 is part of the spacetime, and the relative locations of the 9 points give the 8 mass parameters in the 4d effective theory. 

\subsubsection*{Back to M-theory}
A duality between F-theory on $S^1$ and M-theory allows one to realize the $S^1$-compactification of the E-string theory to 5d as M-theory on the same $\mathrm{ECY}_3$ \cite{Witten:1996qb}. The E-strings (now BPS strings in the 5d theory) can be engineered via M5-branes wrapping the $\frac{1}{2}$K3 as a divisor in $\mathrm{ECY}_3$, which degenerates to zero volume at the origin of the Coulomb branch. This M-theory picture is dual to F-theory on $\mathrm{ECY}_3\times S^1$ where the radius of $S^1$ is inversely proportional to the volume of the elliptic fibers in the M-theory setup. 

On the F-theory side, the E-string is engineered as D3-branes wrapping the base 2-cycle of the $\frac{1}{2}$K3 geometry. When also wrapping the compact $S^1$ direction, it gives rise to 5d BPS particles, which are characterized by the winding number $n$ and KK-momentum $p$. They are dual to M2-branes wrapping around a curve in $\mathrm{ECY}_3$ characterized by $n[\mathbb{P}^1]+p[T^2]$. Therefore, these BPS particles can be counted using topological string amplitudes on $\frac{1}{2}\mathrm{K3}$. Now consider the theory compactified down on another circle, i.e.~M-theory $(S')^1\times \mathrm{ECY}_3$ which is dual to F-theory on $S^1\times (S')^1\times \mathrm{ECY}_3$. In this dual frame, one has these BPS particles given by $n$ M5-branes wrapping $\frac{1}{2}\mathrm{K3}\times (S')^1$. Then, the counting of such BPS particles is given by the partition function of the $n$ M5-branes on $\frac{1}{2}\mathrm{K3}\times T^2$, with KK-momentum $p$.  The latter can be computed using the topologically twisted partition function for $\mathcal{N}=4$ SYM on $\frac{1}{2}$K3 in the instanton number $p$ sector \cite{Vafa:1994tf,Bershadsky:1995qy,Minahan:1998vr}.

This is the setup of the computations for the coefficients of the Seiberg--Witten curve in \cite{Eguchi:2002fc}, which we will review in Appendix~\ref{app:SWdetail}.

\subsubsection*{Rank-$Q$ E-string theories}
In all of the above discussions, we have been implicitly constructing the rank-1 E-string theory, i.e.~a 6d $(1,0)$ SCFT with $E_8$ flavor symmetry and $\frak{sp}(1)$ ``gauge algebra.'' However, there exist rank-$Q$ E-string theories in which the flavor symmetry is $E_8\times SU(2)$ where the gauge algebra is $\frak{sp}(Q)$. These can also be engineered in various brane configurations or stringy settings. In particular, we can make the following modifications to the above constructions regarding the rank-1 E-string theory:
\begin{itemize}
    \item \textbf{M-theory}: $Q$ M5-branes probing the M9-brane;
    \item \textbf{Heterotic string theory}: $Q$ instantons collapsing at the same point;
    \item \textbf{Type IIA string theory}: $Q$ NS5-branes probing the 8 D8/O8$^{-}$-plane;
    \item \textbf{F-theory}: $B_2$ becomes the total space of the bundle of the $\mathcal{O}(-1)$-curve linked to a chain of $(Q-1)$ $\mathcal{O}(-2)$-curves over $\mathbb{P}^1$ \cite{Morrison:1996pp,Gadde:2015tra}. Equivalently, in terms of the notation widely used in current literature for 6d conformal matter \cite{Heckman:2018jxk}, the rank-$Q$ E-string theory is described as $[E_8],\underbrace{1,2,\dots,2}_{Q}$.
\end{itemize}

\subsection{Details on the Seiberg--Witten geometry}\label{app:SWdetail}
With the above constructions, we take the rank-1 E-string theory and compactify it on a torus. The resulting theory is a 4d $\mathcal{N}=2$ theory which, at low energy, is described by its Seiberg--Witten geometry.

Now, to calculate the periods of this fibration near a singularity in the Coulomb branch, we can consider recasting the elliptic curve into the following form with the modulus $\tau_s=\omega_2/\omega_1$, which is a local parameter around a singularity, see e.g.,  \cite[eq~(5.8)]{Minahan:1998vr}\footnote{Notice that there are different conventions as there is freedom in scaling (and shifting) $x$ and $y$ without changing the elliptic curve. In \cite{Minahan:1998vr} the elliptic curve was expressed as
\begin{equation}
\label{eq:local elliptic curve}
    y^2=x^3-\frac{E_4(\tau_s)}{3\omega_1^4}x+\frac{2E_6(\tau_s)}{27\omega_1^6}\,.
\end{equation}
While the $j$-invariant of each curve correctly reproduces $j(\tau_s)=1728E_4^3/(E_4^3-E_6^2)$, the differential $dx/y$ differs by a factor, which needs to be taken into account when relating $\omega_1$ to physical quantities.}
\begin{equation}
\label{eq:tildetauellipticcurve}
    y^2=4x^3-\frac{4\pi^4E_4(\tau_s)}{3\omega_1^4}x-\frac{8\pi^6E_6(\tau_s)}{27\omega_1^6}\,,
\end{equation}
such that by comparing \eqref{eq:tildetauellipticcurve} with \eqref{eq:ellipticCurveES}, the periods can be readily read off as
\begin{equation}
\label{eq:periods}
    \omega_1=i\sqrt{2}\pi\left(\frac{E_4(\tau_s)}{3f(u;\sigma,\{m_i\})}\right)^{1/4},\qquad \omega_2=\tau_s\omega_1\, .
\end{equation}
Furthermore, we have the $j$-invariant for elliptic curves given by
\begin{equation}
\label{eq:jinvariant}
    j(\tau_s)=1728 \frac{f(u;\sigma,\{m_i\})^3}{f(u;\sigma,\{m_i\})^3-27g(u;\sigma,\{m_i\})^2}\, .
\end{equation}
Then, we can compute the prepotential, term by term, according to
\begin{equation}
    \frac{\partial^2 \mathcal{F}}{\partial a^2}=\tau_s\, ,
\end{equation}
where $\tau_s$ can be obtained by inverting the expansion of the $j$-invariant defined by \eqref{eq:jinvariant} in the $1/u\to 0$ limit. By integrating the periods in terms of this $1/u$ expansion, we can relate $1/u$ and $a$ such that $\tau_s$ becomes a polynomial in terms of $e^{2\pi ia}$. Integrating $\tau_s(e^{2\pi ia})$ with respect to $a$ twice gives the expression of $\mathcal{F}$. Note, explicitly, the prepotential has the following instanton expansion in terms of its BPS partition functions~\cite{Minahan:1998vr,Eguchi:2002fc}
\begin{equation}
\label{eq:prepotential}
    \mathcal{F}(a,\sigma,\{m_i\})=\mathcal{F}_{\rm pert}-\frac{1}{(2\pi i)^3}\sum_{n=1}^{\infty}q^{n/2}Z_n(\sigma,\{m_i\})e^{2\pi i  n a}\,,
\end{equation}
such that by power matching in $e^{2\pi in a}$, one can calculate the $Z_n(\sigma,\{m_i\})$ in terms of the coefficients $a_j,b_j$'s appearing in the elliptic curve.

Alternatively, the prepotential can be expressed in terms of instanton amplitudes calculated in the 4d $\mathcal{N}=4$ topological Yang--Mills theory obtained from the dual perspective by considering toroidally compactified M5-brane on $\frac{1}{2}$K3 as outlined in \cite{Minahan:1998vr} and discussed in the beginning of this section. There exists a recursion relation of the partition functions $Z_n$ (or instanton amplitudes) of $U(n)$ gauge theories on $\frac{1}{2}$K3 given by the holomorphic anomaly equation
\begin{equation}
\label{eq:holomorphicGap}
    \frac{\partial Z_n}{\partial E_2}=\frac{1}{24}\sum_{m=1}^{n-1}m(n-m)Z_mZ_{n-m}\, .
\end{equation}
In addition, there is also a gap condition,
\begin{equation}
\label{eq:Zn0}
    Z_{n}(\sigma,\{m_i\})=\frac{1}{n^3}q^{-n/2}+\mathcal{O}\left(q^{n/2}\right)\,  ,
\end{equation}
where the partition functions are $W(E_8)$-invariant quasi-Jacobi forms when $n\ge2$. The gap condition can be interpreted as the following statement: there are no BPS states of 6d E-strings with momentum $k$ less than their winding number $n$. Knowing the elliptic genus of a single E-string as
\begin{equation}
    Z_1=\frac{P(\sigma,\{m_i\})}{\eta(\sigma)^{12}}\, ,
\end{equation}
and by carefully matching the $Z_n$'s obtained from both methods, we can calculate the coefficients in the elliptic curve \eqref{eq:ellipticCurveES}. The details and the explicit expressions for the coefficients can be found in \cite{Eguchi:2002fc,Eguchi:2002nx}.

\subsection{Special limits of the Seiberg--Witten geometry}
\label{sec:specialMass}
Let us consider the consequences of turning on certain holonomies in the $E_8$ symmetry along the $T^2$ in terms of the SW geometry of the effective theory in 4d. Some special limits have been studied in \cite{Eguchi:2002nx}. Here we examine those by also identifying the singularities arising in these limits.  This would be the first step towards a better understanding of how the partition function of the E-string theory encodes other theories in four and five dimensions.

\subsubsection*{$D_8$:}
 The non-trivial holonomy breaking the $E_8$ flavor symmetry down to $D_8$ was discussed in \cite{Ganor:1996mu} and can be understood through the Type IIA construction. The $SO(16)$ flavor symmetry originates from the gauge symmetry on the worldvolume theory of 8 D8/O8$^-$ as mentioned previously. In particular, the O8$^{-}$-plane, by definition, has a $\mathbb{Z}_2$-involution symmetry such that the configuration of 8 D8-branes is symmetric under $\mathbb{Z}_2$ actions.

Another way of obtaining the $SO(16)$ symmetry from $E_8$ has been recently explored in the context of chiral $(E_8)_1$ CFTs in \cite{Burbano:2021loy,BoyleSmith:2023xkd}. By utilizing Kac's theorem \cite{Kac:1969}, one can determine the automorphisms of any simple Lie algebra. In particular, the automorphisms for vertex operator algebra consist of two distinct $\mathbb{Z}_2$'s where the subalgebras corresponding to either $\mathbb{Z}_2$ are $SO(16)$ and $E_7\times SU(2)$.

The value of the holonomy that breaks the $E_8$ flavor symmetry into $D_8$ is \cite{Eguchi:2002nx,Ganor:1996mu}
\begin{equation}
\label{eq:SO16 mass parameters}
    \{m\}_{D_8}=(0,0,0,0,0,0,0,2\pi)\, .
\end{equation}
It is clear that this mass parameter is invariant under $D_7$. Viewed as an element of the $E_8$ Lie group, its stabilizer is in fact $D_8$. More precisely, it is Spin$(16)/\Z_2$---the ``semi-Spin'' group---as opposed to the true $SO(16)$. In Figure~\ref{fig:KK compactification of E-string}, we refer to $\{m\}_{D_8}$ as the ``$-1$'' holonomy, as it is the non-trivial element of the $\Z_2$  center of the Spin$(16)/\Z_2$ subgroup of $E_8$. With this choice, we have $P(\sigma,\{m\}_{D_8})=(-\vartheta_2(\sigma)^8+\vartheta_3(\sigma)^8+\vartheta_4(\sigma)^8)/2$. Thus, we can see that with the choice of $\{m\}_{D_8}$, the coefficients appearing in the Seiberg--Witten curve are no longer invariant under $\vec{\alpha}_{\rm II}$.\footnote{This can be seen by sending $\{m\}_{D_8}\to \{m\}_{D_8}-\{0,\dots,\pi\}$ such that $P(\sigma,\{0,\dots,\pi\})=\vartheta_3(\sigma)\vartheta_4(\sigma)(\vartheta_3(\sigma)^6+\vartheta_4(\sigma)^6)/2\neq P(\sigma,\{m\}_{D_8})$.} 
Therefore, the system reduces to a $W(D_8)$-invariant one as $\{m\}_{D_8}$ is only invariant under $\Gamma_{D_8}$ which is generated by $\alpha_{\rm I}$.
The Seiberg--Witten curve reduces to
\begin{equation}
    y^2=4x^3-\tilde{f}x-\tilde{g}\, ,
\end{equation}
where
\begin{align}
    \tilde{f}&=\frac{1}{12}(u-u_0)^2\left(E_4(u+u_0)^2-\frac{12288}{w}\right)\,,\\
    \tilde{g}&=\frac{1}{216}(u-u_0)^3\left(E_6u^3-\frac{48N_2}{w}u^2-\frac{1536(\vartheta_3^4+\vartheta_4^4)N_1}{w^2}u-\frac{16384N_0}{w^3}\right)\,,
\end{align}
where $w=\vartheta_3^4\vartheta_4^4E_4$, $u_0=32(\vartheta_3^4+\vartheta_4^4)/w$, the modular forms are evaluated at $\sigma$, and, with $T=\vartheta_3^4$ and $Y=\vartheta_4^4$,
\begin{align*}
    N_2&=2T^4-T^3Y+12T^2Y^2-TY^3+2Y^4\,,\\
    N_1&=2T^4-13T^3Y+6T^2Y^2-13TY^3+2Y^4\,,\\
    N_0&=(T^2-4TY+Y^2)(2T^4-25T^3Y+18T^2Y^2-25TY^3+2Y^4)\,.
\end{align*}
The curve has an I$_4^*$ singularity at $u=u_0$ and two I$_1$ singularities (see the table below). Here, in the $\tau\to i\infty$ limit, we recover results discussed in~\cite{Eguchi:2002nx} such that the coefficients of the SW curve become
\begin{align}
    f&=\frac{1}{12}(u-64)^2(u^2+128u-8192)\,,\\
    g&=\frac{1}{216}(u-64)^3(u-32)(u^2-640u+28672)\,.
\end{align}
However, in this limit, only one I$_1$ singularity remains within finite distance.

While this ``$-1$'' holonomy is unique, there are various versions of this holonomy (i.e., $\{m\}_{D_8}$ in \cite{Sakai:2012ik} and \cite{Kim:2014dza,Eguchi:2002fc})  from different parametrizations. An invariant way of characterizing this holonomy is  $\prod_i\vartheta_1(m_i,\sigma)=0$ and $\prod_{i}\vartheta_2(m_i,\sigma)=-\vartheta_2(\sigma)^8$. With our chosen convention reviewed in the previous appendix, \eqref{eq:SO16 mass parameters} is the correct choice. Therefore, the singularities of the Seiberg--Witten curve with this choice of holonomy parameters are

\begin{center}
    \centering
    \begin{tabular}[t]{lcccc}
    \toprule
         &  $\text{ord}_{\Sigma}(f)$ & $\text{ord}_{\Sigma}(g)$ & $\text{ord}_{\Sigma}(\Delta_{\Sigma})$ & Kodaira type\\
    \midrule
         $u_{s,0}$& 2 & 3 & 10& I$_4^*$\\
         $u_{s,1\leq i\leq 2}$& 0& 0& 1& I${}_1$\\
    \bottomrule
    \end{tabular}
\end{center}

\subsubsection*{$E_6$:}
    One can also break the $E_8$ symmetry down to $E_6\times SU(2)\times U(1)$ or $E_6\times U(1)^2$. The latter is achieved by the following holonomy,
    \begin{equation}
        \{m\}_{E_6}=(0,0,0,0,0,\pi,\pi+\pi\sigma,\pi\sigma)\,.
    \end{equation}
     Note that $m_{5,6,7,8}$ are the zeros of $\vartheta_{1,2,3,4}$, respectively, leading to $P(\sigma,\{m\}_{E_6})=0$. 
     
     For this special value of holonomy, the only three coefficients that survive are
    \begin{align}
        a_0&=\frac{1}{12}E_4\, ,\\
        b_0&=\frac{1}{216}E_6\, ,\\
        b_2&=\frac{4}{q^{1/2}\eta^{12}}\, ,
    \end{align}
    making the elliptic curve 
    \begin{equation}
        y^2=4x^3-a_0u^4x-b_0u^6-b_2u^4\, .
    \end{equation}
    The discriminant of this curve is then 
    \begin{equation}
        \Delta_{\Sigma_{E_6}}(\tau,u)=\frac{1}{1728}E_4^3u^{12}-27\left(\frac{1}{216}E_6u^6+\frac{4}{q^{1/2}\eta^{12}}u^4\right)^2\, .
    \end{equation}
    Computing the zeros of $\Delta_{\Sigma_{E_6}}$ gives us the singularities of $\Sigma_{E_6}$
    \begin{equation}
    \begin{cases}
    u_{s,0}=0\, ,\\
    u_{s,1}=12 \sqrt{6} \sqrt{\frac{1}{q^{1/2}\eta^{12}(E_4^{3/2}-E_6)}}\, ,\\
    u_{s,2}=-u_{s,1}\, ,\\
    u_{s,3}=12 \sqrt{6} \sqrt{-\frac{1}{q^{1/2}\eta^{12}(E_4^{3/2}+E_6)}}\, ,\\
    u_{s,4}=-u_{s,3}\, ,
    \end{cases}
    \end{equation}
    and their singularity types for generic $\sigma$ are \begin{center}
    \centering
    \begin{tabular}[t]{lcccc}
    \toprule
         &  $\text{ord}_{\Sigma}(f)$ & $\text{ord}_{\Sigma}(g)$ & $\text{ord}_{\Sigma}(\Delta_{\Sigma})$ & Kodaira type\\
    \midrule
         $u_{s,0}$& 4 & 4 & 8& IV$^*$\\
         $u_{s,1\leq i\leq 4}$& 0& 0& 1& I${}_1$\\
    \bottomrule
    \end{tabular}
    \end{center}
    \noindent The IV$^*$ singularity is also known as the $E_6$ singularity where a Minahan--Nemeschansky $E_6$ SCFT lives. The remaining $E_6$ subgroup of the global symmetry acts on its Higgs branch. Furthermore, there exists a $\mathbb{Z}_2$ action $u\mapsto -u$ that leaves the Coulomb branch invariant as $u_{s,1}=-u_{s,2}$, $u_{s,3}=-u_{s,4}$, and $u_{s,0}$ is a fixed point of this action.

\subsubsection*{$D_4\oplus D_4$:}
    Here, we also review the case where we can realize two $D_4$ singularities on the $u$-plane as discussed in~\cite{Eguchi:2002nx}. This is achieved with the special holonomy
    \begin{equation}
        \{m\}_{D_4\oplus D_4}=(0,0,\pi,\pi,\pi+\pi\sigma,\pi+\pi\sigma,\pi\sigma,\pi\sigma)\,.
    \end{equation}
    With this, the Seiberg--Witten curve becomes
    \begin{equation}
    \label{eq:Nf4Estring}   
        y^2=4x^3-\frac{E_4}{12}\bigg(u^2-\frac{4}{q\eta^{24}}\bigg)^2x-\frac{E_6}{216}\bigg(u^2-\frac{4}{q\eta^{24}}\bigg)^3\,.
    \end{equation}
    We can also write this in a manifestly modular way via a redefinition of $u$ as $\tilde{u}=u\sqrt{q\eta^{24}/4}$. At the same time, we must also rescale $x,y$ as $\tilde{x}=x\,q\eta^{24}/4$ and $\tilde{y}=y\,(q\eta^{24}/4)^{3/2}$. Then, the Seiberg--Witten curve becomes
    \begin{equation}
        \tilde{y}^2=4\tilde{x}^3-\frac{E_4}{12}\big(\tilde{u}^2-1\big)^2\tilde{x}-\frac{E_6}{216}\big(\tilde{u}^2-1\big)^3\,.
    \end{equation}
    Nevertheless, using \eqref{eq:Nf4Estring}, the associated discriminant simply becomes
    \begin{equation}
        \Delta_{\Sigma_{D_4\oplus D_4}}=\eta^{24}\big(u-u_{s,-}\big)^6\big(u-u_{s,+}\big)^6\, ,
    \end{equation}
    where we have the two singularities on the Coulomb branch as
    \begin{equation}
        u_{s,\pm}=\pm\sqrt{\frac{4}{q\eta^{24}}}\,.
    \end{equation}
    We have the following order of vanishing of $f,g,\Delta$ at each $u_{s,\pm}$. 
        \begin{center}
    \centering
    \begin{tabular}[t]{lccccc}
    \toprule
         &  $\text{ord}_{\Sigma}(f)$ & $\text{ord}_{\Sigma}(g)$ & $\text{ord}_{\Sigma}(\Delta_{\Sigma})$ & Kodaira type \\
    \midrule
         $u_{s,\pm}$& 2 & 3 & 6& I$_0^*$\\
    \bottomrule
    \end{tabular}
    \end{center}

\subsection{Coefficients of $\Delta_{\Sigma}$}
\label{sec:modularity of coefficients}
Examining 
\begin{equation*}
    \sum_{i=0}^{12}D_iu^{12-i}=f(u;\sigma,\{m_i\})^3-27g(u;\sigma,\{m_i\})^2\,,
\end{equation*}
where $f(u;\sigma,\{m_i\})$ and $g(u;\sigma,\{m_i\})$ are given in \cite{Eguchi:2002fc}, we have, with the ``gauge choice'' $a_1=0$, 
\begin{subequations}
\begin{align}
    D_0&=a_0^3-27b_0^2\,,\\
    D_1&=-54b_0b_1\,,\\
    D_2&=3a_0^2a_2-27b_1^2-54b_0b_2\,,\\
    D_3&=3a_0^2a_3-54b_1b_2-54b_0b_3\,,\\
    D_4&=3a_0a_2^2+3a_0^2a_4-27b_2^2-54b_1b_3-54b_0b_4\,,\\
    D_5&=6a_0a_2a_3-54b_2b_3-54b_1b_4-54b_0b_5\,,\\
    D_6&=a_2^3+3a_0a_3^2+6a_0a_2a_4-27b_3^2-54b_2b_4-54b_1b_5-54b_0b_6\,,\\
    D_7&=3 a_2^2 a_3 + 6 a_0 a_3 a_4 - 54 b_3 b_4 - 54 b_2 b_5 - 54 b_1 b_6\,,\\
    D_8&=3 a_2 a_3^2 + 3 a_2^2 a_4 + 3 a_0 a_4^2 - 27 b_4^2 - 54 b_3 b_5 - 54 b_2 b_6\,,\\
    D_9&=a_3^3 + 6 a_2 a_3 a_4 - 54 b_4 b_5 - 54 b_3 b_6\,,\\
    D_{10}&=3 a_3^2 a_4 + 3 a_2 a_4^2 - 27 b_5^2 - 54 b_4 b_6\,,\\
    D_{11}&=3 a_3 a_4^2 - 54 b_5 b_6\,,\\
    D_{12}&=a_4^3 - 27 b_6^2\,.
\end{align}
\end{subequations}
Furthermore, by Vieta's formula and Newton's identities, we have
\begin{equation}
\begin{cases}
    kD_k+\sum_{i=1}^{k}D_{k-i}P_{i}=0&1\leq k\leq 12,\\
    \sum_{i=k-12}^{k}D_{k-i}P_{i}=0&k>12\,,
\end{cases}
\end{equation}
where $P_k=\sum_{i=1}^{12}u_{s,i}^k$ and $u_s$ denote the singularities/roots of $\Delta$. Explicitly, we have
\begin{subequations}
\begin{align}
    P_1=\sum_{i=1}^{12}u_{s,i}&=-\frac{D_{1}}{D_{0}}\,,\\
    P_2=\sum_{i=1}^{12}u_{s,i}^2&=\frac{D_1^2}{D_0^2}-2\frac{D_2}{D_0}\,,\\
    \vdots\nonumber
\end{align}
\end{subequations}
Furthermore, as we will encounter summation of negative powers of singularities, it is beneficial to demonstrate the inverse sums of roots of polynomials via Vieta's formula. To this end, we can invert the polynomial above such that we obtain
\begin{subequations}
\begin{align}
    \sum_su_s^{-1}&=-\frac{D_{11}}{D_{12}}\,,\\
    \sum_su_s^{-2}&=\frac{D_{11}^2-2D_{10}D_{12}}{D_{12}^2}\,,\\
    \sum_su_s^{-3}&=\frac{-D_{11}^3+3D_{10}D_{11}D_{12}-3D_9D_{12}^2}{D_{12}^3}\,,\\
    \sum_su_s^{-4}&=\frac{D_{11}^4-4D_{10}D_{11}^2D_{12}+4D_9D_{11}D_{12}^2+2D_{10}^2D_{12}^2-4D_8D_{12}^3}{D_{12}^4}\,,\\
    \sum_s u_s^{-5}&=\frac{1}{D_{12}^5}\Big(-D_{11}^5+5D_{10}D_{11}^3D_{12}-5D_{10}^2D_{11}D_{12}^2-5D_9D_{11}^2D_{12}^2\nonumber\\ &\qquad\quad+5D_9D_{10}D_{12}^3+5D_8D_{11}D_{12}^3-5D_7D_{12}^4\Big)\,.
\end{align}
\end{subequations}
From these expressions, we can also deduce how $u$ scales with $m$ near the small-$m$ limit. Namely, using~\eqref{eq:small-m expansion of ai bi}, we have
\begin{equation}
\label{eq:u inverse sum}
    -\frac{D_{11}}{D_{12}}=-\frac{3\tilde{a_3}(\tau)\tilde{a_4}(\tau)^2m^{22}-54\tilde{b_5}(\tau)\tilde{b_6}(\tau)m^{18}+\dots}{\tilde{a_4}(\tau)^3m^{24}-27\tilde{b_6}(\tau)^2m^{20}+\dots}\sim-2\frac{\tilde{b_5}(\tau)}{\tilde{b_6}(\tau)}\,m^{-2}\,.
\end{equation}
Therefore, for small masses, we have $u_s^{-1}\sim m^{-2}$ and the 10 I$_1$ singularities are roughly at distance $m^2$ away from the origin.

In this form, we have analyzed the first two coefficients in the main text, and now we will analyze the modularity of the rest of the coefficients. Prior to doing so, let us examine the modularity of some crucial components of the coefficients, namely the following specialized functions of $E_{4},E_6,\vartheta_{i}(\sigma,m)$, where $i=1,\dots,4$, detailed in \cite{Eguchi:2002fc},\footnote{In what follows, we only list the relevant terms with those that have identical weights and compatible levels omitted. The level listed here might only be a subgroup of the ``true level.''}

\begin{center}
\begin{tabular}{c|c|c}
    function &  level & weight\\
     \hline
     $P(\sigma,\{m_i\})$ & $\SL(2,\mathbb{Z})$ & $4$\\
    $h_0(\sigma)$ & $\Gamma_1(3)$ & $1$\\
    $h_2(\sigma)$ & $\Gamma_1(3)$ & $3$ \\
    $h_3(\sigma)$ & $\Gamma_1(3)$ & $3$\\
    $f_{a_2,0}(\sigma)$ & $\Gamma_1(2)$ & $4$ \\
    $f_{b_2,0}(\sigma)$ & $\Gamma_1(2)$ & $10$\\
    $f_{a_3,0}(\sigma)$ & $\Gamma_1(3)$ & $14$\\
    $f_{b_3,0}(\sigma)$ & $\Gamma_1(3)$ & $20$\\
    $f_{b_4,0}(\sigma)$ & $\Gamma_1(4)$ & $6$\\
    $f_{b_5,0}(\sigma)$ & $\Gamma_1(5)$ & $16$\\
    $g_{b_5,0}(\sigma)$ & $\Gamma_1(2)$ & $12$\\
    $f_{b_6,0}(\sigma)$ & $\Gamma_1(6)$ & $6$\\
    $g_{b_6,0}(\sigma)$ & $\Gamma_0(2)$ & $2$\\
\end{tabular}
\end{center}
where $f_{a_n,1},f_{b_n,1},g_{b_n,1}$ are $S$-transformations of $f_{a_n,0},f_{b_n,0},g_{b_n,0}$ and $f_{b_n,2},g_{b_n,2}$ are $S$-transformations of $f_{b_n,1},g_{b_n,1}$. Using this data, we can now see the modularity of the original coefficients appearing in the (relative) Weierstrass form of the Seiberg--Witten curve $\Sigma$, namely of the $a_i,b_i$:

\begin{minipage}{.5\textwidth}
\begin{center}
\begin{tabular}{c|c|c}
    coefficient & weight & index\\
    \hline
    $a_0$ & $4$ & $0$\\
    $b_0$ & $6$ & $0$\\
    $a_1$ & --- & ---\\
    $b_1$ & $0$ & $1$\\
    $a_2$ & $-8$ & $2$\\
    $b_2$ & $-6$ & $2$
\end{tabular}
\end{center}
\end{minipage}
\begin{minipage}{.5\textwidth}
\begin{center}
\begin{tabular}{c|c|c}
    coefficient & weight& index\\
    \hline
    $a_3$ & $-14$ & $3$\\
    $b_3$ & $-12$ & $3$\\
    $a_4$ & $-20$ & $4$\\
    $b_4$ & $-18$ & $4$\\
    $b_5$ & $-24$ & $5$\\
    $b_6$ & $-30$ & $6$
\end{tabular}
\end{center}
\end{minipage}

\noindent These are now $E_8$ Jacobi forms under the full $\SL(2,\mathbb{Z})$. Now, we can assemble them into $D_i$ which are also $E_8$ Jacobi forms under the full modular group $\SL(2,\mathbb{Z})$, whose weights and indices are listed below:
\begin{table}[!htp]
\begin{minipage}[t]{.5\textwidth}
\centering
    \begin{tabular}{c|c|c}
        coefficient & weight& index\\
        \hline
        $D_0$ &$12$ & $0$\\
        $D_1$ & $6$ & $1$\\
        $D_2$ &$0$ & $2$\\
        $D_3$ & $-6$ & $3$\\
        $D_4$&$-12$ & $4$\\
        $D_5$& $-18$  & $5$\\
        $D_6$ & $-24$ & $6$
    \end{tabular}
\end{minipage}
\begin{minipage}[t]{.5\textwidth}
\centering
        \begin{tabular}{c|c|c}
            coefficient  & weight & index\\
            \hline
            $D_7$&$-30$ & $7$\\
            $D_8$ & $-36$ & $8$\\
            $D_9$& $-42$ &$9$\\
            $D_{10}$ & $-48$ & $10$\\
            $D_{11}$ & $-54$ & $11$\\
            $D_{12}$ & $-60$ & $12$
        \end{tabular}
\end{minipage}
\end{table}

\noindent This full modularity is expected from the 6d perspective as the E-string theory is absolute and the 4d effective theory is automatically an absolute theory without choosing a polarization on $T^2$ that can break full modularity. In particular, $D_i$ is an $E_8$ Jacobi form of weight $12-6i$ and index $i$. Knowing this, we have that $f(u_s;\sigma,\{m_i\})$ and $g(u_s;\sigma,\{m_i\})$ are (weak) $E_8$ Jacobi forms with weight $-20,-30$ and index $4,6$, respectively.

\section{Details on the partition function}
\label{sec:details on the partition function}
From the Weierstrass form of the elliptic curve near a singularity \eqref{eq:tildetauellipticcurve}, we can gather the following information regarding the non-vanishing period $\omega_1$ in the local coordinates $\tau_s=\omega_2/\omega_1$
\begin{equation}
    \omega_1^4=\frac{4\pi^4E_4(\tau_s)}{3f(u;\sigma,\{m_i\})}\,,\qquad \omega_1^6=\frac{8\pi^6E_6(\tau_s)}{27g(u;\sigma,\{m_i\})}\,.
\end{equation}
Therefore, as computed in the previous appendix, we obtained \eqref{eq:periods} and also, by taking the ratio between the two expressions above, we have
\begin{equation}
    \omega_1^2=2\left(\frac{\pi}{3}\right)^2\frac{E_6(\tau_s)}{E_4(\tau_s)}\frac{f(u;\sigma,\{m_i\})}{g(u;\sigma,\{m_i\})}\,.
\end{equation}
Comparing this expression to that used in \cite{Eguchi:2002fc} to compute the Seiberg--Witten curve for the E-string theory, we have that the relative normalization between $da_s/du$ and $\omega_1$ around a singularity $u_s$ is
\begin{equation}
    \frac{\dd a_s}{\dd u}=\varrho\cdot\omega_1\,.
\end{equation}
Here, the value of $\varrho$ depends on the normalization convention for $\omega_1$, and is chosen to be $1$ in~\eqref{eq:Seiberg--Witten coordinates}. Another common choice in the literature is $\varrho=i/4\pi^2$. From Section~\ref{sec:3} on, we use $\varrho=i/2\pi$, e.g.~in~\eqref{eq:dasdu relation to period}. Here we will keep $\varrho$ in the formulas to make it easier to switch between different conventions. 

Furthermore, using~\eqref{eq:tildetauellipticcurve}, the discriminant of the Weierstrass form becomes
\begin{equation}
    \Delta_{0}(\tau_s)=\left(\frac{4\pi^4E_4(\tau_s)}{3\omega_1^4}\right)^3-27\left(\frac{8\pi^6E_6(\tau_s)}{27\omega_1^6}\right)^2=(2\pi)^{12}\omega_1^{-12}\eta(\tau_s)^{24}\,,
\end{equation}
where we have made use of $(E_4^3-E_6^2)/1728=\eta^{24}$. Using this, we can reproduce the desired relation given in \eqref{eq:local discriminant}, enabling us to compute the expansions \eqref{eq:uexpansions} and \eqref{eq:aexpansions}.

\label{qs-expansion}
Now, let us derive the $q_s$-expansion of $u$ and $a_s$ near a given $u_s$ singularity in the Coulomb branch explicitly.

\subsubsection*{$u$:}
Examining \eqref{eq:useries}, as explained in the main text, this can be derived from the $j$-invariant of a given elliptic curve\footnote{Note, sometimes the elliptic curves may be expressed in their regular form which usually takes on the following expression
\begin{equation}
    (\tilde{y})^2=(\tilde{x})^3-\tilde{f}\cdot (\tilde{x})-\tilde{g}\,.
\end{equation}
This is related to the one given here via a rescaling, i.e.,
\begin{equation}
    y=2\tilde{y}\,,\qquad x=\tilde{x}\,,\qquad  \tilde{f}=f/4\,,\qquad \tilde{g}=g/4\,.
\end{equation}
of which the $j$-invariant associated to these coefficients is
\begin{equation}
    j=12^3\frac{4\tilde{f}^3}{4\tilde{f}^3-27\tilde{g}^2}\,.
\end{equation}}
\begin{equation}
    y^2=4x^3-fx-g\,,
\end{equation}
via the following identification
\begin{equation}
    12^3\frac{f^3}{f^3-27g^2}=12^3\frac{E_4(\tau_s)^3}{E_4(\tau_s)^3-E_6(\tau_s)^2}\,.
\end{equation}
Assuming the ansatz $u=u_s+\sum_{n>0}\mu_nq_s^n$ and expanding $f,g,\Delta$ in terms of $\delta:=u-u_s$, we can write
\begin{equation}
f=\sum_{k\geq 0} \frac{f^{(k)}}{k!}\delta^k\,,\qquad g=\sum_{k\geq 0} \frac{g^{(k)}}{k!}\delta^k\,,\qquad  \Delta=f^3-27g^2=\sum_{k\geq 1}\frac{\Delta_k}{k!}\delta^k\,.
\end{equation}
where $f^{(n)}:=\frac{\partial^n f}{\partial u^n}\vert_{u=u_s}$, and $g^{(n)}:=\frac{\partial^n g}{\partial u^n}\vert_{u=u_s}$. Then, we have
\begin{subequations}
\begin{align}
    \Delta_1&=3f^2f'-54gg'\,,\\
    \Delta_2&=3f^2f''+6f(f')^2-54(gg''+(g')^2)\,,\\
    \Delta_3&=3f^2f^{(3)}+18ff'f''+6(f')^3-54gg^{(3)}-162g'g''\,,
\end{align}
\end{subequations}
where $f,g$ are evaluated at $u_s$.
With this, we can expand $j^{-1}$ as
\begin{equation}
j^{-1}=\frac{f^3-27g^2}{12^3f^3}=\frac{\Delta_1}{12^3f^3}\delta+\frac{\frac12\Delta_2-\frac{3f'}{f}\Delta_1}{12^3f^3}\delta^2+\frac{\frac16\Delta_3-\frac{3f'}{2f}\Delta_2+\frac{12(f')^2-3ff''}{2f^2}\Delta_1}{12^3f^3}\delta^3+\dots\,.
\end{equation}
On the RHS of~\eqref{eq:useries}, we can obtain $q_s$ via the $j$-invariant as
\[
q_s=j^{-1}+744 j^{-2}+750420 j^{-3}+\dots\,.
\]
Now, combining with the above expansion from the LHS, we have
\begin{multline}
    q_s=\frac{\Delta_1}{1728f^3}\delta+\left[\frac{\frac12\Delta_2-\frac{3f'}{f}\Delta_1}{12^3f^3}+744 \left(\frac{\Delta_1}{12^3f^3}\right)^2\right]\delta^2\\
    +\left[\frac{\frac16\Delta_3-\frac{3f'}{2f}\Delta_2+\frac{12(f')^2-3ff''}{2f^2}\Delta_1}{12^3f^3}\right.\\
    \left.
    +1488 \left(\frac{\Delta_1}{12^3f^3}\right)\left(\frac{\frac12\Delta_2-\frac{3f'}{f}\Delta_1}{12^3f^3}\right)+750420\left(\frac{\Delta_1}{12^3f^3}\right)^3\right]\delta^3+\dots
\end{multline}
For convenience, let us rewrite the above $q_s$ expansion as $q_s=\sum_{k\geq 1}s_k\delta^k$. Now, inverting the series, we can obtain
\begin{subequations}
\begin{align}
    \mu_1&=\frac{1}{s_1}\,,\\
    \mu_2&=-\frac{s_2}{s_1^3}\,,\\
    \mu_3&=\frac{2s_2^2}{s_1^5}-\frac{s_3}{s_1^4}\,,\\
    &\vdots\nonumber
\end{align}
\end{subequations}

In the massless limit of the E-string theory, these reduce to
\begin{subequations}
\begin{align}
    \mu_1&=\frac{E_4^3u_s^3}{D}\,,\\
    \label{eq:mu2} 
    \mu_2&=\frac{E_4^3u_s^3[E_4^3u_s^2(1296-E_6u_s)-744D^2]}{D^3}\,,\\
    \mu_3&=\frac{E_4^3u_s^3}{D^5}\left\{2(744D^2-E_4^3u_s^2(1296-E_6u_s))^2\nonumber\right.\\
    &\quad \left.-D(750420D^3-1488DE_4^3u_s^2(1296-E_6u_s)+E_4^6u_s^4(1728-E_6u_s))\right\}\,,\\
    &\vdots\nonumber
\end{align}
\end{subequations}
where $D=864-E_6u_s$ and we have made use of the relation
\begin{equation}
\frac{E_4^3-E_6^2}{1728}u_s^2=432-E_6u_s\,
\end{equation}
near the I$_1$ singularity in the massless limit.

\subsubsection*{$a_s$:}
Knowing $u$'s expansion around $u_s$, we have
\begin{align}
\label{eq:dudqsEisenstein}
\begin{split}
    \frac{\dd u}{\dd q_s}=\sum_{n>0}n\mu_nq_s^{n-1}\,.
\end{split}
\end{align}
With this, we can solve for the coefficients in the $q_s$-expansion of $a_s$'s, i.e.,~the expansion presented in \eqref{eq:aexpansions}, by similarly expanding both sides of \eqref{eq:dasdu relation to period} as $q_s$-series.
To do so, let us consider the similar expansion of $f,g,\Delta$ used for the $u$-expansion above. Let us also rewrite~\eqref{eq:dasdu relation to period} here for convenience
\[
\left(\frac{\dd a_s}{\dd q_s}\cdot\frac{\dd q_s}{\dd u}\right)^2=2\left(\frac{\pi\varrho}{3}\right)^2\frac{E_6(\tau_s)}{E_4(\tau_s)}\cdot\frac{f(u;\sigma,\{m_i\})}{g(u;\sigma,\{m_i\})}\,.
\]
Now, the expansion in the ratio of $f/g$ is $\frac{f(u)}{g(u)}=\sum_{k\geq 0}\frac{b_k}{k!}\delta^k$ with the coefficients 
\begin{subequations}
\begin{align}
    b_0&=\frac{f}{g}\,,\\
    b_1&=\frac{f'g-fg'}{g^2}\,,\\
    b_2&=\frac{f''}{g}-\frac{2f'g'+fg''}{g^2}+\frac{2f(g')^2}{g^3}\,,\\
    \vdots \nonumber
\end{align}
\end{subequations}
Similarly, we can expand $E_6/E_4=1+\sum_{k\geq 1}r_kq_s^k$.
Let us now further introduce the following $q_s$-series
\begin{equation}
B:=2\left(\frac{\pi\varrho}{3}\right)^2\frac{E_6(\tau_s)}{E_4(\tau_s)}\cdot \frac{f(u)}{g(u)}\,,\quad C:=\left(\frac{\dd u}{\dd q_s}\right)^2\,,\quad T=BC=\left(\frac{\dd a_s}{\dd q_s}\right)^2\,.
\end{equation}
These each have a $q_s$-expansion which we can write as $B=\sum_{k\geq 0}B_kq_s^k$, $C=\sum_{k\geq 0}C_kq_s^k$, and $T=\sum_{k\geq 0}T_kq_s^k$. Then, we can derive the coefficients (shown only up to the first three leading terms)
\begin{alignat*}{3}
    B_0&=\frac{2\pi^2\varrho^2}{9}b_0\,,\quad &&B_1=\frac{2\pi^2\varrho^2}{9}(b_1\mu_1+r_1b_0)\,,\quad &&B_2=\frac{2\pi^2\varrho^2}{9}(b_1\mu_2+\frac12b_2\mu_1^2+r_1b_1\mu_1+r_2b_0)\,,\\
    C_0&=\mu_1^2\,,\quad && C_1=4\mu_1\mu_2\,,\quad &&C_2=4\mu_2^2+6\mu_1\mu_3\,,\\
    T_0&=B_0C_0\,,\quad &&T_1=B_1C_0+B_0C_1\,,\quad &&T_2=B_2C_0+B_1C_1+B_0C_2\,.
\end{alignat*}
Then, bringing this back to the form of interest, we have
$
\frac{\dd a_s}{\dd q_s}=\sqrt{T}=\sum_{k\geq 0}\alpha_kq_s^k
$, where the coefficients are
\begin{equation}
\alpha_0=\sqrt{T_0}\,,\quad \alpha_1=\frac{T_1}{2\sqrt{T_0}}\,,\quad \alpha_2=\frac{T_2}{2\sqrt{T_0}}-\frac{T_1^2}{8T_0^{3/2}}\dots
\end{equation}
Now, bringing this back into the expression for $a_s$, we see that the first term is
\begin{equation}
    \kappa_1^2=2\left(\frac{\pi\varrho}{3}\right)^2\frac{f}{g}\left(\frac{12^3f^3}{\Delta_1}\right)^2=2\left(\frac{\pi\varrho}{3}\right)^2\frac{f}{g}\mu_1^2\,.
\end{equation}
Then, iteratively, we can solve for the remaining ratios which take on the following explicit forms,
\begin{subequations}
\begin{align}
    \frac{\kappa_2}{\kappa_1}&=\frac14\left(-744-4\frac{s_2}{s_1^2}+\frac{b_1}{b_0s_1}\right)\,,\\
    \frac{\kappa_3}{\kappa_1}&=\frac{2s_2^2}{s_1^4}-\frac{r_1}{3}\frac{s_2}{s_1^2}-\frac{b_1}{2b_0s_1}\frac{s_2}{s_1^2}-\frac{s_3}{s_1^3}-\frac{744^2}{24}+\frac{159768}{6}-744\frac{b_1}{12b_0s_1}+\frac{b_2}{12b_0s_1^2}-\frac{b_1^2}{24b_0^2s_1^2}\,.
\end{align}
\end{subequations}

In the massless limit, these become
\begin{subequations}
\begin{align}
\label{eq:kappa2}
    \frac{\kappa_2}{\kappa_1}&=-930+\frac{E_4^3u^2}{2D}+\frac{648E_4^3u^2}{D^2}\,,\\
    \frac{\kappa_3}{\kappa_1}&=\frac{D^2-876D+218952}{3}\,.
\end{align}
\end{subequations}
when $u^2\to u_s^2$.

Therefore, we can compute the expansion of $a_s$ around $u_s$ as $a_s=\sum_{n>0}\kappa_nq_s^n$. Similar to \cite{Moore:2017cmm}, let us introduce
\begin{equation}
\label{eq:checkE1}
    \check{E}_1(\tau_s)=\sum_{n\geq 0}e_nq_s^n=\frac{a_s/\kappa_1}{q_s}=\sum_{n\geq 0}\frac{\kappa_{n+1}}{\kappa_1}q_s^n\, ,
\end{equation}
where this is, in principle, generated by $\kappa_1 \frac{\dd}{\dd q_s}q_s\check{E}_1(\tau_s)=\frac{\dd a_s}{\dd q_s}$.

\section{Normalization of the partition function}
\label{app:normalization}

For the integrality and TMF-divisibility test, it is important that we completely fix the overall coefficient $\tilde c(\chi,\sigma)$ given by~\eqref{eq:overall coefficient}. We explain here in some detail how it is determined. 

We will achieve this in three steps in this appendix, with the first being a recap of the topological string computation presented in the main text. The gravitational couplings $\alpha$ and $\beta$ in Section~\ref{sec:effective gravitational couplings} are sensitive only to the coefficients of $\chi$ and $\sigma$, and what is left to be determined is an overall constant and a normalization for the Coulomb parameter $a$. Therefore, two more steps are needed. But before that we first review in more detail the constants that we are aiming to fix.

\subsubsection*{What has to be fixed}

The contribution~\eqref{eq:Zsw} of an I$_1$ singularity is a product of the Seiberg--Witten invariant $\SW(\lambda)$, which is canonically normalized, and the ``$u$-plane measure.'' The latter includes local data of the Seiberg--Witten geometry near the singularity (e.g.~$\mu_l$ and $\kappa_l$ in~\eqref{eq:uexpansions} and~\eqref{eq:aexpansions}) which has several numerical constants. The constants have three origins: the gravitational couplings $\alpha$ and $\beta$ of Section~\ref{sec:effective gravitational couplings}, the relative normalization between the Coulomb parameter and the period of the curve, e.g.~the  $\varrho$ in
\begin{equation}
\label{eq:app rho}
    \frac{\dd a_s}{\dd u}=\varrho\,\omega_1\, ,
\end{equation}
and a possible overall normalization constant $K_0$ of the path integral.

Another way of thinking about the present task is the following. The $u$-plane measure is assembled from local densities on $M_4$, and the only two independent integrals of local curvature densities are $\chi$ and $\sigma$, giving $K_0\,A^{\chi}B^{\sigma}$. The part $A^\chi B^\sigma$ can be changed by adding constant multiples of the Euler and Pontryagin densities to the low-energy effective action, while $K_0$ does not depend on $M_4$ at all. It turns out to be convenient to reparametrize the normalization constant as
\begin{equation}
\label{eq:app general ctilde}
    \tilde c(\chi,\sigma)=K_0\cdot 18^{d_T/4}\cdot X^{\chi_h}\,Y^{\sigma}\,,\qquad d_T=2(\chi_h+\sigma)\,,
\end{equation} 
where we have separated the factor $18^{d_T/4}$ because, for the normalization~\eqref{eq:E8curve} of the curve used here, it is produced by the Seiberg--Witten curve itself, as we will show momentarily.

As the partition function of the UV theory is canonically normalized, the values of $K_0$, $X$ and $Y$ are all uniquely determined, and our job is to perform three independent computations of ``protected quantities'' to find their values. This is done by first using the Nekrasov partition function (equivalently topological string amplitude) as we did in Section~\ref{sec:effective gravitational couplings} (reviewed as Step 1 below), giving $Y=1$ and $X=2\pi\varrho$, then the central charge of BPS particles (Step 2), fixing $\varrho$ to be $i/2\pi$, and finally the Witten index of the E-string theory (Step 3), confirming the overall normalization $K_0$ is just 1.

\subsubsection*{Step 1: the local factors near an I$_1$ singularity}

For a basic class with $n(\lambda)=0$, only the leading terms of the $q_s$-expansions in~\eqref{eq:Zsw} matter. As $q_s\to0$,
\begin{equation}
    \frac{a_s}{q_s}\to\kappa_1\,,\qquad \frac{\dd u}{\dd q_s}\to\mu_1\,,\qquad \frac{\dd a_s}{\dd u}\to\frac{\kappa_1}{\mu_1}\,,\qquad \frac{\eta(\tau_s)^{24}}{q_s}\to1\,,
\end{equation}
so the bracket in~\eqref{eq:Zsw} becomes
\begin{equation}
    \label{eq:app bracket}
    \kappa_1^{\chi_h-1}\,\mu_1\left(\frac{\kappa_1}{\mu_1}\right)^{1-2\chi_h-\sigma}=\left(\frac{\kappa_1}{\mu_1}\right)^{-d_T/2}\mu_1^{\chi_h}\,,
\end{equation}
where the ratio $\kappa_1/\mu_1$ is the value of $\dd a_s/\dd u$ at the singularity. Combining~\eqref{eq:app rho} with~\eqref{eq:omega1 expression}, where $E_6(\tau_s)/E_4(\tau_s)\to1$ as $q_s\to0$, gives
\begin{equation}
\label{eq:app kappa mu}
    \left(\frac{\kappa_1}{\mu_1}\right)^2=2\left(\frac{\pi\varrho}{3}\right)^2\frac{f(u_s)}{g(u_s)}=\frac{(2\pi\varrho)^2}{18}\,\frac{f(u_s)}{g(u_s)}\,,
\end{equation}
which is also the relation for $\kappa_1^2$ in Appendix~\ref{sec:details on the partition function}. The $1/18$ here is the origin of the factor $18^{d_T/4}$ as the bracket in~\eqref{eq:Zsw} at leading order equals $18^{d_T/4}(2\pi\varrho)^{-d_T/2}\left(f(u_s)/g(u_s)\right)^{-d_T/4}\mu_1^{\chi_h}$. Summed over the two massless I$_1$ singularities, the last two factors give~\eqref{eq:Ztest}, whose leading coefficients are listed in Table~\ref{tab:original grav coef}.

Two more quantities depend on $\varrho$. The first is the local discriminant, given by
\begin{equation}
    \Delta_0(\tau_s)=(2\pi)^{12}\omega_1^{-12}\eta(\tau_s)^{24}=(2\pi\varrho)^{12}\left(\frac{\dd a_s}{\dd u}\right)^{-12}\eta(\tau_s)^{24}\,,
\end{equation}
so the factor
\begin{equation*}
    \left(\frac{\Delta_\Sigma}{D_0}\right)^{\sigma/8}=\eta(\sigma)^{-3\sigma}\,\Delta_0(\tau_s)^{\sigma/8}
\end{equation*}
of the measure contributes $(2\pi\varrho)^{3\sigma/2}\eta^{-3\sigma}$. For $\varrho=i/2\pi$ this is the prefactor $i^{3\sigma/2}\eta^{-3\sigma}$ in~\eqref{eq:Zsw}. 
The second is the gravitational coupling $\alpha$. At large $u$ the period behaves as $\omega_1\simeq2\pi/u$, so $\dd u/\dd a\simeq u/(2\pi\varrho)$, and the Seiberg--Witten side gives 
\begin{equation}
    b(u)=\log\alpha-\frac12\log(2\pi\varrho)+\frac12\log u+\CO(u^{-1}).
\end{equation}
 Matching with results from topological strings gives $\alpha=(2\pi\varrho)^{1/2}$, while $\beta=D_0^{1/8}$ cancels $\eta^{-3\sigma}$ as explained in the main text.

Collecting all factors, we find
\begin{equation}
\label{eq:app chain}
    \tilde c=K_0\cdot\alpha^{\chi}\cdot(2\pi\varrho)^{3\sigma/2}\cdot18^{d_T/4}(2\pi\varrho)^{-d_T/2}=K_0\cdot18^{d_T/4}\cdot(2\pi\varrho)^{\chi_h}\,.
\end{equation}
In the language of~\eqref{eq:app general ctilde}, we have determined $X=2\pi\varrho$ and $Y=1$.

\subsubsection*{Step 2: the normalization of the Coulomb parameter}

As the Coulomb parameter $a$ and the $E_8$ masses $m_i$ both enter the central charges of BPS particles in the 4d effective theory \cite{Ganor:1996pc}, they should be measured in the same units. A convenient place to compare them is the weakly coupled region of $u\to\infty$. There,
\begin{equation*}
    a\simeq2\pi\varrho\log u\,,
\end{equation*} 
and going once around $u=\infty$ shifts $a$ by $4\pi^2i\varrho$. This monodromy is the shift of the Kaluza--Klein momentum by one unit. 

On the other hand, the masses take values in $\mathbb{C}^8/(2\pi\Gamma_{E_8}+2\pi\sigma\Gamma_{E_8})$, and shifting $m$ by $2\pi(e_1+e_2)$, a vector of $\Gamma_{E_8}$, is a large gauge transformation, which also shifts the Kaluza--Klein momentum of a state of flavor weight $w$ by $\langle w,e_1+e_2\rangle$ units, e.g.\ by one unit for $w=e_1+e_3$. Hence, in the units of the masses, the Kaluza--Klein shift is the real number $\pm2\pi$, and requiring $4\pi^2i\varrho=\mp2\pi$ gives
\begin{equation}
    \varrho=\pm\frac{i}{2\pi}\,,\qquad 2\pi\varrho=\pm i\,.
\end{equation}
The upper sign is the normalization~\eqref{eq:dasdu relation to period} used in the main text. The other sign changes $\tilde c$ by $(-1)^{\chi_h}$, which seems to be related to changing the homology orientation. With this choice, we have only the overall normalization to be determined,
\begin{equation}
    \tilde c=K_0\cdot i^{\chi_h}\,18^{(\chi_h+\sigma)/2}\,,
\end{equation}
and no factors of $\pi$ remain. By contrast, the other common normalization $\varrho=i/4\pi^2$ would give $\tilde c\propto(i/2\pi)^{\chi_h}$, so that every partition function with $\chi_h\neq0$ would have coefficients proportional to $\pi^{-\chi_h}$.

\subsubsection*{Step 3: counting BPS states}

Now we need to fix the overall normalization constant $K_0$, and we achieve this by computing a canonically normalized partition function on $T^6$---the Witten index of the E-string theory. 

For $M_4=T^4$ one has $\chi=\sigma=0$, hence $\chi_h=d_T=0$ and $\tilde c=K_0$. The bracket in~\eqref{eq:Zsw} is then $(\dd a_s/\dd q_s)/(a_s/q_s)$, whose constant term is exactly $1$. Each of the 12 type I$_1$ singularities present for generic masses therefore contributes $K_0$ times the Seiberg--Witten invariant of the trivial class, which is $1$ for $T^4$. Hence $Z[T^4\times T^2]=12K_0$.\footnote{The formula~\eqref{eq:Zsw} was written for simply-connected $M_4$. For $\chi=\sigma=0$ the additional zero modes from $b_1(T^4)=4$ do not change this coefficient (see the discussion of $b_1>0$ in~\cite{Moore:1997pc}).} On the other hand, reducing first on a $T^3$ gives a sigma model with target being the 3d $\CN=4$ Coulomb branch (the total space of the elliptic fibration over the $u$-plane, cf.~Section~\ref{sec:moduli space}). This is given by the rational elliptic surface with the fiber at infinity removed. The Witten index of this 3d sigma model is expected to be the Euler characteristic of its target, which is $12$. Therefore $K_0=1$. Alternatively, one can use the $E_8\times E_8$ model, whose target is a (now compact) K3 surface whose Euler characteristic is 24.

\subsubsection*{A check}

 As the normalization of $\tilde c$ underlies much of our discussion in Section~\ref{sec:examples}, we want to gain confidence in our choice. Here we give an independent argument that supports our choice of the normalization. 
 
 Consider now a basic class with $\SW(\lambda)=\pm1$ and $n(\lambda)=0$ on a four-manifold with $b_2^+>1$, for example the canonical class of a minimal surface of general type with $c_1^2>5\chi_h$. 

In the spectrum of the theory, from the low-energy point of view, one would have towers of states coming from the two I$_1$ singularities, and the one with lowest energy is from $u_{s,+}$. This part can be captured by the 5d $\CN=1$ theory with U(1) gauge group and a charge-1 hypermultiplet. However, when considering this theory on a small $M_4$ with a topological twist, the low-energy theory is captured by the 1d $\CN=1$ sigma model with target the Seiberg--Witten moduli space. Therefore, the $q$-series should always start with $\pm 1$ in such cases.

This leading coefficient is, up to a phase, $K_0|X|^{\chi_h}|Y|^\sigma$ times the leading coefficient of $18^{d_T/4}\mu_1^{\chi_h}\left(f(u_{s,+})/g(u_{s,+})\right)^{-d_T/4}$, and the latter is exactly $\pm1$ whenever $c_1^2-5\chi_h\geq2$ for all cases we have checked. Requiring $K_0|X|^{\chi_h}|Y|^\sigma=1$ for all these values of $(\chi_h,\sigma)$ gives $K_0=|X|=|Y|=1$, in agreement with $K_0=1$ from $T^6$ and with $X=i$ and $Y=1$ from the computation above. 

This also explains why on the line $c_1^2=5\chi_h$, the leading coefficient is $\pm2$---there, both I$_1$ singularities contribute at the same order with two states of lowest energy.

\section{Seiberg--Witten invariants}
\label{appendix:SW}

For a more detailed review on the Seiberg--Witten (SW) invariants, we refer the readers to~\cite{MorgenBook,Hutchings-Taubes-1997,Nicolaescu-1999,Moore-2010} and references therein. Here, we will briefly summarize some key definitions and properties that are most relevant for our study of the E-string theory partition function. 

Consider a compact, connected, oriented and smooth  4-manifold $M_4$ equipped with a spin$^c$ structure. Recall that the spin$^c$ spinor bundle $S^+_c$ (or $S^-_c$) is a rank-2 vector bundle associated with the defining representation of the $U(2)_+\simeq (SU(2)_+\times U(1))/\Z_2$ (or $U(2)_-$) subgroup of Spin$^{c}(4)\simeq (SU(2)_+\times SU(2)_-\times U(1))/\Z_2$. One can informally write $S^\pm_c=S^{\pm}\otimes L^{1/2}$, except that the spinor bundle $S^\pm$ and the line bundle $L^{1/2}$ do not exist individually when $w_2(M_4)$ is non-trivial. Let $A$ be a unitary connection on the Hermitian line bundle $L\simeq\Lambda^2S_c^{\pm}$, $M$---the ``monopole field''---be a section of $S_c^+$, and $D_A: \Gamma(S_c^+)\rightarrow \Gamma(S_c^-)$ be the spin$^c$ Dirac operator. The definition of $D_A$ uses the Clifford multiplication, which is usually written in the physics context via the gamma matrices $\gamma^\mu_{\alpha\dot\alpha}$. Then the monopole/SW equations are 
\begin{subequations}
\begin{align}
\label{eq:Monopoles}
    D_AM&=0,\\
\label{eq:Monopoles2}
F_A^+&=\sigma(M)+i\eta
\end{align}
\end{subequations}
where $\eta\in\Omega^{2,+}(M_4,\R)$ is a self-dual 2-form that can be regarded as a small perturbation to the original monopole equations, $F^+_A=(dA+\star d A)/2$ denotes the self-dual part of the curvature of $A$, and
\begin{equation*}
    \sigma:S_c^+\to i\Lambda^{2,+}T^*M_4,
\end{equation*}
is the squaring map. 

One key property of the moduli space $\mathcal{M}_{\SW}(M_4)$ of the SW equations is its compactness. The moduli space can be identified, without the perturbation by $\eta$, with the minima of the following action (coming from the 4d $\CN=2$ ``monopole theory''),
\begin{equation}
\label{eq:SW action}
    S(A,\psi)=\int_{M_4}dV|D_AM|^2+|F_A^+-\sigma(M)|^2.
\end{equation}
In fact, the moduli space $\mathcal{M}_{\SW}(M_4)$ is also the space of supersymmetric configurations of the monopole theory. After choosing an orientation of the moduli space, one can define the $\SW$ invariants,  $$\SW:\text{spin}^c(M_4)\to \mathbb{Z}$$
which are in fact topological when $b_2^+(M_4)>1$.

The invariant depends on a choice of a $\text{spin}^c$ structure, which can be labeled by $s=c_1(L)\in H^2(M_4,\Z)$ satisfying the condition $s\equiv w_2\pmod2$. In the physics literature, the spin$^c$ structure is sometimes labeled by $\lambda=c_1(L^{1/2})=s/2$, which should be viewed as living in ``$H^2(M_4,\Z)+w_2/2$.'' In other words, upon subtracting half of (the integral lift of) $w_2$, it is a well-defined cohomology class.

As one goal of the present work is to find new constraints on the SW invariants, we list below several well-known properties of the Seiberg--Witten invariants (see e.g.~\cite{Hutchings-Taubes-1997} for more detailed discussions).

\paragraph{Properties of SW$(s)$.} For $M_4$ a connected, smooth, compact and oriented four-dimensional manifold, we have:
\begin{enumerate}
    \item $\SW(s)=0$ for all $s$, if $b^+_2-b_1\in 2\mathbb{Z}$. In other words, $\chi_h:=(\chi+\sigma)/4$ has to be integral in order for any SW$(s)$ to be non-vanishing.
    \item $\SW(s)=0$ if the virtual dimension of $\mathcal{M}_{\SW}(M_4)$, given by $2n(s)=\frac{1}{4}(s^2 -2\chi -3\sigma)$,  is negative. (We use $n(s)$ for the complex virtual dimension, which is given by one half of this formula.)

\item $\SW(s)=0$ for all $s$, if $M_4$ admits a metric with positive scalar curvature.
    
    \item $\SW(s)=(-1)^{\chi_h}\SW(-s)$. As a consequence, non-vanishing SW invariants come in pairs (unless $s=0$), and such a pair is often known as a \emph{basic class}.
\end{enumerate}
If we further restrict to symplectic four-manifolds, we have \cite{Taubes-1994,Taubes-1995} 
\begin{enumerate}[resume]
    \item $\SW(s)=\pm 1$, where $s=c_1(K)$ is the first Chern class of the canonical bundle of the almost complex structure, $K=\Lambda^2T^{1,0}$, where $T^{1,0}$ is the holomorphic component of $T^*M_4\otimes \mathbb{C}$, with $c_1(K)^2=2\chi+3\sigma$ being another topological/homotopy invariant useful in describing the geography of 4-manifolds.
\end{enumerate}
    The second constraint also has a counterpart for $\dim \mathcal{M}_{\SW}(M_4)>0$ when $M_4$ is simply-connected, known as the SW simple-type conjecture (discussed in more detail in Section~\ref{sec:examples}), 
\begin{enumerate}[resume]
    \item $\SW(s)=0$, unless $n(s)=\frac{1}{8}(s^2 -2\chi -3\sigma)=0$.
\end{enumerate}
Therefore, there can only be finitely many basic classes when $b_2^+>1$. When $b_2^+=1$, the SW invariants have wall-crossings and there are typically different chambers in the space of metrics. We briefly review this phenomenon below. 

\paragraph{Wall-crossing for SW$(s)$.} The location of the wall is characterized by the existence of a reducible solution $(A,M)=(A_0,0)$ to the SW equations, where $F^+(A_0)=0$. Now consider a perturbation to the metric, giving the following deformation of the solution, 
\begin{equation*}
    \begin{cases}
        M=mM_0\\
        A=A_0+\epsilon \delta A
    \end{cases}
\end{equation*}
parametrized by $\epsilon\in\mathbb{R}$, where $m\in\mathbb{C}$ is a complex number and $DM_0=0$. Note that, after perturbation, $M=m M_0$ is still a solution to $DM=0$. Plugging these into \eqref{eq:Monopoles2}, we have
\begin{equation*}
    \epsilon(d\delta A)^+-mm^*(M\overline{M})^+=0.
\end{equation*}
We can choose $\delta A$ where $(d\delta A)^+=c(M\overline{M})^+$ such that the above equation becomes
\begin{equation}
\label{eq:c-good}
    c\epsilon-m m^*=0.
\end{equation}
Now, depending on the sign of $c$ and $\epsilon$, there only exists a single solution for $m$ as $mm^*$ is positive definite. Therefore, assuming $n(s)=0$, the Seiberg--Witten invariants experience a jump as we vary $\epsilon$ crossing over $\epsilon=0$,
\begin{equation}\label{SWwallcrossing}
    \mathrm{SW}^+(s)-\mathrm{SW}^-(s)=1\,,
\end{equation}
where $\mathrm{SW}^{\pm}$ are in the chamber where the self-dual part of $c_1(L)$ is negative and positive with respect to the chosen orientation of $H^{2,+}$. A similar formula holds for positive ${n(s)}$.\footnote{One relevant fact, which we won't elaborate on here, is that the orientation of $\CM_{\SW}$, and hence the Seiberg--Witten invariant, also depends on a choice of orientation of $H^1\oplus H^{2,+}$. See \cite{Salamon-2014} for a detailed discussion of this point.} In other words, given a metric $g$, there exists a unique self-dual harmonic 2-form $\omega_g$ with $\int_{M_4}\omega_g\^\omega_g=1$ in the positive orientation, and the sign of $-\int_{M_4}c_1(L)\^ \omega_g$ determines which chamber $g$ lives in. For the $\beta$-twisted SW invariant, $\eta$ is no longer a small perturbation but can be large. Then the two different chambers correspond to the two different signs of  
\begin{equation}
    \epsilon(g,\eta):=\int_{M_4} (\eta-\pi c_1(L))\^\omega_g\,.
\end{equation}
Although the above is a good conceptual explanation of the wall-crossing phenomenon, we have suppressed some technical subtleties (e.g.~$n(s)$ can be non-zero and $M_4$ can be non-simply-connected) that slightly complicate the situation. In general, one has $\mathrm{SW}^+(s)-\mathrm{SW}^-(s)=w(M_4,s)$ with the jump being a function of the spin$^c$ structure (see e.g.,~\cite{Salamon-2014} for details). One special example in the $b_2^+=1$ world is $\mathbb{CP}^2$, which we turn to next.

\subsubsection*{SW invariants of $\mathbb{CP}^2$}

Using a Weitzenb\"{o}ck-like formula, one can show that the SW invariants for $\mathbb{CP}^2$ vanish due to the positive scalar curvature~\cite{Witten:1994cg}. Hence, $\SW_{\mathbb{CP}^2}(s)=0$ for any $s=c_1(L)$. Furthermore, the Kähler form is self-dual, and there are no walls in the positive cone as the only abelian anti-self-dual connection is the flat one. However, the $\beta$-twisted version of the Seiberg--Witten invariants, prevalent in the mathematical literature, can be non-trivial for $\mathbb{CP}^2$ and exhibit interesting wall-crossing behaviors (see e.g.~\cite{Okonek-Teleman-1996}). This is in fact not in contradiction to the statement made above. 

Given the local coordinates $(z_1,z_2)$ on $\mathbb{CP}^2$, we consider the following Fubini--Study metric, 
\begin{equation}
    g_{ij}=\frac{\delta_{ij}(1+|z|^2)-\bar{z}_iz_j}{(1+|z|^2)^2}dz_i\otimes d\bar{z}_j.
\end{equation}
The Kähler form
\begin{equation}
    \omega=\frac{i}{2\pi}\partial\bar{\partial}\log(1+|z|^2)
\end{equation}
is a self-dual harmonic 2-form with unit volume, $\int_{\mathbb{CP}^2}\omega\wedge \omega=1$, and the class $[\omega]$ generates $H^2(M_4;\mathbb{Z})=\mathbb{Z}$. Furthermore, we have $c_1(K)=-3 [\omega]$ for the canonical bundle $K$. 

When $\eta$ is infinitesimal, we have
\begin{equation}
    \lim_{\eta\to 0}\epsilon(g,\eta)=-\pi [\omega]\cdot c_1(L)=-\pi\cdot s
\end{equation}
where $s=2\lambda$ is odd due to the spin$^c$ condition. Therefore, the positive curvature argument only tells us that $\SW^-(\lambda)=0$ for $\lambda> 0$ and $\SW^+(\lambda)=0$ when $\lambda< 0$. 

While there are no other chambers for the original SW invariant in this case, there is a more interesting story when the ``deformation'' $\eta$ is allowed to be large. This is known as the $\beta$-twisted Seiberg--Witten invariant, which is defined with the same equations \eqref{eq:Monopoles} and \eqref{eq:Monopoles2} except that $\eta$ is now allowed to be any closed self-dual 2-form. Then, for each $\lambda$, there will be another chamber, and one can obtain the value of the invariant there via the same wall-crossing formula \eqref{SWwallcrossing}. To apply this formula, one also needs to make sure that the virtual dimension of the SW moduli space given by
\begin{equation}
    \dim \mathcal{M}_{\rm SW}(\mathbb{CP}^2)=\lambda^2 - \frac{9}{4}
\end{equation}
is non-negative. Thus, the twisted invariants are given by
\begin{equation}
    \SW^{\beta,+}_{\mathbb{CP}^2}(\lambda)=\begin{cases}
        + 1&\lambda\geq 3/2\\
        0&\lambda<3/2
    \end{cases},\qquad 
    \SW^{\beta,-}_{\mathbb{CP}^2}(\lambda)=\begin{cases}
        -1&\lambda\leq -3/2\\
        0&\lambda>-3/2
    \end{cases}
\end{equation}
which exactly recover the results in \cite{Okonek-Teleman-1996,Salamon-2014}.




\bibliography{papers}
\bibliographystyle{JHEPbib}

\end{document}